\documentclass[12pt]{article}
\usepackage{amssymb}
\usepackage{amsmath}
\usepackage{sw20elba}
\usepackage{hyperref}

\numberwithin{equation}{subsection}
\input{tcilatex}
\begin{document}

\title{Wave-corpuscle mechanics for elementary charges}
\author{Anatoli Babin and Alexander Figotin \\
University of California at Irvine}
\maketitle

\begin{abstract}
It is well known that the concept of a point charge interacting with the
electromagnetic (EM) field has a problem. To address that problem we
introduce the concept of \emph{wave-corpuscle} to describe spinless
elementary charges interacting with the classical EM field. Every charge
interacts only with the EM field and is described by a complex valued wave
function over the 4-dimensional space time continuum. A system of many
charges interacting with the EM field is defined by a local, gauge and
Lorentz invariant Lagrangian with a key ingredient - a nonlinear
self-interaction term providing for a cohesive force assigned to every
charge. An ideal wave-corpuscle is an exact solution to the Euler-Lagrange
equations describing both free and accelerated motions. It carries
explicitly features of a point charge and the de Broglie wave. A system of
well separated charges moving with nonrelativistic velocities are
represented accurately as wave-corpuscles governed by the Newton equations
of motion for point charges interacting with the Lorentz forces. In this
regime the nonlinearities are "stealthy" and don't show explicitly anywhere,
but they provide for the binding forces that keep localized every individual
charge.
\end{abstract}

\tableofcontents

\section{Introduction}

We all know from textbooks that if there is a point charge $q$ of a mass $m$
in an external electromagnetic (EM) field its non-relativistic dynamics is
governed by the equation%
\begin{equation}
\frac{\mathrm{d}}{\mathrm{d}t}\left[ m\mathbf{v}\left( t\right) \right] =q%
\left[ \mathbf{E}\left( \mathbf{r}\left( t\right) ,t\right) +\frac{1}{%
\mathrm{c}}\mathbf{v}\left( t\right) \times \mathbf{B}\left( \mathbf{r}%
\left( t\right) ,t\right) \right]   \label{pchar1}
\end{equation}%
where $\mathbf{r}$ and $\mathbf{v}=\mathbf{\dot{r}=}\frac{\mathrm{d}\mathbf{r%
}}{\mathrm{d}t}$ are respectively the charge's time-dependent position and
velocity, $\mathbf{E}\left( t,\mathbf{r}\right) $ and $\mathbf{B}\left( t,%
\mathbf{r}\right) $ are the electric field and the magnetic induction, and
the right-hand side of the equation (\ref{pchar1}) is the Lorentz force. We
also know that if the charge's time-dependent position and velocity are $%
\mathbf{r}$ and $\mathbf{v}$ then there is associated with them an EM field
described by the equations%
\begin{equation}
\frac{1}{\mathrm{c}}\frac{\partial \mathbf{B}}{\partial t}+\nabla \times 
\mathbf{E}=\boldsymbol{0},\ \nabla \cdot \mathbf{B}=0,  \label{psys1}
\end{equation}%
\begin{equation}
\frac{1}{\mathrm{c}}\frac{\partial \mathbf{E}}{\partial t}-\nabla \times 
\mathbf{B}=-\frac{4\pi }{\mathrm{c}}q\delta \left( \mathbf{x}-\mathbf{r}%
\left( t\right) \right) \mathbf{v}\left( t\right) ,\ \nabla \cdot \mathbf{E}%
=4\pi q\delta \left( \mathbf{x}-\mathbf{r}\left( t\right) \right) ,\ \mathbf{%
v}\left( t\right) =\mathbf{\dot{r}}\left( t\right) ,  \label{psys2}
\end{equation}%
where $\delta $ is the Dirac delta-function. But if naturally we would like
to consider the equation (\ref{pchar1}) and (\ref{psys1})-(\ref{psys2}) as a
closed system "charge-EM field" there is a problem. The origin of the
problems is in the divergence of the EM field exactly at the position of the
point charge, as, for instance, for the electrostatic field $\mathbf{E}$
with the Coulomb's potential $\frac{q}{\left\vert \mathbf{x}-\mathbf{r}%
\right\vert }$ with a singularity at $\mathbf{x}=\mathbf{r}$. If (\ref%
{pchar1}) is replaced by a relativistic equation 
\begin{equation}
\frac{\mathrm{d}}{\mathrm{d}t}\left[ \gamma m\mathbf{v}\left( t\right) %
\right] =q\left[ \mathbf{E}\left( \mathbf{r}\left( t\right) ,t\right) +\frac{%
1}{\mathrm{c}}\mathbf{v}\left( t\right) \times \mathbf{B}\left( \mathbf{r}%
\left( t\right) ,t\right) \right] ,  \label{psys3}
\end{equation}%
where $\gamma =1/\sqrt{1-\mathbf{v}^{2}\left( t\right) /\mathrm{c}^{2}}$ is
the Lorentz factor, the system constituted by (\ref{psys3}) and (\ref{psys1}%
)-(\ref{psys2}) becomes Lorentz invariant and has a Lagrangian that yields
it via the variational principle \cite[(4.21)]{Barut}, \cite[(2.36)]{Spohn},
but the problem still persists. Some studies indicate that, \cite{Yaghjian1}%
, "a fully consistent classical equation of motion for a point charge,
unlike that of an extended charge, does not exist". If one wants to stay
within the classical theory of electromagnetism a possible remedy is the
introduction of an extended charge which, though very small, is not a point.
There are two most well known models for such an extended charge: the
Abraham rigid charge model and the Lorentz relativistically covariant model.
These models are considered, studied and advanced in \cite[Sections 16]%
{Jackson}, \cite{Pearle1}, \cite[Sections 2, 6]{Rohrlich}, \cite{Schwinger}, 
\cite{Spohn}, \cite{Yaghjian}. Importantly for what we do here, Poincar\'{e}
suggested in 1905-1906, \cite{Poincare} (see also \cite[Sections 16.4-16.6]%
{Jackson}, \cite[Sections 2.3, 6.1- 6.3]{Rohrlich}, \cite[Section 63]{Pauli
RT}, \cite{Schwinger}, \cite[Section 4.2]{Yaghjian} and references there
in), to add to the Lorentz-Abraham model non-electromagnetic cohesive forces
which balance the charge internal repulsive electromagnetic forces and
remarkably restore also the covariance of the entire model. W. Pauli argues
very convincingly based on the relativity principle in \cite[Section 63]%
{Pauli RT} the necessity to introduce for the electron an energy of
non-electromagnetic origin.

An alternative approach to deal with the above-mentioned divergences goes
back to G. Mie who proposed to modify the Maxwell equations making them
nonlinear, \cite[Section 64]{Pauli RT}, \cite[Section 26]{Weyl STM}, and a
particular example of the Mie approach is the Born-Infeld theory, \cite{Born
Infeld 1}. M. Kiessling showed that, \cite{Kiessling 1}, "a relativistic
Hamilton--Jacobi type law of point charge motion can be consistently coupled
with the nonlinear Maxwell--Born--Infeld field equations to obtain a
well-defined relativistic classical electrodynamics with point charges".

A substantially different approach to elementary charges was pursued by E.
Schr\"{o}dinger who tried to develop a concept of \emph{wavepacket} as a
model for spatially localized charge. The Schr\"{o}dinger wave theory, \cite%
{Schrodinger ColPap}, was inspired by de Broglie ideas, \cite{de Broglie 2}, 
\cite[Section II.1]{Barut}. The theory was very successful in describing
quantum phenomena in the hydrogen atom, but it had great difficulties in
treating the elementary charge as the material wave as it moves and
interacts with other elementary charges. M. Born commented on this, \cite[%
Chapter IV.7]{Born1}: "To begin with, Schr\"{o}dinger attempted to interpret
corpuscles and particularly electrons, as \emph{wave packets}. Although his
formulae are entirely correct, his interpretation cannot be maintained,
since on the one hand, as we have already explained above, the wave packets
must in course of time become dissipated, and on the other hand the
description of the interaction of two electrons as a collision of two wave
packets in ordinary three-dimensional space lands us in grave difficulties."

We develop here a concept of wave-corpuscle, which is understood as a
spatially localized excitation in a dispersive medium, and which is to
substitute for the point charge concept. Our approach to a spatially
distributed but localized elementary charge has some features in common with
the above discussed concepts of extended charge, but it differs from any of
them substantially. In particular, our approach provides for an
electromagnetic theory in which (i) a "bare" elementary charge and the EM
field described by the Maxwell equations form an inseparable entity; (ii)
every elementary "bare" charge interacts directly only with the EM field;
(iii) the EM field is a single entity providing for the interaction between
"bare" elementary charges insuring the maximum speed of interaction not to
ever exceed the speed of light. To emphasize the inseparability of the
"bare" elementary charge from the EM field we refer to their entity as to 
\emph{dressed charge}.

The best way to describe our concept of a spatially distributed but
localized dressed charge in one word is by the name \emph{wave-corpuscle,}
since it is a stable localized excitation of a dispersive medium propagating
in the three-dimensional space. An instructive example of a wave-corpuscle
is furnished by our nonrelativistic charge model. In that model, in the
simplest case, \ an \emph{ideal wave-corpuscle} is described by a
complex-valued wave function $\psi $ of the form%
\begin{equation}
\psi =\psi \left( t,\mathbf{x}\right) =\exp \left\{ \frac{\mathrm{i}}{\hbar }%
\left[ \mathbf{p}\left( t\right) \cdot \mathbf{x}-\dint_{0}^{t}\frac{\mathbf{%
p}\left( t^{\prime }\right) ^{2}}{2m}\,\mathrm{d}t^{\prime }\right] \right\} 
\mathring{\psi}\left( \left\vert \mathbf{x}-\mathbf{r}\left( t\right)
\right\vert \right) ,  \label{wavco1}
\end{equation}%
where $\mathring{\psi}\left( s\right) $, $s\geq 0$, is a non negative,
monotonically decaying function which vanishes at infinity at a sufficiently
fast rate. Importantly, for the above wave function $\psi $ to be an exact
solution of corresponding field equations, the parameters $\mathbf{r}\left(
t\right) $ and $\mathbf{p}\left( t\right) $ satisfy the Newton's equations
which in this simplest case have the form%
\begin{equation}
m\frac{\mathrm{d}^{2}\mathbf{r}\left( t\right) }{\mathrm{d}t^{2}}=q\mathbf{E}%
_{\mathrm{ex}},\ \mathbf{p}\left( t\right) =m\frac{\mathrm{d}\mathbf{r}%
\left( t\right) }{\mathrm{d}t},  \label{wavco2}
\end{equation}%
where $m$ and $q$ are respectively its mass and the charge and $\mathbf{E}_{%
\mathrm{ex}}\left( t\right) $ is an external homogeneous electric field. We
would like to emphasize that the Newton's equations are not postulated as in
(\ref{pchar1}) or (\ref{psys3}) but rather are derived from the field
equations. \emph{The ideal wave-corpuscle wave function }$\psi \left( t,%
\mathbf{x}\right) $\emph{\ defined by (\ref{wavco1}), (\ref{wavco2})
together with the corresponding EM field forms an exact solution to the
relevant Euler-Lagrange field equations describing an accelerating dressed
charge. The point charge momentum }$\mathbf{p}\left( t\right) $\emph{\ turns
out to be exactly equal to the total momentum of the charge as a
wave-corpuscle and its electromagnetic field. Remarkably the point charge
features appear in the phase and amplitude of the ideal wave-corpuscle in a
transparent and direct way without any limit process. The wave-corpuscle is
a material wave, the quantity }$q\left\vert \psi \left( t,\mathbf{x}\right)
\right\vert ^{2}$\emph{\ corresponds to the charge density and the density }$%
\left\vert \psi \left( t,\mathbf{x}\right) \right\vert ^{2}$\emph{\ is not
given a probabilistic interpretation. The wave-corpuscle provides we believe
an alternative resolution to the wave-particle duality problem. In one word
our theory can be characterized as neo-classical as suggested by Michael
Kiessling.}

\section{Sketch of the wave-corpuscle mechanics\label{ssketch}}

We describe a bare single elementary charge by a complex-valued scalar field 
$\psi =\psi \left( x\right) =\psi \left( t,\mathbf{x}\right) $, where $%
x=\left( t,\mathbf{x}\right) \in \mathbb{R}^{4}$ is the space-time variable.
The charge is coupled at all times with the classical EM field as described
by its potentials $A^{\mu }=\left( \varphi ,\mathbf{A}\right) $ related to
the EM field by the standard formulas 
\begin{equation}
\mathbf{E}=-\nabla \varphi -\frac{1}{\mathrm{c}}\partial _{t}\mathbf{A},\ 
\mathbf{B}=\nabla \times \mathbf{A},  \label{afi1}
\end{equation}%
where $\mathrm{c}$ is the speed of light. The dynamics of the system of a
single charge and the EM field is described via its Lagrangian 
\begin{gather}
L_{0}\left( \psi ,A^{\mu }\right) =\frac{\chi ^{2}}{2m}\left\{ \frac{1}{%
\mathrm{c}^{2}}\left\vert \tilde{\partial}_{t}\psi \right\vert
^{2}-\left\vert \tilde{\nabla}\psi \right\vert ^{2}-\kappa
_{0}^{2}\left\vert \psi \right\vert ^{2}-G\left( \psi ^{\ast }\psi \right)
\right\}  \label{afi3} \\
+\frac{1}{8\pi }\left[ \left( \nabla \varphi +\frac{1}{\mathrm{c}}\partial
_{t}\mathbf{A}\right) ^{2}-\left( \nabla \times \mathbf{A}\right) ^{2}\right]
,  \notag
\end{gather}%
where $\tilde{\partial}_{t}$ and $\tilde{\nabla}$ are the covariant
differentiation operators defined by 
\begin{equation}
\tilde{\partial}_{t}=\partial _{t}+\frac{\mathrm{i}q\varphi }{\chi },\ 
\tilde{\nabla}=\nabla -\frac{\mathrm{i}q\mathbf{A}}{\chi \mathrm{c}},\ 
\tilde{\partial}_{t}^{\ast }=\partial _{t}-\frac{\mathrm{i}q\varphi }{\chi }%
,\ \tilde{\nabla}^{\ast }=\nabla +\frac{\mathrm{i}q\mathbf{A}}{\chi \mathrm{c%
}},  \label{afi4}
\end{equation}%
$m>0$ is the \emph{charge mass}, $\chi >0$ is a constant similar to the
Planck constant $\hbar =\frac{h}{2\pi }$ and it might be dependent on the
charge; $q$ is the total charge of the particle.

Let us take a closer look at the components of the Lagrangian (\ref{afi3}).
It involves constants $\kappa _{0}$, $\mathrm{c}$, $\chi $ and $m$ and,
acting similarly to the case of the Klein-Gordon equation for a relativistic
particle (see \cite[Sections 1, 18, 19]{Pauli PWM} and Section \ref%
{skleingordon}), we introduce a fundamental frequency $\omega _{0}$ relating
it to the above constants by the following formulas%
\begin{equation}
\omega _{0}=\frac{m\mathrm{c}^{2}}{\chi },\ \kappa _{0}=\frac{\omega _{0}}{%
\mathrm{c}}=\frac{m\mathrm{c}}{\chi }.  \label{afi2}
\end{equation}%
A key component of the Lagrangian in (\ref{afi3}) is a real-valued nonlinear
function $G\left( s\right) $, $s\geq 0$, providing for the charge cohesive
self-interaction. The second part of the expression (\ref{afi3}) is the
standard Lagrangian of the EM field coupled to the charge via the covariant
derivatives. Observe that the Lagrangian $L_{0}$ defined by (\ref{afi3})-(%
\ref{afi4}) is manifestly (i) local; (ii) Lorentz and gauge invariant, and
(iii) it has a local nonlinear term providing for a cohesive self-force
similar to the Poincar\'{e} force for the Lorentz-Poincar\'{e} model of an
extended charge.

Since a single charge is coupled at all times to the EM field we always deal
with the \emph{system "charge-EM field"}, $\left\{ \psi ,\psi ^{\ast
},A^{\mu }\right\} $, and call it for short \emph{dressed charge}. The
dressed charge motion is governed by the relevant Euler-Lagrange field
equations (see (\ref{fpar4}), (\ref{fpar5})), and when the charge is at rest
in the origin $\mathbf{x}=\mathbf{0}$ it is described by the fields 
\begin{equation}
\psi \left( t,\mathbf{x}\right) =\mathrm{e}^{-\mathrm{i}\omega _{0}t}%
\mathring{\psi}\left( \left\vert \mathbf{x}\right\vert \right) ,\ \varphi
\left( t,\mathbf{x}\right) =\mathring{\varphi}\left( \left\vert \mathbf{x}%
\right\vert \right) ,\ \mathbf{A}\left( t,\mathbf{x}\right) =\mathbf{0},
\label{psfi1}
\end{equation}%
where $\psi ^{\ast }$ is the complex conjugate to $\psi $ and the real
valued functions $\mathring{\psi}$ and $\mathring{\varphi}$ satisfy the
following system of equations%
\begin{gather}
-\Delta \mathring{\varphi}=4\pi \mathring{\rho},\ \mathring{\rho}=q\left( 1-%
\frac{q\mathring{\varphi}}{m\mathrm{c}^{2}}\right) \mathring{\psi}^{2},
\label{psfi1a} \\
-\Delta \mathring{\psi}+\frac{m\mathring{\varphi}}{\chi ^{2}}q\left( 2-\frac{%
q\mathring{\varphi}}{m\mathrm{c}^{2}}\right) \mathring{\psi}+G^{\prime
}\left( |\mathring{\psi}|^{2}\right) \mathring{\psi}=0.  \label{psfi1b}
\end{gather}%
where $\Delta =\nabla ^{2}$ \ is Laplace operator. We refer to the state of
the dressed charge of the form (\ref{psfi1}) as $\omega _{0}$\emph{-static}.
The functions $\mathring{\psi}$ and $\mathring{\varphi}$ in the above
formulas are instrumental for our constructions and we refer to them
respectively as the \emph{charge form factor} and \emph{form factor potential%
}. Using Green's function to solve equation (\ref{psfi1a}) we see that the
charge form factor $\mathring{\psi}$ determines the Coulomb-like potential $%
\mathring{\varphi}=\mathring{\varphi}_{\mathring{\psi}}$ by the formula%
\begin{equation}
\mathring{\varphi}=\mathring{\varphi}_{\mathring{\psi}}=4\pi q\left( -\Delta
+\frac{4\pi q^{2}}{m\mathrm{c}^{2}}\mathring{\psi}^{2}\right) ^{-1}\mathring{%
\psi}^{2}.\   \label{psfi1c}
\end{equation}%
Consequently, plugging the above expression into the equation (\ref{psfi1b})
we get the following nonlinear equation%
\begin{equation}
-\Delta \mathring{\psi}+\frac{m\mathring{\varphi}_{\mathring{\psi}}}{\chi
^{2}}q\left( 2-\frac{q\mathring{\varphi}_{\mathring{\psi}}}{m\mathrm{c}^{2}}%
\right) \mathring{\psi}+G^{\prime }\left( |\mathring{\psi}|^{2}\right) 
\mathring{\psi}=0.  \label{psfi1d}
\end{equation}%
The above equation (\ref{psfi1d}) signifies a complete balance of the three
forces acting upon the resting charge: (i) internal elastic deformation
force associated with the term $-\Delta \mathring{\psi}$; (ii) charge's
electromagnetic self-interaction force associated with the term $\frac{m%
\mathring{\varphi}_{\mathring{\psi}}}{\chi ^{2}}\left( 2q-\frac{q^{2}%
\mathring{\varphi}_{\mathring{\psi}}}{m\mathrm{c}^{2}}\right) \mathring{\psi}
$; (iii) internal nonlinear self-interaction of the charge associated with
the term $G^{\prime }\left( |\mathring{\psi}|^{2}\right) \mathring{\psi}$.
In what follows we refer to the equation (\ref{psfi1d}) as \emph{charge
equilibrium equation}. Importantly, the static charge equilibrium equation (%
\ref{psfi1d}) establishes an explicit relation between the form factor $%
\mathring{\psi}$ and the self-interaction nonlinearity $G$. Hence being
given the form factor $\mathring{\psi}$ we can find from the equilibrium
equation (\ref{psfi1d}) the self-interaction nonlinearity $G$ which exactly
produces this factor under an assumption that $\mathring{\psi}\left(
r\right) $, $r\geq 0,$ is a nonnegative, monotonically decaying and
sufficiently smooth function. \emph{The later is a key feature of our
approach: it allows to choose the form factor }$\mathring{\psi}$\emph{\ and
then to determine matching self-interaction nonlinearity }$G$\emph{\ rather
than to deal with solving a nontrivial nonlinear partial differential
equation}.

\emph{Thus, to summarize an important point of our method: we pick the form
factor }$\mathring{\psi}$\emph{\ and then the nonlinear self interaction
function }$G$\emph{\ is determined based on a physically sound ground: the
charge equilibrium\ equation (\ref{psfi1d})}. Needless to say that under
this approach the nonlinearity $G$ is not expected to be a simple polynomial
function but rather a function with properties that ought to be established.
Then having fixed the nonlinear self-interaction $G$ based on the charge
equilibrium equation (\ref{psfi1d}) the challenge is to figure out the
dynamics of the charge as it interacts with other charges or is acted upon
by an external EM field and hence accelerates. The nonlinear
self-interaction $G$ evidently brings into the charge model
non-electromagnetic forces, the necessity of which for a consistent
relativistic electromagnetic theory was argued convincingly by W. Pauli in 
\cite[Section 63]{Pauli RT}. It is worth to point out that the nonlinearity $%
G$ introduced via the charge equilibrium equation (\ref{psfi1d}) differs
significantly from nonlinearities considered in similar problems in
literature including attempts to introduce nonlinearity in the quantum
mechanics, \cite{Bialyanicki}, \cite{Holland}, \cite{Weinberg}. Important
features of our nonlinearity include: (i) the boundedness of its derivative $%
G^{\prime }\left( s\right) $ for $s\geq 0$ with consequent boundedness from
below of the wave energy; (ii) non analytic behavior for small $s$ that is
for small wave amplitudes.

We would like to mention that an idea to use concept of a solitary wave in
nonlinear dispersive media for modelling wave-particles was quite popular.
Luis de Broglie tried to use it in his pursuit of the material wave
mechanics. G. Lochak wrote in his preface to the de Broglie's monograph, 
\cite[page XXXIX]{de Broglie 2}: "...The first idea concerns the solitons,
which we would call \emph{ondes \`{a} bosses} (humped waves) at the \emph{%
Institut Henri Poincar\'{e}}. This idea of de Broglie's used to be
considered as obsolete and too classical, but it is now quite well known, as
I mentioned above, and is likely to be developed in the future, but only
provided we realize what the obstacle is and has been for twenty-five years:
It resides in the lack of a general principle in the name of which we would
be able to choose one nonlinear wave equation from among the infinity of
possible equations. If we succeed one day in finding such an equation, a new
microphysics will arise." G. Lochak raised an interesting point of the
necessity of a general principle that would allow to choose one nonlinearity
among infinitely many. We agree to G. Lochak to the extend that there has to
be an important physical principle that would allow to choose the
nonlinearity but whether it has to be unique is different matter. In our
approach such a principle is the exact balance of all forces for the resting
dressed charge via the static charge equilibrium equation (\ref{psfi1d}). As
to a possibility of spatially localized excitations such as wave-packets to
maintain their basic properties when propagate in a dispersive medium with a
nonlinearity we refer to our work \cite{BabFig1}-\cite{BabFig3}.

The gauge invariance of the Lagrangian $L_{0}$ allows us to introduce in a
standard fashion the \emph{microcharge density} $\rho $ and the \emph{%
microcurrent density} $\mathbf{J}$ by 
\begin{equation}
\rho =-\frac{\chi q}{2m\mathrm{c}}\mathrm{i}\left( \tilde{\partial}%
_{t}^{\ast }\psi ^{\ast }\psi -\psi ^{\ast }\tilde{\partial}_{t}\psi \right)
,\ \mathbf{J}=\frac{\chi q}{2m}\mathrm{i}\left( \tilde{\nabla}^{\ast }\psi
^{\ast }\psi -\psi ^{\ast }\tilde{\nabla}\psi \right) .  \label{psfi4}
\end{equation}%
They satisfy the conservation (continuity) equation 
\begin{equation}
\partial _{t}\rho +\nabla \cdot \mathbf{J}=0,  \label{psfi5}
\end{equation}%
and, consequently, the total charge\ is conserved: 
\begin{equation}
\dint\nolimits_{\mathbb{R}^{3}}\rho \left( t,\mathbf{x}\right) \,\mathrm{d}%
\mathbf{x}=const.  \label{psfi6}
\end{equation}%
For the fundamental pair $\left\{ \mathring{\psi},\mathring{\varphi}\right\} 
$ the corresponding microcharge density defined by (\ref{psfi4}) turns into 
\begin{equation}
\rho =\rho \left( \left\vert \mathbf{x}\right\vert \right) =q\left( 1-\frac{q%
\mathring{\varphi}\left( \left\vert \mathbf{x}\right\vert \right) }{m\mathrm{%
c}^{2}}\right) \mathring{\psi}^{2}\left( \left\vert \mathbf{x}\right\vert
\right) .  \label{psfi5a}
\end{equation}%
Note that equation (\ref{psfi1a}) turns into the classical equation for the
Coulomb's potential if $\rho $ is replaced by $q\delta \left( \mathbf{x}%
\right) $ where delta function has standard property $\dint \delta \left( 
\mathbf{x}\right) \,\mathrm{d}\mathbf{x}=1$. Since we want $\mathring{\varphi%
}$ to behave as Coulomb's electrostatic potential at large distances and $q$
to be the charge, we introduce the following \emph{charge normalization
condition} imposed on the form factor $\mathring{\psi}$%
\begin{equation}
\dint\nolimits_{\mathbb{R}^{3}}\left( 1-\frac{q\mathring{\varphi}\left(
\left\vert \mathbf{x}\right\vert \right) }{m\mathrm{c}^{2}}\right) \mathring{%
\psi}^{2}\left( \left\vert \mathbf{x}\right\vert \right) \,\mathrm{d}\mathbf{%
x}=1.  \label{psfi7}
\end{equation}%
Notice that we introduced the above terms \emph{microcharge} and \emph{%
microcurrent} densities to emphasize their relation to the internal
structure of elementary charges and the difference from commonly used charge
and the current densities as macroscopic quantities. It is worth noticing
though that if it comes to the interaction with the electromagnetic field
the "micro" charges and microcurrents densities behave exactly the same way
as the macroscopic charges and densities, but microcharges are also subjects
to the internal elastic and nonlinear self-interaction forces of
non-electromagnetic nature.

\subsection{Energy considerations}

Let us denote by $\mathcal{E}_{0}\left( \psi ,A^{\mu }\right) $ the energy
of the dressed charge $\left\{ \psi ,A^{\mu }\right\} $ derived from the
Lagrangian $L_{0}$. We found that for the fundamental pair $\left\{ 
\mathring{\psi},\mathring{\varphi}\right\} $ the energy $\mathcal{E}_{0}$
can be written in the following form%
\begin{gather}
\mathcal{E}_{0}\left( \mathring{\psi},\mathring{\varphi}\right) =\mathcal{E}%
_{0}\left( \mathring{\psi}\right) =m\mathrm{c}^{2}+\mathcal{E}_{0}^{\prime
}\left( \mathring{\psi}\right) ,  \label{psi1e} \\
\mathcal{E}_{0}^{\prime }\left( \mathring{\psi}\right) =\frac{2}{3}\int_{%
\mathbb{R}^{3}}\left[ \frac{\chi ^{2}}{2m}\nabla \mathring{\psi}^{\ast
}\cdot \nabla \mathring{\psi}-\frac{\left( \nabla \mathring{\varphi}\right)
^{2}}{8\pi }\right] \,\mathrm{d}x,  \notag
\end{gather}%
where we use the relation $\mathring{\varphi}=\mathring{\varphi}_{\mathring{%
\psi}}$ from (\ref{psfi1c}) to emphasize an important fact that the above
energy $\mathcal{E}_{0}$ is a functional of $\mathring{\psi}$ and the model
constants only. We refer to the energy $\mathcal{E}_{0}^{\prime }\left( 
\mathring{\psi}\right) $ defined in (\ref{psi1e}) as the relative energy.
The significance of the representation (\ref{psi1e}) for the energy $%
\mathcal{E}_{0}$ is in the fact that it does not explicitly involve the
nonlinear self-interaction $G$.

Applying to the energy $\mathcal{E}_{0}$ the Einstein principle of
equivalence of mass and energy, namely $E=m\mathrm{c}^{2}$, \cite[Section 41]%
{Pauli RT}, we define the \emph{dressed charge mass} $\tilde{m}=\tilde{m}%
\left( \mathring{\psi}\right) $ by the equality 
\begin{equation}
\mathcal{E}_{0}\left( \mathring{\psi}\right) =\tilde{m}\mathrm{c}^{2}=\tilde{%
m}\left( \mathring{\psi}\right) \mathrm{c}^{2}.  \label{psi2a}
\end{equation}%
Combining the relation (\ref{psi2a}) with (\ref{psi1e}) we readily obtain%
\begin{equation}
\left( \tilde{m}-m\right) \mathrm{c}^{2}=\mathcal{E}_{0}^{\prime }\left( 
\mathring{\psi}\right) =\frac{2}{3}\int_{\mathbb{R}^{3}}\left[ \frac{\chi
^{2}}{2m}\nabla \mathring{\psi}^{\ast }\cdot \nabla \mathring{\psi}-\frac{%
\left( \nabla \mathring{\varphi}\right) ^{2}}{8\pi }\right] \,\mathrm{d}x.
\label{psi2e}
\end{equation}%
We also want the fundamental frequency $\omega _{0}$ to satisfy the Einstein
relation $\mathcal{E}_{0}=\hbar \omega _{0}$, which would determine $\omega
_{0}$ as a function of $\mathring{\psi}$, constants $\mathrm{c}$, $m$, $q$
and, importantly, $\chi $, namely 
\begin{gather}
\mathcal{E}_{0}\left( \mathring{\psi}\right) =\hbar \omega _{0}\left( 
\mathring{\psi},\chi \right) ,\text{ or}  \label{psfi3} \\
\omega _{0}=\omega _{0}\left( \mathring{\psi},\chi \right) =\frac{1}{\hbar }%
\left\{ m\mathrm{c}^{2}+\frac{2}{3}\int_{\mathbb{R}^{3}}\left[ \frac{\chi
^{2}}{2m}\nabla \mathring{\psi}^{\ast }\cdot \nabla \mathring{\psi}-\frac{%
\left( \nabla \mathring{\varphi}\right) ^{2}}{8\pi }\right] \,\mathrm{d}%
x\right\} .  \notag
\end{gather}%
Then to be consistent with the earlier relation (\ref{afi2}) we have to set
the value of $\chi $ so that%
\begin{equation}
\chi \omega _{0}\left( \mathring{\psi},\chi \right) =m\mathrm{c}^{2},
\label{psfi3e}
\end{equation}%
which, in view of the representation (\ref{psfi3}) for $\omega _{0}\left( 
\mathring{\psi},\chi \right) $, is equivalent to the requirement for $\chi
=\chi \left( \mathring{\psi}\right) $ to be a positive solution to the
following cubic equation 
\begin{gather}
\chi \left\{ c_{2}\chi ^{2}+c_{1}\right\} =\hbar \text{, where}
\label{psi2b} \\
c_{2}=\frac{1}{3m^{2}\mathrm{c}^{2}}\int_{\mathbb{R}^{3}}\nabla \mathring{%
\psi}^{\ast }\cdot \nabla \mathring{\psi}\,\mathrm{d}x,\ c_{1}=1-\frac{2}{3m%
\mathrm{c}^{2}}\int_{\mathbb{R}^{3}}\frac{\left( \nabla \mathring{\varphi}%
\right) ^{2}}{8\pi }\,\mathrm{d}x,  \notag
\end{gather}%
where $\mathring{\varphi}$, in view of defining it equation (\ref{psfi1a}),
depends only on $\mathring{\psi}$ and constants $\mathrm{c}$, $m$, $q$. Note
also that we can choose $\mathring{\psi}$ independently of $\chi $, and then
determine $G$. Notice that cubic equation (\ref{psi2b}) always has a
positive solution, and if, in addition to that, we know that $c_{1}\geq 0,$
then the left-hand side of the equation (\ref{psi2b}) is a monotonically
increasing function for $\chi \geq 0$ implying that the solution is unique.

In the case of a generic form factor $\mathring{\psi}$ the relative energy $%
\mathcal{E}_{0}^{\prime }\left( \mathring{\psi}\right) $ does not
necessarily have to vanish and, in view of the formula (\ref{psi2e}), the
mass $\tilde{m}$ \ may be different from $m$. Then, as it follows from the
relation (\ref{psi2a})-(\ref{psi2b}) $\chi \neq \hbar $. The very same
relations also readily imply that%
\begin{equation}
\chi =\hbar \text{ if }\tilde{m}=m.  \label{psi2c}
\end{equation}%
For a number of reasons, mainly for the perfect agreement between the
relativistic energy-momentum and its nonrelativistic approximation
constructed below, it is attractive to have $\tilde{m}=m$ implying also, in
view of (\ref{psi2c}), $\chi =\hbar $. The question though is if that is
possible. The answer is affirmative and the equality $\tilde{m}=m$ of the
two masses can be achieved as follows. We introduce for the form factor its 
\emph{size representation} involving size parameter $a$ and normalization
constant $\mathring{C}$: \ \ 
\begin{equation}
\mathring{\psi}\left( s\right) =\mathring{\psi}_{a}\left( s\right) =\frac{%
\mathring{C}}{a^{3/2}}\mathring{\psi}_{1}\left( \frac{s}{a}\right) ,\ a>0,\
s\geq 0,  \label{psfi8}
\end{equation}%
where the function $\mathring{\psi}_{1}\left( \left\vert \mathbf{x}%
\right\vert \right) $ satisfies the normalization condition%
\begin{equation}
\int\nolimits_{\mathbb{R}^{3}}\mathring{\psi}_{1}^{2}\left( \left\vert 
\mathbf{x}\right\vert \right) \,\mathrm{d}\mathbf{x}=1.  \label{psfi9}
\end{equation}%
We consider then values of $\mathring{C}$ and $a$ in the representation (\ref%
{psfi8}), (\ref{psfi9}) that satisfy two conditions: (i) the charge
normalization condition (\ref{psfi7}), namely%
\begin{equation}
\mathring{C}^{2}\dint\nolimits_{\mathbb{R}^{3}}\left( 1-\frac{q\mathring{%
\varphi}_{a}\left( \left\vert \mathbf{x}\right\vert \right) }{m\mathrm{c}^{2}%
}\right) \mathring{\psi}_{a}^{2}\left( \left\vert \mathbf{x}\right\vert
\right) \,\mathrm{d}\mathbf{x}=1,  \label{psixi2}
\end{equation}%
and the energy normalization condition 
\begin{equation}
\mathcal{E}_{0}^{\prime }\left( \mathring{\psi}_{a_{0}}\right) =0.
\label{psi2d}
\end{equation}%
We provide arguments in Subsection \ref{simnorm} \ based on the smallness of
the Sommerfeld fine structure constant showing that there exist such $%
\mathring{C}$, $a=a_{0}$ for which both the normalization conditions (\ref%
{psixi2}) and (\ref{psi2d}) can hold. In view of (\ref{psi1e}) and (\ref%
{psi2c}) the above equality implies 
\begin{equation}
\mathcal{E}_{0}\left( \mathring{\psi}_{a_{0}}\right) =m\mathrm{c}^{2}\text{
and }\tilde{m}=m,\ \chi =\hbar .  \label{psfi2}
\end{equation}%
In other words, the requirement $\tilde{m}=m$ fixes the charge size $a=a_{0}$
as well as the constant $\chi =\hbar $, the magnitude of $a_{0}$ is of the
same order as Bohr's radius. One might ask if it is necessary to require the
exact equality $\tilde{m}=m$? For a good agreement between the relativistic
energy-momentum and its nonrelativistic approximation constructed below it
would be sufficient for $\tilde{m}-m$ to be small enough rather than zero.
For this reason and because of the scope of this paper we decided not to
impose here the exact mass equality $\tilde{m}=m$ leaving this decision for
future work. \emph{So from now on we assume that the value of the constant }$%
\chi $\emph{\ to be set by equations (\ref{psfi3e}), (\ref{psi2b}).}

\subsection{Moving free charge}

Using the Lorentz invariance of the system we can obtain, as it is often
done, a representation for the dressed charge moving with a constant
velocity $\mathbf{v}$ simply by applying to the rest solution (\ref{psfi1})
the Lorentz transformation from the original "rest frame" to the frame in
which the "rest frame" moves with the constant velocity $\mathbf{v}$ as
described by the formulas (\ref{rkin3}), (\ref{maxw8}) (so $x^{\prime }$ and 
$x$ correspond, respectively, to the "rest" and "moving" frames). Namely,
introducing%
\begin{equation}
\mathbf{\beta }=\frac{\mathbf{v}}{\mathrm{c}},\ \beta =\left\vert \mathbf{%
\beta }\right\vert ,\ \gamma =\left( 1-\left( \frac{v}{\mathrm{c}}\right)
^{2}\right) ^{-1/2},  \label{wcem1}
\end{equation}%
we obtain the following representation for the dressed charge moving with
velocity $\mathbf{v}$ 
\begin{equation}
\psi \left( t,\mathbf{x}\right) =\mathrm{e}^{-\mathrm{i}\left( \omega t-%
\mathbf{k}\cdot \mathbf{x}\right) }\mathring{\psi}\left( \mathbf{x}^{\prime
}\right) ,\ \varphi \left( t,\mathbf{x}\right) =\gamma \mathring{\varphi}%
\left( \left\vert \mathbf{x}^{\prime }\right\vert \right) ,\ \mathbf{A}%
\left( t,\mathbf{x}\right) =\gamma \mathbf{\beta }\mathring{\varphi}\left(
\left\vert \mathbf{x}^{\prime }\right\vert \right) ,  \label{wcem1a}
\end{equation}%
\begin{equation}
\mathbf{x}^{\prime }=\mathbf{x}+\frac{\gamma -1}{\beta ^{2}}\left( \mathbf{%
\beta }\cdot \mathbf{x}\right) \mathbf{\beta }-\gamma \mathbf{v}t,\text{ or }%
\mathbf{x}_{\Vert }^{\prime }=\gamma \left( \mathbf{x}_{\Vert }-\mathbf{v}%
t\right) ,\ \mathbf{x}_{\bot }^{\prime }=\mathbf{x}_{\bot },  \label{wcem1d}
\end{equation}%
\begin{equation}
\omega =\gamma \omega _{0},\ \mathbf{k}=\gamma \mathbf{\beta }\frac{\omega
_{0}}{\mathrm{c}},  \label{wcem1c}
\end{equation}%
where $\mathbf{x}_{\Vert }$ and $\mathbf{x}_{\bot }$ refer, respectively, to
the components of $\mathbf{x}$ parallel and perpendicular to the velocity $%
\mathbf{v}$, with the fields given by 
\begin{equation}
\mathbf{E}\left( t,\mathbf{x}\right) =-\gamma \nabla \mathring{\varphi}%
\left( \left\vert \mathbf{x}^{\prime }\right\vert \right) +\frac{\gamma ^{2}%
}{\gamma +1}\left( \mathbf{\beta }\cdot \nabla \mathring{\varphi}\left(
\left\vert \mathbf{x}^{\prime }\right\vert \right) \right) \mathbf{\beta ,\ B%
}\left( t,\mathbf{x}\right) =\gamma \mathbf{\beta }\times \nabla \mathring{%
\varphi}\left( \left\vert \mathbf{x}^{\prime }\right\vert \right) .
\label{wcem1b}
\end{equation}%
The above formulas (which provide a solution to the field equations (\ref%
{fpar4}), (\ref{fpar5}) indicate that the fields of the dressed charge
contract by the factor $\gamma $ as it moves with the velocity $\mathbf{v}$
compared to their rest state. The first oscillatory exponential factor in (%
\ref{wcem1}) is the de Broglie plane wave of frequency $\omega =\omega
\left( \mathbf{k}\right) $ and a wave-vector $\mathbf{k}$\textbf{,}
satisfying%
\begin{equation}
\omega ^{2}-\mathrm{c}^{2}\mathbf{k}^{2}=\omega _{0}^{2},\ \chi \omega _{0}=m%
\mathrm{c}^{2}.  \label{wcem2}
\end{equation}%
Based on the Lagrangian $L_{0}$ we found the symmetric energy-momentum
tensor, which shows that the dressed charge moving with a constant velocity $%
\mathbf{v}$ and described by (\ref{wcem1})-(\ref{wcem1d}) has energy $%
\mathsf{E}$ and momentum $\mathsf{P}$, which satisfy the \emph{Einstein-de
Broglie relations} 
\begin{equation}
\mathsf{E}=\hbar \omega ,\ \mathsf{P}=\hbar \mathbf{k}.  \label{wcem3}
\end{equation}%
In addition to that, the charge velocity $\mathbf{v}$ and its de Broglie
wave vector $\mathbf{k}$ satisfy the following relation%
\begin{equation}
\mathbf{v}=\nabla _{\mathbf{k}}\omega \left( \mathbf{k}\right) ,
\label{wcem4}
\end{equation}%
signifying that the velocity $\mathbf{v}$ of the dressed charge is the group
velocity of the linear medium with the dispersion relation (\ref{wcem2}).

The second factor in the formula (\ref{wcem1a}) for $\psi $ involves the
form factor $\mathring{\psi}\left( r\right) $, $r\geq 0$, which is a
monotonically decreasing function of $r$ decaying at infinity. For such a
form factor the form factor potential $\mathring{\varphi}\left( r\right) $
decays at infinity as the Coulomb's potential as it follows from (\ref%
{psfi1c}), i.e. $\mathring{\varphi}\left( r\right) \sim qr^{-1}$ for large $%
r $. Consequently, the dressed charge moving with constant velocity $\mathbf{%
v} $ as described by equations (\ref{wcem1a})-(\ref{wcem1d}) remains well
localized and does not disperse in the space at all times justifying its
characterization as a wave-corpuscle.

\subsection{Nonrelativistic approximation for the charge in an external EM
field\label{nrapr}}

Our nonrelativistic Lagrangian for a single charge in external EM field $%
\mathbf{\ }$with potentials $\varphi _{\mathrm{ex}},\mathbf{A}_{\mathrm{ex}}$%
\ has the form 
\begin{equation}
\hat{L}_{0}\left( \psi ,\psi ^{\ast },\varphi \right) =\frac{\chi }{2}%
\mathrm{i}\left[ \psi ^{\ast }\tilde{\partial}_{t}\psi -\psi \tilde{\partial}%
_{t}^{\ast }\psi ^{\ast }\right] -\frac{\chi ^{2}}{2m}\left\{ \tilde{\nabla}%
\psi \tilde{\nabla}^{\ast }\psi ^{\ast }+G\left( \psi ^{\ast }\psi \right)
\right\} +\frac{\left\vert \nabla \varphi \right\vert ^{2}}{8\pi },
\label{nop1}
\end{equation}%
where%
\begin{equation}
\tilde{\partial}_{t}=\partial _{t}+\frac{\mathrm{i}q\bar{\varphi}}{\chi },\ 
\tilde{\nabla}=\nabla -\frac{\mathrm{i}q\mathbf{A}_{\mathrm{ex}}}{\chi c},\ 
\tilde{\partial}_{t}^{\ast }=\partial _{t}-\frac{\mathrm{i}q\bar{\varphi}}{%
\chi },\ \tilde{\nabla}^{\ast }=\nabla +\frac{\mathrm{i}q\mathbf{A}_{\mathrm{%
ex}}}{\chi c},\ \bar{\varphi}=\varphi +\varphi _{\mathrm{ex}}.  \label{nop1a}
\end{equation}

Derivation of (\ref{nop1}) from relativistic Lagrangian (\ref{afi3}) is
discussed in Section \ref{cheqdimf}. For simplicity, we consider at first
the case where an external magnetic field vanishes, $\mathbf{A}_{\mathrm{ex}%
}=0$. The field equations for this case take the form%
\begin{gather}
\chi \mathrm{i}\partial _{t}\psi -q\left( \varphi +\varphi _{\mathrm{ex}%
}\right) \psi =-\frac{\chi ^{2}}{2m}\left[ \Delta \psi -G^{\prime }\left(
\psi ^{\ast }\psi \right) \psi \right] ,  \label{nop2} \\
-\Delta \varphi =4\pi q\psi \psi ^{\ast },\ \text{where }G^{\prime }\left(
s\right) =\partial _{s}G,  \label{nop3}
\end{gather}%
and we refer to the pair $\left\{ \psi ,\varphi \right\} $ as \emph{dressed
charge}. Recall that $\psi ^{\ast }$ is complex conjugate to $\psi $.

Let us consider now the\ case of resting charge with no external EM field
described by a static time independent solution to the equations (\ref{nop2}%
)-(\ref{nop3}). These equations under the assumption $\mathbf{E}_{\mathrm{ex}%
}\left( t\right) =0$ turn into the following \emph{rest charge equations }
for a static state, as described by time independent real-valued radial
functions $\mathring{\psi}=\mathring{\psi}\left( \left\vert \mathbf{x}%
\right\vert \right) $ and $\mathring{\varphi}=\mathring{\varphi}\left(
\left\vert \mathbf{x}\right\vert \right) $:%
\begin{gather}
-\Delta \mathring{\varphi}=4\pi q\left\vert \mathring{\psi}\right\vert ^{2},
\label{nop4} \\
-\Delta \mathring{\psi}+\frac{2m}{\chi ^{2}}q\mathring{\varphi}\mathring{\psi%
}+G^{\prime }\left( \left\vert \mathring{\psi}\right\vert ^{2}\right) 
\mathring{\psi}=0.  \label{nop5}
\end{gather}%
The quantities $\mathring{\psi}$ and $\mathring{\varphi}$ are fundamental
for our theory and we refer to them, respectively, as \emph{form factor} and 
\emph{form factor potential}. In view of the equation (\ref{nop4}) the
charge form factor $\mathring{\psi}$ determines the form factor potential $%
\mathring{\varphi}$ by the formula%
\begin{equation}
\mathring{\varphi}\left( \left\vert \mathbf{x}\right\vert \right) =\mathring{%
\varphi}_{\mathring{\psi}}\left( \left\vert \mathbf{x}\right\vert \right)
=4\pi q\dint_{\mathbb{R}^{3}}\frac{\mathring{\psi}^{2}\left( \left\vert 
\mathbf{y}\right\vert \right) }{\left\vert \mathbf{y}-\mathbf{x}\right\vert }%
\,\mathrm{d}\mathbf{y},  \label{nop6}
\end{equation}%
and if we plug in the above expression into equation (\ref{nop5}), we get
the following nonlinear equation%
\begin{equation}
-\Delta \mathring{\psi}+\frac{2mq}{\chi ^{2}}\mathring{\varphi}_{\mathring{%
\psi}}\mathring{\psi}+G^{\prime }(\left\vert \mathring{\psi}\right\vert ^{2})%
\mathring{\psi}=0.  \label{nop7}
\end{equation}%
The equation (\ref{nop7}) signifies a complete balance of the three forces
acting upon the resting charge: (i) internal elastic deformation force
associated with the term $-\Delta \mathring{\psi}$; (ii) the charge's
electromagnetic self-interaction force associated with the term $\frac{2mq}{%
\chi ^{2}}\mathring{\varphi}_{\mathring{\psi}}\mathring{\psi}$; (iii)
internal nonlinear self-interaction of the charge associated with the term $%
G^{\prime }\left( \left\vert \mathring{\psi}\right\vert ^{2}\right) 
\mathring{\psi}$. We refer to the equation (\ref{nop7}), which establishes
an explicit relation between the form factor $\mathring{\psi}$ and the
self-interaction nonlinearity $G,$ as the \emph{charge equilibrium equation}%
. Hence, being given the form factor $\mathring{\psi}$ we can find from the
equilibrium equation (\ref{nop7}) the self-interaction nonlinearity $G$
which exactly produces this factor under the assumption that $\mathring{\psi}%
\left( r\right) $, $r\geq 0$ is a nonnegative, monotonically decaying and
sufficiently smooth function. \emph{Thus, we pick the form factor }$%
\mathring{\psi},$\emph{\ and then the nonlinear self interaction function }$%
G $\emph{\ is determined based on the charge equilibrium\ equation (\ref%
{nop7})}. One of the advantages of determining $G$ in terms of $\mathring{%
\psi}$ is that we more often use properties of $\mathring{\psi}$\ in our
analysis rather than properties of $G.$ Note that after the nonlinearity $G$
\ is determined, it is fixed forever, and while solutions of equations (\ref%
{nop2})-(\ref{nop3}) may evolve in time, they do not need to coincide with $%
\{\mathring{\psi},\mathring{\varphi}\}.$ Details and examples of the
construction of the nonlinear self-interaction function $G$ based on the
form factor are provided in the following sections.

The 4-microcurrent density $J^{\mu }$ related to the Lagrangian $\hat{L}_{0}$
is 
\begin{equation}
J^{\mu }=\left( \mathrm{c}\rho ,\mathbf{J}\right) ,\ \rho =q\psi \psi ^{\ast
},\ \mathbf{J}=\frac{\mathrm{i}\chi q}{2m}\left[ \psi \nabla ^{\ast }\psi
^{\ast }-\psi ^{\ast }\nabla \psi \right] =\frac{\chi q}{m}\func{Im}\frac{%
\nabla \psi }{\psi }\left\vert \psi \right\vert ^{2},  \label{nop8}
\end{equation}%
and it satisfies the conservation/continuity equations%
\begin{equation}
\partial _{\nu }J^{\nu }=0,\ \partial _{t}\rho +\nabla \cdot \mathbf{J}=0.
\label{nop9}
\end{equation}%
For the fundamental pair $\{\mathring{\psi},\mathring{\varphi}\}$ the
corresponding microcharge density defined by (\ref{nop8}) turns into 
\begin{equation}
\rho =\rho \left( \left\vert \mathbf{x}\right\vert \right) =q\left\vert 
\mathring{\psi}\right\vert ^{2}\left( \left\vert \mathbf{x}\right\vert
\right) .  \label{nop10}
\end{equation}%
The \emph{charge normalization condition} (\ref{psfi7}) which, in
particular, ensures that $\mathring{\varphi}\left( \left\vert \mathbf{x}%
\right\vert \right) $ is close to the Coulomb's potential with the charge $q$
for large $\left\vert \mathbf{x}\right\vert $ now takes the simpler form 
\begin{equation}
\dint\nolimits_{\mathbb{R}^{3}}\mathring{\psi}^{2}\left( \left\vert \mathbf{x%
}\right\vert \right) \,\mathrm{d}\mathbf{x}=1.  \label{nop11}
\end{equation}%
Interestingly the momentum and the current densities of the dressed charge
derived from $\hat{L}_{0}$ are identical up to a factor $\frac{m}{q}$, namely%
\begin{equation}
\mathbf{P}=\frac{m}{q}\mathbf{J}=\frac{\mathrm{i}\chi }{2}\left[ \psi \nabla
^{\ast }\psi ^{\ast }-\psi ^{\ast }\nabla \psi \right] =\chi \func{Im}\frac{%
\nabla \psi }{\psi }\left\vert \psi \right\vert ^{2}.  \label{nop12}
\end{equation}

It turns out that the field equations (\ref{nop2})-(\ref{nop3}) have a
closed form solution in terms of (i) the static dressed charge state $%
\left\{ \mathring{\psi},\mathring{\varphi}\right\} $ satisfying (\ref{nop4}%
)-(\ref{nop5}) and (ii) the basic quantities related to the point charge of
mass $m$ moving in external homogeneous electric field $\mathbf{E}_{\mathrm{%
ex}}\left( t\right) $ with the electric potential $\varphi _{\mathrm{ex}%
}\left( t,\mathbf{x}\right) =\varphi _{\mathrm{ex}}^{0}\left( t\right) -%
\mathbf{E}_{\mathrm{ex}}\left( t\right) \cdot \mathbf{x}$. Indeed, let $%
\mathbf{r}$, $\mathbf{v}$, $\mathbf{p}$, $L_{\mathrm{p}}\left( \mathbf{v},%
\mathbf{r}\right) $, $s_{\mathrm{p}}$ and $H_{\mathrm{p}}\left( \mathbf{p},%
\mathbf{r},t\right) $ be, respectively, the point charge position, velocity,
momentum, Lagrangian, action and Hamiltonian satisfying the following
familiar relations:%
\begin{gather}
L_{\mathrm{p}}\left( \mathbf{v},\mathbf{r},t\right) =\frac{m\mathbf{v}^{2}}{2%
}-q\varphi _{\mathrm{ex}}\left( t,\mathbf{x}\right) ,  \label{exac2} \\
\frac{\mathrm{d}s_{\mathrm{p}}}{\mathrm{d}t}=L_{\mathrm{p}}\left( \mathbf{v},%
\mathbf{r,}t\right) =\frac{\mathrm{d}}{\mathrm{d}t}\left( \mathbf{p}\cdot 
\mathbf{r}\right) -\frac{\mathbf{p}^{2}}{2m}-\varphi _{\mathrm{ex}%
}^{0}\left( t\right) ,  \notag \\
H_{\mathrm{p}}\left( \mathbf{p},\mathbf{r},t\right) =\frac{\mathbf{p}^{2}}{2m%
}+q\varphi _{\mathrm{ex}}\left( t,\mathbf{r}\right) ,\ \mathbf{p}=m\mathbf{v}%
,  \notag
\end{gather}%
with point charge equations of motion%
\begin{equation}
\mathbf{v}=\frac{\mathrm{d}\mathbf{r}}{\mathrm{d}t},\ \frac{\mathrm{d}%
\mathbf{p}}{\mathrm{d}t}=q\mathbf{E}_{\mathrm{ex}}\left( t\right) \text{ or
equivalently }m\frac{\mathrm{d}^{2}\mathbf{r}}{\mathrm{d}t^{2}}=q\mathbf{E}_{%
\mathrm{ex}}.  \label{exac3}
\end{equation}%
One can recognize in the expression $q\mathbf{E}_{\mathrm{ex}}\left(
t\right) $ in (\ref{exac3}) the Lorentz force due the external electric
field $\mathbf{E}_{\mathrm{ex}}\left( t\right) $. We refer to $L_{\mathrm{p}%
}\left( \mathbf{v},\mathbf{r}\right) $ and the equations (\ref{exac3})
respectively as \emph{the complementary point charge Lagrangian and the
equations of motion}.

Then the field equations (\ref{nop2})-(\ref{nop3}) have the following closed
form solution%
\begin{gather}
\psi =\psi \left( t,\mathbf{x}\right) =\mathrm{e}^{\mathrm{i}S/\chi }%
\mathring{\psi}\left( \left\vert \mathbf{x}-\mathbf{r}\right\vert \right) ,\
S=\mathbf{p}\cdot \left( \mathbf{x}-\mathbf{r}\right) +s_{\mathrm{p}},
\label{exac1} \\
\varphi \left( t,\mathbf{x}\right) =\mathring{\varphi}_{0}\left( \left\vert 
\mathbf{x}-\mathbf{r}\right\vert \right) ,  \notag
\end{gather}%
where in view of relations (\ref{exac2}) $\psi $ can be also represented as%
\begin{equation}
\psi =\mathrm{e}^{\mathrm{i}S/\chi }\mathring{\psi}\left( \left\vert \mathbf{%
x}-\mathbf{r}\right\vert \right) ,\ S=\mathbf{p}\cdot \mathbf{x}%
-\dint_{0}^{t}\frac{\mathbf{p}^{2}}{2m}\,\mathrm{d}t^{\prime
}-q\dint_{0}^{t}\varphi _{\mathrm{ex}}^{0}\left( t^{\prime }\right) \,%
\mathrm{d}t^{\prime }.  \label{exac1a}
\end{equation}%
A similar exact solution is given for the class of EM fields with non-zero,
spatially constant magnetic field $\mathbf{B}_{\mathrm{ex}}\left( t\right) .$
Of course, in latter case the Lorentz force involves $\mathbf{B}_{\mathrm{ex}%
}$ as in (\ref{pchar1}) (see Section \ref{exelmag} for details). Looking at
the exact solution (\ref{exac1}), (\ref{exac1a}) to the field equations that
describes the \emph{accelerating charge} we would like to acknowledge the 
\emph{truly remarkable simplicity and transparency of the relations between
the two concepts of the charge}: \emph{charge as a field} $\left\{ \psi
,\varphi \right\} $ in (\ref{exac1}), (\ref{exac1a}) and \emph{charge as a
point} described by equations (\ref{exac2}) and (\ref{exac3}). Indeed, the
wave amplitude $\mathring{\psi}\left( \left\vert \mathbf{x}-\mathbf{r}\left(
t\right) \right\vert \right) $ in (\ref{exac1}) is a soliton-like field
moving exactly as the point charge described by its position $\mathbf{r}%
\left( t\right) $. \emph{The exponential factor }$\mathrm{e}^{\mathrm{i}%
S/\chi }$\emph{\ is just a plane wave with the phase }$S$ \emph{that depends
only on the point charge position }$\mathbf{r}$\emph{\ and momentum }$%
\mathbf{p}$\emph{\ and a time dependent gauge term, and it does not depend
on the nonlinear self-interaction}. The phase $S$ has a term in which we
readily recognize the de Broglie wave-vector $\mathbf{k}\left( t\right) $
described exactly in terms of the point charge quantities, namely%
\begin{equation}
\chi \mathbf{k}\left( t\right) =\mathbf{p}\left( t\right) =m\mathbf{v}\left(
t\right) .  \label{exac4}
\end{equation}%
Notice that the dispersion relation $\omega =\omega \left( \mathbf{k}\right) 
$ of the linear kinetic part of the field equations (\ref{nop2}) for $\psi $
is 
\begin{equation}
\omega \left( \mathbf{k}\right) =\frac{\chi \mathbf{k}^{2}}{2m},\text{
implying that the group velocity }\nabla _{\mathbf{k}}\omega \left( \mathbf{k%
}\right) =\frac{\chi \mathbf{k}}{m}.  \label{exac5}
\end{equation}%
Combining the expression (\ref{exac5}) for the group velocity $\nabla _{%
\mathbf{k}}\omega \left( \mathbf{k}\right) $ with the expression (\ref{exac4}%
) for wave vector $\mathbf{k}\left( t\right) $ we establish another exact
relation, namely 
\begin{equation}
\mathbf{v}\left( t\right) =\nabla _{\mathbf{k}}\omega \left( \mathbf{k}%
\left( t\right) \right) ,  \label{exac6}
\end{equation}%
signifying the equality between the point charge velocity $\mathbf{v}\left(
t\right) $ and the group velocity $\nabla _{\mathbf{k}}\omega \left( \mathbf{%
k}\left( t\right) \right) $ at the de Broglie wave vector $\mathbf{k}\left(
t\right) $. Using the relations (\ref{nop8}) and (\ref{nop12}) we readily
obtain the following representations for the micro-charge, the micro-current
and momentum densities 
\begin{gather}
\rho \left( t,\mathbf{x}\right) =q\mathring{\psi}^{2}\left( \left\vert 
\mathbf{x}-\mathbf{r}\left( t\right) \right\vert \right) ,\text{ }\mathbf{J}%
\left( t,\mathbf{x}\right) =q\mathbf{v}\left( t\right) \mathring{\psi}%
^{2}\left( \left\vert \mathbf{x}-\mathbf{r}\left( t\right) \right\vert
\right) ,  \label{exac6a} \\
\mathbf{P}\left( t,\mathbf{x}\right) =\frac{m}{q}\mathbf{J}\left( t,\mathbf{x%
}\right) =\mathbf{p}\left( t\right) \mathring{\psi}^{2}\left( \left\vert 
\mathbf{x}-\mathbf{r}\left( t\right) \right\vert \right) .  \label{exac6b}
\end{gather}%
We also found that the total dressed charge field momentum $\mathsf{P}\left(
t\right) $ and the total current $\mathsf{J}\left( t\right) $ for the
solution (\ref{exac1}) are expressed exactly in terms of point charge
quantities, namely%
\begin{equation}
\mathsf{P}\left( t\right) =\frac{m}{q}\mathsf{J}\left( t\right) =\dint_{%
\mathbb{R}^{3}}\frac{\chi q}{m}\func{Im}\frac{\nabla \psi }{\psi }%
\,\left\vert \psi \right\vert ^{2}\,\mathrm{d}\mathbf{x}=\mathsf{p}\left(
t\right) =m\mathbf{v}\left( t\right) .  \label{exac7}
\end{equation}

To summarize the above analysis we may state that even when the charge
accelerates it perfectly combines the properties of a wave and a corpuscle,
justifying the name wave-corpuscle mechanics. Its wave nature shows, in
particular, in the de Broglie exponential factor and the equality (\ref%
{exac6}), indicating the wave origin the charge motion. The corpuscle
properties are manifested in the factor $\mathring{\psi}\left( \left\vert 
\mathbf{x}-\mathbf{r}\left( t\right) \right\vert \right) $ and soliton like
propagation with $\mathbf{r}\left( t\right) $ satisfying the classical point
charge evolution equation (\ref{exac3}). Importantly, the introduced
nonlinearities are stealthy in the sense that they don't show in the
dynamics and kinematics of what appears to be soliton-like waves propagating
like classical test point charges.

\subsection{Many interacting charges\label{nrapr1}}

A qualitatively new physical phenomenon in the theory of two or more charges
compared with the theory of a single charge is obviously the interaction
between them. In our approach \emph{any individual "bare" charge interacts
directly only with the EM} field and consequently different charges interact
with each other only indirectly through the EM field. We develop the theory
for many interacting charges for both the relativistic and nonrelativistic
cases, and in this section we provide key elements of the nonrelativistic
theory. The primary focus of our studies on many charges is the
correspondence between our wave theory and the point charge mechanics in the
regime of remote interaction when the charges are separated by large
distances compared to their sizes.

Studies of charge interactions at short distances are not within the scope
of this paper, though our approach allows to study short distance
interactions and we provide as an example the hydrogen atom model in Section %
\ref{shydro}. The main purpose of that exercise is to show a similarity
between our and Schr\"{o}dinger's hydrogen atom models and to contrast it to
any kind of Kepler's model. Another point we can make based on our hydrogen
atom model is that our theory does provide a basis for a regime of close
interaction between two charges which differs significantly from the regime
of remote interaction, which is the primary focus of this paper.

Let us consider a system of $N$ charges described by complex-valued fields $%
\psi ^{\ell }$, $\ell =1,\ldots ,N$. The system's nonrelativistic Lagrangian 
$\mathcal{\hat{L}}$ is constructed based on individual charge Lagrangians $%
\hat{L}^{\ell }$ of the form (\ref{nop1})-(\ref{nop1a}) and an assumption
that every charge interacts directly only with the EM field, including the
external one with potentials $\varphi _{\mathrm{ex}}\left( t,\mathbf{x}%
\right) $, $\mathbf{A}_{\mathrm{ex}}\left( t,\mathbf{x}\right) $, namely%
\begin{gather}
\mathcal{\hat{L}}=\sum_{\ell }\hat{L}^{\ell }+\frac{\left\vert \nabla
\varphi \right\vert ^{2}}{8\pi },\ \text{where}  \label{minor1} \\
\hat{L}^{\ell }=\frac{\chi }{2}\mathrm{i}\left[ \psi ^{\ell \ast }\tilde{%
\partial}_{t}^{\ell }\psi ^{\ell }-\psi ^{\ell }\tilde{\partial}_{t}^{\ell
\ast }\psi ^{\ell \ast }\right] -\frac{\chi ^{2}}{2m}\left\{ \tilde{\nabla}%
\psi ^{\ell }\cdot \tilde{\nabla}^{\ast }\psi ^{\ell \ast }+G^{\ell }\left(
\psi ^{\ell \ast }\psi ^{\ell }\right) \right\} ,  \notag \\
\tilde{\partial}_{t}^{\ell }=\partial _{t}+\frac{\mathrm{i}q^{\ell }\left(
\varphi +\varphi _{\mathrm{ex}}\right) }{\chi },\ \tilde{\nabla}=\nabla -%
\frac{\mathrm{i}q\mathbf{A}_{\mathrm{ex}}}{\chi c}.  \notag
\end{gather}%
The nonlinear self-interaction terms $G^{\ell }$ in (\ref{minor1}) are
determined through the charge equilibrium equation (\ref{nop7}), and they
can be the same for different $\ell $, or there may be several types of
charges, for instance, protons and electrons. The field equations for this
Lagrangian are%
\begin{gather}
\mathrm{i}\chi \partial _{t}\psi ^{\ell }=-\frac{\chi ^{2}\nabla ^{2}\psi
^{\ell }}{2m^{\ell }}-\frac{\chi q^{\ell }\mathbf{A}_{\mathrm{ex}}\cdot
\nabla \psi ^{\ell }}{m^{\ell }\mathrm{c}i}+  \label{NLSj0} \\
+q^{\ell }\left( \varphi +\varphi _{\mathrm{ex}}\right) \psi ^{\ell }+\frac{%
q^{\ell 2}\mathbf{A}_{\mathrm{ex}}^{2}\psi ^{\ell }}{2m^{\ell }\mathrm{c}^{2}%
}+\left[ G_{a}^{\ell }\right] ^{\prime }\left( \left\vert \psi ^{\ell
}\right\vert ^{2}\right) \psi ^{\ell },  \notag \\
-\nabla ^{2}\varphi =4\pi \sum_{\ell =1}^{N}q^{\ell }\left\vert \psi ^{\ell
}\right\vert ^{2},\ \ell =1,...,N.  \notag
\end{gather}%
and $\psi ^{\ell \ast }$ is complex conjugate of $\psi ^{\ell }.$ Obviously,
the equations with different $\ell $ are coupled only through the potential $%
\varphi $ which is responsible for the charge interaction. The Lagrangian $%
\mathcal{\hat{L}}$ is gauge invariant with respect to the first and the
second gauge transformations (\ref{lagco3}), (\ref{lagco4}), and
consequently every $\ell $-th charge has a conserved current 
\begin{equation}
J^{\ell \mu }=\left( \mathrm{c}\rho ^{\ell },\mathbf{J}^{\ell }\right) ,\
\rho ^{\ell }=q\left\vert \psi ^{\ell }\right\vert ^{2},\ \mathbf{J}^{\ell
}=\left( \frac{\chi q^{\ell }}{m^{\ell }}\func{Im}\frac{\nabla \psi ^{\ell }%
}{\psi ^{\ell }}-\frac{q^{\ell 2}\mathbf{A}_{\mathrm{ex}}}{m^{\ell }\mathrm{c%
}}\right) \left\vert \psi ^{\ell }\right\vert ^{2},  \label{jco1}
\end{equation}%
satisfying the conservation/continuity equations%
\begin{equation}
\partial _{t}\rho ^{\ell }+\nabla \cdot \mathbf{J}^{\ell }=0\text{ or }%
m^{\ell }\partial _{t}\left\vert \psi ^{\ell }\right\vert ^{2}+\nabla \cdot
\left( \chi \func{Im}\frac{\nabla \psi ^{\ell }}{\psi ^{\ell }}\left\vert
\psi ^{\ell }\right\vert ^{2}-\frac{q^{\ell }}{\mathrm{c}}\mathbf{A}_{%
\mathrm{ex}}\left\vert \psi ^{\ell }\right\vert ^{2}\right) =0.  \label{jco2}
\end{equation}%
The differential form (\ref{jco2}) of the charge conservation implies the
conservation of every total $\ell $-th charge, and we require every total $%
\ell $-th conserved charge to be exactly $q^{\ell }$, rather than just any
constant, implying the following \emph{charge normalization} conditions%
\begin{equation}
\int_{\mathbb{R}^{3}}\left\vert \psi ^{\ell }\right\vert ^{2}{}\mathrm{d}%
\mathbf{x}=1,\ t\geq 0,\ \ell =1,...,N.  \label{norm10}
\end{equation}%
We attribute to every $\ell $-th charge its potential $\varphi ^{\ell }$ by
the formula%
\begin{equation}
\varphi ^{\ell }\left( t,\mathbf{x}\right) =q^{\ell }\dint_{\mathbb{R}^{3}}%
\frac{\left\vert \psi ^{\ell }\right\vert ^{2}\left( t,\mathbf{y}\right) }{%
\left\vert \mathbf{y}-\mathbf{x}\right\vert }\,\mathrm{d}\mathbf{y}.\text{ }
\label{jco3}
\end{equation}%
Hence we can write the equation for $\varphi $ in (\ref{NLSj0}) in the form 
\begin{equation}
\nabla ^{2}\varphi ^{\ell }=-4\pi q^{\ell }\left\vert \psi ^{\ell
}\right\vert ^{2},\ \varphi =\sum_{\ell =1}^{N}\varphi ^{\ell }.
\label{delfi}
\end{equation}%
An analysis of the system of the charges energy-momentum and its partition
between individual charges shows another important property of the
Lagrangian: \emph{the charge gauge invariant momentum density }$\mathbf{P}$ $%
^{\ell }$\emph{\ equals exactly the microcurrent density }$\mathbf{J}^{\ell
} $\textbf{\ }\emph{multiplied by the constant }$m^{\ell }/q^{\ell }$,
namely: 
\begin{equation}
\mathbf{P}^{\ell }=\frac{m^{\ell }}{q^{\ell }}\mathbf{J}^{\ell }=\frac{%
\mathrm{i}\chi }{2}\left[ \psi ^{\ell }\tilde{\nabla}^{\ell \ast }\psi
^{\ell \ast }-\psi ^{\ell \ast }\tilde{\nabla}^{\ell }\psi ^{\ell }\right]
=\left( \chi \func{Im}\frac{\nabla \psi ^{\ell }}{\psi ^{\ell }}-\frac{%
q^{\ell }\mathbf{A}_{\mathrm{ex}}}{\mathrm{c}}\right) \left\vert \psi ^{\ell
}\right\vert ^{2},  \label{jco5}
\end{equation}%
\emph{that can be viewed as the momentum density kinematic representation}: 
\begin{equation}
\mathbf{P}^{\ell }\left( t,\mathbf{x}\right) =m\mathbf{v}^{\ell }\left( t,%
\mathbf{x}\right) ,\ \text{where }\mathbf{v}^{\ell }\left( t,\mathbf{x}%
\right) =\mathbf{J}^{\ell }\left( t,\mathbf{x}\right) /q^{\ell }\text{ is
the velocity density.}  \label{jco6}
\end{equation}

To quantify the conditions of the remote interaction we make use of the
explicit dependence on the size parameter $a$ \ of the nonlinearity $G^{\ell
}=G_{a}^{\ell }$ as in (\ref{totgkap}) and take the size parameter as a
spatial scale characterizing sizes of the fields $\psi _{a}^{\ell }$ and $%
\varphi _{a}^{\ell }$. The charges separation is measured roughly by a
minimal distance $R_{\min }$ between any two charges. Another relevant
spatial scale $R_{\mathrm{ex}}$ measures a typical scale of spatial
inhomogeneity of external fields near charges. Consequently, conditions of
remote interaction are measured by the dimensionless ratio $a/R$ where the
characteristic spatial scale $R=\min \left( R_{\min },R_{\mathrm{ex}}\right) 
$ with the \emph{condition }$a/R\ll 1$\emph{\ to define the regime of remote
interaction}.

Next we would like to take a look on how the point charge mechanics is
manifested in our wave mechanics governed by the Lagrangian $\mathcal{\hat{L}%
}$. There are two distinct ways to correspond our field theory to the
classical point charge mechanics in the case when all charges are well
separated satisfying the condition $a/R\ll 1$. The first way is via averaged
quantities in the spirit of the well known in quantum mechanics \emph{%
Ehrenfest\ Theorem, }\cite[Sections 7, 23]{Schiff}, and the second one via a
construction of approximate solutions to the field equations (\ref{NLSj0})
when every charge is represented as wave-corpuscle similar to one from (\ref%
{exac1}).

We construct the correspondence to the point charges mechanics via averaged
quantities by defining first the $\ell $-th charge location $\mathbf{r}%
_{a}^{\ell }\left( t\right) $ and its potential $\varphi _{a}^{\ell }$ by
the relations 
\begin{equation}
\mathbf{r}^{\ell }\left( t\right) =\mathbf{r}_{a}^{\ell }\left( t\right)
=\int_{\mathbb{R}^{3}}\mathbf{x}\left\vert \psi _{a}^{\ell }\left( t,\mathbf{%
x}\right) \right\vert ^{2}\ \mathrm{d}\mathbf{x},\ \varphi _{a}^{\ell
}\left( t,\mathbf{x}\right) =q^{\ell }\dint_{\mathbb{R}^{3}}\frac{\left\vert
\psi _{a}^{\ell }\right\vert ^{2}\left( t,\mathbf{y}\right) }{\left\vert 
\mathbf{y}-\mathbf{x}\right\vert }\,\mathrm{d}\mathbf{y}.  \label{xloc}
\end{equation}%
Then, for the start, we consider a simpler case when the external magnetic
field vanishes, i.e. $\mathbf{A}_{\mathrm{ex}}=0$, and introduce a potential 
$\varphi _{\mathrm{ex},a}^{\ell }$ describing the interaction of $\ell $-th
charge with all other charges: 
\begin{equation}
\varphi _{\mathrm{ex},a}^{\ell }=\varphi +\varphi _{\mathrm{ex}}-\varphi
_{a}^{\ell }=\varphi _{\mathrm{ex}}+\sum_{\ell ^{\prime }\neq \ell }\varphi
_{a}^{\ell ^{\prime }}.  \label{fiaa}
\end{equation}%
In this case, based on the momentum balance equation for every $\ell $-th
charge combined with the momentum density kinematic representation (\ref%
{delfi}), (\ref{jco5}) and the representation (\ref{fiaa}), we obtain the
following exact equations of motion 
\begin{equation}
m^{\ell }\frac{\mathrm{d}^{2}\mathbf{r}^{\ell }}{\mathrm{d}t^{2}}=-\int_{%
\mathbb{R}^{3}}q^{\ell }\left\vert \psi _{a}^{\ell }\right\vert ^{2}\nabla
\varphi _{\mathrm{ex},a}^{\ell }\ \mathrm{d}\mathbf{x},\ \ell =1,...,N,
\label{Couf}
\end{equation}%
where we recognize in the integrand the Lorentz force density exerted on the 
$\ell $-th charge by other charges and the external field. Suppose now that
fields $\psi _{a}^{\ell }\left( t,\mathbf{x}\right) $ are localized about
point $\mathbf{r}_{a}^{\ell }\left( t\right) $ with the localization radius
of the order $a$ and observe that the normalization (\ref{norm10}) combined
with the localization implies that 
\begin{equation}
\left\vert \psi _{a}^{\ell }\left( t,\mathbf{x}\right) \right\vert
^{2}\rightarrow \delta \left( \mathbf{x}-\mathbf{r}_{0}^{\ell }\left(
t\right) \right) ,\ a\rightarrow 0,\text{ implying }\mathbf{r}_{0}^{\ell
}\left( t\right) =\lim_{a\rightarrow 0}\mathbf{r}_{a}^{\ell }\left( t\right)
.  \label{psitodel0}
\end{equation}%
\ \ The relations (\ref{xloc}) and (\ref{psitodel0}) imply consequently 
\begin{gather}
\varphi _{a}^{\ell }\left( t,\mathbf{x}\right) \rightarrow \varphi
_{0}^{\ell }=\frac{q^{\ell }}{\left\vert \mathbf{x}-\mathbf{r}_{0}^{\ell
}\right\vert }\text{ as}\ a\rightarrow 0,\text{ and}  \label{fiexl0} \\
\varphi _{\mathrm{ex},0}^{\ell }\left( t,\mathbf{x}\right) =\varphi _{%
\mathrm{ex}}\left( t,\mathbf{x}\right) +\sum_{\ell ^{\prime }\neq \ell }%
\frac{q^{\ell ^{\prime }}}{\left\vert \mathbf{x}-\mathbf{r}_{0}^{\ell
^{\prime }}\left( t\right) \right\vert }.  \notag
\end{gather}%
Then combining relations (\ref{xloc}), (\ref{psitodel0}) we can pass in (\ref%
{Couf}) to the limit $a\rightarrow 0$, obtaining Newton's equations of
motion for the system of $N$ point charges 
\begin{equation}
m^{\ell }\frac{\mathrm{d}^{2}\mathbf{r}_{0}^{\ell }}{\mathrm{d}t^{2}}\left(
t\right) =-q^{\ell }\nabla \varphi _{\mathrm{ex},a}^{\ell }\left( \mathbf{r}%
_{0}^{\ell }\left( t\right) ,t\right) ,\ \ell =1,...,N,  \label{Newtel}
\end{equation}%
where in the right-hand side of equation (\ref{Newtel}) we see the Lorentz
force acting on the $\ell $-th point charge interacting with other charges
via the point charges Coulomb's potentials (\ref{fiexl0}).

In the case when the external magnetic field potential $\mathbf{A}_{\mathrm{%
ex}}\left( t,\mathbf{x}\right) $ does not vanish, based on the exact
equations of motion similar to (\ref{Couf}) we establish that in the limit
as $a\rightarrow 0$ the following point charges equation of motion hold: 
\begin{equation}
m^{\ell }\frac{\mathrm{d}^{2}\mathbf{r}_{0}^{\ell }}{\mathrm{d}t^{2}}=%
\mathsf{f}^{\ell },\ \mathsf{f}^{\ell }=q^{\ell }\mathbf{E}^{\ell }+\frac{%
q^{\ell }}{\mathrm{c}}\frac{\mathrm{d}\mathbf{r}_{0}^{\ell }}{\mathrm{d}t}%
\times \mathbf{B^{\ell },}\ \ell =1,...N\mathbf{,}  \label{Newt}
\end{equation}%
with the Lorentz force $\mathsf{f}^{\ell }$ involving electric and magnetic
fields defined by the classical formulas: 
\begin{equation}
\mathbf{E^{\ell }}=-\left[ \nabla \varphi _{\mathrm{ex},0}^{\ell }\left( 
\mathbf{r}_{0}^{\ell }\right) +\frac{1}{\mathrm{c}}\partial _{t}\mathbf{A}_{%
\mathrm{ex}}\left( \mathbf{r}_{0}^{\ell }\right) \right] ,\ \mathbf{B^{\ell }%
}=\nabla \times \mathbf{A}_{\mathrm{ex}}\left( t,\mathbf{r}_{0}^{\ell
}\right) .  \label{Flor}
\end{equation}%
Obviously, the above force $\mathsf{f}^{\ell }$ coincides with the Lorentz
force acting on a point charge\ as in (\ref{pchar1}). To summarize our first
way of correspondence, we observe that the exact equations of motion (\ref%
{Couf}) form a basis for relating the field \ and point mechanics under an
assumption that charge fields remain localized during time interval of
interest. \emph{Notice the equations of motion (\ref{Couf}) do not involve
the nonlinear interaction terms }$G^{\ell }$\emph{\ in an explicit way
justifying their characterization as "stealthy" in the regime of remote
interactions}. As to the assumption that the charge fields remain localized,
it has to be verified based on the field equations (\ref{NLSj0}) where the
nonlinear interaction terms $G^{\ell }$ provide for cohesive forces for
individual charges. The fact that they can do just that is demonstrated for
a single charge represented as wave-corpuscle (\ref{exac1})-(\ref{exac1a})
as it accelerates in an external EM field.

The second way of correspondence between the charges as fields and points is
based on established by us fact that all charge fields $\psi _{a}^{\ell }$
can be represented as wave-corpuscles (\ref{exac1})-(\ref{exac1a}) which,
though do not satisfy the field equations (\ref{NLSj0}) exactly, but they
satisfy them with small discrepancies in the regime of remote interaction
when $a/R\ll 1$. More detailed presentation of that idea is as follows.
Consider for simplicity a simpler case when $\mathbf{A}_{\mathrm{ex}}=0$ and
introduce the following wave-corpuscle representation similar to (\ref{exac1}%
)-(\ref{exac1a}): 
\begin{gather}
\psi _{a}^{\ell }\left( t,\mathbf{x}\right) =\mathrm{e}^{\mathrm{i}S^{\ell
}/\chi }\mathring{\psi}^{\ell }\left( \left\vert \mathbf{x}-\mathbf{r}%
_{0}^{\ell }\right\vert \right) ,\ \varphi _{a}^{\ell }\left( t,\mathbf{x}%
\right) =\mathring{\varphi}^{\ell }\left( \mathbf{x}-\mathbf{r}_{0}^{\ell
}\right) ,\text{ where}  \label{apsol} \\
S^{\ell }\left( t,\mathbf{x}\right) =\mathbf{p}_{0}^{\ell }\cdot \mathbf{x}%
-\dint_{0}^{t}\frac{\mathbf{p}_{0}^{\ell 2}}{2m^{\ell }}\,\mathrm{d}%
t^{\prime }-q^{\ell }\dint_{0}^{t}\varphi _{\mathrm{ex}}^{0}\left( t^{\prime
},\mathbf{r}_{0}^{\ell }\right) \,\mathrm{d}t^{\prime },\ \mathbf{p}%
_{0}^{\ell }=m^{\ell }\frac{\mathrm{d}\mathbf{r}_{0}^{\ell }}{\mathrm{d}t},
\label{plslfi}
\end{gather}%
and the position functions $\mathbf{r}_{0}^{\ell }\left( t\right) $ satisfy
the Newton's equations of motion (\ref{Newt}). It turns out that
wave-corpuscles $\left\{ \psi _{a}^{\ell },\varphi _{a}^{\ell }\right\} $
defined by (\ref{apsol}), (\ref{plslfi}) and the complementary point charge
Newton's equations of motion (\ref{Newt}) \ solve the field equations (\ref%
{NLSj0}) with a small discrepancy which approaches zero as $a/R$ approaches
zero. The point charge mechanics features are transparently integrated into
the fields $\left\{ \psi _{a}^{\ell },\varphi _{a}^{\ell }\right\} $ in (\ref%
{apsol}), (\ref{plslfi}) via the de Broglie factor phases $S^{\ell }$ and
spatial shifts $\mathbf{r}_{0}^{\ell }$. Comparing with the motion of a
single charge in an external field we observe that now the acceleration of
the wave-corpuscle center $\mathbf{r}_{0}^{\ell }\left( t\right) $ is
determined not only by the Lorentz force due to the external field but also
by electric interactions with the remaining charges $\mathring{\psi}^{\ell
^{\prime }}$, $\ell ^{\prime }\neq \ell $ according to the Coulomb's law (%
\ref{fiexl0}), (\ref{Newt}).

We would like to point out that when analyzing the system of charges in the
regime of remote interaction we do not use any specific form of the
nonlinearities. Note also that solutions to the field equations (\ref{NLSj0}%
) depend on the size parameter $a$, which is proportional to the radius of
the wave-corpuscle, through the nonlinearity $G_{a}^{\ell }$, but the
integral equations (\ref{Couf}) do not involve explicit dependence on $a$.
Equation (\ref{apsol}) which describes the structure of the wave-corpuscle
involves $a$ only through radial shape factors $\mathring{\psi}^{\ell }=%
\mathring{\psi}_{a}^{\ell }\ $and through the electric potential $\mathring{%
\varphi}^{\ell }=\mathring{\varphi}_{a}^{\ell }$. The dependence of $%
\mathring{\psi}_{a}^{\ell }$ on $a$ is explicitly singular at zero, as that
is expected since in the singular limit $a\rightarrow 0$ the wave-corpuscle
should turn into the point charge with the square of amplitude described by
a delta function as in (\ref{psys2}). Nevertheless, for \emph{arbitrary small%
} $a>0$ the wave-corpuscle structure of every charge is preserved including
its principal wave-vector. The dependence of $\mathring{\varphi}_{a}^{\ell }$
on small $a$ can be described as a regular convergence to the classical
singular Coulomb's potential, see (\ref{dca}) for details. That allows to
apply representation (\ref{apsol}) to arbitrary small charges with radius
proportional to $a$ without compromising the accuracy of the description
and, in fact, increasing the accuracy as $a\rightarrow 0$.

\subsection{Comparative summary with the Schr\"{o}dinger wave mechanics\label%
{scompsum}}

The nonrelativistic version of our wave mechanics has many features in
common with the Schr\"{o}dinger wave mechanics. In particular, the charges
wave functions are complex valued, they satisfy equations resembling the Schr%
\"{o}dinger equation, the charge normalization condition is the same as in
the Schr\"{o}dinger wave mechanics. Our theory provides for a hydrogen atom
model which has a lot in common with that of the Schr\"{o}dinger one, but
its detailed study is outside of the scope of this article. There are
features of our wave theory though that distinguish it significantly from
the Schr\"{o}dinger wave mechanics, and they are listed below.

\begin{itemize}
\item Charges are always coupled with and inseparable from the EM field.

\item Every charge has a nonlinear self-interaction term in its Lagrangian
providing for a cohesive force holding it together as it moves freely or
accelerates.

\item A single charge either free or in external EM field is described by a
soliton-like wave function parametrized by the position and the momentum
related to the corresponding point mechanics. It propagates in the space
without dispersion even when it accelerates, and this addresses one of the
above mentioned "grave difficulties" with the Schr\"{o}dinger's
interpretation of the wave function expressed by M. Born.

\item When dressed charges are separated by distances considerably larger
than their sizes their wave functions and the corresponding EM fields
maintain soliton-like representation.

\item The correspondence between the wave mechanics and a point mechanics
comes through the closed form soliton-like representation of wave functions
in which point mechanics positions and momenta enter as parameters. In
particular, the wave function representation includes the de Broglie wave
vector as an exact parameter, it equals (up to the Planck constant) the
point mechanics momentum. In addition to that, the corresponding group
velocity matches exactly the velocity of a soliton-like solution and the
point mechanics velocity.

\item In the case of many interacting charges every charge is described by
its own wave function over the same three dimensional space in contrast to
the Schr\"{o}dinger wave mechanics for many charges requiring
multidimensional configuration space.

\item Our theory has a relativistic version based on a local, gauge and
Lorentz invariant Lagrangian with most of the above listed features.
\end{itemize}

\section{Single free relativistic charge\label{freepar}}

A single free charge is described by a complex scalar field $\psi =\psi
\left( t,\mathbf{x}\right) $ and it is coupled to the EM field described by
its 4-potential $A^{\mu }=\left( \varphi ,\mathbf{A}\right) $. To emphasize
the fact that our charge is always coupled with the EM field we name the
pair $\left\{ \psi ,A^{\mu }\right\} $ \emph{dressed charge}. So whenever we
use the term dressed charge we mean the charge and the EM field as an
inseparable entity. The dressed charge is called free if there are no
external forces acting upon it. The free charge Lagrangian is defined by the
following formula%
\begin{equation}
L_{0}\left( \psi ,A^{\mu }\right) =\frac{\chi ^{2}}{2m}\left\{ \psi _{;\mu
}^{\ast }\psi ^{;\mu }-\kappa _{0}^{2}\psi ^{\ast }\psi -G\left( \psi ^{\ast
}\psi \right) \right\} -\frac{F^{\mu \nu }F_{\mu \nu }}{16\pi },
\label{fpar1}
\end{equation}%
where 
\begin{gather}
\kappa _{0}=\frac{\omega _{0}}{\mathrm{c}}=\frac{m\mathrm{c}}{\chi },\
\omega _{0}=\frac{m\mathrm{c}^{2}}{\chi },\ F^{\mu \nu }=\partial ^{\mu
}A^{\nu }-\partial ^{\nu }A^{\mu }  \label{fpar1a} \\
\psi ^{;\mu }=\tilde{\partial}^{\mu }\psi ,\ \psi ^{;\mu \ast }=\tilde{%
\partial}^{\ast \mu }\psi ^{\ast },\ \tilde{\partial}^{\mu }=\partial ^{\mu
}+\frac{\mathrm{i}q}{\chi \mathrm{c}}A^{\mu },\ \tilde{\partial}^{\ast \mu
}=\partial ^{\mu }-\frac{\mathrm{i}q}{\chi \mathrm{c}}A^{\mu }.
\label{fpar1b}
\end{gather}%
In the above relations $m>0$ is the mass of the charge, $q$ is the charge
value, and $\chi >0$ is a parameter similar to the Planck constant $\hbar $
the value of which will be set later to satisfy the Einstein relation $%
\mathcal{E}=\hbar \omega _{0}$. The term $G\left( \psi ^{\ast }\psi \right) $
corresponds to the nonlinear self-interaction and is to be determined later, 
$\psi ^{\ast }\ $is complex conjugate to $\psi $. The above Lagrangian
involves the so-called \emph{covariant differentiation operators }$\tilde{%
\partial}^{\mu }$ and $\tilde{\partial}^{\ast \mu }$ with abbreviated
notations $\psi ^{;\mu }$ and $\psi ^{;\mu \ast }$ for the corresponding 
\emph{covariant derivatives}. In what follows we use also the following
abbreviations 
\begin{equation}
\partial ^{\mu }\psi =\psi ^{,\mu },\ \partial ^{\mu }\psi ^{\ast }=\psi
^{,\mu \ast }.  \label{fpar3a}
\end{equation}%
We remind the reader also that%
\begin{gather}
\partial _{\mu }=\frac{\partial }{\partial x^{\mu }}=\left( \frac{1}{\mathrm{%
c}}\partial _{t},\nabla \right) ,\ \partial ^{\mu }=\frac{\partial }{%
\partial x_{\mu }}=\left( \frac{1}{\mathrm{c}}\partial _{t},-\nabla \right) ,
\label{fpar3c} \\
A^{\mu }=\left( \varphi ,\mathbf{A}\right) ,\ A_{\mu }=\left( \varphi ,-%
\mathbf{A}\right) ,\ \mathbf{E}=-\nabla \varphi -\frac{1}{\mathrm{c}}%
\partial _{t}\mathbf{A},\ \mathbf{B}=\nabla \times \mathbf{A}.  \notag
\end{gather}

Evidently the Lagrangian $L_{0}$ defined by the formulas (\ref{fpar1})-(\ref%
{fpar1b}) is obtained from the Klein-Gordon Lagrangian, \cite[Section 7.1,
11.2]{Griffiths}, \cite[III.3]{Barut}, by adding to it the nonlinear term $%
G\left( \psi ^{\ast }\psi \right) $. The Lagrangian expression indicates
that the charge is coupled to the EM field through the covariant
derivatives, and such a coupling is well known and called \emph{minimal}.
The Klein-Gordon Lagrangian is a commonly used model for a \emph{%
relativistic spinless charge}, and the introduced nonlinearity $G\left( \psi
^{\ast }\psi \right) $ can provide for a binding self-force. Nonlinear
alterations of the Klein-Gordon Lagrangian were considered in the
literature, see, for instance, \cite[Section 11.7, 11.8]{Griffiths} and \cite%
{Benci 1}, for rigorous mathematical studies. Our way to choose of the
nonlinearity $G\left( \psi ^{\ast }\psi \right) $ differs from those.

Observe that the Lagrangian $L_{0}$ defined by (\ref{fpar1})-(\ref{fpar1b})
is manifestly Lorentz and gauge invariant, and it is a special case of a
general one charge Lagrangian studied in Section \ref{1part}. This allows us
to apply to the Lagrangian $L_{0}$ formulas from there to get the field
equations, the 4-microcurrent\emph{\ }and the energy-momenta tensors.
Consequently, the Euler-Lagrange field equations (\ref{frlag4}) take here
the form%
\begin{gather}
\partial _{\mu }F^{\mu \nu }=\frac{4\pi }{\mathrm{c}}J^{\nu },  \label{fpar4}
\\
\left[ \tilde{\partial}_{\mu }\tilde{\partial}^{\mu }+\kappa
_{0}^{2}+G^{\prime }\left( \psi ^{\ast }\psi \right) \right] \psi =0.
\label{fpar5}
\end{gather}%
The formula (\ref{frlag6}) for the 4-microcurrent density $J^{\mu }$ turns
into%
\begin{equation}
J^{\nu }=-\frac{\chi q}{2m}\mathrm{i}\left( \psi \tilde{\partial}^{\nu \ast
}\psi ^{\ast }-\psi ^{\ast }\tilde{\partial}^{\nu }\psi \right) =-\left( 
\frac{\chi q}{m}\func{Im}\frac{\partial ^{\nu }\psi }{\psi }+\frac{q^{2}}{m%
\mathrm{c}}A^{\nu }\right) \left\vert \psi \right\vert ^{2},  \label{fpar6}
\end{equation}%
or, in the time-space variables,%
\begin{gather}
\rho =-\frac{\chi q}{2m\mathrm{c}}\mathrm{i}\left( \psi \tilde{\partial}%
_{t}^{\ast }\psi ^{\ast }-\psi ^{\ast }\tilde{\partial}_{t}\psi \right)
=-\left( \frac{\chi q}{m\mathrm{c}^{2}}\func{Im}\frac{\partial _{t}\psi }{%
\psi }+\frac{q^{2}\varphi }{m\mathrm{c}^{2}}\right) \left\vert \psi
\right\vert ^{2},  \label{fpar7} \\
\mathbf{J}=\frac{\chi q}{2m}\mathrm{i}\left( \psi \tilde{\nabla}^{\ast }\psi
^{\ast }-\psi ^{\ast }\tilde{\nabla}\psi \right) =\left( \frac{\chi q}{m}%
\func{Im}\frac{\nabla \psi }{\psi }-\frac{q^{2}\mathbf{A}}{m\mathrm{c}}%
\right) \left\vert \psi \right\vert ^{2}.  \notag
\end{gather}%
The above formulas for the 4-microcurrent density $J^{\mu }=\left( \rho 
\mathrm{c},\mathbf{J}\right) $ are well known in the literature, see for
instance, \cite[(11.3)]{Wentzel}, \cite[Section 3.3, (3.3.27), (3.3.34),
(3.3.35)]{Morse Feshbach I}. It satisfies the conservation/continuity
equations%
\begin{equation}
\partial _{\nu }J^{\nu }=0,\ \partial _{t}\rho +\nabla \cdot \mathbf{J}=0,\
J^{\nu }=\left( \rho \mathrm{c},\mathbf{J}\right) .  \label{fpar10}
\end{equation}%
Consequently, the total charge $\int \rho \left( \mathbf{x}\right) \,\mathrm{%
d}\mathbf{x}$ of the elementary charge remains constant in the course of
evolution, and we impose a \emph{charge normalization} condition which
extends (\ref{psfi7}), namely%
\begin{equation}
\int\nolimits_{\mathbb{R}^{3}}\frac{\rho \left( t,\mathbf{x}\right) }{q}\,%
\mathrm{d}\mathbf{x}=-\int\nolimits_{\mathbb{R}^{3}}\left( \frac{\chi }{m%
\mathrm{c}^{2}}\func{Im}\frac{\partial _{t}\psi }{\psi }+\frac{q\varphi }{m%
\mathrm{c}^{2}}\right) \left\vert \psi \right\vert ^{2}\,\mathrm{d}\mathbf{x}%
=1.  \label{fpar11}
\end{equation}%
We would like to stress that the equation (\ref{fpar11}) is perfectly
consistent with the field equations and the conservation laws (\ref{fpar10}%
), and \emph{it constitutes an independent and physically significant
constraint for the total charge to be exactly }$q$\emph{\ as in the
Coulomb's potential, rather than an arbitrary constant}.

Applying the general formulas (\ref{frlag9})-(\ref{frlag10}) to the
Lagrangian $L_{0}$ defined by (\ref{fpar1})-(\ref{fpar1b}) we obtain the
following representations for the symmetric and gauge invariant
energy-momenta tensors $T^{\mu \nu }$ and $\Theta ^{\mu \nu }$ for,
respectively, the charge and the EM field: 
\begin{gather}
T^{\mu \nu }=\frac{\chi ^{2}}{2m}\left\{ \left[ \psi ^{;\mu \ast }\psi
^{;\nu }+\psi ^{;\mu }\psi ^{;\nu \ast }\right] -\left[ \psi _{;\mu }^{\ast
}\psi ^{;\mu }-\kappa _{0}^{2}\psi ^{\ast }\psi -G\left( \psi ^{\ast }\psi
\right) \right] g^{\mu \nu }\right\} ,  \label{thm1} \\
\Theta ^{\mu \nu }=\frac{1}{4\pi }\left( g^{\mu \gamma }F_{\gamma \xi
}F^{\xi \nu }+\frac{1}{4}g^{\mu \nu }F_{\gamma \xi }F^{\gamma \xi }\right) .
\label{thm2}
\end{gather}%
Energy conservation equations which we derive later in (\ref{divten1})-(\ref%
{divten2}) turn here into%
\begin{equation}
\partial _{\mu }T^{\mu \nu }=f^{\nu },\text{ }\partial _{\mu }\Theta ^{\mu
\nu }=-f^{\nu },  \label{thm3}
\end{equation}%
where%
\begin{equation}
f^{\nu }=\frac{1}{\mathrm{c}}J_{\mu }F^{\nu \mu }=\left( \frac{1}{\mathrm{c}}%
\mathbf{J}\cdot \mathbf{E},\rho \mathbf{E}+\frac{1}{\mathrm{c}}\mathbf{J}%
\times \mathbf{B}\right) \text{ }  \label{thm4}
\end{equation}%
is the Lorentz force density.

\subsection{Charge at rest}

We say the dressed charge\ to be at rest at the origin $\mathbf{x}=\mathbf{0}
$ if it is a radial solution to the field equations (\ref{fpar4})-(\ref%
{fpar5}) of the following special form%
\begin{equation}
\psi \left( t,\mathbf{x}\right) =\mathrm{e}^{-\mathrm{i}\omega _{0}t}%
\mathring{\psi}\left( \left\vert \mathbf{x}\right\vert \right) ,\ \varphi
\left( t,\mathbf{x}\right) =\mathring{\varphi}\left( \left\vert \mathbf{x}%
\right\vert \right) ,\ \mathbf{A}\left( t,\mathbf{x}\right) =\mathbf{0},\
\omega _{0}=\frac{m\mathrm{c}^{2}}{\chi },  \label{psif1}
\end{equation}%
and we refer to such a solution as $\omega _{0}$\emph{-static}. Observe that
as it follows from (\ref{fpar6}), (\ref{fpar7}), the micro-density $\rho $
and micro-current $\mathbf{J}$ for the $4$-microcurrent $J^{\nu }=\left(
\rho \mathrm{c},\mathbf{J}\right) $ for the $\omega _{0}$-static solution (%
\ref{psif1}) are 
\begin{equation}
\rho =q\left( 1-\frac{q\mathring{\varphi}}{m\mathrm{c}^{2}}\right) \mathring{%
\psi}^{2},\ \mathbf{J}=\mathbf{0}.  \label{psif6}
\end{equation}%
The charge normalization condition (\ref{fpar11}) then turns into (\ref%
{psfi7}).

For the charge at rest as described by relations (\ref{psif1}) the field
equations (\ref{fpar4})-(\ref{fpar5}) turn into the following system of two
equations for the real-valued functions $\mathring{\psi}$ and $\mathring{%
\varphi}$ which we call \emph{rest charge equations}:%
\begin{gather}
-\Delta \mathring{\varphi}=4\pi \mathring{\rho},\ \mathring{\rho}=q\left( 1-%
\frac{q\mathring{\varphi}_{a}}{m\mathrm{c}^{2}}\right) \mathring{\psi}^{2},
\label{psif2} \\
-\Delta \mathring{\psi}+\frac{mq\mathring{\varphi}}{\chi ^{2}}\left( 2-\frac{%
q\mathring{\varphi}}{m\mathrm{c}^{2}}\right) \mathring{\psi}+G^{\prime
}\left( |\mathring{\psi}|^{2}\right) \mathring{\psi}=0.  \label{psif3}
\end{gather}

The radial functions $\mathring{\psi}$ and $\mathring{\varphi}$ play
instrumental role in our constructions, and we name them respectively \emph{%
charge form factor} and \emph{form factor potential}. As it follows from the
equation (\ref{psif2}), the charge form factor $\mathring{\psi}=\mathring{%
\varphi}_{\mathring{\psi}}$ determines the form factor potential $\mathring{%
\varphi}$ by the formula (\ref{psfi1c}). \ Consequently, plugging in the
above expression into the equation (\ref{psfi1b}) we get the nonlinear
equation (\ref{psfi1d}) as follows: 
\begin{equation}
-\Delta \mathring{\psi}+\frac{m\mathring{\varphi}_{\mathring{\psi}}}{\chi
^{2}}q\left( 2-\frac{q\mathring{\varphi}_{\mathring{\psi}}}{m\mathrm{c}^{2}}%
\right) \mathring{\psi}+G^{\prime }\left( |\mathring{\psi}|^{2}\right) 
\mathring{\psi}=0.  \label{psif5}
\end{equation}%
As it is shown in the next section, the equation (\ref{psif5}) signifies a
complete balance (equilibrium) of the three forces acting upon the resting
charge: (i) internal elastic deformation force associated with the term $%
-\Delta \mathring{\psi}$; (ii) charge's electromagnetic self-interaction
force associated with the term $\frac{m\mathring{\varphi}_{\mathring{\psi}}}{%
\chi ^{2}}\left( 2q-\frac{q^{2}\mathring{\varphi}_{\mathring{\psi}}}{m%
\mathrm{c}^{2}}\right) \mathring{\psi}$; (iii) internal nonlinear
self-interaction of the charge associated with the term $G^{\prime }\left( |%
\mathring{\psi}|^{2}\right) \mathring{\psi}$. We refer to the equation (\ref%
{psif5}) as \emph{charge equilibrium equation} or just the \emph{equilibrium
equation}. Importantly, the charge equilibrium equation (\ref{psif5})
establishes an explicit relation between the form factor $\mathring{\psi}$
and the self-interaction nonlinearity $G$.

\emph{Now we come to a key point of our construction: determination of the
nonlinearity }$G$\emph{\ from the equilibrium equation (\ref{psif5}). First
we pick and fix a form factor }$\mathring{\psi}\left( r\right) $\emph{, }$%
r\geq 0$\emph{, which is assumed to be a nonnegative, monotonically decaying
and sufficiently smooth function. Then we determine consequently }$G^{\prime
}$\emph{\ and }$G$\emph{\ from the equilibrium equation (\ref{psif5}). This
gives us at once the desired state of the resting charge }$\{\mathring{\psi}%
\emph{,}\mathring{\varphi}\}$\emph{\ without solving any nontrivial
nonlinear partial differential equation, which is a stumbling block in most
theories involving nonlinearities. Of course such a benefit of our approach
comes at a cost of dealing with a nontrivial nonlinearity }$G$\emph{\ at all
further steps, but it turns out that the definition of the nonlinearity via
the equilibrium equation (\ref{psif5}) is constructive enough for
representing many important physical quantities in terms of }$\mathring{\psi}
$\emph{, }$\mathring{\varphi}$\emph{\ and }$G$\emph{\ without explicit
formulas for them. Curiously, for certain choices of }$\mathring{\psi}$\emph{%
\ one can find explicit formulas for }$\mathring{\varphi}$\emph{, }$G$\emph{%
\ and other important physical quantities, as we show in Section \ref%
{Examnon}.}

\subsection{Energy-momentum tensor, forces and equilibrium\label{senmomforce}%
}

In any classical field theory over the four-dimensional continuum of
space-time the energy-momentum tensor is of a fundamental importance. It
provides the density of the energy, the momentum and the surface forces as
well as the conservation laws that govern the energy and momentum transport.
It is worth to point out that it is the differential form of the
energy-momentum conservation which involve the densities of energy, momentum
and forces rather than the original field equations\ are more directly
related to corpuscular properties of the fields. In particular, for the
charge model we study here the Lorentz force density arises in the
differential form of the energy-momentum conservation equations and not in
the original field equations. For detailed considerations of the structure
and properties of the energy-momentum tensor including its symmetry, gauge
invariance and conservation laws we refer the reader to Section \ref%
{sclasfield}. Here, using the results of that section, we compute and
analyze the energy-momentum tensor for the Lagrangian $L_{0}$ defined by the
formulas (\ref{fpar1})-(\ref{fpar1b}) and for the $\omega _{0}$-static state
defined by (\ref{psif1}).

Using the interpretation form (\ref{emten2})-(\ref{emten3}) of the
energy-momentum $T^{\mu \nu }$ and formulas (\ref{thm1})-(\ref{thm2}) we
find that the energy-momentum tensor takes the following form 
\begin{equation}
T^{\mu \nu }=\left[ 
\begin{array}{cccc}
u & \mathrm{c}p_{1} & \mathrm{c}p_{2} & \mathrm{c}p_{3} \\ 
\mathrm{c}^{-1}s_{1} & -\sigma _{11} & -\sigma _{12} & -\sigma _{13} \\ 
\mathrm{c}^{-1}s_{2} & -\sigma _{21} & -\sigma _{22} & -\sigma _{23} \\ 
\mathrm{c}^{-1}s_{3} & -\sigma _{31} & -\sigma ^{32} & -\sigma _{33}%
\end{array}%
\right] ,  \label{psif7}
\end{equation}%
where the energy density $u$, the momentum and the energy flux components $%
p_{j}$ and $s_{j}$ are as follows:%
\begin{equation}
u=\frac{\chi ^{2}}{2m}\left[ \left( \nabla \mathring{\psi}\right)
^{2}+G\left( \mathring{\psi}^{2}\right) \right] +\left( m\mathrm{c}^{2}-q%
\mathring{\varphi}+\frac{q\mathring{\varphi}^{2}}{2m\mathrm{c}^{2}}\right) 
\mathring{\psi}^{2},  \label{psif8}
\end{equation}%
\begin{equation}
p^{j}=0,\ s^{j}=0,\ j=1,2,3,  \label{psif9}
\end{equation}%
and the stress tensor components $\sigma _{ij}$ are represented by the
formulas%
\begin{gather}
\sigma _{ij}=-\frac{\chi ^{2}}{m}\left[ \partial _{i}\mathring{\psi}\partial
_{j}\mathring{\psi}-\frac{1}{2}\left( \nabla \mathring{\psi}\right)
^{2}\delta _{ij}\right] +  \label{psif10} \\
\left[ q\left( \mathring{\varphi}-\frac{q}{2m\mathrm{c}^{2}}\mathring{\varphi%
}^{2}\right) \mathring{\psi}^{2}+\frac{\chi ^{2}}{2m}G\left( \mathring{\psi}%
^{2}\right) \right] \delta _{ij},\ i,j=1,2,3.  \notag
\end{gather}%
Notice that the vanishing of the momentum $\mathbf{p}$ and the energy flux $%
\mathbf{s}$ in (\ref{psif9}) is yet another justification for the name $%
\omega _{0}$-static solution. Observe also that for the $\omega _{0}$-static
state defined by (\ref{psif1}) the EM field is%
\begin{equation}
\mathbf{E}=-\nabla \mathring{\varphi},\ \mathbf{B}=\mathbf{0}.
\label{psif11}
\end{equation}%
Using the representation (\ref{maxw12})-(\ref{maxw13}) for the EM
energy-momentum $\Theta ^{\mu \nu }$ combined with the formulas (\ref{psif11}%
) for the EM field we obtain the following representation of $\Theta ^{\mu
\nu }$ for the $\omega _{0}$-static solution (\ref{psif1}): 
\begin{equation}
\Theta ^{\mu \nu }=%
\begin{bmatrix}
\left( \nabla \mathring{\varphi}\right) ^{2}/\left( 8\pi \right) & 0 \\ 
0 & -\tau _{ij}%
\end{bmatrix}%
,\text{ }  \label{psif12}
\end{equation}%
where%
\begin{equation}
-\tau _{ij}=\Theta ^{ij}=-\frac{1}{4\pi }\left[ \partial _{i}\mathring{%
\varphi}\partial _{j}\mathring{\varphi}-\frac{\left( \nabla \mathring{\varphi%
}\right) ^{2}}{2}\delta _{ij}\right] ,\ i,j=1,2,3.  \label{psif13}
\end{equation}%
Combining the conservation law (\ref{thm3}) with the general representation (%
\ref{psif7}) of the charge energy-momentum tensor $T^{\mu \nu }$ we obtain 
\begin{equation}
\partial _{t}p_{i}=\dsum_{j=1,2,3}\partial _{j}\sigma _{ji}+\left[ \rho
E_{i}+\frac{1}{\mathrm{c}}\left( \mathbf{J}\times \mathbf{B}\right) _{i}%
\right] =0,\ i=1,2,3.  \label{psif15a}
\end{equation}%
Notice that for the $\omega _{0}$-static solution (\ref{psif1}), in view of (%
\ref{psif11}), (\ref{psif12}), (\ref{psif13}), the equation (\ref{psif15a})
turns into the equilibrium equations%
\begin{equation}
\dsum_{j=1,2,3}\partial _{j}\sigma _{ji}-\rho \partial _{i}\mathring{\varphi}%
=0,\ i=1,2,3.  \label{psif15b}
\end{equation}

Observe now that the stress tensor (str.t.) $\sigma _{ij}$ defined in (\ref%
{psif10}) can be naturally decomposed into the three components which we
name as follows%
\begin{equation}
\sigma _{ij}=\sigma _{ij}^{\mathrm{el}}+\sigma _{ij}^{\mathrm{em}}+\sigma
_{ij}^{\mathrm{nl}},\ i,j=1,2,3,  \label{strc1}
\end{equation}%
where%
\begin{equation}
\sigma _{ij}^{\mathrm{el}}=-\frac{\chi ^{2}}{m}\left[ \partial _{i}\mathring{%
\psi}\partial _{j}\mathring{\psi}-\frac{1}{2}\left( \nabla \mathring{\psi}%
\right) ^{2}\delta _{ij}\right]  \label{strc2}
\end{equation}
is the elastic deformation stress tensor,%
\begin{equation}
\sigma _{ij}^{\mathrm{em}}=-p^{\mathrm{em}}\delta _{ij},\ p^{\mathrm{em}%
}=-q\left( \mathring{\varphi}-\frac{q\mathring{\varphi}^{2}}{2m\mathrm{c}^{2}%
}\right) \mathring{\psi}^{2}\text{ }  \label{strc3}
\end{equation}%
is the EM interaction stress tensor,%
\begin{equation}
\sigma _{ij}^{\mathrm{nl}}=-p^{\mathrm{nl}}\delta _{ij},\ p^{\mathrm{nl}}=-%
\frac{\chi ^{2}G\left( \mathring{\psi}^{2}\right) }{2m}  \label{strc4}
\end{equation}%
is the nonlinear self-interaction stress tensor, and consequently the
respective volume force densities are%
\begin{gather}
\dsum_{j=1,2,3}\partial _{j}\sigma _{ij}^{\mathrm{el}}=f_{i}^{\mathrm{el}}=-%
\frac{\chi ^{2}}{m}\Delta \mathring{\psi}\partial _{i}\mathring{\psi},\
i=1,2,3  \label{strc5} \\
\dsum_{j=1,2,3}\partial _{j}\sigma _{ij}^{\mathrm{em}}=f_{i}^{\mathrm{em}%
}+\rho \partial _{i}\mathring{\varphi},  \label{strc6}
\end{gather}%
\begin{gather}
\ f_{i}^{\mathrm{em}}=q\left( 2\mathring{\varphi}-\frac{q}{m\mathrm{c}^{2}}%
\mathring{\varphi}^{2}\right) \mathring{\psi}\partial _{i}\mathring{\psi}, \\
\dsum_{j=1,2,3}\partial _{j}\sigma _{ij}^{\mathrm{nl}}=f_{i}^{\mathrm{nl}}=%
\frac{\chi ^{2}}{m}G^{\prime }\left( \mathring{\psi}^{2}\right) \mathring{%
\psi}\partial _{i}\mathring{\psi}.  \label{strc7}
\end{gather}%
Notice that the volume force density for the electromagnetic interaction
stress in (\ref{strc6}) has two parts: $f_{i}^{\mathrm{em}}$, which we call 
\emph{internal electromagnetic force}, and $\rho \partial _{i}\mathring{%
\varphi}$ which is the negative of the Lorentz force. Observe that the
stress tensor $\sigma _{ij}^{\mathrm{el}}$ has a structure similar to the
one for compressional waves, see Section \ref{scompress} and (\ref{compr6}),
whereas the both stress tensors $\sigma _{ij}^{\mathrm{em}}$ and $\sigma
_{ij}^{\mathrm{nl}}$ have the structure typical for perfect fluids, \cite[%
Section 6.6]{Moller}, with respective hydrostatic pressures $p^{\mathrm{em}}$
and $p^{\mathrm{nl}}$ defined by the relations (\ref{strc3})-(\ref{strc4}). 
\emph{Notice that the formula (\ref{strc4}) provides an interpretation of
the nonlinearity }$G\left( \psi ^{2}\right) $\emph{: }$p^{\mathrm{nl}}=-\chi
^{2}G\left( \psi ^{2}\right) /(2m)$\emph{\ is the hydrostatic pressure when
the charge as at rest.}

Based on the equalities (\ref{strc5})-(\ref{strc7}) we can recast the
equilibrium equation (\ref{psif15b}) as%
\begin{equation}
f_{i}^{\mathrm{el}}+f_{i}^{\mathrm{em}}+f_{i}^{\mathrm{nl}}=0,\ i=1,2,3,%
\text{ }  \label{strc8}
\end{equation}%
or%
\begin{equation*}
\left[ -\frac{\chi ^{2}}{m}\Delta \mathring{\psi}+q\left( 2-\frac{q}{m%
\mathrm{c}^{2}}\mathring{\varphi}\right) \mathring{\varphi}\mathring{\psi}+%
\frac{\chi ^{2}}{m}G^{\prime }\left( \mathring{\psi}^{2}\right) \mathring{%
\psi}\right] \partial _{i}\mathring{\psi}=0.
\end{equation*}%
\emph{The equation (\ref{strc8}) signifies the ultimate equilibrium for the $%
\omega _{0}$-static charge}. It is evident from equation (\ref{strc8}) that
the scalar expression in the brackets before $\nabla \mathring{\psi}$ up to
the factor $\frac{m}{\chi ^{2}}$ is exactly the left-hand side of the
equilibrium equation (\ref{psif5}). In fact if $\nabla \mathring{\psi}\neq 0$
then the equilibrium equation (\ref{strc8}) is equivalent to the scalar
equilibrium equation (\ref{psif5}).

Notice that since the $\mathring{\psi}\left( \left\vert \mathbf{x}%
\right\vert \right) $ and $\mathring{\varphi}\left( \left\vert \mathbf{x}%
\right\vert \right) $ are radial and monotonically decaying functions of $%
\left\vert \mathbf{x}\right\vert $ we readily have 
\begin{equation}
\nabla \mathring{\psi}\left( \left\vert \mathbf{x}\right\vert \right)
=-\left\vert \nabla \mathring{\psi}\right\vert \mathbf{\hat{x}},\ \nabla 
\mathring{\varphi}\left( \left\vert \mathbf{x}\right\vert \right)
=-\left\vert \nabla \mathring{\varphi}\right\vert \mathbf{\hat{x}},\ \mathbf{%
\hat{x}}=\frac{\mathbf{x}}{\left\vert \mathbf{x}\right\vert }=\left( \hat{x}%
_{1},\hat{x}_{2},\hat{x}_{3}\right) .  \label{strc9}
\end{equation}%
The relations (\ref{strc9}) combined with (\ref{strc5})-(\ref{strc7}) imply
that for the resting charge all the forces $\mathbf{f}^{\mathrm{el}}$, $%
\mathbf{f}^{\mathrm{em}}$ and $\mathbf{f}^{\mathrm{nl}}$ and radial, i.e.
they are functions of $\left\vert \mathbf{x}\right\vert $ and point toward
or outward the origin.

Notice that it follows from the relations (\ref{psif9}) and (\ref{psif12})
that \emph{the total momentum }$P$\emph{\ and the energy flux }$S$\emph{\ of
the resting dressed charge, i.e. the charge and the EM field together, vanish%
}, and hence we have $P^{\nu }=\left( u,\mathbf{P}\right) $ \ with 
\begin{equation}
\mathbf{P}=\mathbf{0},\ \mathbf{S}=\mathbf{0},\text{ }  \label{fshif6}
\end{equation}%
where the energy density $u$ is represented by the formula (\ref{psif8}). We
would like to point out that the vanishing for the resting charge of the
micro-current $\mathbf{J}$ in (\ref{psif6}) as well as the momentum $\mathbf{%
P}$ and the energy flux $\mathbf{S}$ in (\ref{fshif6}) justifies the name $%
\omega _{0}$-static solution.

Using the general formulas (\ref{ttbe4}) for the angular momentum density $%
M^{\mu \nu \gamma }$ and combining them with the relations (\ref{psif7})-(%
\ref{psif9}) for the energy-momentum tensor $T^{\mu \nu }$ we readily obtain
that the \emph{total angular momentum }$J^{\nu \gamma }$\emph{\ vanishes},
namely $\ M^{0\nu \gamma }=0$ implying 
\begin{equation}
J^{\nu \gamma }=\dint_{\mathbb{R}^{3}}M^{0\nu \gamma }\left( x\right) \,%
\mathrm{d}\mathbf{x}=0.  \label{fshif7}
\end{equation}

\subsection{Frequency shifted Lagrangian and the reduced energy\label{sfreq}}

The time harmonic factor $\mathrm{e}^{-\mathrm{i}\omega _{0}t}$ which
appears in $\omega _{0}$-static states as in (\ref{psif1}) plays a very
important role in this theory, including the nonrelativistic case. To
reflect that we introduce a change of variables%
\begin{equation}
\psi \left( t,\mathbf{x}\right) \rightarrow \mathrm{e}^{-\mathrm{i}\omega
_{0}t}\psi \left( t,\mathbf{x}\right)  \label{fshif1}
\end{equation}%
and substitute it in the Lagrangian $L_{0}$ defined by (\ref{fpar1}) to
obtain the Lagrangian $L_{\omega _{0}}$, which we call \emph{frequency
shifted}, namely%
\begin{gather}
L_{\omega _{0}}\left( \psi ,A^{\mu }\right) =\frac{\chi }{2}\mathrm{i}\left(
\psi ^{\ast }\tilde{\partial}_{t}\psi -\psi \tilde{\partial}_{t}^{\ast }\psi
^{\ast }\right) +  \label{fshif2} \\
+\frac{\chi ^{2}}{2m}\left\{ \frac{1}{\mathrm{c}^{2}}\tilde{\partial}%
_{t}\psi \tilde{\partial}_{t}^{\ast }\psi ^{\ast }-\tilde{\nabla}\psi \tilde{%
\nabla}^{\ast }\psi ^{\ast }-G\left( \psi ^{\ast }\psi \right) \right\} -%
\frac{F^{\mu \nu }F_{\mu \nu }}{16\pi },  \notag
\end{gather}%
where we use notation 
\begin{equation*}
\tilde{\partial}_{t}=\partial _{t}+\frac{\mathrm{i}q\varphi }{\chi },\ 
\tilde{\nabla}=\nabla -\frac{\mathrm{i}q\mathbf{A}}{\chi \mathrm{c}},\ 
\tilde{\partial}_{t}^{\ast }=\partial _{t}-\frac{\mathrm{i}q\varphi }{\chi }%
,\ \tilde{\nabla}^{\ast }=\nabla +\frac{\mathrm{i}q\mathbf{A}}{\chi \mathrm{c%
}}.
\end{equation*}%
If we use the relation (\ref{flagr7a}) we can rewrite it in the form%
\begin{gather}
L_{\omega _{0}}\left( \psi ,\psi ^{\ast },A^{\mu }\right) =\frac{\chi }{2}%
\mathrm{i}\left( \psi ^{\ast }\tilde{\partial}_{t}\psi -\psi \tilde{\partial}%
_{t}^{\ast }\psi ^{\ast }\right) +  \label{fshif3} \\
\frac{\chi ^{2}}{2m}\left\{ \frac{1}{\mathrm{c}^{2}}\tilde{\partial}_{t}\psi 
\tilde{\partial}_{t}^{\ast }\psi ^{\ast }-\tilde{\nabla}\psi \tilde{\nabla}%
^{\ast }\psi ^{\ast }-G\left( \psi ^{\ast }\psi \right) \right\} +  \notag \\
+\frac{1}{8\pi }\left[ \left( \nabla \varphi +\frac{1}{\mathrm{c}}\partial
_{t}\mathbf{A}\right) ^{2}-\left( \nabla \times \mathbf{A}\right) ^{2}\right]
\notag
\end{gather}%
The Lagrangian $L_{\omega _{0}}$ defined by the formula (\ref{fshif2}) is
manifestly gauge and space-time translation invariant, it also invariant
with respect to space rotations but it is not invariant with respect to the
entire group of Lorentz transformations. Notice also that $\omega _{0}$%
-static states for the original Lagrangian defined by (\ref{fpar1}) turn
into regular static states for the Lagrangian $L_{\omega _{0}}$, and that
was one of the reasons to introduce it.

For a $\omega _{0}$-static state $\left\{ \mathrm{e}^{-\mathrm{i}\omega
_{0}t}\mathring{\psi},\mathring{\varphi}\right\} $ satisfying the field
equations (\ref{psif2})-(\ref{psif3}) its canonical density of energy $%
\mathring{u}_{L_{0}}\left( \mathring{\psi},\ \mathring{\varphi}\right) $ as
defined by (\ref{flagr5}) can be simply related to the canonical energy $%
\mathring{u}_{L_{\omega _{0}}}\left( \mathring{\psi},\ \mathring{\varphi}%
\right) $ of the frequency shifted Lagrangian $L_{\omega _{0}}$. Indeed
applying the arguments provided in Section \ref{stharm}, particularly
relations (\ref{kab10})-(\ref{kab13}), and combined with the representation (%
\ref{fshif3}) we find that%
\begin{gather}
\mathring{u}_{L_{0}}\left( \mathring{\psi},\ \mathring{\varphi}\right) =%
\frac{m\mathrm{c}^{2}}{q}\rho -L_{\omega _{0}}\left( \mathring{\psi},%
\mathring{\psi}^{\ast },\mathring{\varphi}\right) =  \label{fshif4} \\
=\frac{m\mathrm{c}^{2}}{q}\rho +\frac{\chi ^{2}}{2m}\left[ \left( \nabla 
\mathring{\psi}\right) ^{2}+\left( \frac{\omega _{0}}{\mathrm{c}}-\frac{q}{%
\chi \mathrm{c}}\mathring{\varphi}\right) ^{2}\mathring{\psi}^{2}+G\left( 
\mathring{\psi}^{2}\right) \right] -\frac{\left( \nabla \mathring{\varphi}%
\right) ^{2}}{8\pi },  \notag
\end{gather}%
and that the total energy in this state can be represented in the form (\ref%
{psi1e}) using results of Section \ref{senrgypart}. The energy
representation (\ref{psi1e}) is important to us since it does not involve
explicitly the nonlinear self-interaction $G$.

\subsection{Moving charge}

As it is often done in the literature we use the Lorentz invariance of the
system to obtain the state of the dressed charge moving with a constant
velocity $\mathbf{v}$. Namely, we apply to the rest solution described by (%
\ref{psif1})-(\ref{psif3}) the Lorentz transformation from the original
"rest frame" to the frame in which the "rest frame" moves with the constant
velocity $\mathbf{v}$ as described by the formulas (\ref{rkin3}), (\ref%
{maxw8}) (so $\mathbf{x}^{\prime }$ and $\mathbf{x}$ correspond respectively
to the "rest" and "moving" frames) yielding 
\begin{equation}
\psi \left( t,\mathbf{x}\right) =\mathrm{e}^{-\mathrm{i}\left( \omega t-%
\mathbf{k}\cdot \mathbf{x}\right) }\mathring{\psi}\left( \mathbf{x}^{\prime
}\right) ,\ \varphi \left( t,\mathbf{x}\right) =\gamma \mathring{\varphi}%
\left( \left\vert \mathbf{x}^{\prime }\right\vert \right) ,\ \mathbf{A}%
\left( t,\mathbf{x}\right) =\gamma \mathbf{\beta }\mathring{\varphi}\left(
\left\vert \mathbf{x}^{\prime }\right\vert \right) ,  \label{mvch1}
\end{equation}%
\begin{equation}
\mathbf{E}\left( t,\mathbf{x}\right) =-\gamma \nabla \mathring{\varphi}%
\left( \left\vert \mathbf{x}^{\prime }\right\vert \right) +\frac{\gamma ^{2}%
}{\gamma +1}\left( \mathbf{\beta }\cdot \nabla \mathring{\varphi}\left(
\left\vert \mathbf{x}^{\prime }\right\vert \right) \right) \mathbf{\beta ,\ B%
}\left( t,\mathbf{x}\right) =\gamma \mathbf{\beta }\times \nabla \mathring{%
\varphi}\left( \left\vert \mathbf{x}^{\prime }\right\vert \right) ,
\label{mvch2}
\end{equation}%
where%
\begin{equation}
\omega =\gamma \omega _{0},\ \mathbf{k}=\gamma \mathbf{\beta }\frac{\omega
_{0}}{\mathrm{c}},\ \mathbf{\beta }=\frac{\mathbf{v}}{\mathrm{c}},\ \beta
=\left\vert \mathbf{\beta }\right\vert ,\ \gamma =\left( 1-\left( \frac{v}{%
\mathrm{c}}\right) ^{2}\right) ^{-1/2},  \label{mvch3}
\end{equation}%
\begin{equation}
\mathbf{x}^{\prime }=\mathbf{x}+\frac{\gamma -1}{\beta ^{2}}\left( \mathbf{%
\beta }\cdot \mathbf{x}\right) \mathbf{\beta }-\gamma \mathbf{v}t,\text{ or }%
\mathbf{x}_{\Vert }^{\prime }=\gamma \left( \mathbf{x}_{\Vert }-\mathbf{v}%
t\right) ,\ \mathbf{x}_{\bot }^{\prime }=\mathbf{x}_{\bot },  \label{mvch4}
\end{equation}%
where $\mathbf{x}_{\Vert }$ and $\mathbf{x}_{\bot }$ refer respectively to
the components of $\mathbf{x}$ parallel and perpendicular to the velocity $%
\mathbf{v}$. The above formulas provide a solution to field equations (\ref%
{fpar4}), (\ref{fpar5}) and indicate that the fields of the dressed charge
contract by the factor $\gamma $ as it moves with the velocity $\mathbf{v}$
compared to their rest state. The first oscillatory exponential factor in (%
\ref{mvch1}) is the \emph{de Broglie plane wave} of the frequency $\omega $
and the \emph{de Broglie wave-vector} $\mathbf{k}$. Notice that the
equalities (\ref{mvch3}) readily imply the following relations between $%
\omega $, $\mathbf{k}$ and $\mathbf{v}$ 
\begin{equation}
\omega =\omega \left( \mathbf{k}\right) =\sqrt{\omega _{0}^{2}+\mathrm{c}^{2}%
\mathbf{k}^{2}},\ \mathbf{v}=\nabla _{\mathbf{k}}\omega \left( \mathbf{k}%
\right) ,  \label{mvch5}
\end{equation}%
$\ $where%
\begin{equation*}
\text{ }\omega _{0}=\frac{\mathcal{E}_{0}\left( \mathring{\psi}\right) }{%
\hbar }=\frac{m\mathrm{c}^{2}}{\chi },
\end{equation*}%
and we refer to Section \ref{ssketch}, formulas (\ref{psfi3})-(\ref{psi2b}),
for the values of the frequency $\omega _{0}$ and the constant $\chi $.

Notice that the above relations show, in particular, that for the freely
moving dressed charge defined by equalities (\ref{mvch1})-(\ref{mvch4}) its
velocity $\mathbf{v}$ equals exactly the group velocity $\nabla _{\mathbf{k}%
}\omega \left( \mathbf{k}\right) $ computed for the de Broglie wave vector $%
\mathbf{k}$. This fact clearly points to the wave origin of the charge
kinematics as it moves in the three dimensional space continuum with the
dispersion relation $\omega =\sqrt{\omega _{0}^{2}+\mathrm{c}^{2}\mathbf{k}%
^{2}}$. Notice that this dispersion relation is identical to the dispersion
relation of the Klein-Gordon equation as a model for a free charge, \cite[%
Section 18]{Pauli PWM}.

Now we consider the total 4-momentum $\mathsf{P}$ obtained from its density
by integration over the space $\mathbb{R}^{3}$. Since the dressed charge is
a closed system, its total 4-momentum $\mathsf{P}^{\nu }=\left( \mathsf{E},%
\mathrm{c}\mathsf{P}\right) $ is 4-vector, see the end of Section \ref%
{senergymom}. Using this vector property and the value $\mathsf{P}^{\nu
}=\left( \mathcal{E}_{0}\left( \mathring{\psi}\right) ,\mathbf{0}\right) $
for the resting dressed charge we find, by applying the relevant Lorentz
transformation, that the dressed charge 4-momentum $\mathsf{P}^{\nu }$
satisfies%
\begin{equation}
\mathsf{P}^{\nu }=\left( \mathsf{E},\mathrm{c}\mathsf{P}\right) ,\ \mathsf{E}%
=\hbar \omega ,\ \mathsf{P}=\hbar \mathbf{k},  \label{mvch6}
\end{equation}%
showing that \emph{the Einstein-de Broglie relations hold for the moving
charge}. We would like to point out that, though the above argument used to
obtain the relations (\ref{mvch6}) is rather standard, in our case relations
(\ref{mvch6}) are deduced rather than rationally imposed.

Observe that our relations (\ref{mvch5}) under the assumption that $\chi
=\hbar $ are identical to those of a free charge as described by the
Klein-Gordon equation, \cite[Sections 1, 18]{Pauli PWM}, (see also Section %
\ref{skleingordon}) but there are several significant differences between
the two models which are as follows. First of all, our charge is a dressed
charge described by the pair $\left\{ \psi ,A^{\mu }\right\} $. From the
very outset it includes the EM field as its inseparable part whereas the
Klein-Gordon model describes a free charge by a complex-valued wave function 
$\psi $ which is not coupled to its own EM field (not to be confused with an
external EM field). Second, our free dressed charge when it moves, evidently
preserves its shape up to the natural Lorentz construction whereas any
wavepacket satisfying Klein-Gordon equation spreads out in the course of
time.

\subsection{Correspondence with the point charge mechanics}

The free dressed charge as described by equalities (\ref{mvch1})-(\ref{mvch4}%
) allows for a certain reduction to the model of point charge (mass). Notice
that combining the relations (\ref{mvch6}) with (\ref{mvch3}) we obtain the
well known point mass kinematic representations (\ref{rkin6}) for the total
energy $\mathsf{E}$ and the momentum $\mathsf{P}$ of the dressed charge,
namely 
\begin{gather}
\mathsf{P}=\hbar \mathbf{k}=\gamma \mathbf{\beta }\frac{\hbar \omega _{0}}{%
\mathrm{c}}=\gamma \tilde{m}\mathbf{v,}\gamma =\left( 1-\left( \frac{v}{%
\mathrm{c}}\right) ^{2}\right) ^{-1/2},  \label{mvch7} \\
\mathsf{E}=\hbar \omega =\hbar \gamma \omega _{0}=\gamma \tilde{m}\mathrm{c}%
^{2}=\mathrm{c}\sqrt{\mathsf{P}^{2}+\tilde{m}^{2}\mathrm{c}^{2}},
\label{mvch8}
\end{gather}%
where $\tilde{m}$ is the dressed charge mass defined by (\ref{psi2a}). We
can also reasonably assign to the dressed charge described by equalities (%
\ref{mvch1})-(\ref{mvch4}) a location $\mathbf{r}\left( t\right) $ at any
instant $t$ of time which is obtained from the (\ref{mvch4}) by setting
there $\mathbf{x}^{\prime }$ and solving it for $\mathbf{x}$, $\mathbf{r}%
\left( t\right) =\mathbf{x}\left( \mathbf{x}^{\prime },t\right) \mathbf{.}$
Not surprisingly, its solution is%
\begin{equation}
\mathbf{r}\left( t\right) =\mathbf{v}t.  \label{mvch9}
\end{equation}%
An elementary examination confirms that $\left( \mathrm{c}t,\mathbf{v}%
t\right) $ transforms as a 4-vector implying that the definition (\ref{mvch9}%
) is both natural and relativistically consistent. From (\ref{mvch9}) we
readily obtain another fundamental relation for the point charge%
\begin{equation}
\mathbf{v}=\frac{\mathrm{d}\mathbf{r}\left( t\right) }{\mathrm{d}t}.
\label{mvch10}
\end{equation}

\section{Single nonrelativistic free and resting charge\label{snfree}}

The nonrelativistic case, i.e. the case when a charge moves with a velocity
much smaller than the velocity of light, is important for our studies for at
least two reasons. First of all, we need it to relate the wave-corpuscle
mechanics to the Newtonian mechanics for point charges in EM fields. Second
of all, in the nonrelativistic case we can carry out rather detailed
analytical studies of many physical quantities in a closed form. With that
in mind, we would like to treat the nonrelativistic case not just as an
approximation to the relativistic theory but rather as a case on its own,
and we do it by constructing a certain \emph{nonrelativistic Lagrangian }$%
\hat{L}_{0}$ intimately related to the relativistic Lagrangian defined in (%
\ref{fpar1})-(\ref{fpar1b}). This nonrelativistic Lagrangian constitutes a
fundamental basis for our nonrelativistic studies including the construction
of the nonlinear self-interaction. The relation between the relativistic and
nonrelativistic Lagrangians is considered in Section \ref{cheqdimf}.

The nonrelativistic Lagrangian $\hat{L}_{0}$ is constructed as a certain
nonrelativistic modification of the frequency shifted Lagrangian $L_{\omega
_{0}}\left( \psi ,A^{\mu }\right) $ introduced in Section \ref{sfreq}. The
first step in this modification is the change of variables (\ref{fshif1}),
namely%
\begin{equation}
\psi \left( t,\mathbf{x}\right) \rightarrow \mathrm{e}^{-\mathrm{i}\omega
_{0}t}\psi \left( t,\mathbf{x}\right)  \label{norel}
\end{equation}%
which was the initial step in the construction of the frequency-shifted
Lagrangian $L_{\omega _{0}}$ defined by (\ref{fshif2})-(\ref{fshif3}). Then
a gauge invariant and nonrelativistic Lagrangian $\hat{L}_{0}$ is obtained
from the Lagrangian $L_{\omega _{0}}$ by omitting in (\ref{fshif3}) the term 
$\frac{\chi ^{2}}{2m\mathrm{c}^{2}}\tilde{\partial}_{t}\psi \tilde{\partial}%
_{t}^{\ast }\psi ^{\ast }$ and setting $\mathbf{A}=0$, namely 
\begin{gather}
\hat{L}_{0}\left( \psi ,\psi ^{\ast },\varphi \right) =\frac{\chi \mathrm{i}%
}{2}\left[ \psi ^{\ast }\tilde{\partial}_{t}\psi -\psi \tilde{\partial}%
_{t}^{\ast }\psi ^{\ast }\right] -\frac{\chi ^{2}}{2m}\left[ \nabla \psi
\nabla \psi ^{\ast }+G\left( \psi ^{\ast }\psi \right) \right] +\frac{%
\left\vert \nabla \varphi \right\vert ^{2}}{8\pi },  \label{nore1b} \\
\tilde{\partial}_{t}=\partial _{t}+\frac{\mathrm{i}q\varphi }{\chi },\ 
\tilde{\partial}_{t}^{\ast }=\partial _{t}-\frac{\mathrm{i}q\varphi }{\chi }
\notag
\end{gather}%
where, we recall, the term $G\left( \psi ^{\ast }\psi \right) $ corresponds
to the charge nonlinear self-interaction. Observe that the assumption $%
\mathbf{A}=0$ in view of (\ref{maxw3a}) readily implies%
\begin{equation}
\mathbf{E}=-\nabla \varphi ,\ \mathbf{B}=\mathbf{0}.  \label{nore1a}
\end{equation}%
Hence, the EM field tensor $F^{\mu \nu }$ defined by (\ref{maxw3}) takes
here a simpler form 
\begin{equation}
F^{\mu \nu }=\left[ 
\begin{array}{cccc}
0 & -E_{1} & -E_{2} & -E_{3} \\ 
E_{1} & 0 & 0 & 0 \\ 
E_{2} & 0 & 0 & 0 \\ 
E_{3} & 0 & 0 & 0%
\end{array}%
\right] .  \label{nore1c}
\end{equation}%
The mentioned gauge invariance is understood with respect first to the gauge
transformation of the first kind (global) as in (\ref{flagr8}) and of the
second (local) types as in (\ref{flagr8})-(\ref{flagr8a}), namely 
\begin{equation}
\psi \rightarrow \mathrm{e}^{\mathrm{i}\gamma }\psi ,\ \psi ^{\ast
}\rightarrow \mathrm{e}^{-\mathrm{i}\gamma }\psi ^{\ast },\text{ }
\label{nore2}
\end{equation}%
where $\gamma $ is any real constant, and with respect to a reduced version
of the second type gauge transformation 
\begin{equation}
\psi \rightarrow \mathrm{e}^{-\frac{\mathrm{i}q\lambda \left( t\right) }{%
\chi }}\psi ,\ \psi ^{\ast }\rightarrow \mathrm{e}^{\frac{\mathrm{i}q\lambda
\left( t\right) }{\chi }}\psi ^{\ast },\ \varphi \rightarrow \varphi
+\partial _{t}\lambda \left( t\right) ,  \label{nore3}
\end{equation}%
which is similar to (\ref{flagr8a}) but the function $\lambda \left(
t\right) $ may depend only on time.

Evidently the EM field of the charge is represented in the above Lagrangian $%
\hat{L}_{0}$ only by its scalar potential $\varphi $ and the corresponding
electric field $\mathbf{E}=-\nabla \varphi $ , since $\mathbf{A}=0$. The
charge's magnetic field is identically zero in view of the equalities (\ref%
{nore1a}), and, consequently, any radiation phenomena are excluded in this
model. The Lagrangian $\hat{L}_{0}$ can be viewed as a field version of the
point charges model (\ref{minor4}) that neglects all retardation effects in
the static limit (zeroth order in $\frac{v}{\mathrm{c}}$) with the
"instantaneous" interaction Lagrangian $-\frac{q_{1}q_{2}}{\left\vert 
\mathbf{r}_{1}-\mathbf{r}_{2}\right\vert }$ between two charges, \cite[%
Section 12.6]{Jackson}. More detailed discussions on the relations between
relativistic and nonrelativistic Lagrangians and the corresponding
Euler-Lagrange equations are provided in Section \ref{cheqdimf}.

The Euler-Lagrange field equations for this Lagrangian are%
\begin{gather}
\chi \mathrm{i}\tilde{\partial}_{t}\psi =\frac{\chi ^{2}}{2m}\left[ -\Delta
\psi +G^{\prime }\left( \psi ^{\ast }\psi \right) \psi \right] ,\ 
\label{nore3a} \\
-\Delta \varphi =4\pi q\psi \psi ^{\ast },\   \label{nore3b}
\end{gather}%
where $G^{\prime }\left( s\right) =\partial _{s}G\left( s\right) $, \ and we
refer to the pair $\left\{ \psi ,\varphi \right\} $ as \emph{dressed charge}%
. Taking into account the form of the covariant time derivative from (\ref%
{nore1b}) we can recast the field equations (\ref{nore3a})-(\ref{nore3b})
for the dressed charge as%
\begin{equation}
\chi \mathrm{i}\partial _{t}\psi =\frac{\chi ^{2}}{2m}\left[ -\Delta +\frac{%
2mq}{\chi ^{2}}\varphi +G^{\prime }\left( \left\vert \psi \right\vert
^{2}\right) \right] \psi ,\ -\Delta \varphi =4\pi q\left\vert \psi
\right\vert ^{2}.  \label{nore5}
\end{equation}%
which imply (\ref{nop2}), (\ref{nop3}).

Applying \ the general formulas (\ref{lagco9})-(\ref{lagco12}) for the
charge and current densities to the Lagrangian $\hat{L}_{0}$ we obtain
expressions (\ref{nop8}) for the densities and since external fields are
absent, the current $J^{\mu }$ satisfies the conservation/continuity
equations (\ref{nop9}). Consequently, the total charge remains constant in
the course of evolution, and as always we set this constant charge to be
exactly $q$, namely \ we impose \emph{charge normalization} condition (\ref%
{nop11}) 
\begin{equation}
\dint\nolimits_{\mathbb{R}^{3}}\rho \left( x\right) \,\mathrm{d}\mathbf{x}%
=q\dint\nolimits_{\mathbb{R}^{3}}\psi \psi ^{\ast }\mathrm{d}\mathbf{x}=q%
\text{ or }\dint\nolimits_{\mathbb{R}^{3}}\left\vert \psi \right\vert ^{2}\,%
\mathrm{d}\mathbf{x}=1.  \label{nrac4}
\end{equation}%
As in the relativistic case the equation (\ref{nop9}) follows from the field
equations, therefore (\ref{nrac4}) is preserved for all times.

\subsection{Symmetries and conservation laws}

To carry out a systematic analysis of conservation laws associated with the
Lagrangian $\hat{L}_{0}$ defined by (\ref{nore1b}) via Noether theorem, see
Section \ref{snoether}, we need to find a Lie group of transformations which
preserve it. The Lagrangian $\hat{L}_{0}$ is not invariant with respect to
either the Lorentz or the Galilean groups of transformations. But a
straightforward examination shows that $\hat{L}_{0}$ is invariant with
respect to the following \emph{Galilean-gauge group of transformations}%
\begin{equation}
t^{\prime }=t,\ \mathbf{x}^{\prime }=\mathbf{x}-\mathbf{v}t,  \label{gal1}
\end{equation}%
or%
\begin{equation*}
x^{0\prime }=x^{0},\ \mathbf{x}^{\prime }=\mathbf{x}-\frac{\mathbf{v}}{%
\mathrm{c}}x^{0},
\end{equation*}%
\begin{equation}
\psi \left( t,\mathbf{x}\right) =\mathrm{e}^{\mathrm{i}\frac{m}{2\chi }%
\left( \mathbf{v}^{2}t^{\prime }+2\mathbf{v}\cdot \mathbf{x}^{\prime
}\right) }\psi ^{\prime }\left( t^{\prime },\mathbf{x}^{\prime }\right) ,
\label{gal2}
\end{equation}%
or%
\begin{equation*}
\psi ^{\prime }\left( t^{\prime },\mathbf{x}^{\prime }\right) =\mathrm{e}^{%
\mathrm{i}\frac{m}{2\chi }\left( \mathbf{v}^{2}t-2\mathbf{v}\cdot \mathbf{x}%
\right) }\psi \left( t,\mathbf{x}\right) ,
\end{equation*}%
with $\varphi \left( t,\mathbf{x}\right) =\varphi ^{\prime }\left( t^{\prime
},\mathbf{x}^{\prime }\right) $. One can also verify that the above
transformations form an Abelian (commutative) group of transformations
parametrized by the velocity parameter $\mathbf{v}$. It is curious to
observe that according to the Galilean-gauge transformations (\ref{gal1}), (%
\ref{gal2}) the charge wave function does not transform as a scalar as in
the relativistic case. These transformations are known, \cite[Section 7.3]%
{Gottfried}, and were used, in particular, in studies of nonlinear Schr\"{o}%
dinger equations, \cite[Section 2.3]{Sulem}.

The above defined Galilean-gauge group is naturally extended to the \emph{%
general inhomogeneous Galilean-gauge group} by adding to it the group of
spacial rotations $O$ and space-time translations $a^{\mu }$, namely%
\begin{equation}
t^{\prime }=t+\tau ,\ \mathbf{x}^{\prime }=O\mathbf{x}-\mathbf{v}t+\mathbf{a}%
,  \label{gal2a}
\end{equation}%
\begin{equation}
\psi ^{\prime }\left( t^{\prime },\mathbf{x}^{\prime }\right) =\mathrm{e}^{%
\mathrm{i}\frac{m}{2\chi }\left[ \mathbf{v}^{2}\left( t+\tau \right) -2%
\mathbf{v}\cdot \left( O\mathbf{x}+\mathbf{a}\right) \right] }\psi \left( t,%
\mathbf{x}\right) ,\varphi ^{\prime }\left( t^{\prime },\mathbf{x}^{\prime
}\right) =\varphi \left( t,\mathbf{x}\right) .  \label{gal2b}
\end{equation}%
The\ \emph{infinitesimal form} of the above group of transformations is as
follows%
\begin{equation}
t^{\prime }=t+\tau ,\ \mathbf{x}^{\prime }=\mathbf{x}+\mathbf{\xi }\times 
\mathbf{x}-\mathbf{v}t+\mathbf{a},\ a^{\mu }=\left( \mathrm{c}\tau ,\mathbf{a%
}\right) ,\ a^{0}=\mathrm{c}\tau ,\ x^{0}=\mathrm{c}t,  \label{gal3}
\end{equation}%
or, equivalently,%
\begin{equation}
x^{\prime \mu }=x^{\mu }+\xi ^{\mu \nu }x_{\nu }+a^{\mu },\ \xi ^{\mu \nu }= 
\left[ 
\begin{array}{cccc}
0 & 0 & 0 & 0 \\ 
-\frac{v^{1}}{\mathrm{c}} & 0 & -\xi ^{3} & \xi ^{2} \\ 
-\frac{v^{2}}{\mathrm{c}} & \xi ^{3} & 0 & -\xi ^{1} \\ 
-\frac{v^{3}}{\mathrm{c}} & -\xi ^{2} & \xi ^{1} & 0%
\end{array}%
\right] ,  \label{gal4}
\end{equation}%
where the real number $\tau $ and the coordinates of the three
three-dimensional vectors $\mathbf{v}$, $\mathbf{\xi }$, $\mathbf{a}$
provide the total of ten real parameters as in the case of the infinitesimal
inhomogeneous Lorentz group defined by (\ref{ttbe2}). The infinitesimal form
of the transformations (\ref{gal2b}) is 
\begin{equation}
\delta \psi =-\mathrm{i}\frac{m}{\chi }\left( \sum_{j=1,2,3}x^{j}\cdot
\delta v^{j}\right) \psi ,\ \ \delta \varphi =0.  \label{gal5}
\end{equation}%
Coming back to the analysis of basic features of our model we acknowledge
the use in this section of the relativistic conventions for upper and lower
indices and the summation as in Section \ref{srelkin}, including%
\begin{equation}
x^{\mu }=\left( x^{0},\mathbf{x}\right) ,\ x_{\mu }=\left( x^{0},-\mathbf{x}%
\right) ,\ x^{0}=\mathrm{c}t.  \label{nore4a}
\end{equation}%
Carrying out the Noether currents analysis as in Section \ref{snoether} for
the Lagrangian $\hat{L}_{0}$ we obtain 10 conservation laws which, as it
turns out, can be formulated in terms of the canonical energy-momentum
tensor $\mathcal{\mathring{T}}^{\mu \nu }$, which in turn is obtained from
the general formula (\ref{flagr5}): 
\begin{equation}
\mathcal{\mathring{T}}^{\mu \nu }=\frac{\partial \hat{L}_{0}}{\partial \psi
_{,\mu }}\psi _{,\nu }+\frac{\partial \hat{L}_{0}}{\partial \psi _{,\mu
}^{\ast }}\psi _{,\nu }^{\ast }+\frac{\partial \hat{L}_{0}}{\partial \varphi
_{,\mu }}\varphi _{,\nu }-\hat{L}_{0}g^{\mu \nu }.  \label{gal6}
\end{equation}%
Namely, we get the total of ten conservation laws:%
\begin{gather}
\partial _{\mu }\mathcal{\mathring{T}}^{\mu \nu }=0\text{ - energy-momentum
conserv.,}  \label{gal7} \\
\mathcal{\mathring{T}}^{ij}=\mathcal{\mathring{T}}^{ji},\ i,j=1,2,3\text{ -
space angular momentum conserv.,}  \label{gal8} \\
P^{i}=\mathcal{\mathring{T}}^{0i}=\frac{m}{q}J^{i},\ i=1,2,3\text{ -
time-space angular momentum conserv.}  \label{gal9}
\end{gather}%
The first four standard conservation laws (\ref{gal7}) are associated with
the Noether's currents with respect to space-time translations $a^{\mu }$.
The second three conservation laws in (\ref{gal8}) are associated with space
rotation parameters $\mathbf{\xi }$, and they turn into the symmetry of the
energy-momentum tensor $\mathcal{\mathring{T}}^{\mu \nu }$ for the spatial
indices similarly to relations (\ref{ttbe4})-(\ref{ttbe5}). The form of the 
\emph{last three conservation laws (\ref{gal9}) is special to the
nonrelativistic Lagrangian }$\hat{L}_{0}$\emph{, and it is due to the
Galilean-gauge invariance (\ref{gal2}), (\ref{gal2b})}. These relations
indicate that the \emph{total momentum density }$P^{i}$\emph{\ is
identically equal up to the factor }$\frac{m}{q}$\emph{\ to the microcurrent
density }$J^{i}$ defined by (\ref{nop9}). This important identity is
analogous to the kinematic representation $\mathbf{p}=m\mathbf{v}$ of the
momentum $\mathbf{p}$ of a point charge. It is related to the velocity
components $\mathbf{v}$ in the Galilean-gauge transformations (\ref{gal2}), (%
\ref{gal2b}), and can be traced to the infinitesimal transformation (\ref%
{gal5}) and the phases in (\ref{gal2b}). The proportionality of the momentum
and the current is known to occur for systems governed by the nonlinear Schr%
\"{o}dinger equations, \cite[Section 2.3]{Sulem}.

The issue of fundamental importance of studies of the energy-momentum tensor
has been already addressed in the beginning of Section \ref{senmomforce}. To
find the energy-momentum tensor for the Lagrangian $\hat{L}_{0}$ we apply to
it the general formulas from Section \ref{1part}. The canonical
energy-momentum $\mathring{\Theta}^{\mu \nu }$ for the EM field is obtained
by applying the general formula (\ref{flagr5}) to the Lagrangian $\hat{L}%
_{0} $ yielding%
\begin{equation}
\mathring{\Theta}^{\mu \nu }=\left[ 
\begin{array}{cccc}
\mathring{w} & 0 & 0 & 0 \\ 
\mathrm{c}^{-1}\mathring{s}_{1} & -\mathring{\tau}_{11} & -\mathring{\tau}%
_{12} & -\mathring{\tau}_{13} \\ 
\mathrm{c}^{-1}\mathring{s}_{2} & -\mathring{\tau}_{21} & -\mathring{\tau}%
_{22} & -\mathring{\tau}_{23} \\ 
\mathrm{c}^{-1}\mathring{s}_{3} & -\mathring{\tau}_{31} & -\mathring{\tau}%
_{32} & -\mathring{\tau}_{33}%
\end{array}%
\right] ,  \label{emfr1}
\end{equation}%
\begin{gather}
\mathring{w}=-\frac{\left\vert \nabla \varphi \right\vert ^{2}}{8\pi },\ 
\mathring{g}_{j}=0,\ \mathring{s}_{j}=\mathrm{c}\frac{\partial _{j}\varphi
\partial _{0}\varphi }{4\pi },  \label{emfr2} \\
\mathring{\tau}_{jj}=\frac{\partial _{j}^{2}\varphi }{4\pi }-\frac{%
\left\vert \nabla \varphi \right\vert ^{2}}{8\pi }=\frac{\partial
_{j}^{2}\varphi }{4\pi }+\mathring{w},\ \mathring{\tau}_{ij}=\frac{\partial
_{i}\varphi \partial _{j}\varphi }{4\pi }.  \notag
\end{gather}%
The gauge invariant energy-momentum of the EM field takes the form%
\begin{equation}
\Theta ^{\mu \nu }=\left[ 
\begin{array}{cccc}
w & 0 & 0 & 0 \\ 
0 & -\tau _{11} & -\tau _{12} & -\tau _{13} \\ 
0 & -\tau _{21} & -\tau _{22} & -\tau _{23} \\ 
0 & -\tau _{31} & -\tau _{32} & -\tau _{33}%
\end{array}%
\right] ,  \label{emfr3}
\end{equation}%
with matrix entries%
\begin{gather}
\partial _{0}w=\frac{\mathbf{J}\cdot \nabla \varphi }{\mathrm{c}}=-\frac{%
\mathbf{J}\cdot \mathbf{E}}{\mathrm{c}},\ g_{j}=0,\ s_{j}=0,  \label{emfr4}
\\
\tau _{jj}=\frac{\partial _{j}^{2}\varphi }{4\pi }-\frac{\left\vert \nabla
\varphi \right\vert ^{2}}{8\pi },\ \tau _{ij}=\frac{\partial _{i}\varphi
\partial _{j}\varphi }{4\pi }.  \notag
\end{gather}%
As we can see the relation (\ref{emfr4}) involves the time derivative $%
\partial _{0}w$ for the energy density $w$ rather then the density itself,
and it follows from it that 
\begin{equation}
w=w\left( t,\mathbf{x}\right) =w_{0}\left( \mathbf{x}\right) +\int_{-\mathbb{%
\infty }}^{t}\frac{\mathbf{J}\left( t^{\prime },\mathbf{x}\right) \cdot
\nabla \varphi \left( t^{\prime },\mathbf{x}\right) }{\mathrm{c}}\,\mathrm{d}%
t^{\prime },  \label{emfr5}
\end{equation}%
where $w_{0}\left( \mathbf{x}\right) $ is a time independent energy density.
Notice also that combining the relations (\ref{nop8}), (\ref{nop9}), (\ref%
{nore3b}) and (\ref{emfr5}) we obtain the following identity%
\begin{gather}
\partial _{0}\int_{\mathbb{R}^{3}}w\,\mathrm{d}\mathbf{x}=\int_{\mathbb{R}%
^{3}}\frac{\mathbf{J}\cdot \nabla \varphi }{\mathrm{c}}\,\mathrm{d}\mathbf{x}%
=-\int_{\mathbb{R}^{3}}\frac{\varphi \left( \nabla \cdot \mathbf{J}\right) }{%
\mathrm{c}}\,\mathrm{d}\mathbf{x}=\frac{1}{4\pi \mathrm{c}}\int_{\mathbb{R}%
^{3}}\varphi \partial _{t}\rho \,\mathrm{d}\mathbf{x}=  \label{emfr11a} \\
-\frac{1}{4\pi \mathrm{c}}\int_{\mathbb{R}^{3}}\varphi \partial _{t}\nabla
^{2}\varphi \,\mathrm{d}\mathbf{x}=\frac{1}{4\pi \mathrm{c}}\int_{\mathbb{R}%
^{3}}\nabla \varphi \cdot \partial _{t}\nabla \varphi \,\mathrm{d}\mathbf{x}%
=\partial _{0}\int_{\mathbb{R}^{3}}\frac{\left( \nabla \varphi \right) ^{2}}{%
8\pi }\,\mathrm{d}\mathbf{x}.  \notag
\end{gather}%
A choice for $w_{0}\left( \mathbf{x}\right) $ in (\ref{emfr5}), consistent
with the canonical energy-momentum, is 
\begin{equation*}
w_{0}\left( \mathbf{x}\right) =\frac{\left( \nabla \varphi \right) ^{2}}{%
8\pi }.
\end{equation*}

The canonical energy-momentum tensor $\mathring{T}^{\mu \nu }$ is not gauge
invariant, but the following decomposition holds for it 
\begin{equation}
\mathring{T}^{\mu \nu }=\tilde{T}^{\mu \nu }+\frac{1}{\mathrm{c}}J^{\mu
}A^{\nu },\ A^{\nu }=\left( \varphi ,0\right)  \label{emfr6}
\end{equation}%
where $\tilde{T}^{\mu \nu }$ is a gauge invariant energy-momentum obtained
by applying formula (\ref{frlag9}) to the Lagrangian $\hat{L}_{0}$, namely%
\begin{equation}
\tilde{T}^{\mu \nu }=\left[ 
\begin{array}{cccc}
\tilde{u} & \mathrm{c}\tilde{p}_{1} & \mathrm{c}\tilde{p}_{2} & \mathrm{c}%
\tilde{p}_{3} \\ 
\mathrm{c}^{-1}\tilde{s}_{1} & -\tilde{\sigma}_{11} & -\tilde{\sigma}_{12} & 
-\tilde{\sigma}_{13} \\ 
\mathrm{c}^{-1}\tilde{s}_{2} & -\tilde{\sigma}_{21} & -\tilde{\sigma}_{22} & 
-\tilde{\sigma}_{23} \\ 
\mathrm{c}^{-1}\tilde{s}_{3} & -\tilde{\sigma}_{31} & -\tilde{\sigma}_{32} & 
-\tilde{\sigma}_{33}%
\end{array}%
\right] ,\text{ }  \label{emfr7}
\end{equation}%
where%
\begin{gather}
\tilde{u}=\frac{\chi ^{2}}{2m}\left[ \left\vert \nabla \psi \right\vert
^{2}+G\left( \left\vert \psi \right\vert ^{2}\right) \right] ,  \label{emfr8}
\\
\tilde{p}_{j}=\frac{\chi \mathrm{i}}{2}\left( \psi \partial _{j}\psi ^{\ast
}-\psi ^{\ast }\partial _{j}\psi \right) ,\ \tilde{s}_{j}=-\frac{\chi ^{2}%
\mathrm{i}}{2m}\left( \tilde{\partial}_{t}\psi \partial _{j}\psi ^{\ast }+%
\tilde{\partial}_{t}^{\ast }\psi ^{\ast }\partial _{j}\psi \right) ,\
j=1,2,3,  \label{emfr9}
\end{gather}%
and the stress tensor components $\sigma _{ij}$ are represented by the
formulas 
\begin{gather}
\tilde{\sigma}_{ii}=\tilde{u}-\frac{\chi ^{2}}{m}\partial _{i}\psi \partial
_{i}\psi ^{\ast }+\frac{\chi \mathrm{i}}{2}\left( \psi \tilde{\partial}%
_{t}\psi ^{\ast }-\psi ^{\ast }\tilde{\partial}_{t}\psi \right) ,
\label{emfr10} \\
\tilde{\sigma}_{ij}=\tilde{\sigma}_{ji}=-\frac{\chi ^{2}}{2m}\left( \partial
_{i}\psi \partial _{j}\psi ^{\ast }+\partial _{j}\psi \partial _{i}\psi
^{\ast }\right) \text{ for }i\neq j,\ i,j=1,2,3.  \notag
\end{gather}%
One can verify using the field equations (\ref{nore3a}), (\ref{nore3b}) and
the current conservation law (\ref{nop9}) that the canonical and gauge
invariant energy-momentum tensors satisfy the the following relations%
\begin{equation}
\mathring{\Theta}^{\mu j}=\Theta ^{\mu j},\ \mathring{T}^{\mu j}=\tilde{T}%
^{\mu j},\ j=1,2,3\text{ }  \label{emfr10a}
\end{equation}%
and 
\begin{equation*}
\partial _{\mu }\left[ \left( \Theta ^{\mu 0}+\tilde{T}^{\mu 0}\right)
-\left( \mathring{\Theta}^{\mu 0}+\mathring{T}^{\mu 0}\right) \right] =0,
\end{equation*}%
and that the conservation laws in view of the representation (\ref{nore1c})
take the following form 
\begin{equation}
\partial _{\mu }\tilde{T}^{\mu \nu }=f^{\nu },\ \partial _{\mu }\Theta ^{\mu
\nu }=-f^{\nu },  \label{emfr11}
\end{equation}%
where 
\begin{equation*}
\ f^{\nu }=\frac{1}{\mathrm{c}}J_{\mu }F^{\nu \mu }=\left( \frac{1}{\mathrm{c%
}}\mathbf{J}\cdot \mathbf{E},\rho \mathbf{E}\right) ,
\end{equation*}%
and we recognize in $f^{\nu }$ the Lorentz force density.

\subsection{Resting charge}

For a resting charge the representations (\ref{emfr3})-(\ref{emfr5}) and (%
\ref{emfr7})-(\ref{emfr10}) for the energy-momentum tensors $\Theta ^{\mu
\nu }$ and $\tilde{T}^{\mu \nu }$ turn into%
\begin{equation}
\Theta ^{\mu \nu }=\left[ 
\begin{array}{cccc}
w & 0 & 0 & 0 \\ 
0 & -\tau _{11} & -\tau _{12} & -\tau _{13} \\ 
0 & -\tau _{21} & -\tau _{22} & -\tau _{23} \\ 
0 & -\tau _{31} & -\tau _{32} & -\tau _{33}%
\end{array}%
\right] ,\ \tilde{T}^{\mu \nu }=\left[ 
\begin{array}{cccc}
\tilde{u} & 0 & 0 & 0 \\ 
0 & -\tilde{\sigma}_{11} & -\tilde{\sigma}_{12} & -\tilde{\sigma}_{13} \\ 
0 & -\tilde{\sigma}_{21} & -\tilde{\sigma}_{22} & -\tilde{\sigma}_{23} \\ 
0 & -\tilde{\sigma}_{31} & -\tilde{\sigma}_{32} & -\tilde{\sigma}_{33}%
\end{array}%
\right] ,  \label{emfr12}
\end{equation}%
showing, in particular, that the momentum and flux densities for the charge
and for the EM field are all identically zero. Consequently, the total
momentum $\mathsf{P}$\ and the energy flux $\mathsf{S}$\ of the resting
dressed charge vanish and we have $\mathsf{P}^{\nu }=\left( \tilde{u},%
\mathsf{P}\right) $ \ with 
\begin{equation}
\mathsf{P}=\mathbf{0},\ \mathsf{S}=\mathbf{0}\text{ and}\ \tilde{u}=\frac{%
\chi ^{2}}{2m}\left[ \left( \nabla \mathring{\psi}\right) ^{2}+G\left( 
\mathring{\psi}^{2}\right) \right] .  \label{emfr13}
\end{equation}%
Observe now that the stress tensor (str. t.) $\sigma _{ij}$ defined by
relations (\ref{emfr10}) in the case of resting charge can be naturally
decomposed into three components which we name as follows%
\begin{equation}
\sigma _{ij}=\sigma _{ij}^{\mathrm{el}}+\sigma _{ij}^{\mathrm{em}}+\sigma
_{ij}^{\mathrm{nl}},\ i,j=1,2,3,\text{ }  \label{nstrc1a}
\end{equation}%
where%
\begin{equation}
\sigma _{ij}^{\mathrm{el}}=-\frac{\chi ^{2}}{m}\left[ \partial _{i}\mathring{%
\psi}\partial _{j}\mathring{\psi}-\frac{1}{2}\left( \nabla \mathring{\psi}%
\right) ^{2}\delta _{ij}\right]  \label{nstrc2}
\end{equation}%
is elastic deformation stress tensor,%
\begin{equation}
\sigma _{ij}^{\mathrm{em}}=-p^{\mathrm{em}}\delta _{ij},\ p^{\mathrm{em}}=-q%
\mathring{\varphi}\mathring{\psi}^{2}\text{ }  \label{nstrc3}
\end{equation}%
is EM interaction stress tensor,%
\begin{equation}
\sigma _{ij}^{\mathrm{nl}}=-p^{\mathrm{nl}}\delta _{ij},\ p^{\mathrm{nl}}=-%
\frac{\chi ^{2}G\left( \mathring{\psi}^{2}\right) }{2m}  \label{nstrc4}
\end{equation}%
is nonlinear self-interaction stress tensor. \ Consequently, the respective
volume force densities are%
\begin{eqnarray}
\dsum_{j=1,2,3}\partial _{j}\sigma _{ij}^{\mathrm{el}} &=&f_{i}^{\mathrm{el}%
}=-\frac{\chi ^{2}}{m}\Delta \mathring{\psi}\partial _{i}\mathring{\psi},\
i=1,2,3,  \label{nstrc5} \\
\dsum_{j=1,2,3}\partial _{j}\sigma _{ij}^{\mathrm{em}} &=&f_{i}^{\mathrm{em}%
}+\rho \partial _{i}\mathring{\varphi},\ f_{i}^{\mathrm{em}}=2q\mathring{%
\varphi}\mathring{\psi}\partial _{i}\mathring{\psi},  \label{nstrc6} \\
\dsum_{j=1,2,3}\partial _{j}\sigma _{ij}^{\mathrm{nl}} &=&f_{i}^{\mathrm{nl}%
}=\frac{\chi ^{2}}{m}G^{\prime }\left( \mathring{\psi}^{2}\right) \mathring{%
\psi}\partial _{i}\mathring{\psi}.  \label{nstrc7}
\end{eqnarray}%
Notice that the volume force density for the electromagnetic interaction
stress in (\ref{nstrc6}) has two parts: $f_{i}^{\mathrm{em}}$, which we call 
\emph{internal electromagnetic force}, and $\rho \partial _{i}\mathring{%
\varphi}$ which is the negative of the Lorentz force. Observe that the
stress tensor $\sigma _{ij}^{\mathrm{el}}$ has the structure similar to the
one for compressional waves, see Section \ref{scompress} and (\ref{compr6}),
whereas the both stress tensors $\sigma _{ij}^{\mathrm{em}}$ and $\sigma
_{ij}^{\mathrm{em}}$ have the structure typical for perfect fluids, \cite[%
Section 6.6]{Moller}, with respective hydrostatic pressures $p^{\mathrm{em}}$
and $p^{\mathrm{nl}}$ defined by the relations (\ref{nstrc3})-(\ref{nstrc4}).

Based on the equalities (\ref{nstrc5})-(\ref{nstrc7}) we can recast the
equilibrium equation (\ref{psif15b}) as%
\begin{gather}
f_{i}^{\mathrm{el}}+f_{i}^{\mathrm{em}}+f_{i}^{\mathrm{nl}}=0,\ i=1,2,3,%
\text{ or}  \label{nstrc8} \\
\left[ -\frac{\chi ^{2}}{m}\Delta \mathring{\psi}+2q\mathring{\varphi}%
\mathring{\psi}+\frac{\chi ^{2}}{m}G^{\prime }\left( \mathring{\psi}%
^{2}\right) \mathring{\psi}\right] \partial _{i}\mathring{\psi}=0.  \notag
\end{gather}%
\emph{The equation (\ref{nstrc8}) signifies the ultimate equilibrium for the
static charge}. It is evident from equation (\ref{nstrc8}) that the scalar
expression in the brackets before $\nabla \mathring{\psi}$ up to the factor $%
\frac{m}{\chi ^{2}}$ is exactly the left-hand side of the equilibrium
equation (\ref{nop5}). In fact if $\nabla \mathring{\psi}\neq 0$ then the
equilibrium equation (\ref{nstrc8}) is equivalent to the scalar equilibrium
equation (\ref{nop5}).

Notice that since the $\mathring{\psi}\left( \left\vert \mathbf{x}%
\right\vert \right) $ and $\mathring{\varphi}\left( \left\vert \mathbf{x}%
\right\vert \right) $ are radial and monotonically decaying functions of $%
\left\vert \mathbf{x}\right\vert $ we readily have%
\begin{equation}
\nabla \mathring{\psi}\left( \left\vert \mathbf{x}\right\vert \right) =%
\mathbf{\hat{x}}-\left\vert \nabla \mathring{\psi}\right\vert \mathbf{\hat{r}%
},\ \nabla \mathring{\varphi}\left( \left\vert \mathbf{r}\right\vert \right)
=-\left\vert \nabla \mathring{\varphi}\right\vert \mathbf{\hat{x}},\ \mathbf{%
\hat{x}}=\frac{\mathbf{x}}{\left\vert \mathbf{x}\right\vert }=\left( \hat{x}%
_{1},\hat{x}_{2},\hat{x}_{3}\right) .  \label{nstrc9}
\end{equation}%
The relations (\ref{nstrc9}) combined with (\ref{nstrc5})-(\ref{nstrc7})
imply that for the resting charge all the forces $\mathbf{f}^{\mathrm{el}}$, 
$\mathbf{f}^{\mathrm{em}}$ and $\mathbf{f}^{\mathrm{nl}}$ and radial, i.e.
they are functions of $\left\vert \mathbf{x}\right\vert $ and point toward
or outward the origin.

The total energy of the resting dressed charge $\mathcal{E}\left( \mathring{%
\psi}\right) $ can be estimated based on either canonical energy-momentum or
the gauge invariant one. If we use the canonical energy-momentum tensors
defined by (\ref{emfr1}), (\ref{emfr2}) and (\ref{emfr6})-(\ref{emfr8}) we
find the following expressions for the respectively the charge energy
density $\mathring{u}$, the EM field energy density $\mathring{w}$ and the
total energy density of the dressed charge $\mathring{u}+\mathring{w}$:%
\begin{equation}
\mathring{u}+\mathring{w}=\left( \tilde{u}+\mathring{\psi}^{2}\mathring{%
\varphi}\right) +\mathring{w}=\frac{\chi ^{2}}{2m}\left[ \left( \nabla 
\mathring{\psi}\right) ^{2}+G\left( \mathring{\psi}^{2}\right) \right] +%
\mathring{\psi}^{2}\mathring{\varphi}-\frac{\left( \nabla \mathring{\varphi}%
\right) ^{2}}{8\pi }.  \label{nstrc9a}
\end{equation}%
Using now the results of Section \ref{senrgypart} including the relation (%
\ref{epar12}) we obtain the following representation for the total energy of
the resting dressed charge 
\begin{equation}
\mathcal{E}\left( \mathring{\psi}\right) =\int_{\mathbb{R}^{3}}\left( 
\mathring{u}+\mathring{w}\right) \,\mathrm{d}\mathbf{x}=\frac{2}{3}\int_{%
\mathbb{R}^{3}}\left[ \frac{\chi ^{2}\left( \nabla \mathring{\psi}\right)
^{2}}{2m}-\frac{\left( \nabla \mathring{\varphi}\right) ^{2}}{8\pi }\right]
\,\mathrm{d}\mathbf{x}.  \label{nstrc10}
\end{equation}%
If we wanted to use the gauge invariant energy-momentum tensors (\ref{emfr3}%
)-(\ref{emfr10}) for the same evaluation, a choice for $w_{0}\left( \mathbf{x%
}\right) $ in (\ref{emfr5}) consistent with the canonical energy-momentum
would be $w_{0}\left( \mathbf{x}\right) =\mathring{w}=-\frac{\left( \nabla
\varphi \right) ^{2}}{8\pi }$.

\subsection{Freely moving charge\label{snfreemov}}

We can use the invariance of the Lagrangian $\hat{L}_{0}$ with respect to
Galilean-gauge transformations (\ref{gal1})-(\ref{gal2}) to obtain a freely
moving charge solution to the field equations (\ref{nore3a})-(\ref{nore3b})
based on the resting charge solution (\ref{nop4})-(\ref{nop5}) similarly to
what is done in the relativistic case where we obtain a freely moving charge
solution applying Lorentz transformation to the resting one. Namely, the
field equations (\ref{nop2})-(\ref{nop3}) have the following closed form
solution%
\begin{gather}
\psi =\psi \left( t,\mathbf{x}\right) =\mathrm{e}^{\mathrm{i}S/\chi }%
\mathring{\psi}\left( \left\vert \mathbf{x}-\mathbf{v}t\right\vert \right) ,
\label{nexac1} \\
S=\frac{m}{2}\left[ \mathbf{v}^{2}t+2\mathbf{v}\cdot \left( \mathbf{x}-%
\mathbf{v}t\right) \right] ,\ \varphi \left( t,\mathbf{x}\right) =\mathring{%
\varphi}_{0}\left( \left\vert \mathbf{x}-\mathbf{v}t\right\vert \right) , 
\notag
\end{gather}%
where $\psi $ in view of relations (\ref{exac2}) can be also represented as%
\begin{equation}
\psi =\psi \left( t,\mathbf{x}\right) =\mathrm{e}^{\mathrm{i}S/\chi }%
\mathring{\psi}\left( \left\vert \mathbf{x}-\mathbf{v}t\right\vert \right)
,\ S=\mathbf{p}\cdot \mathbf{x}-\frac{\mathbf{p}^{2}t}{2m},\ \mathbf{p}=m%
\mathbf{v}.  \label{nexac1a}
\end{equation}%
Solutions of a similar form propagating with a constant speed are well-known
in the theory of Nonlinear Schr\"{o}dinger equations, see \cite{Sulem} and
references therein. In what follows we refer to a wave function represented
by the formulas (\ref{nexac1}), (\ref{nexac1a}) as a \emph{wave-corpuscle}.
Looking at the exact solution (\ref{nexac1}), (\ref{nexac1a}) to the field
equations describing the \emph{freely moving charge} we observe that it
harmoniously integrates the features of the point charge. Indeed, the wave
amplitude $\mathring{\psi}\left( \left\vert \mathbf{x}-\mathbf{v}%
t\right\vert \right) $ in (\ref{nexac1}) is a soliton-like field moving
exactly as a free point charge described by its position $\mathbf{r}=\mathbf{%
v}t$. \emph{The exponential factor }$\mathrm{e}^{\mathrm{i}S/\chi }$\emph{\
is a plane wave with the phase }$S$ \emph{that depends only on the point
charge position }$\mathbf{v}t$\emph{\ and momentum }$\mathbf{p}=m\mathbf{v}$%
\emph{, and it does not depend on the nonlinear self-interaction}. The phase 
$S$ has a term in which we readily recognize the de Broglie wave-vector $%
\mathbf{k}$ described exactly in terms of the point charge quantities, namely%
\begin{equation}
\mathbf{k}=\frac{\mathbf{p}}{\chi }=\frac{m}{\chi }\mathbf{v}.
\label{nexac4}
\end{equation}%
Notice that the dispersion relation $\omega =\omega \left( \mathbf{k}\right) 
$ of the linear part of the field equations (\ref{nore3a}) for $\psi $ is 
\begin{equation}
\omega \left( \mathbf{k}\right) =\frac{\chi \mathbf{k}^{2}}{2m},\text{
implying that the group velocity }\nabla _{\mathbf{k}}\omega \left( \mathbf{k%
}\right) =\frac{\chi \mathbf{k}}{m}.  \label{nexac5}
\end{equation}%
Combining the expression (\ref{nexac5}) for the group velocity $\nabla _{%
\mathbf{k}}\omega \left( \mathbf{k}\right) $ with the expression (\ref%
{nexac4}) for wave vector $\mathbf{k}$ we establish another exact relation 
\begin{equation}
\mathbf{v}=\nabla _{\mathbf{k}}\omega \left( \mathbf{k}\right) ,
\label{nexac6}
\end{equation}%
signifying the equality between the point charge velocity $\mathbf{v}$ and
the group velocity $\nabla _{\mathbf{k}}\omega \left( \mathbf{k}\right) $ at
the de Broglie wave vector $\mathbf{k}$. Using the relations (\ref{nop8})
and (\ref{gal9}) we readily obtain the following representations for the
micro-charge, the micro-current and momentum densities 
\begin{gather}
\rho \left( t,\mathbf{x}\right) =q\mathring{\psi}^{2}\left( \left\vert 
\mathbf{x}-\mathbf{v}t\right\vert \right) ,\text{ }\mathbf{J}\left( t,%
\mathbf{x}\right) =q\mathbf{v}\mathring{\psi}^{2}\left( \left\vert \mathbf{x}%
-\mathbf{v}t\right\vert \right) ,  \label{nexac6a} \\
\mathbf{P}\left( t,\mathbf{x}\right) =\frac{m}{q}\mathbf{J}\left( t,\mathbf{x%
}\right) =\mathbf{p}\mathring{\psi}^{2}\left( \left\vert \mathbf{x}-\mathbf{v%
}t\right\vert \right) .  \label{nexac6b}
\end{gather}%
The above expressions and the charge normalization condition (\ref{nrac4})
readily imply the following representations for the total dressed charge
field momentum $\mathsf{P}$ and the total current $\mathsf{J}$ for the
solution (\ref{exac1}) in terms of point charge quantities, namely%
\begin{equation}
\mathsf{P}=\frac{m}{q}\mathsf{J}=\dint_{\mathbb{R}^{3}}\frac{\chi q}{m}\func{%
Im}\frac{\nabla \mathring{\psi}}{\mathring{\psi}}\,\mathring{\psi}^{2}\,%
\mathrm{d}\mathbf{x}=\mathbf{p}=m\mathbf{v}.  \label{nexac7}
\end{equation}

\subsection{Nonlinear self-interaction and its basic properties}

As we have already explained in the beginning of Section \ref{snfree} the 
\emph{nonlinear self interaction function }$G$\emph{\ is determined from the
charge equilibrium\ equation} (\ref{nop7}) based on the form factor $%
\mathring{\psi}$ and the form factor potential $\mathring{\varphi}$. It is
worth to point out that such a nonlinearity differs significantly from
nonlinearities considered in similar problems in literature. Important
features of our nonlinearity include: (i) the boundedness of its derivative $%
G^{\prime }\left( s\right) $ for $s\geq 0$ with consequent boundedness from
below of the wave energy; (ii) non analytic behavior for small $s$ that is
for small wave amplitudes.

In this section we consider the construction of the function $G$, study its
properties and provide examples for which the construction of $G$ is carried
out explicitly. Throughout this section we have 
\begin{equation*}
\psi ,\mathring{\psi}\geq 0\text{ and hence }\left\vert \psi \right\vert
=\psi .
\end{equation*}%
We introduce explicitly the \emph{size parameter} $a>0$ through the
following representation of the fundamental functions $\mathring{\psi}\left(
r\right) $ and $\mathring{\varphi}\left( r\right) $%
\begin{gather}
\mathring{\psi}\left( r\right) =\mathring{\psi}_{a}\left( r\right) =a^{-3/2}%
\mathring{\psi}_{1}\left( a^{-1}r\right) ,\   \label{nrac5} \\
\mathring{\varphi}\left( r\right) =\mathring{\varphi}_{a}\left( r\right)
=a^{-1}\mathring{\varphi}_{1}\left( a^{-1}r\right) ,\   \notag
\end{gather}%
where $\mathring{\psi}_{1}\left( \mathsf{r}\right) $ and $\mathring{\varphi}%
_{1}\left( \mathsf{r}\right) $ are functions of the dimensionless parameter $%
\mathsf{r}$, and, as a consequence of (\ref{nrac4}), the function $\mathring{%
\psi}_{a}\left( r\right) $ satisfies the charge normalization condition%
\begin{equation}
\dint\nolimits_{\mathbb{R}^{3}}\mathring{\psi}_{a}^{2}\left( \left\vert 
\mathbf{x}\right\vert \right) \,\mathrm{d}\mathbf{x}=1\text{ for all }a>0.
\label{nrac6}
\end{equation}%
The \emph{size parameter} $a$ naturally has the dimension of length, but we
do not yet identify it with the size. Indeed, any properly defined spatial
size of $\mathring{\psi}_{a}$, based, for instance, on the variance or on an
energy-based scale as in (\ref{apsi}), is proportional to $a$\ with a
coefficient depending on $\mathring{\psi}_{1}$. The charge equilibrium\
equation (\ref{nop7}) can be written in the following form 
\begin{gather}
-\frac{\chi ^{2}}{2m}\nabla ^{2}\mathring{\psi}_{a}+q\mathring{\varphi}_{a}%
\mathring{\psi}_{a}+\frac{\chi ^{2}}{2m}G_{a}^{\prime }\left( \mathring{\psi}%
_{a}^{2}\right) \mathring{\psi}_{a}=0,  \label{stp} \\
-\nabla ^{2}\mathring{\varphi}_{a}=4\pi q\left\vert \mathring{\psi}%
_{a}\right\vert ^{2}.  \label{stpfi}
\end{gather}%
The function $\mathring{\psi}_{a}\left( r\right) $ is assumed to be a
positive, monotonically decreasing function of $r\geq 0$, and to satisfy the
charge normalization condition (\ref{nrac6}). Recall that $\mathring{\psi}%
_{a}\left( \left\vert \mathbf{x}\right\vert \right) $ and $\mathring{\varphi}%
_{a}\left( \left\vert \mathbf{x}\right\vert \right) $ are radial functions
and consequently, when solving the equation (\ref{stpfi}) for $\mathring{%
\varphi}_{a},$ \ we obtain (see Section \ref{scoulomb} and (\ref{ficr}) for
details) the formula 
\begin{equation}
\mathring{\varphi}_{a}\left( r\right) =\frac{q}{r}\left[ 1-\frac{4\pi }{a}%
\int_{r/a}^{\infty }\left( r_{1}-\frac{r}{a}\right) r_{1}\mathring{\psi}%
_{1}^{2}\left( r_{1}\right) \,\mathrm{d}r_{1}\right] .  \label{fiexp}
\end{equation}%
Obviously, if $\mathring{\psi}_{1}^{2}\left( r\right) $ decays sufficiently
fast as $r\rightarrow \infty $ and $a$ is \ sufficiently small then the
potential $\mathring{\varphi}_{a}\left( r\right) $ is very close to the
Coulomb's potential $q/r,$ as we show in Section \ref{scoulomb}.

Let us look first at the case $a=1$,\ $\mathring{\psi}_{a}=\mathring{\psi}%
_{1}$,\ $\mathring{\varphi}_{a}=\mathring{\varphi}_{1}$, for which the
equation (\ref{stp}) yields the following representation for $G^{\prime }(%
\mathring{\psi}_{1}^{2})$ from (\ref{stp})%
\begin{equation}
G^{\prime }\left( \mathring{\psi}_{1}^{2}\left( r\right) \right) =\frac{%
(\nabla ^{2}\mathring{\psi}_{1})\left( r\right) }{\mathring{\psi}_{1}\left(
r\right) }-\frac{2m}{\chi ^{2}}q\mathring{\varphi}_{1}\left( r\right) .
\label{gg}
\end{equation}%
Since $\mathring{\psi}_{1}^{2}\left( r\right) $ is a monotonic function,\ we
can find its inverse $r=r\left( \psi ^{2}\right) ,$ yielding 
\begin{equation}
G^{\prime }\left( s\right) =\left[ \frac{\nabla ^{2}\mathring{\psi}_{1}}{%
\mathring{\psi}_{1}}-\frac{2m}{\chi ^{2}}q\mathring{\varphi}_{1}\right]
\left( r\left( s\right) \right) ,\ 0=\mathring{\psi}_{1}^{2}\left( \infty
\right) \leq s\leq \mathring{\psi}_{1}^{2}\left( 0\right) .  \label{intps}
\end{equation}%
We extend $G^{\prime }\left( s\right) $ for $s\geq \mathring{\psi}%
_{1}^{2}\left( 0\right) $ to be a constant, namely 
\begin{equation}
G^{\prime }\left( s\right) =G^{\prime }\left( \mathring{\psi}_{1}^{2}\left(
\infty \right) \right) \text{ if }s\geq \mathring{\psi}_{1}^{2}\left( \infty
\right) .  \label{intps1}
\end{equation}%
\emph{Observe that the positivity and the monotonicity of the form factor }$%
\mathring{\psi}_{1}$\emph{\ was instrumental for recovering the function }$%
G^{\prime }\left( s\right) $\emph{\ from the charge balance equation (\ref%
{stp})}.

Using the representation (\ref{intps}) for the function $G^{\prime }\left(
s\right) $ we decompose it naturally into two components: 
\begin{equation}
G^{\prime }\left( s\right) =G_{\nabla }^{\prime }\left( s\right) -\frac{2}{%
a_{\chi }}G_{\varphi }^{\prime }\left( s\right) ,\text{ }  \label{gdet}
\end{equation}%
where 
\begin{equation}
a_{\chi }=\frac{\chi ^{2}}{mq^{2}}  \label{achi}
\end{equation}%
and for all $r\geq 0$ 
\begin{equation}
G_{\nabla }^{\prime }\left( \mathring{\psi}_{1}^{2}\right) \left( r\right) =%
\frac{\left( \nabla ^{2}\mathring{\psi}_{1}\right) }{\mathring{\psi}_{1}}%
\left( r\right) ,\ G_{\varphi }^{\prime }\left( \mathring{\psi}%
_{1}^{2}\right) \left( r\right) =\frac{\mathring{\varphi}_{1}\left( r\right) 
}{q}=\mathring{\phi}_{1}\left( r\right) .  \label{gdet2}
\end{equation}%
We refer to $G_{\nabla }^{\prime }\left( s\right) $ and $G_{\varphi
}^{\prime }\left( s\right) $ as\emph{\ elastic and EM components}
respectively. In the case of arbitrary size parameter $a$ we find first that%
\begin{equation}
G_{\nabla ,a}^{\prime }\left( s\right) =a^{-2}G_{\nabla ,1}^{\prime }\left(
a^{3}s\right) ,\ G_{\varphi ,a}^{\prime }\left( s\right) =a^{-1}G_{\varphi
,1}^{\prime }\left( a^{3}s\right) ,\ a>0,  \label{gkap}
\end{equation}%
and then combining (\ref{gkap}) with (\ref{gdet}) and (\ref{gdet2}) we
obtain the following representation for the function $G_{a}^{\prime }\left(
s\right) $%
\begin{equation}
G_{a}^{\prime }\left( s\right) =\frac{G_{\nabla ,1}^{\prime }\left(
a^{3}s\right) }{a^{2}}-\frac{2G_{\varphi ,1}^{\prime }\left( a^{3}s\right) }{%
aa_{\chi }}.  \label{totgkap}
\end{equation}

Let us take a look at general properties of $G^{\prime }\left( s\right) $
and its components $G_{\nabla }^{\prime }\left( s\right) $ and $G_{\varphi
}^{\prime }\left( s\right) $ as they follow from defining them relations (%
\ref{intps})-(\ref{totgkap}). Starting with the EM component $G_{\varphi
}^{\prime }\left( s\right) $ we notice that $\mathring{\varphi}_{1}\left(
\left\vert \mathbf{x}\right\vert \right) $ is a radial solution to the
equation (\ref{stpfi}). Combining that with $\psi ^{2}\geq 0$ and using the
Maximum principle we conclude that $\mathring{\varphi}_{1}\left( \left\vert 
\mathbf{x}\right\vert \right) /q$\emph{\ is a positive function without
local minima, implying that it is a monotonically decreasing function of }$%
\left\vert \mathbf{x}\right\vert $. Consequently, $G_{\varphi }^{\prime
}\left( s\right) $ defined by (\ref{gdet2}) is a monotonically increasing
function of $s$, and hence 
\begin{equation}
G_{\varphi }^{\prime }\left( s\right) >0\text{ for all\ }s>0\text{ and }%
G_{\varphi }^{\prime }\left( 0\right) =0.  \label{gfi0}
\end{equation}%
\ Note that $G_{\varphi }^{\prime }\left( s\right) $\emph{\ is not
differentiable at zero}, which can be seen by comparing the behavior of $%
\mathring{\varphi}_{1}\left( r\right) $ and $\mathring{\psi}_{1}\left(
r\right) $ at infinity. Indeed, $\varphi _{1}\left( r\right) /q\sim r^{-1}$
as $r\rightarrow \infty $ and since $\mathring{\psi}^{2}\left( \left\vert 
\mathbf{x}\right\vert \right) $\ is integrable, it has to decay faster than $%
\left\vert \mathbf{x}\right\vert ^{-3}$ as $\left\vert \mathbf{x}\right\vert
\rightarrow \infty $.\ Consequently, $\left\vert G_{\varphi }^{\prime
}\left( s\right) \right\vert $ for small $s$ has to be greater than $s^{1/3}$
which prohibits its the differentiability at zero. One has to notice though
that the nonlinearity $G^{\prime }\left( \left\vert \psi \right\vert
^{2}\right) \psi $ as it enters the field equation (\ref{nore5}) is
differentiable for all $\psi $ including zero, hence it satisfies\ a
Lipschitz condition required for uniqueness of solutions of initial value
problem for (\ref{nore5}).

Let us look at the elastic component $G_{\nabla }^{\prime }\left( s\right) $
defined by the relations (\ref{gdet2}). Since $\mathring{\psi}\left(
\left\vert \mathbf{x}\right\vert \right) >0$ the sign of $G_{\nabla
}^{\prime }\left( \left\vert \psi \right\vert ^{2}\right) $ \ coincides with
the sign of $\nabla ^{2}\mathring{\psi}_{1}\left( \left\vert \mathbf{x}%
\right\vert \right) $. At the origin $\mathbf{x}=\mathbf{0}$ the function $%
\mathring{\psi}_{1}\left( \left\vert \mathbf{x}\right\vert \right) $ has its
maximum and, consequently, $G_{\nabla }^{\prime }\left( s\right) \leq 0$ for
all $s$ close to $s=\mathring{\psi}_{1}^{2}\left( \infty \right) $, implying 
\begin{equation}
G_{\nabla }^{\prime }\left( s\right) \leq 0\text{ for }s\gg 1.  \label{ggr0}
\end{equation}%
The Laplacian applied to radial functions $\mathring{\psi}_{1}$ takes the
form $\frac{1}{r}\frac{\partial ^{2}}{\partial r^{2}}\left( r\mathring{\psi}%
_{1}\left\vert \mathbf{x}\right\vert \right) $. Consequently, if \ $r%
\mathring{\psi}_{1}\left( r\right) $ is convex at $r=\left\vert \mathbf{x}%
\right\vert $ we have $\nabla ^{2}\mathring{\psi}_{1}\left( \left\vert 
\mathbf{x}\right\vert \right) \geq 0$. Since $r^{2}\mathring{\psi}\left(
r\right) $ is integrable we can naturally assume that $\left\vert \mathbf{x}%
\right\vert \mathring{\psi}_{1}\left( \left\vert \mathbf{x}\right\vert
\right) \rightarrow 0$ as $\left\vert \mathbf{x}\right\vert \rightarrow
\infty $. Then if the second derivative of $r\mathring{\psi}_{1}\left(
r\right) $ has a constant sign near infinity, it must be non-negative. For
an exponentially decaying $\mathring{\psi}_{1}\left( r\right) $ the second
derivative of $r\mathring{\psi}_{a}\left( r\right) $ is positive implying 
\begin{equation}
G_{\nabla }^{\prime }\left( s\right) >0\text{ for }s\ll 1.  \label{ggr1}
\end{equation}%
Combining this with the equality $G_{\varphi }^{\prime }\left( 0\right) =0$
from (\ref{gfi0})\ we readily obtain 
\begin{equation}
G^{\prime }\left( s\right) >0\text{ for }s\ll 1.  \label{ggr2}
\end{equation}%
>From the relations (\ref{gdet}), (\ref{gfi0}), (\ref{ggr0}) we also obtain%
\begin{equation}
G^{\prime }\left( s\right) <0\ \text{if }s\gg 1.  \label{gg0}
\end{equation}%
We remind the reader that the sign of the $G^{\prime }\left( s\right) $
according to the representation (\ref{nstrc7}) for nonlinear
self-interaction force density $f_{i}^{\mathrm{nl}}$ controls its direction.

\subsection{Examples of nonlinearities\label{Examnon}}

In this section we provide two examples of the form factor $\mathring{\psi}$
for which the form factor potential $\mathring{\varphi}_{0}$ and the
corresponding nonlinear self-interaction function $G$ can be constructed
explicitly. The first example is for the form factor $\mathring{\psi}\left(
r\right) $ decaying as a power law as $r\rightarrow \infty $. In this case
both $\mathring{\varphi}_{0}$ and $G$ are represented by rather simple,
explicit formulas, but some properties of these functions are not as
appealing. Namely, the variance of the function $\mathring{\psi}$ is
infinite and the rate of approximation of the exact Coulomb's potential by $%
\mathring{\varphi}_{a}\left( \mathbf{x}\right) $ for small $a\ $is not as
fast. The second example is for the form factor $\mathring{\psi}\left(
r\right) $ decaying exponentially as $r\rightarrow \infty $. In this case
the representations for $\mathring{\varphi}_{0}$ and $G$ are more involved
compared with the power law form factor but all the properties of $\mathring{%
\psi}$ and $\mathring{\varphi}$ are satisfactory in any regard.

\subsubsection{Nonlinearity for the form factor decaying as a power law}

We introduce here a form factor $\mathring{\psi}_{1}\left( r\right) $
decaying as a power law of the form 
\begin{equation}
\mathring{\psi}_{1}\left( r\right) =\frac{c_{\mathrm{pw}}}{\left(
1+r^{2}\right) ^{5/4}},\   \label{expsi1}
\end{equation}%
where $c_{\mathrm{pw}}$ \ is the normalization factor, 
\begin{equation*}
c_{\mathrm{pw}}=\frac{3^{1/2}}{\left( 4\pi \right) ^{1/2}}\text{.}
\end{equation*}%
This function evidently is positive and monotonically decreasing as
required. Let us find now $G_{\nabla }^{\prime }\left( s\right) $ and $%
G_{\varphi }^{\prime }$ based on the relations (\ref{gdet2}). An elementary
computation shows that%
\begin{equation*}
\nabla ^{2}\mathring{\psi}_{1}=\frac{15}{4c_{\mathrm{pw}}^{4/5}}\left( 1-%
\frac{3}{c_{\mathrm{pw}}^{4/5}}\mathring{\psi}_{1}^{4/5}\right) \mathring{%
\psi}_{1}^{1+4/5},\text{ }
\end{equation*}%
implying%
\begin{equation}
G_{\nabla }^{\prime }\left( s\right) =\frac{15s^{2/5}}{4c_{\mathrm{pw}}^{4/5}%
}-\frac{45s^{4/5}}{4c_{\mathrm{pw}}^{8/5}},\ G_{\nabla }\left( s\right) =%
\frac{75s^{7/5}}{28c_{\mathrm{pw}}^{4/5}}-\frac{25s^{9/5}}{4c_{\mathrm{pw}%
}^{8/5}},\text{ for }0\leq s\leq c_{\mathrm{pw}}^{2}.  \label{exGd1}
\end{equation}%
To determine $G_{\varphi }^{\prime }$ we find by a straightforward
examination that function 
\begin{equation}
\mathring{\varphi}_{1}=\frac{q}{c_{\mathrm{pw}}^{2/5}}\mathring{\psi}%
_{1}^{2/5}\text{ }  \label{exfi1}
\end{equation}%
solves \ equation 
\begin{equation*}
\nabla ^{2}\mathring{\varphi}_{1}=-4\pi q\mathring{\psi}_{1}^{2},
\end{equation*}
\ and that together with (\ref{gdet2}) yields 
\begin{equation}
G_{\varphi }^{\prime }\left( s\right) =\frac{s^{1/5}}{c_{\mathrm{pw}}^{2/5}}%
,\ G_{\varphi }\left( s\right) =\frac{5s^{6/5}}{6c_{\mathrm{pw}}^{2/5}},%
\text{\ for }0\leq s\leq c_{\mathrm{pw}}^{2}.  \label{exGfi1}
\end{equation}%
Observe that the both components $G_{\nabla }^{\prime }\left( s\right) $ and 
$G_{\varphi }^{\prime }\left( s\right) $ in (\ref{exGd1}), (\ref{exGfi1}) of
the total nonlinearity $G_{\nabla }^{\prime }\left( s\right) $ defined by (%
\ref{gdet}) \emph{are not differentiable} at $s=0$.

If we explicitly introduce size parameter $a$ into the form factor, namely 
\begin{equation}
\psi _{a}\left( r\right) =\frac{ac_{\mathrm{pw}}}{\left( a^{2}+r^{2}\right)
^{5/4}},  \label{exGfi2}
\end{equation}%
then combining (\ref{exGd1}), (\ref{exGfi1}) with (\ref{gkap}) we obtain the
following representation for the nonlinearity components%
\begin{gather}
G_{\nabla ,a}^{\prime }\left( s\right) =\frac{15s^{2/5}}{4a^{4/5}c_{\mathrm{%
pw}}^{4/5}}-\frac{45a^{2/5}s^{4/5}}{4c_{\mathrm{pw}}^{8/5}},
\label{exGd1kap} \\
G_{\nabla ,a}\left( s\right) =\frac{75s^{7/5}}{28a^{4/5}c_{\mathrm{pw}}^{4/5}%
}-\frac{25a^{2/5}s^{9/5}}{4c_{\mathrm{pw}}^{8/5}}\text{\ for }0\leq s\leq c_{%
\mathrm{pw}}^{2}a^{-3},  \notag
\end{gather}%
\ 
\begin{equation}
G_{\varphi ,a}^{\prime }\left( s\right) =\frac{s^{1/5}}{a^{2/5}c_{\mathrm{pw}%
}^{2/5}},\ G_{\varphi }\left( s\right) =\frac{5s^{6/5}}{6a^{2/5}c_{\mathrm{pw%
}}^{2/5}}\text{\ for }0\leq s\leq c_{\mathrm{pw}}^{2}a^{-3}.
\label{exfi1cap}
\end{equation}%
Notice that the variance of the form factor $\mathring{\psi}_{1}^{2}\left(
\left\vert \mathbf{x}\right\vert \right) $ decaying as a power law (\ref%
{expsi1}) is infinite, i.e.%
\begin{equation}
\int_{\mathbb{R}^{3}}\left\vert \mathbf{x}\right\vert ^{2}\mathring{\psi}%
_{1}^{2}\left( \left\vert \mathbf{x}\right\vert \right) \,\mathrm{d}\mathbf{x%
}=4\pi \int_{0}^{\infty }\frac{c_{\mathrm{pw}}^{2}}{\left( 1+r^{2}\right)
^{5/2}}r^{4}\,\mathrm{d}r=3\int_{0}^{\infty }\frac{1}{\left( 1+r^{2}\right)
^{5/2}}r^{4}\,\mathrm{d}r=\infty .  \label{sol2}
\end{equation}

\subsubsection{Nonlinearity for the form factor decaying exponentially}

We introduce here an exponentially decaying form factor $\mathring{\psi}_{1}$
of the form 
\begin{equation}
\mathring{\psi}_{1}\left( r\right) =c_{\mathrm{e}}\mathrm{e}^{-\left(
r^{2}+1\right) ^{1/2}}\text{ }  \label{grs}
\end{equation}%
where $c_{\mathrm{e}}$ is the normalization factor, 
\begin{equation*}
c_{\mathrm{e}}=\left( 4\pi \int_{0}^{\infty }r^{2}\mathrm{e}^{-2\left(
r^{2}+1\right) ^{1/2}}\,\mathrm{d}r\right) ^{-1/2}\simeq 0.79195,
\end{equation*}%
Evidently $\mathring{\psi}_{1}\left( r\right) $ is positive and
monotonically decreasing as required. The dependence $r\left( s\right) $
defined by the relation (\ref{grs})\ is as follows:\ 
\begin{equation}
r=\left[ \ln ^{2}\left( c_{\mathrm{e}}/\sqrt{s}\right) -1\right] ^{1/2},%
\text{ if }\sqrt{s}\leq \mathring{\psi}_{1}\left( 0\right) =c_{\mathrm{e}}%
\mathrm{e}^{-1}.  \label{rofp}
\end{equation}%
An elementary computation shows that $\ \nabla ^{2}\mathring{\psi}_{1}=-W%
\mathring{\psi}_{1}$ where \ 
\begin{equation}
\text{ }W=\frac{2}{\left( r^{2}+1\right) ^{\frac{1}{2}}}+\frac{1}{\left(
r^{2}+1\right) }+\frac{1}{\left( r^{2}+1\right) ^{\frac{3}{2}}}-1,
\label{rofp1}
\end{equation}%
implying%
\begin{equation}
G_{\nabla }^{\prime }\left( \psi _{1}^{2}\left( r\right) \right) =-W\left(
r\right) =1-\frac{2}{\left( r^{2}+1\right) ^{\frac{1}{2}}}-\frac{1}{\left(
r^{2}+1\right) }-\frac{1}{\left( r^{2}+1\right) ^{\frac{3}{2}}}.
\label{rofp2}
\end{equation}%
Combining (\ref{rofp}) with (\ref{rofp2}) we readily obtain the following
function \ for $\sqrt{s}\leq c_{\mathrm{e}}\mathrm{e}^{-1}\simeq 0.29134$ 
\begin{equation}
G_{\nabla ,1}^{\prime }\left( s\right) =\left[ 1-\frac{4}{\ln \left( c_{%
\mathrm{e}}^{2}/s\right) }-\frac{4}{\ln ^{2}\left( c_{\mathrm{e}%
}^{2}/s\right) }-\frac{8}{\ln ^{3}\left( c_{\mathrm{e}}^{2}/s\right) }\right]
,\text{ }  \label{ggkap}
\end{equation}%
which is evidently monotonically decreasing. We extend it for larger $s$ as
follows:%
\begin{equation}
G_{\nabla ,1}^{\prime }\left( s\right) =G_{\nabla ,1}^{\prime }\left( c_{%
\mathrm{e}}^{2}\mathrm{e}^{-2}\right) =-3\text{ if }\ \sqrt{s}\geq c_{%
\mathrm{e}}\mathrm{e}^{-1}.  \label{ggkap1}
\end{equation}%
The relations (\ref{ggkap}) and (\ref{ggkap1}) imply $G_{\nabla ,1}^{\prime
}\left( s\right) $ takes values in the interval $\left[ 1,-3\right] $. It
also follows from (\ref{ggkap}) that%
\begin{equation}
G_{\nabla ,1}^{\prime }\left( s\right) \cong 1-\frac{4}{\ln 1/s}\text{ as }%
s\rightarrow 0,  \label{ggkap2}
\end{equation}%
implying that the function $G_{\nabla ,1}^{\prime }\left( s\right) $ \emph{%
is not differentiable} at $s=0$ and consequently is not analytic about $0$.

\ To determine the second component $G_{\varphi }^{\prime }$ we need to
solve (\ref{gdet2}). Using the fact that $\mathring{\varphi}_{1}$, $%
\mathring{\psi}_{1}$ are radial functions we obtain the following equation
for $\mathring{\varphi}_{1}\left( r\right) $ 
\begin{equation}
-\frac{1}{4\pi r}\partial _{r}^{2}\left( r\mathring{\varphi}_{1}\right) =qc_{%
\mathrm{e}}^{2}\mathrm{e}^{-2\sqrt{r^{2}+1}}.  \label{ggkap3}
\end{equation}%
We seek a solution of equation (\ref{ggkap3}) that is regular at zero and
behaves like the Coulomb's potential $\frac{q}{r}$ for large $r$. Taking
that into account we obtain after the first integration of (\ref{ggkap3}) 
\begin{equation}
\partial _{r}\left( r\mathring{\varphi}_{1}\right) =\pi qc_{\mathrm{e}}^{2}%
\left[ 1+2\left( r^{2}+1\right) ^{1/2}\right] \mathrm{e}^{-2\left(
r^{2}+1\right) ^{1/2}},  \label{drrfi}
\end{equation}%
and integrating (\ref{drrfi}) yields the ultimate formula for the form
factor potential: 
\begin{eqnarray}
\mathring{\varphi}_{1}\left( r\right) &=&\frac{\pi qc_{\mathrm{e}}^{2}}{r}%
\int_{0}^{r}\left[ 1+2\left( r_{1}^{2}+1\right) ^{1/2}\right] \mathrm{e}%
^{-2\left( r_{1}^{2}+1\right) ^{1/2}}\,\mathrm{d}r_{1}  \label{drrfi1} \\
&=&\frac{q}{r}-\frac{\pi qc_{\mathrm{e}}^{2}}{r}\int_{r}^{\infty }\left[
1+2\left( r_{1}^{2}+1\right) ^{1/2}\right] \mathrm{e}^{-2\left(
r_{1}^{2}+1\right) ^{1/2}}\,\mathrm{d}r_{1}.  \notag
\end{eqnarray}%
The above formula shows that the form factor potential $\mathring{\varphi}%
_{1}\left( r\right) $ is exponentially close to the Coulomb's potential $q/r$
for large $r$. But if we use the substitution $\left( r_{1}^{2}+1\right)
^{1/2}=u$ in the second integral in (\ref{drrfi1}) we can recast $\mathring{%
\varphi}_{1}\left( r\right) $ in even more convenient form for estimations
of its proximity\ to the Coulomb's potential $q/r$, namely 
\begin{align}
\mathring{\varphi}_{1}\left( r\right) & =\frac{q}{r}-\frac{\pi qc_{\mathrm{e}%
}^{2}\left[ 1+2\left( r^{2}+1\right) ^{1/2}\right] }{2r}\mathrm{e}^{-2\left(
r^{2}+1\right) ^{1/2}}  \label{drrfi1a} \\
& -\frac{\pi qc_{\mathrm{e}}^{2}}{r}\int_{\left( r^{2}+1\right)
^{1/2}}^{\infty }\left[ \frac{\left( 2u+1\right) }{\left( 1-u^{-2}\right)
^{1/2}}-2u\right] \mathrm{e}^{-2u}\,\mathrm{d}u.  \notag
\end{align}%
Then, based on the relation (\ref{gdet2}) and (\ref{drrfi1}), we find
consequently 
\begin{equation}
G_{\varphi ,1}^{\prime }\left( \psi ^{2}\right) =\frac{1}{q}\mathring{\varphi%
}_{1}\left( r\left( \psi \right) \right) =\frac{\pi c_{\mathrm{e}}^{2}}{%
r\left( \psi \right) }\int_{0}^{r\left( \psi \right) }\left[ 1+2\left(
r_{1}^{2}+1\right) ^{1/2}\right] \mathrm{e}^{-2\left( r_{1}^{2}+1\right)
^{1/2}}\,\mathrm{d}r_{1},\text{ }  \label{drrfi3}
\end{equation}%
where%
\begin{equation*}
r\left( \psi \right) =\left[ \ln ^{2}\left( c_{\mathrm{e}}/\psi \right) -1%
\right] ^{1/2}\text{\ for }\psi \leq c_{\mathrm{e}}\mathrm{e}^{-1}.
\end{equation*}%
We extend $G_{\varphi ,1}^{\prime }\left( \psi ^{2}\right) $ for larger
values of $\psi $ as a constant: 
\begin{equation}
G_{\varphi ,1}^{\prime }\left( \psi ^{2}\right) =\lim_{r\rightarrow 0}\frac{%
\mathring{\varphi}\left( r\right) }{q}=3\pi c_{\mathrm{e}}^{2}\mathrm{e}%
^{-2}\simeq 0.79998,\text{ for }\psi \geq c_{\mathrm{e}}\mathrm{e}^{-1}.
\label{drrfi4}
\end{equation}%
Using the representation (\ref{drrfi1a}) we obtain the following formula for 
$G_{\varphi ,1}^{\prime }$ 
\begin{align}
G_{\varphi ,1}^{\prime }\left( \psi ^{2}\right) & =\frac{1}{\left[ \ln
^{2}\left( c_{\mathrm{e}}/\psi \right) -1\right] ^{1/2}}-\frac{\pi \left[
1+2\ln \left( c_{\mathrm{e}}/\psi \right) \right] \psi ^{2}}{2\left[ \ln
^{2}\left( c_{\mathrm{e}}/\psi \right) -1\right] ^{1/2}}  \label{drrfi5} \\
& -\frac{\pi c_{\mathrm{e}}^{2}}{\left[ \ln ^{2}\left( c_{\mathrm{e}}/\psi
\right) -1\right] ^{1/2}}\int_{\ln \left( c_{\mathrm{e}}/\psi \right)
}^{\infty }\left[ \frac{\left( 2u+1\right) }{\left( 1-u^{-2}\right) ^{1/2}}%
-2u\right] \mathrm{e}^{-2u}\,\mathrm{d}u.  \notag
\end{align}%
To find $G_{a}^{\prime }\left( s\right) $ for arbitrary $a$ we use its
representation (\ref{totgkap}), i.e.%
\begin{equation}
G_{a}^{\prime }\left( s\right) =\frac{G_{\nabla ,1}^{\prime }\left(
a^{3}s\right) }{a^{2}}-\frac{2G_{\varphi ,1}^{\prime }\left( a^{3}s\right) }{%
aa_{\chi }}  \label{drrfi6}
\end{equation}%
and combine with the formulas (\ref{ggkap}) and (\ref{drrfi5}). We don't
write the final formula since it is quite long but it is clear from formulas
(\ref{ggkap}) and (\ref{drrfi5}) that $G_{a}^{\prime }\left( s\right) $ 
\emph{does not depend analytically on }$s$ at $s=0$, and that the following
asymptotic formula holds%
\begin{equation}
G_{a}^{\prime }\left( s\right) =\frac{1}{a^{2}}-\left( \frac{1}{a^{2}}+\frac{%
1}{aa_{\chi }}\right) \frac{4}{\ln \left( c_{\mathrm{e}}^{2}/\left(
a^{3}s\right) \right) }\text{ for }s\rightarrow 0.  \label{drrfi6a}
\end{equation}%
The variance of the exponential form factor $\mathring{\psi}_{1}\left(
r\right) $ is%
\begin{equation}
\int_{\mathbb{R}^{3}}\left\vert \mathbf{x}\right\vert ^{2}\mathring{\psi}%
_{1}^{2}\left( \left\vert \mathbf{x}\right\vert \right) \,\mathrm{d}\mathbf{x%
}=4\pi c_{\mathrm{e}}^{2}\int_{0}^{\infty }r^{4}\mathrm{e}^{-2\left(
r^{2}+1\right) ^{1/2}}\,\mathrm{d}r\simeq 3.826\,8.  \label{drrfi7}
\end{equation}

\subsection{Form factor potential proximity to the Coulomb's potential\label%
{scoulomb}}

In this subsection we study the proximity of the potential form factor $%
\mathring{\varphi}_{a}\left( \left\vert \mathbf{x}\right\vert \right) $ to
the Coulomb's potential $q/\left\vert \mathbf{x}\right\vert $ for small $a$.
This is an important issue since it is a well known experimental fact that
the Coulomb's potential $q/\left\vert \mathbf{x}\right\vert $ represents the
electrostatic field of the charge very accurately even for very small values
of $\left\vert \mathbf{x}\right\vert $.

According to the rest charge equation (\ref{nop7}) and the equation (\ref%
{nop6}) the potential $\mathring{\phi}_{a}\left( \left\vert \mathbf{x}%
\right\vert \right) =\mathring{\varphi}_{a}\left( \left\vert \mathbf{x}%
\right\vert \right) /q$ satisfies%
\begin{equation*}
\nabla ^{2}\mathring{\phi}_{a}\left( \left\vert \mathbf{x}\right\vert
\right) =-4\pi \mathring{\psi}_{a}^{2}\left( \left\vert \mathbf{x}%
\right\vert \right) ,
\end{equation*}%
hence 
\begin{equation}
\text{ }\mathring{\phi}_{a}\left( \left\vert \mathbf{x}\right\vert \right)
=\dint\nolimits_{\mathbb{R}^{3}}\frac{\mathring{\psi}_{a}^{2}\left(
\left\vert \mathbf{y}\right\vert \right) }{\left\vert \mathbf{x}-\mathbf{y}%
\right\vert }\,\mathrm{d}\mathbf{y}>0.  \label{vph1}
\end{equation}%
In view of the relations (\ref{nrac5}) the dependence of the potential $%
\mathring{\phi}_{a}\left( r\right) $ on the size parameter $a$ is of the
form 
\begin{equation}
\mathring{\phi}_{a}\left( r\right) =a^{-1}\phi _{1}\left( a^{-1}r\right) ,
\label{vph2}
\end{equation}%
and consequently its behavior for small $a$ is determined by the behavior of 
$\phi _{1}\left( r\right) $ for large $r$. To find the latter, consider the
radial solution $\zeta $ $\left( r\right) $ to the Poisson equation%
\begin{equation}
\frac{1}{r}\left( \frac{d}{dr}\right) ^{2}\zeta \left( r\right) =-4\pi 
\mathring{\psi}_{1}^{2}\left( r\right) ,\ \zeta \left( r\right) =r\phi
_{1}\left( r\right) ,\ r\geq 0.  \label{polfi}
\end{equation}%
We seek a solution $\zeta \left( r\right) $ to the above equation that is
close to the Coulomb's potential $1/r$ and hence satisfies the following
condition 
\begin{equation}
\zeta \left( r\right) =r\phi _{1}\left( r\right) \rightarrow 1\text{ as }%
r\rightarrow \infty .  \label{polfi1}
\end{equation}%
Taking into account (\ref{polfi1}) when integrating equation (\ref{polfi})
twice yields 
\begin{equation}
\zeta \left( r\right) =1-4\pi \int_{r}^{\infty }\int_{r_{2}}^{\infty }r_{1}%
\mathring{\psi}_{1}^{2}\left( r_{1}\right) \,\mathrm{d}r_{1}\mathrm{d}%
r_{2}=1-4\pi \int_{r}^{\infty }\left( r_{1}-r\right) r_{1}\mathring{\psi}%
_{1}^{2}\left( r_{1}\right) \,\mathrm{d}r_{1},  \label{rfi1}
\end{equation}%
where the second equality in (\ref{rfi1}) is obtained by rewriting the
preceding repeated integral as as a double integral and changing the order
of integration, namely 
\begin{gather*}
\int_{r}^{\infty }\int_{r_{2}}^{\infty }r_{1}\mathring{\psi}_{1}^{2}\left(
r_{1}\right) \,\mathrm{d}r_{1}\mathrm{d}r_{2}=\int_{r}^{\infty
}\int_{r}^{r_{1}}r_{1}\mathring{\psi}_{1}^{2}\left( r_{1}\right) \,\mathrm{d}%
r_{2}\mathrm{d}r_{1}= \\
=\int_{r}^{\infty }\left( r_{1}-r\right) r_{1}\mathring{\psi}_{1}^{2}\left(
r_{1}\right) \,\mathrm{d}r_{1}.
\end{gather*}%
In view of the charge normalization condition (\ref{nrac6}) we readily
obtain from (\ref{rfi1})%
\begin{equation}
\zeta \left( 0\right) =1-4\pi \int_{0}^{\infty }r_{1}^{2}\mathring{\psi}%
_{1}^{2}\left( r_{1}\right) \,\mathrm{d}r_{1}=1-\int_{\mathbb{R}^{3}}%
\mathring{\psi}_{1}^{2}\left( \left\vert \mathbf{x}\right\vert \right) d%
\mathbf{x}=0.  \label{rfi2}
\end{equation}%
The representation (\ref{rfi1}) for $\zeta $ $\left( r\right) =r\phi
_{1}\left( r\right) $ readily implies the following representation for the
potential $\phi _{1}\left( r\right) $: 
\begin{equation}
\phi _{1}\left( r\right) =\frac{1}{r}\left[ 1-4\pi \int_{r}^{\infty }\left(
r_{1}-r\right) r_{1}\mathring{\psi}_{1}^{2}\left( r_{1}\right) \,\mathrm{d}%
r_{1}\right] .  \label{ficr}
\end{equation}%
Combining (\ref{ficr}) with (\ref{rfi2}) we conclude that $\phi _{1}\left(
r\right) $ is regular for small $r\geq 0$. Using (\ref{ficr}) once more we
obtain the following expression for the difference $D_{C}$ between $\phi
_{1}\left( r\right) $ and $1/r$ 
\begin{equation}
D_{C}\left( \phi _{1}\right) =\phi _{1}\left( r\right) -\frac{1}{r}=-\frac{%
4\pi }{r}\int_{r}^{\infty }\left( r_{1}-r\right) r_{1}\mathring{\psi}%
_{1}^{2}\left( r_{1}\right) \,\mathrm{d}r_{1}.  \label{dc1}
\end{equation}%
The relation (\ref{dc1}) together with (\ref{vph2}) implies 
\begin{equation}
D_{C}\left( \phi _{a}\right) =\phi _{a}\left( r\right) -\frac{1}{r}=-\frac{%
4\pi }{r}\int_{a^{-1}r}^{\infty }\left( r_{1}-a^{-1}r\right) r_{1}\mathring{%
\psi}_{1}^{2}\left( r_{1}\right) \,\mathrm{d}r_{1},  \label{dca}
\end{equation}%
showing in particular that the difference $D_{C}$ becomes small for small $a$%
. More precisely, if $\mathring{\psi}_{1}^{2}$ decays exponentially as in (%
\ref{grs}) then 
\begin{gather}
\left\vert D_{C}\left( \phi _{a}\right) \left( r\right) \right\vert
=\left\vert \phi _{a}\left( r\right) -\frac{1}{r}\right\vert \leq
\label{dca1} \\
\leq \frac{4\pi }{r}\int_{a^{-1}r}^{\infty }\left( r_{1}-a^{-1}r\right)
r_{1}c_{\mathrm{e}}^{2}\mathrm{e}^{-2\left( r_{1}^{2}+1\right) ^{1/2}}\,%
\mathrm{d}r_{1}  \notag \\
\leq \frac{4\pi c_{\mathrm{e}}^{2}\mathrm{e}^{-2a^{-1}r}}{r}\int_{0}^{\infty
}\left( r_{1}+a^{-1}r\right) r_{1}\mathrm{e}^{-2r_{1}}\,\mathrm{d}r_{1} 
\notag \\
=\frac{\pi c_{\mathrm{e}}^{2}\left( a^{-1}r+1\right) }{r}\mathrm{e}%
^{-2a^{-1}r}.  \notag
\end{gather}%
For instance, for $r\geq 10a$ the difference $D_{C}$ between the potential $%
\phi _{a}\left( r\right) $ and the Coulomb's potential $1/r$ is extremely
small: 
\begin{equation}
\left\vert D_{C}\left( \phi _{a}\right) \left( r\right) \right\vert =\pi c_{%
\mathrm{e}}^{2}\left( a^{-1}r+1\right) \mathrm{e}^{-2a^{-1}r}\lesssim
4.4674\times 10^{-8}\text{.}  \label{dca2}
\end{equation}%
Similar estimates for the power law decaying $\mathring{\psi}_{1}^{2}$ as in
(\ref{expsi1}) yields%
\begin{align}
\left\vert D_{C}\left( \phi _{a}\right) \left( r\right) \right\vert &
=\left\vert \phi _{a}\left( r\right) -\frac{1}{r}\right\vert \leq \frac{1}{r}%
\int_{a^{-1}r}^{\infty }\frac{3\left( r_{1}-a^{-1}r\right) r_{1}}{r_{1}^{5}}%
\,\mathrm{d}r_{1}  \label{dca3} \\
& =\frac{1}{r}\int_{a^{-1}r}^{\infty }\left( \frac{3}{r_{1}^{3}}-\frac{%
3a^{-1}r}{r_{1}^{4}}\right) \,\mathrm{d}r_{1}=\frac{1}{r}\left( \frac{3a^{2}%
}{2r^{2}}-\frac{3a^{3}}{5r^{3}}\right) ,\text{ }  \notag
\end{align}%
implying%
\begin{equation*}
\left\vert D_{C}\left( \phi _{a}\right) \right\vert \leq \frac{0.009}{r},%
\text{ for }r\geq 10a.
\end{equation*}%
Notice that if we would take $\mathring{\psi}_{1}\left( r\right) =0$ for all 
$r\geq \mathsf{r}_{0}$, as it is the case in the Abraham-Lorentz model, the
formula (\ref{ficr}) would imply that $\mathring{\varphi}_{a}\left( r\right) 
$ would be exactly the Coulomb's potential for $r\geq a\mathsf{r}_{0}$. But
for such a $\mathring{\psi}_{1}\left( r\right) $ we would not be able to
construct the nonlinear self-interaction component $G_{\varphi }^{\prime }$\
which would satisfy (\ref{gdet2}) since it requires $\mathring{\psi}%
_{1}^{2}\left( r\right) $ to be strictly positive for all $r\geq 0$.

\subsection{Energy related spacial scale\label{senergyscale}}

An attractive choice for the spacial scale can be obtained based on the
requirement that the total energy $\mathcal{E}\left( \mathring{\psi}\right) $
of the resting dressed charge defined by the expression (\ref{nstrc10}) be
exactly 0, which readily reduces to the requirement%
\begin{equation}
\mathcal{E}_{1}\left( \mathring{\psi}\right) =\mathcal{E}_{2}\left( 
\mathring{\varphi}\right) \text{,}  \label{ensp1}
\end{equation}
where%
\begin{equation*}
\mathcal{E}_{1}\left( \mathring{\psi}\right) =\frac{\chi ^{2}}{2m}\int_{%
\mathbb{R}^{3}}\left\vert \nabla \mathring{\psi}\right\vert ^{2}\,\mathrm{d}%
\mathbf{x},\ \mathcal{E}_{2}\left( \mathring{\varphi}\right) =\frac{1}{8\pi }%
\int_{\mathbb{R}^{3}}\left( \nabla \mathring{\varphi}\right) ^{2}\,\mathrm{d}%
\mathbf{x}
\end{equation*}%
(this condition is similar to (\ref{psi2d})). Plugging $\mathring{\psi}=%
\mathring{\psi}_{a}$ and $\mathring{\varphi}=\mathring{\varphi}_{a}$ defined
by (\ref{nrac5}) into the equalities (\ref{ensp1}) we obtain 
\begin{equation}
\mathcal{E}_{1}\left( \mathring{\psi}_{a}\right) =a^{-2}\mathcal{E}%
_{1}\left( \mathring{\psi}_{1}\right) ,\ \mathcal{E}_{2}\left( \mathring{%
\varphi}_{a}\right) =a^{-1}\mathcal{E}_{2}\left( \mathring{\varphi}%
_{1}\right) =a^{-1}q^{2}\mathcal{E}_{2}\left( \mathring{\phi}_{1}\right) ,\ 
\mathring{\phi}_{1}=q^{-1}\mathring{\varphi}_{1}.  \label{ensp2}
\end{equation}%
Hence, the requirement $\mathcal{E}_{1}\left( \mathring{\psi}\right) =%
\mathcal{E}_{2}\left( \mathring{\varphi}\right) $ in view of the relations (%
\ref{ensp2}) is equivalent to the following choice $a=a_{\psi }$ of size
parameter $a,$ \ with 
\begin{equation}
a_{\psi }=\frac{\mathcal{E}_{1}\left( \mathring{\psi}_{1}\right) }{\mathcal{E%
}_{2}\left( \mathring{\varphi}_{1}\right) }=\frac{4\pi \chi ^{2}}{m}\frac{%
\int_{\mathbb{R}^{3}}\left\vert \nabla \mathring{\psi}_{1}\right\vert ^{2}\,%
\mathrm{d}\mathbf{x}}{\int_{\mathbb{R}^{3}}\left( \nabla \mathring{\varphi}%
_{1}\right) ^{2}\,\mathrm{d}\mathbf{x}}=a_{\chi }\theta _{\psi },
\label{apsi}
\end{equation}%
\begin{equation*}
\theta _{\psi }=\frac{4\pi \int_{\mathbb{R}^{3}}\left\vert \nabla \mathring{%
\psi}_{1}\right\vert ^{2}\,\mathrm{d}\mathbf{x}}{\int_{\mathbb{R}^{3}}\left(
\nabla \mathring{\phi}_{1}\right) ^{2}\,\mathrm{d}\mathbf{x}},\ a_{\chi }=%
\frac{\chi ^{2}}{mq^{2}}.
\end{equation*}%
Since the functions $\mathring{\psi}_{1},\mathring{\phi}_{1}$ in the above
relations are radial, the Dirichlet integrals in (\ref{apsi}) can be recast
as 
\begin{equation}
\int_{\mathbb{R}^{3}}\left\vert \nabla \mathring{\psi}_{1}\right\vert ^{2}\,%
\mathrm{d}\mathbf{x}=4\pi \int_{0}^{\infty }\left( \partial _{r}\left( r%
\mathring{\psi}_{1}\left( r\right) \right) \right) ^{2}\,\mathrm{d}r\text{ }
\label{grint}
\end{equation}%
with the similar formula for $\mathring{\phi}_{1}.$ We refer the space scale 
$a_{\psi }$ in (\ref{apsi}) obtained based on the equality $\mathcal{E}%
_{1}\left( \mathring{\psi}\right) =\mathcal{E}_{2}\left( \mathring{\varphi}%
\right) $ as \emph{energy-based spacial scale.}

The energy-based spatial scale $a_{\psi }$ defined by (\ref{apsi}) for the
power law form factor (\ref{expsi1}) $a_{\psi }=\theta _{\psi }a_{\chi },$ \
with 
\begin{equation}
\ \theta _{\psi }=\frac{4\pi \int_{\mathbb{R}^{3}}\left( \nabla \mathring{%
\psi}_{1}\right) ^{2}\,\mathrm{d}\mathbf{x}}{\int_{\mathbb{R}^{3}}\left(
\nabla \mathring{\phi}_{1}\right) ^{2}\,\mathrm{d}\mathbf{x}}=\frac{40}{7\pi 
}\simeq 1.\,8189.  \label{sol3}
\end{equation}%
For the exponentially decaying form factor we get $\ a_{\psi }=\theta _{\psi
}a_{\chi }$ with 
\begin{equation*}
\theta _{\psi }\simeq 1.\,2473
\end{equation*}%
Note that the energy based spatial scales for power law form factor and
exponential form factor are of the same order, though their variances are
absolutely different (infinite variance for the power law as in (\ref{expsi1}%
)).

\section{Accelerated motion of a single nonrelativistic charge in an
external EM field\label{ssaccel}}

The key objective of this section is an extension of the wave-corpuscle
representation defined by formulas (\ref{nexac1}), (\ref{nexac1a}) to the
case of a single nonrelativistic charge accelerating in an external EM
field. Here we discuss in detail results sketched in Sections \ref{nrapr}, %
\ref{nrapr1}. Recall that as in (\ref{nore1b}) \ we neglect the charge's own
magnetic field and set $\mathbf{A}=0$ taking into account only external
magnetic field. In the case of a general external EM field no exact closed
form solution to the field equations seems to be available, but there is an
approximate wave-corpuscle solution and its accuracy is a subject of our
studies in this case; this solution is exact for special external fields.

The total EM fields are described by their potentials $\bar{\varphi},\mathbf{%
\bar{A}}$ \ which involve potentials \ of the external field and particle's
own field, namely 
\begin{equation}
\bar{\varphi}=\varphi _{\mathrm{ex}}+\varphi ,\ \mathbf{\bar{A}=A}_{\mathrm{%
ex}}.  \label{expo1}
\end{equation}%
The nonrelativistic Lagrangian $\hat{L}_{0}$ for the charge in external
field is obtained from the one for the free charge in (\ref{nore1b}) by
modifying there the covariant derivative to include the external potential,
namely 
\begin{equation}
\hat{L}_{0}\left( \psi ,\psi ^{\ast },\varphi \right) =\frac{\chi }{2}%
\mathrm{i}\left[ \psi ^{\ast }\tilde{\partial}_{t}\psi -\psi \tilde{\partial}%
_{t}^{\ast }\psi ^{\ast }\right] -\frac{\chi ^{2}}{2m}\left\{ \tilde{\nabla}%
\psi \tilde{\nabla}^{\ast }\psi ^{\ast }+G\left( \psi ^{\ast }\psi \right)
\right\} -\frac{\left\vert \nabla \varphi \right\vert ^{2}}{8\pi },
\label{expo2}
\end{equation}%
where%
\begin{equation*}
\tilde{\partial}_{t}=\partial _{t}+\frac{\mathrm{i}q\bar{\varphi}}{\chi },\ 
\tilde{\nabla}=\nabla -\frac{\mathrm{i}q\mathbf{A}_{\mathrm{ex}}}{\chi c},\ 
\tilde{\partial}_{t}^{\ast }=\partial _{t}-\frac{\mathrm{i}q\bar{\varphi}}{%
\chi },\ \tilde{\nabla}^{\ast }=\nabla +\frac{\mathrm{i}q\mathbf{A}_{\mathrm{%
ex}}}{\chi c}.
\end{equation*}%
This modified Lagrangian remains to be gauge invariant with respect to the
transformations (\ref{nore2}) and the general formulas (\ref{lagco9})-(\ref%
{lagco12}) for the charge and current densities applied to the Lagrangian $%
\hat{L}_{0}$ yield $\ J^{\mu }=\left( \mathrm{c}\rho ,\mathbf{J}\right) $ \
with 
\begin{gather}
\ \rho =q\psi \psi ^{\ast },\ \mathbf{J}=\frac{\mathrm{i}\chi q}{2m}\left[
\psi \tilde{\nabla}^{\ast }\psi ^{\ast }-\psi ^{\ast }\tilde{\nabla}\psi %
\right] =  \label{expo3} \\
=\frac{\chi q}{2m}\mathrm{i}\left( \nabla \psi ^{\ast }\psi -\psi ^{\ast
}\nabla \psi \right) -\frac{q^{2}\mathbf{A}_{\mathrm{ex}}}{m\mathrm{c}}\psi
^{\ast }\psi =\left( \frac{\chi q}{m}\func{Im}\frac{\nabla \psi }{\psi }-%
\frac{q^{2}\mathbf{A}_{\mathrm{ex}}}{m\mathrm{c}}\right) \left\vert \psi
\right\vert ^{2}.  \notag
\end{gather}%
This current satisfies the conservation/continuity equations $\partial _{\nu
}J^{\nu }=0$ \ which take the form 
\begin{equation}
\partial _{t}\rho +\nabla \cdot \mathbf{J}=0.  \label{expo4}
\end{equation}%
The Euler-Lagrange field equations in this case are%
\begin{gather}
\chi \mathrm{i}\tilde{\partial}_{t}\psi =\frac{\chi ^{2}}{2m}\left[ -\tilde{%
\nabla}^{2}\Delta \psi +G^{\prime }\left( \psi ^{\ast }\psi \right) \psi %
\right] ,  \label{expo5} \\
-\Delta \varphi =4\pi q\psi \psi ^{\ast },  \label{expo6}
\end{gather}%
where $G^{\prime }\left( s\right) =\partial _{s}G$ \ \ and, as always, $\psi
^{\ast }$ is complex conjugate to $\psi $. \ Then the field equations (\ref%
{expo5})-(\ref{expo6}) can be recast as the following \emph{field equations}%
\begin{gather}
\mathrm{i}\chi \partial _{t}\psi =-\frac{\chi ^{2}\nabla ^{2}\psi }{2m}-%
\frac{\chi q\mathbf{A}_{\mathrm{ex}}\cdot \nabla \psi }{m\mathrm{ci}}%
+q\left( \varphi +\varphi _{\mathrm{ex}}+\frac{q\mathbf{A}_{\mathrm{ex}}^{2}%
}{2m\mathrm{c}^{2}}\right) \psi +\frac{\chi ^{2}G^{\prime }\psi }{2m},
\label{NLS1} \\
\nabla ^{2}\varphi =-4\pi q\left\vert \psi \right\vert ^{2}.  \label{fis}
\end{gather}%
As in the case of a free charge we set the total conserved charge to be
exactly $q$ and, similarly to (\ref{nrac4}), we have the following \emph{%
charge normalization} condition%
\begin{equation}
\text{ }\dint\nolimits_{\mathbb{R}^{3}}\left\vert \psi \right\vert ^{2}\,%
\mathrm{d}\mathbf{x}=1.  \label{expo8}
\end{equation}

The presence of the external EM field turns the dressed charge into an open
system with consequent subtleties in the treatment of the energy-momentum.
All elements of the proper treatment of the energy and momentum densities in
such a situation are provided in Section \ref{1partext} and we apply them to
the Lagrangian $\hat{L}_{0}$ defined by (\ref{expo2}). \emph{An instrumental
element in the analysis of the energy-momentum tensor is its partition
between the charge and the EM field. In carrying out such a partition we are
guided by two principles: }(i)\emph{\ \ both the energy-momenta tensors and
the forces have to be gauge invariant; }(ii)\emph{\ the forces must be of
the Lorentz form. The second principle is evidently special to the EM system
consisting of the charge and the EM field.}

\subsection{Wave-corpuscle concept for an accelerating charge\label{swcacel}}

In Section \ref{snfreemov} we introduced a wave-corpuscle by the relations (%
\ref{nexac1}), (\ref{nexac1a}) for a free moving dressed charge. In this
section we study wave-corpuscles in an external EM field. Recall that the
wave-corpuscle (\ref{nexac1}), (\ref{nexac1a}) for a free moving dressed
charge is an exact solution to the fields equations (\ref{nop2}), (\ref{nop3}%
), and when constructing the wave-corpuscle for a dressed charge in external
EM field we also want it to be an exact solutions to the field equation (\ref%
{NLS1})-(\ref{fis}). It turns out that it is possible \ if the external EM
field is a homogeneous electric field, but no closed form solution seems to
be available for a general external EM field as defined by its potentials $%
\varphi _{\mathrm{ex}}$, $\mathbf{A}_{\mathrm{ex}}$. To describe explicitly
the class of external EM fields \ for which wave-corpuscles are exact
solutions, we introduce \emph{auxiliary field equations}. If the
coefficients of the auxiliary field equations are linear functions of
spacial variables, we write explicit solutions. \emph{\ }The coefficients of
the auxiliary field equations are in a simple, explicit correspondence with
the coefficients of the original field equations, therefore we obtain
wave-corpuscles which exactly solve original ones (\ref{NLS1})-(\ref{fis})
for certain classes of external EM fields. If external EM fields \ are
general, \emph{we construct a wave-corpuscle so that it exactly solves
properly defined auxiliary field equations which differ from the original
ones (\ref{NLS1})-(\ref{fis}) by an explicitly defined discrepancy }$D$\emph{%
, based on which we judge the accuracy of the approximation}. Executing this
plan we introduce the following system of \emph{auxiliary field equations} 
\begin{gather}
\mathrm{i}\chi \partial _{t}\psi =-\frac{\chi ^{2}\nabla ^{2}\psi }{2m}-%
\frac{\chi q\mathbf{\tilde{A}}_{\mathrm{ex}}\cdot \nabla \psi }{m\mathrm{ci}}%
+q\left( \varphi +\tilde{\varphi}_{\mathrm{ex}}\right) \psi +\frac{\chi
^{2}G^{\prime }\psi }{2m},\   \label{Schmag1} \\
\nabla ^{2}\varphi =-4\pi q\left\vert \psi \right\vert ^{2},  \notag
\end{gather}%
where the auxiliary potentials $\tilde{\varphi}_{\mathrm{ex}}\left( t,%
\mathbf{x}\right) $, $\mathbf{\tilde{A}}_{\mathrm{ex}}\left( t,\mathbf{x}%
\right) $ are \ linear in $\mathbf{x}$ \ and generally may differ from the
original potentials $\varphi _{\mathrm{ex}}$, $\mathbf{A}_{\mathrm{ex}}$.
Evidently, in addition to the alteration of the potentials $\varphi _{%
\mathrm{ex}}$, $\mathbf{A}_{\mathrm{ex}}$, the auxiliary field equations
differ from the original ones (\ref{NLS1}), (\ref{fis}) only by a single
term $\chi q^{2}\mathbf{A}_{\mathrm{ex}}^{2}\psi /\left( 2m\mathrm{c}%
^{2}\right) $ (which, in fact, is absorbed in $\tilde{\varphi}_{\mathrm{ex}}$%
).

We define the wave-corpuscle $\psi ,\varphi $ by the formula similar to (\ref%
{nexac1}), (\ref{nexac1a}), namely 
\begin{gather}
\psi \left( t,\mathbf{x}\right) =\mathrm{e}^{\mathrm{i}S/\chi }\hat{\psi},\
S=m\mathbf{v}\left( t\right) \cdot \left[ \mathbf{x}-\mathbf{r}\left(
t\right) \right] +s_{\mathrm{p}}\left( t\right) \text{,}\ \hat{\psi}=%
\mathring{\psi}\left( \left\vert \mathbf{x}-\mathbf{r}\left( t\right)
\right\vert \right) ,\   \label{psil0} \\
\varphi =\mathring{\varphi}\left( \left\vert \mathbf{x}-\mathbf{r}\left(
t\right) \right\vert \right) .  \notag
\end{gather}%
In the above formula $\mathring{\psi}$ and $\mathring{\varphi}$ are,
respectively, the form factor and the form factor potential satisfying (\ref%
{nop4}), (\ref{nop5}); the function $\mathbf{r}\left( t\right) $ is
determined by the following \emph{complementary point charge equations}%
\begin{gather}
m\frac{\mathrm{d}^{2}\mathbf{r}}{\mathrm{d}^{2}t}=q\mathbf{E}_{\mathrm{ex}%
}\left( t,\mathbf{r}\right) +\frac{q}{\mathrm{c}}\frac{\mathrm{d}\mathbf{r}}{%
\mathrm{d}t}\times \mathbf{B}_{\mathrm{ex}}\left( t,\mathbf{r}\right) ,\ 
\label{Lor1} \\
\mathbf{r}\left( 0\right) =\mathbf{r}_{0},\ \frac{\mathrm{d}\mathbf{r}}{%
\mathrm{d}t}\left( 0\right) =\mathbf{\dot{r}}_{0}\text{ }  \notag
\end{gather}%
where 
\begin{equation*}
\mathbf{E}_{\mathrm{ex}}=-\nabla \varphi _{\mathrm{ex}}-\frac{\partial _{t}%
\mathbf{A}_{\mathrm{ex}}}{\mathrm{c}},\ \mathbf{B}_{\mathrm{ex}}=\nabla
\times \mathbf{A}_{\mathrm{ex}},
\end{equation*}%
are based on the EM potentials of \ the original \ equations (\ref{NLS1})-(%
\ref{fis}) and $\mathbf{r}_{0},\mathbf{\dot{r}}_{0}$ are initial data. \ The
functions $\mathbf{v}\left( t\right) ,s_{\mathrm{p}}\left( t\right) $ \ in (%
\ref{psil0}) are determined based on the solution $\mathbf{r}\left( t\right) 
$ by the formulas 
\begin{equation}
\mathbf{v}\left( t\right) =\frac{\mathrm{d}\mathbf{r}}{\mathrm{d}t}\left(
t\right) +\frac{q}{m\mathrm{c}}\mathbf{A}_{\mathrm{ex}}\left( \mathbf{r}%
\left( t\right) \right) ,\ \frac{\mathrm{d}s_{\mathrm{p}}}{\mathrm{d}t}=%
\frac{m\mathbf{v}^{2}\left( t\right) }{2}-q\varphi _{\mathrm{ex}}\left( 
\mathbf{r}\left( t\right) \right) .  \label{Lor2}
\end{equation}%
We readily recognize in the equation (\ref{Lor1}) the point charge motion in
the external EM field equation. Notice that the \emph{function }$\mathbf{v}%
\left( t\right) $\emph{\ defined by the first equation in (\ref{Lor2}) if }$%
\mathbf{A}_{\mathrm{ex}}\neq 0$\emph{\ is not the charge velocity} $\mathbf{%
\dot{r}}\left( t\right) $, but it is simply related to canonical momentum $%
\mathbf{\mathring{p}}$ (see (\ref{rkin8})-(\ref{rkin9})) by the formula 
\begin{equation}
\mathbf{v}\left( t\right) =\frac{\mathbf{\mathring{p}}\left( t\right) }{m},\ 
\mathbf{\mathring{p}}=\mathbf{p}+\frac{q}{\mathrm{c}}\mathbf{A}_{\mathrm{ex}%
},\ \text{where }\mathbf{p}=m\frac{\mathrm{d}\mathbf{r}}{\mathrm{d}t}\text{
is the kinetic momentum.}  \label{Lor3}
\end{equation}%
\emph{We refer to the function }$\mathbf{r}\left( t\right) $\emph{\ as
wave-corpuscle center or wave-corpuscle position. \ }Note that since\emph{\ }%
$\hat{\psi}$ \ is center-symmetric, \ this definition agrees with (\ref{peu4}%
).

Now we define the auxiliary linear in $\mathbf{x}$ potentials $\tilde{\varphi%
}_{\mathrm{ex}}\left( t,\mathbf{x}\right) $, $\mathbf{\tilde{A}}_{\mathrm{ex}%
}\left( t,\mathbf{x}\right) $ by the following formulas 
\begin{equation}
\tilde{\varphi}_{\mathrm{ex}}=\varphi _{0,\mathrm{ex}}\left( t\right)
+\varphi _{0,\mathrm{ex}}^{\prime }\left( t\right) \cdot \left( \mathbf{x}-%
\mathbf{r}\left( t\right) \right) ,\   \label{fiexlin}
\end{equation}%
where%
\begin{equation*}
\varphi _{0,\mathrm{ex}}\left( t\right) =\varphi _{\mathrm{ex}}\left( t,%
\mathbf{r}\left( t\right) \right) ,\ \varphi _{0,\mathrm{ex}}^{\prime
}\left( t\right) =\nabla \varphi _{\mathrm{ex}}\left( t,\mathbf{r}\left(
t\right) \right) ,
\end{equation*}%
and 
\begin{equation}
\mathbf{\tilde{A}}_{\mathrm{ex}}=\mathbf{A}_{\mathrm{ex},0}\left( t\right) +%
\frac{1}{2}\mathbf{B}_{0}\left( t\right) \times \left[ \mathbf{x}-\mathbf{r}%
\left( t\right) \right] ,\   \label{Aexlin}
\end{equation}%
where%
\begin{equation*}
\mathbf{A}_{\mathrm{ex},0}\left( t\right) =\mathbf{A}_{\mathrm{ex}}\left( t,%
\mathbf{r}\left( t\right) \right) ,\ \mathbf{B}_{0}\left( t\right) =\left[
\nabla \times \mathbf{A}_{\mathrm{ex}}\right] \left( t,\mathbf{r}\left(
t\right) \right) ,
\end{equation*}

The verification of the fact that the wave-corpuscle defined by the
relations (\ref{Lor1})-(\ref{Lor3}) is either exact or an approximate
solution to the field equations (\ref{NLS1})-(\ref{fis}) with estimated
accuracy is carried out in the following sections.

\subsection{Energy-momentum tensor\label{senmomnonrel}}

The canonical energy-momentum $\mathring{\Theta}^{\mu \nu }$ for the EM
field is obtained by applying the general formula (\ref{flagr5}) to the
Lagrangian $\hat{L}_{0},$ yielding%
\begin{equation}
\mathring{\Theta}^{\mu \nu }=\left[ 
\begin{array}{cccc}
\mathring{w} & \mathrm{c}\mathring{g}_{1} & \mathrm{c}\mathring{g}_{2} & 
\mathrm{c}\mathring{g}_{3} \\ 
\mathrm{c}^{-1}\mathring{s}_{1} & -\mathring{\tau}_{11} & -\mathring{\tau}%
_{12} & -\mathring{\tau}\mathring{\sigma}_{13} \\ 
\mathrm{c}^{-1}\mathring{s}_{2} & -\mathring{\tau}_{21} & -\mathring{\tau}%
_{22} & -\mathring{\tau}\mathring{\sigma}_{23} \\ 
\mathrm{c}^{-1}\mathring{s}_{3} & -\mathring{\tau}_{31} & -\mathring{\tau}%
^{32} & -\mathring{\tau}\mathring{\sigma}_{33}%
\end{array}%
\right] =\left[ 
\begin{array}{cccc}
\mathring{w} & 0 & 0 & 0 \\ 
\mathrm{c}^{-1}\mathring{s}_{1} & -\mathring{\tau}_{11} & -\mathring{\tau}%
_{12} & -\mathring{\tau}_{13} \\ 
\mathrm{c}^{-1}\mathring{s}_{2} & -\mathring{\tau}_{21} & -\mathring{\tau}%
_{22} & -\mathring{\tau}_{23} \\ 
\mathrm{c}^{-1}\mathring{s}_{3} & -\mathring{\tau}_{31} & -\mathring{\tau}%
^{32} & -\mathring{\tau}_{33}%
\end{array}%
\right] ,  \label{emna1}
\end{equation}%
with%
\begin{equation}
\mathring{w}=-\frac{\left\vert \nabla \varphi \right\vert ^{2}}{8\pi },\ 
\mathring{g}_{j}=0,\ \mathring{s}_{j}=\mathrm{c}\frac{\partial _{j}\varphi
\partial _{0}\varphi }{4\pi },\ \mathring{\tau}_{jj}=\frac{\partial
_{j}^{2}\varphi }{4\pi }-\frac{\left\vert \nabla \varphi \right\vert ^{2}}{%
8\pi },\ \mathring{\tau}_{ij}=\frac{\partial _{i}\varphi \partial
_{j}\varphi }{4\pi }.  \label{emna2}
\end{equation}%
The gauge invariant energy-momentum of the EM field takes the form%
\begin{equation}
\Theta ^{\mu \nu }=\left[ 
\begin{array}{cccc}
w & \mathrm{c}g_{1} & \mathrm{c}g_{2} & \mathrm{c}g_{3} \\ 
\mathrm{c}^{-1}s_{1} & -\tau _{11} & -\tau _{12} & -\tau _{13} \\ 
\mathrm{c}^{-1}s_{2} & -\tau _{21} & -\tau _{22} & -\tau _{23} \\ 
\mathrm{c}^{-1}s_{3} & -\tau _{31} & -\tau ^{32} & -\tau _{33}%
\end{array}%
\right] =\left[ 
\begin{array}{cccc}
w & 0 & 0 & 0 \\ 
0 & -\tau _{11} & -\tau _{12} & -\tau _{13} \\ 
0 & -\tau _{21} & -\tau _{22} & -\tau _{23} \\ 
0 & -\tau _{31} & -\tau ^{32} & -\tau _{33}%
\end{array}%
\right] ,  \label{emna3}
\end{equation}%
\begin{gather}
w\left( t,\mathbf{x}\right) =w_{0}\left( \mathbf{x}\right) +\int_{-\mathbb{%
\infty }}^{t}\frac{\mathbf{J}\left( t^{\prime },\mathbf{x}\right) \cdot
\nabla \varphi \left( t^{\prime },\mathbf{x}\right) }{\mathrm{c}}\,\mathrm{d}%
t^{\prime },\ g_{j}=0,\ s_{j}=0,  \label{emna4} \\
\tau _{ij}=\frac{\partial _{i}\varphi \partial _{j}\varphi }{4\pi },\ \ \tau
_{jj}=\frac{\partial _{j}^{2}\varphi }{4\pi }-\frac{\left\vert \nabla
\varphi \right\vert ^{2}}{8\pi }.  \label{emna5}
\end{gather}%
The canonical energy-momentum tensor $\mathring{T}^{\mu \nu }$ is not gauge
invariant, but the following decomposition holds for it 
\begin{equation}
\mathring{T}^{\mu \nu }=\tilde{T}^{\mu \nu }+\frac{1}{\mathrm{c}}J^{\mu }%
\bar{A}^{\nu },\ \bar{A}^{\nu }=\left( \varphi _{\mathrm{ex}}+\varphi ,%
\mathbf{A}_{\mathrm{ex}}\right) ,  \label{emna6}
\end{equation}%
where $\tilde{T}^{\mu \nu }$ is a gauge invariant energy-momentum obtained
from formula (\ref{tsing1}) applied to the Lagrangian $\hat{L}_{0}$ yielding%
\begin{equation}
\tilde{T}^{\mu \nu }=\left[ 
\begin{array}{cccc}
\tilde{u} & \mathrm{c}\tilde{p}_{1} & \mathrm{c}\tilde{p}_{2} & \mathrm{c}%
\tilde{p}_{3} \\ 
\mathrm{c}^{-1}\tilde{s}_{1} & -\tilde{\sigma}_{11} & -\tilde{\sigma}_{12} & 
-\tilde{\sigma}_{13} \\ 
\mathrm{c}^{-1}\tilde{s}_{2} & -\tilde{\sigma}_{21} & -\tilde{\sigma}_{22} & 
-\tilde{\sigma}_{23} \\ 
\mathrm{c}^{-1}\tilde{s}_{3} & -\tilde{\sigma}_{31} & -\tilde{\sigma}_{32} & 
-\tilde{\sigma}_{33}%
\end{array}%
\right] ,\text{ where}  \label{emna7}
\end{equation}%
\begin{equation}
\tilde{u}=\frac{\chi ^{2}}{2m}\left[ \tilde{\nabla}\psi \cdot \tilde{\nabla}%
^{\ast }\psi ^{\ast }+G\left( \psi ^{\ast }\psi \right) \right] ,
\label{emna8}
\end{equation}%
\begin{equation}
\tilde{p}_{j}=\frac{\chi \mathrm{i}}{2}\left( \psi \tilde{\partial}%
_{j}^{\ast }\psi ^{\ast }-\psi ^{\ast }\tilde{\partial}_{j}\psi \right) ,\ 
\tilde{s}_{j}=-\frac{\chi ^{2}\mathrm{i}}{2m}\left( \tilde{\partial}_{t}\psi 
\tilde{\partial}_{j}^{\ast }\psi ^{\ast }+\tilde{\partial}_{t}^{\ast }\psi
^{\ast }\tilde{\partial}_{j}\psi \right) ,\ j=1,2,3,  \label{emna9}
\end{equation}%
and the stress tensor components $\sigma _{ij}$ are represented by the
formulas 
\begin{gather}
\tilde{\sigma}_{ii}=\tilde{u}-\frac{\chi ^{2}}{m}\tilde{\partial}_{i}\psi 
\tilde{\partial}_{i}^{\ast }\psi ^{\ast }+\frac{\chi \mathrm{i}}{2}\left(
\psi \tilde{\partial}_{t}\psi ^{\ast }-\psi ^{\ast }\tilde{\partial}_{t}\psi
\right) ,\   \label{emna10} \\
\tilde{\sigma}_{ij}=\tilde{\sigma}_{ji}=-\frac{\chi ^{2}}{2m}\left( \tilde{%
\partial}_{i}\psi \tilde{\partial}_{j}^{\ast }\psi ^{\ast }+\tilde{\partial}%
_{j}\psi \tilde{\partial}_{i}^{\ast }\psi ^{\ast }\right) \text{ for }i\neq
j,\ i,j=1,2,3.  \notag
\end{gather}%
It follows from (\ref{expo3}) and (\ref{emna9}) that \emph{the charge gauge
invariant momentum }$\mathbf{P}$ \emph{\ equals exactly the microcurrent
density }$\mathbf{J}$\textbf{\ }\emph{multiplied by the constant }$m/q$,
namely the following identity holds 
\begin{equation}
\mathbf{P}=\frac{m}{q}\mathbf{J}=\frac{\mathrm{i}\chi }{2}\left[ \psi \tilde{%
\nabla}^{\ast }\psi ^{\ast }-\psi ^{\ast }\tilde{\nabla}\psi \right] =\left(
\chi \func{Im}\frac{\nabla \psi }{\psi }-\frac{q\mathbf{\bar{A}}}{c}\right)
\left\vert \psi \right\vert ^{2},  \label{emna11}
\end{equation}%
\emph{which can be viewed as the kinematic representation of the momentum
density }%
\begin{equation}
\mathbf{P}=m\mathbf{v},\ \mathbf{v}=\mathbf{J}/q.  \label{emna12}
\end{equation}%
So we refer to the identities (\ref{emna11})-(\ref{emna12}) as \emph{%
momentum density kinematic representation}.

Using the field equations we can also verify that the following
conservations laws for the charge and its EM field hold: 
\begin{equation}
\partial _{\mu }\tilde{T}^{\mu \nu }=f^{\nu }+f_{\mathrm{ex}}^{\nu },\
\partial _{\mu }\Theta ^{\mu \nu }=-f^{\nu },\ \partial _{\mu }\mathcal{T}%
^{\mu \nu }=\partial _{\mu }\left( \tilde{T}^{\mu \nu }+\Theta ^{\mu \nu
}\right) =f_{\mathrm{ex}}^{\nu },\text{ }  \label{emna13}
\end{equation}%
where%
\begin{gather}
f^{\nu }=\frac{1}{\mathrm{c}}J_{\mu }F^{\nu \mu }=\left( \frac{1}{\mathrm{c}}%
\mathbf{J}\cdot \mathbf{E},\rho \mathbf{E}\right) ,  \label{emna14} \\
f_{\mathrm{ex}}^{\nu }=\frac{1}{\mathrm{c}}J_{\mu }F_{\mathrm{ex}}^{\nu \mu
}=\left( \frac{1}{\mathrm{c}}\mathbf{J}\cdot \mathbf{E}_{\mathrm{ex}},\rho 
\mathbf{E}_{\mathrm{ex}}+\frac{1}{\mathrm{c}}\mathbf{J}\times \mathbf{B}_{%
\mathrm{ex}}\right) ,  \notag
\end{gather}%
We readily recognize in $f^{\nu }$ and $f_{\mathrm{ex}}^{\nu }$ in equations
(\ref{emna13}) respectively the Lorentz force densities for the charge \ in
its own and the external EM fields. We also see to our satisfaction from the
first two equations in (\ref{emna13}) that the Newton's principle "action
equals reaction" does manifestly hold for all involved densities at every
point of the space-time.

\subsection{Point charge mechanics via averaged quantities}

Combining now the conservations laws (\ref{emna13}) with energy-momentum
tensors representations (\ref{emna3})-(\ref{emna5}) and (\ref{emna7})-(\ref%
{empa10}) we obtain the following equations for the total dressed charge
momentum density $\mathbf{P}=\left( P^{1},P^{2},P^{3}\right) $ and the
energy density $U$ 
\begin{equation}
\partial _{t}P^{i}=\partial _{t}\left( \tilde{p}^{i}+g^{i}\right)
=\dsum_{j=1,2,3}\partial _{j}\left( \tilde{\sigma}^{ji}+\tau _{ij}\right)
+\left( \rho \mathbf{E}_{\mathrm{ex}}+\frac{1}{\mathrm{c}}\mathbf{J}\times 
\mathbf{B}_{\mathrm{ex}}\right) ^{i},\ i=1,2,3,  \label{crv1}
\end{equation}%
\begin{equation}
\partial _{t}U=\partial _{t}\left( \tilde{u}+w\right)
=-\dsum_{j=1,2,3}\partial _{j}\left( \tilde{s}_{1}+s_{1}\right) +\mathbf{J}%
\cdot \mathbf{E}_{\mathrm{ex}}.  \label{crv2}
\end{equation}%
Integrating the above conservation laws over the entire space $\mathbb{R}%
^{3} $ we obtain the following equations for the total momentum $\mathsf{P}$
and the total energy $\mathsf{E}$%
\begin{eqnarray}
\frac{\mathrm{d}\mathsf{P}}{\mathrm{d}t} &=&\int_{\mathbb{R}^{3}}\left[ \rho 
\mathbf{E}_{\mathrm{ex}}+\frac{1}{\mathrm{c}}\mathbf{J}\times \mathbf{B}_{%
\mathrm{ex}}\right] \left( t,\mathbf{x}\right) \,\mathrm{d}\mathbf{x},
\label{crv3} \\
\frac{\mathrm{d}\mathsf{E}}{\mathrm{d}t} &=&\int_{\mathbb{R}^{3}}\mathbf{J}%
\cdot \mathbf{E}_{\mathrm{ex}}\left( t,\mathbf{x}\right) \,\mathrm{d}\mathbf{%
x}.  \notag
\end{eqnarray}%
Let us introduce a \emph{charge average position} $\mathbf{r}\left( t\right) 
$ and \emph{average velocity} $\mathsf{v}\left( t\right) $ by the following
relations%
\begin{gather}
\mathbf{r}\left( t\right) =\int_{\mathbb{R}^{3}}\mathbf{x}\left\vert \psi
\left( t,\mathbf{x}\right) \right\vert ^{2}\,\mathrm{d}\mathbf{x},
\label{crv4} \\
\ \mathsf{v}\left( t\right) =\frac{1}{q}\int_{\mathbb{R}^{3}}\mathbf{J}%
\left( t,\mathbf{x}\right) \,\mathrm{d}\mathbf{x}.
\end{gather}%
Then using the charge conservation law (\ref{expo4}) we find%
\begin{equation}
\frac{\mathrm{d}\mathbf{r}\left( t\right) }{\mathrm{d}t}=\int_{\mathbb{R}%
^{3}}\mathbf{x}\partial _{t}\left\vert \psi \right\vert ^{2}\,\mathrm{d}%
\mathbf{x}=\mathbf{-}\frac{1}{q}\int_{\mathbb{R}^{3}}\mathbf{x}\nabla \cdot 
\mathbf{J}\,\mathrm{d}\mathbf{x}=\frac{1}{q}\int_{\mathbb{R}^{3}}\mathbf{J}\,%
\mathrm{d}\mathbf{x}=\mathsf{v}\left( t\right) .  \label{crv5}
\end{equation}%
Utilizing the momentum density kinematic representation (\ref{emna11})-(\ref%
{emna12}) and the fact the momentum density of the charge's EM field is
identically zero according to (\ref{emna4}) we obtain the following
kinematic representation for charge and hence the dressed charge total
momentum:%
\begin{equation}
\mathsf{P}\left( t\right) =\frac{m}{q}\int_{\mathbb{R}^{3}}\mathbf{J}\left(
t,\mathbf{x}\right) \,\mathrm{d}\mathbf{x}=m\mathsf{v}\left( t\right) .
\label{crv6}
\end{equation}%
Notice now that for the spatially homogeneous EM fields $\mathbf{E}_{\mathrm{%
ex}}\left( t\right) $ and $\mathbf{B}_{\mathrm{ex}}\left( t\right) $ the
equations (\ref{crv3}) take a simpler form 
\begin{eqnarray}
\frac{\mathrm{d}\mathsf{P}}{\mathrm{d}t} &=&q\mathbf{E}_{\mathrm{ex}}\left(
t\right) +\frac{q\mathbf{v}\left( t\right) }{\mathrm{c}}\times \mathbf{B}_{%
\mathrm{ex}}\left( t\right) ,  \label{crv7} \\
\ \frac{\mathrm{d}\mathsf{E}}{\mathrm{d}t} &=&q\mathsf{v}\left( t\right)
\cdot \mathbf{E}_{\mathrm{ex}}\left( t\right) .  \notag
\end{eqnarray}%
In addition to that, in this case combining the first equality in (\ref{crv7}%
) with the momentum kinematic representation (\ref{crv6}) we get%
\begin{equation}
\frac{\mathrm{d}m\mathsf{v}\cdot \mathsf{v}}{2\mathrm{d}t}=\mathsf{v}\cdot 
\frac{\mathrm{d}m\mathsf{v}}{\mathrm{d}t}=q\mathsf{v}\cdot \mathbf{E}_{%
\mathrm{ex}}\left( t\right) ,  \label{crv8}
\end{equation}%
and this combined with the second equality in (\ref{crv7}) implies the
following energy kinematic representation:%
\begin{equation}
\mathsf{E}=\frac{m\mathsf{v}\cdot \mathsf{v}}{2}+\limfunc{constant}.
\label{crv9}
\end{equation}%
Combining the relations (\ref{crv5})-(\ref{crv7}) we also obtain%
\begin{equation}
m\frac{\mathrm{d}^{2}\mathbf{r}\left( t\right) }{\mathrm{d}t^{2}}=q\mathbf{E}%
_{\mathrm{ex}}\left( t\right) +\frac{q}{\mathrm{c}}\frac{\mathrm{d}\mathbf{r}%
\left( t\right) }{\mathrm{d}t}\times \mathbf{B}_{\mathrm{ex}}\left( t\right)
,  \label{crv10}
\end{equation}%
in which we recognize the point charge in a homogeneous EM field dynamic
equation with the familiar expression for the Lorentz force. Notice the
above\ found correspondence between field quantities and point mechanics
quantities via the charge position and velocity defined as average values (%
\ref{crv4}) is similar to the well known \emph{Ehrenfest\ Theorem} in
quantum mechanics, \cite[Sections 7, 23]{Schiff}. This is, of course, not
accidental as one can see from the Lagrangian representation of the Schr\"{o}%
dinger wave mechanics briefly discussed in Section \ref{sschrodinger}. \emph{%
The key argument for the Ehrenfest theorem as in our case is the momentum
density kinematic representation (\ref{emna11})-(\ref{emna12})}.

\subsection{Accelerated motion in an external electric field}

In this subsection we consider a purely electric external EM field, i.e.
when $\mathbf{A}_{\mathrm{ex}}=0$, for which the field equations (\ref{NLS1}%
), (\ref{fis}) take the form 
\begin{gather}
\mathrm{i}\chi \partial _{t}\psi =-\frac{\chi ^{2}\nabla ^{2}\psi }{2m}%
+q\left( \varphi +\varphi _{\mathrm{ex}}\right) \psi +\frac{\chi ^{2}}{2m}%
G^{\prime }\left( \left\vert \psi \right\vert ^{2}\right) \psi ,\ 
\label{NLSel} \\
\nabla ^{2}\varphi =-4\pi q\left\vert \psi \right\vert ^{2}.  \notag
\end{gather}%
In this case the wave-corpuscle is defined by the formula (\ref{psil0})" \
with the complementary point charge equations (\ref{Lor1}) taking the form%
\begin{gather}
m\frac{\mathrm{d}^{2}\mathbf{r}\left( t\right) }{\mathrm{d}t^{2}}=q\mathbf{E}%
_{\mathrm{ex}}\left( t,\mathbf{r}\right) ,  \label{nlaw} \\
\ \mathbf{r}\left( 0\right) =\mathbf{r}_{0},\ \frac{\mathrm{d}\mathbf{r}}{%
\mathrm{d}t}\left( 0\right) =\mathbf{\dot{r}}_{0},  \notag
\end{gather}%
where$\ \mathbf{E}_{\mathrm{ex}}\left( t,\mathbf{x}\right) =-\nabla \varphi
_{\mathrm{ex}}\left( t,\mathbf{x}\right) ,$ $\mathbf{r}_{0}$ and $\mathbf{%
\dot{r}}_{0}$ are the initial data and in (\ref{Lor1})%
\begin{equation}
\mathbf{v}\left( t\right) =\frac{\mathrm{d}\mathbf{r}\left( t\right) }{%
\mathrm{d}t},\ \frac{\mathrm{d}s_{\mathrm{p}}\left( t\right) }{\mathrm{d}t}=%
\frac{m\mathbf{v}^{2}\left( t\right) }{2}-q\varphi _{\mathrm{ex}}\left( t,%
\mathbf{r}\left( t\right) \right) .  \label{eel}
\end{equation}%
In the case when the external electric field is homogeneous we show that the
wave-corpuscle is an exact solution to the field equations (\ref{NLSel}),
and if the external electric field is inhomogeneous we show that the
wave-corpuscle is an accurate approximation. Since the electric field
homogeneity plays a role in the wave-corpuscle representation, it is
convenient to extract from the external electric field potential $\varphi _{%
\mathrm{ex}}\left( t,\mathbf{x}\right) $ its linear in $\mathbf{x}$ part $%
\tilde{\varphi}_{\mathrm{ex}}\left( t,\mathbf{x}\right) $ about the
trajectory $\mathbf{r}\left( t\right) $, namely we represent $\varphi _{%
\mathrm{ex}}\left( t,\mathbf{x}\right) $ in the form 
\begin{equation}
\varphi _{\mathrm{ex}}\left( t,\mathbf{x}\right) =\tilde{\varphi}_{\mathrm{ex%
}}\left( t,\mathbf{x}\right) +\varphi _{\mathrm{ex}}^{\left( 1\right)
}\left( t,\mathbf{x}\right) \text{, where}  \label{fiex1}
\end{equation}%
\begin{equation*}
\tilde{\varphi}_{\mathrm{ex}}\left( t,\mathbf{x}\right) =\varphi _{0,\mathrm{%
ex}}\left( t\right) +\varphi _{0,\mathrm{ex}}^{\prime }\left( t\right) \cdot
\left( \mathbf{x}-\mathbf{r}\left( t\right) \right) ,\text{ and}
\end{equation*}%
\begin{equation*}
\varphi _{0,\mathrm{ex}}\left( t\right) =\left. \varphi _{\mathrm{ex}}\left(
t,\mathbf{x}\right) \right\vert _{\mathbf{x}=\mathbf{r}\left( t\right) },\
\varphi _{0,\mathrm{ex}}^{\prime }\left( t\right) =\left. \nabla _{\mathbf{x}%
}\varphi _{\mathrm{ex}}\left( t,\mathbf{x}\right) \right\vert _{\mathbf{x}=%
\mathbf{r}\left( t\right) }.
\end{equation*}%
The remainder $\varphi _{\mathrm{ex}}^{\left( 1\right) }\left( t,\mathbf{x}%
\right) $ in (\ref{fiex1}) is defined by 
\begin{equation}
\varphi _{\mathrm{ex}}^{\left( 1\right) }\left( t,\mathbf{x}\right) =\varphi
_{\mathrm{ex}}\left( t,\mathbf{x}\right) -\varphi _{\mathrm{ex}}\left( t,%
\mathbf{r}\right) -\nabla \varphi _{\mathrm{ex}}\left( t,\mathbf{r}\right)
\left( \mathbf{x}-\mathbf{r}\right) ,\text{ where }\mathbf{r}=\mathbf{r}%
\left( t\right)  \label{figr0}
\end{equation}%
and, consequently, it satisfiies%
\begin{equation*}
\left. \varphi _{\mathrm{ex}}^{\left( 1\right) }\left( t,\mathbf{x}\right)
\right\vert _{\mathbf{x}=\mathbf{r}\left( t\right) }=0,\mathbf{\ }\left.
\nabla \varphi _{\mathrm{ex}}^{\left( 1\right) }\left( t,\mathbf{x}\right)
\right\vert _{\mathbf{x}=\mathbf{r}\left( t\right) }=0.
\end{equation*}%
Notice that in Section \ref{nrapr} we used a slightly different from (\ref%
{fiex1}) form for the linear external potential $\varphi _{\mathrm{ex}%
}\left( t,\mathbf{x}\right) $, namely $\varphi _{\mathrm{ex}}\left( t,%
\mathbf{x}\right) =\varphi _{\mathrm{ex}}^{0}\left( t\right) -\mathbf{E}_{%
\mathrm{ex}}\left( t\right) \cdot \mathbf{x}$, where 
\begin{equation*}
\varphi _{\mathrm{ex}}^{0}\left( t\right) =\varphi _{0,\mathrm{ex}}\left(
t\right) +\mathbf{r}\left( t\right) \cdot \nabla \varphi _{\mathrm{ex}%
}\left( t,\mathbf{r}\left( t\right) \right) .
\end{equation*}

\subsubsection{ Accelerated motion in an external homogeneous electric field 
\label{exel}}

If the external field is a purely electric and homogeneous field $\mathbf{E}%
_{\mathrm{ex}}\left( t\right) $ then its potential $\varphi _{\mathrm{ex}%
}\left( t,\mathbf{x}\right) $ is linear in $\mathbf{x}$ and the
representation (\ref{fiex1}) turns into%
\begin{equation}
\varphi _{\mathrm{ex}}\left( t,\mathbf{x}\right) =\tilde{\varphi}_{\mathrm{ex%
}}\left( t,\mathbf{x}\right) =\varphi _{0,\mathrm{ex}}\left( t\right)
+\varphi _{0,\mathrm{ex}}^{\prime }\left( t\right) \cdot \left( \mathbf{x}-%
\mathbf{r}\left( t\right) \right) ,\   \label{elst}
\end{equation}%
where%
\begin{equation*}
\varphi _{0,\mathrm{ex}}^{\prime }\left( t\right) =\nabla _{\mathbf{x}%
}\varphi _{\mathrm{ex}}\left( \mathbf{r}\left( t\right) ,t\right) =-\mathbf{E%
}_{\mathrm{ex}}\left( t\right) .
\end{equation*}%
The main result of this section is that the \emph{wave-corpuscle as defined
by formula (\ref{psil0}) is an exact solution to the field equation (\ref%
{NLSel}) that can be verified by a straightforward examination}. One can
alternatively establish that result by considering the expression for $\psi $
in (\ref{psil0}) and assuming that the real valued functions $\mathbf{r}%
\left( t\right) $, $\mathbf{v}\left( t\right) $ and $s_{\mathrm{p}}\left(
t\right) $ are unknown and to be found, if possible, from the field
equations (\ref{NLSel}). Indeed, observe that the representation (\ref{psil0}%
) implies 
\begin{equation}
\partial _{t}\psi =\exp \left( \mathrm{i}\frac{S}{\chi }\right) \left\{ %
\left[ \frac{\mathrm{i}m}{\chi }\left( \mathbf{\dot{v}}\cdot \left( \mathbf{x%
}-\mathbf{r}\right) -\mathbf{v}\cdot \mathbf{\dot{r}}\right) +\frac{\mathrm{i%
}\dot{s}_{\mathrm{p}}}{\chi }\right] \hat{\psi}-\mathbf{\dot{r}}\cdot \nabla 
\hat{\psi}\right\} ,  \label{dtpp}
\end{equation}%
\begin{equation*}
\nabla \hat{\psi}=\mathring{\psi}^{\prime }\left( \mathbf{x}-\mathbf{r}%
\right) \frac{\mathbf{x}-\mathbf{r}}{\left\vert \mathbf{x}-\mathbf{r}%
\right\vert },
\end{equation*}%
and by the Leibnitz formula we have 
\begin{equation}
\nabla ^{2}\psi =\exp \left( \mathrm{i}\frac{S}{\chi }\right) \left[ \left( 
\frac{\mathrm{i}m\mathbf{v}}{\chi }\right) ^{2}\hat{\psi}+2\frac{\mathrm{i}m%
}{\chi }\mathbf{v}\cdot \nabla \hat{\psi}+\nabla ^{2}\hat{\psi}\right] .
\label{del2p}
\end{equation}%
To find if the expression (\ref{psil0}) for $\psi $ can solve the field
equations (\ref{NLSel}) we substitute the expression into the field
equations (\ref{NLSel}) obtaining the following equation for functions $%
\mathbf{v}$, $\mathbf{r}$, $s_{\mathrm{p}}$ : 
\begin{gather}
\left[ -m\mathbf{\dot{v}}\cdot \left( \mathbf{x}-\mathbf{r}\right) -\mathbf{v%
}\cdot \mathbf{\dot{r}}-\dot{s}_{\mathrm{p}}\right] \hat{\psi}-\mathrm{i}%
\chi \mathbf{\dot{r}}\cdot \nabla \hat{\psi}  \label{eq2} \\
-\frac{m}{2}\mathbf{v}^{2}\hat{\psi}+\mathrm{i}\chi \mathbf{v}\cdot \nabla 
\hat{\psi}+\frac{\chi ^{2}}{2m}\nabla ^{2}\hat{\psi}-q\left( \tilde{\varphi}%
_{\mathrm{ex}}+\varphi \right) \hat{\psi}-\frac{\chi ^{2}}{2m}G^{\prime }%
\hat{\psi}=0.  \notag
\end{gather}%
Then using the charge equilibrium equation (\ref{nop7}) we eliminate the
nonlinearity $G$ in the above equation (\ref{eq2}) and obtain the following
equation equivalent to it: 
\begin{equation}
-\left\{ m\left[ \mathbf{\dot{v}}\cdot \left( \mathbf{x}-\mathbf{r}\right) -%
\mathbf{v}\cdot \mathbf{\dot{r}}\right] +\frac{m}{2}\mathbf{v}^{2}+\dot{s}_{%
\mathrm{p}}+q\tilde{\varphi}_{\mathrm{ex}}\right\} \hat{\psi}-\mathrm{i}\chi
\left( \mathbf{\dot{r}-v}\right) \nabla \hat{\psi}=0.  \label{eq3}
\end{equation}%
Now to determine if there is a triple of functions $\left\{ \mathbf{r}\left(
t\right) ,\mathbf{v}\left( t\right) ,s_{\mathrm{p}}\left( t\right) \right\} $
for which the equation (\ref{eq3}) holds we equate to zero the coefficients
before $\nabla \hat{\psi}$\ and $\hat{\psi}$ in that equation, resulting in
two equations: 
\begin{equation}
\mathbf{v}=\mathbf{\dot{r}},\ m\left[ \mathbf{\dot{v}}\cdot \left( \mathbf{x}%
-\mathbf{r}\right) -\mathbf{v}\cdot \mathbf{\dot{r}}\right] +\frac{m}{2}%
\mathbf{v}^{2}+\dot{s}_{\mathrm{p}}+q\tilde{\varphi}_{\mathrm{ex}}=0,
\label{ra}
\end{equation}%
where, in view of the representation (\ref{elst}), the second equation in (%
\ref{ra}) can be recast as 
\begin{equation}
m\left[ \mathbf{\dot{v}}\cdot \left( \mathbf{x}-\mathbf{r}\right) -\mathbf{v}%
\cdot \mathbf{\dot{r}}\right] +\dot{s}_{\mathrm{p}}+\frac{m\mathbf{v}^{2}}{2}%
+q\left[ \varphi _{0,\mathrm{ex}}+\varphi _{0,\mathrm{ex}}^{\prime }\cdot
\left( \mathbf{x}-\mathbf{r}\right) \right] =0.  \label{fir}
\end{equation}%
To find out if there is a triple of functions $\left\{ \mathbf{r}\left(
t\right) ,\mathbf{v}\left( t\right) ,s_{\mathrm{p}}\left( t\right) \right\} $
solving the equation (\ref{fir}) we equate to zero the coefficient before $%
\left( \mathbf{x}-\mathbf{r}\right) $ and the remaining coefficient and
obtain the following pair of equations%
\begin{equation}
m\mathbf{\dot{v}}=-q\varphi _{0,\mathrm{ex}}^{\prime }\left( t\right) ,\ 
\dot{s}_{\mathrm{p}}-m\mathbf{v}\cdot \mathbf{\dot{r}}+\frac{m\mathbf{v}^{2}%
}{2}+q\varphi _{0,\mathrm{ex}}\left( t\right) =0.  \label{V1a}
\end{equation}%
Thus, based on the first equation (\ref{ra}) and the equations (\ref{V1a})
we conclude that the wave-corpuscle defined by the formula (\ref{Lor1}) with
the complementary point charge equations (\ref{nlaw}), (\ref{eel}) is indeed
an exact solution to the field equations (\ref{NLSel}).

It is instructive to compare now construction of the exact solutions (\ref%
{psil0}) with the quasi-classical approach based on the WKB theory. The
trajectories of the charges centers as described by our model coincide with
ones obtained based on the well-known quasiclassical asymptotics if one
neglects the nonlinearity. Note though that there are two important effects
of the nonlinearity that are not presented in the formal quasiclassical
approach. First of all, due to the nonlinearity the charge preserves its
shape in the course of evolution whereas in the linear model any wavepacket
disperses over time. Second of all, the quasiclassical asymptotic expansions
produce infinite asymptotic series which provide for a formal solution,
whereas the properly introduced nonlinearity as in (\ref{nop4}), (\ref{nop5}%
) allows one to obtain an exact solution. For a treatment of a nonlinear
wave mechanics based on the WKB asymptotic expansions we refer the reader to 
\cite{Komech05} and references therein.

\subsubsection{ Accelerated motion in an external inhomogeneous electric
field\label{seinhomo}}

In this section we consider a general external electric field $\mathbf{E}_{%
\mathrm{ex}}\left( t,\mathbf{r}\right) $ which can be inhomogeneous with the
corresponding electric potential $\varphi _{\mathrm{ex}}\left( t,\mathbf{x}%
\right) $ as described by relations (\ref{fiex1}), (\ref{figr0}) with
nonzero remainder $\varphi _{\mathrm{ex}}^{\left( 1\right) }\left( t,\mathbf{%
x}\right) $. For an inhomogeneous external electric field $\mathbf{E}_{%
\mathrm{ex}}\left( t,\mathbf{r}\right) $ no closed form solution to the
field equations (\ref{NLSel}) seems to be available but the wave-corpuscle
defined by the relations (\ref{psil0}) with complementary point charge
equations (\ref{nlaw}), (\ref{eel}) turns out to be a good approximation
with the accuracy dependent on (i) the size parameter $a$ defined by
relations (\ref{nrac5}) and (ii) the degree of spatial inhomogeneity of the
electric field measured by the \emph{electric field inhomogeneity length} $%
R_{\mathrm{ex}}$ introduced below. The parameter $R_{\mathrm{ex}}$ is
similar to the radius of curvature of the graph of $\varphi _{\mathrm{ex}%
}\left( t,\mathbf{x}\right) $, and large or small values $R_{\mathrm{ex}}$
correspond, respectively, to almost homogeneous or highly inhomogeneous
electric fields. It turns out that the wave-corpuscle solves the field
equations (\ref{NLSel}) with the discrepancy $D=O\left( \left( a/R_{\mathrm{%
ex}}\right) ^{2}\right) $ for $a\ll R_{\mathrm{ex}}$ as we show below. \emph{%
We fix now for the rest of this section the initial data }$r_{0}$\emph{\ and 
}$v_{0}$\emph{\ in (\ref{nlaw}) and consequently the function }$\mathbf{r}%
\left( t\right) $. We assume here that the factor $\left\vert \mathring{\psi}%
_{1}\left( \left\vert \mathbf{x}\right\vert \right) \right\vert ^{2}$ \emph{%
decays exponentially} as $\left\vert \mathbf{x}\right\vert \rightarrow
\infty $, and, in particular, for some constant $C_{0}$ 
\begin{equation}
\int_{\mathbb{R}^{3}}\left\vert \mathring{\psi}_{1}\left( \left\vert \mathbf{%
x}\right\vert \right) \right\vert ^{2}\left\vert \mathbf{x}\right\vert ^{2}\,%
\mathrm{d}\mathbf{x}\leq C_{0},\mathbf{\ }\int_{\mathbb{R}^{3}}\left\vert 
\mathring{\psi}_{1}\left( \left\vert \mathbf{x}\right\vert \right)
\right\vert ^{2}\left\vert \mathbf{x}\right\vert \,\mathrm{d}\mathbf{x}\leq
C_{0}.  \label{psexp}
\end{equation}%
To assess the accuracy of the wave-corpuscle solution defined by relations (%
\ref{psil0}), (\ref{nlaw}), (\ref{eel}) we follow an approach discussed in
Section \ref{swcacel}. Namely, we introduce \ write auxiliary field equation
(\ref{Schmag1}) \ with $\mathbf{\tilde{A}}_{\mathrm{ex}}=0$ in the form 
\emph{\ }%
\begin{gather}
\mathrm{i}\chi \partial _{t}\psi -\frac{\chi ^{2}\nabla ^{2}\psi }{2m}%
+q\left( \varphi +\tilde{\varphi}_{\mathrm{ex}}\left( t,\mathbf{x}\right)
\right) \psi -\frac{\chi ^{2}G^{\prime }\psi }{2m}=0,\mathbf{\ }
\label{auf1} \\
\nabla ^{2}\varphi =-4\pi q\left\vert \psi \right\vert ^{2},  \notag
\end{gather}%
where $\tilde{\varphi}_{\mathrm{ex}}\left( t,\mathbf{x}\right) $ is defined
by (\ref{fiex1}). In view of the relations (\ref{elst}) the corresponding
external field $\mathbf{E}_{\mathrm{ex}}\left( t\right) =-\nabla _{\mathbf{x}%
}\varphi _{\mathrm{ex}}\left( \mathbf{r}\left( t\right) ,t\right) $ is
homogeneous. A straightforward examination shows that the results of Section %
\ref{exel} apply and the wave-corpuscle defined by (\ref{psil0}) with
complementary point charge equations (\ref{nlaw}), (\ref{eel}) is an exact
solution to the auxiliary field equations (\ref{auf1}). Now notice that the
auxiliary field equations differ from the original ones (\ref{NLSel}) only
by the replacement of $\varphi _{\mathrm{ex}}\left( t,\mathbf{x}\right) $
with $\tilde{\varphi}_{\mathrm{ex}}\left( t,\mathbf{x}\right) $ with the
consequent \emph{discrepancy}%
\begin{equation}
D_{0}\left( t,\mathbf{x}\right) =q\left[ \tilde{\varphi}_{\mathrm{ex}}\left(
t,\mathbf{x}\right) -\varphi _{\mathrm{ex}}\left( t,\mathbf{x}\right) \right]
\psi \left( t,\mathbf{x}\right) =-q\varphi ^{\left( 1\right) }\left( t,%
\mathbf{x}\right) \psi \left( t,\mathbf{x}\right) .  \label{auf2}
\end{equation}%
Based on the above discrepancy and taking into account the dependence on
size parameter $a$ for $\mathring{\psi}=\mathring{\psi}_{a}$ as in (\ref%
{nrac5}) we introduce the \emph{integral discrepancy} 
\begin{gather}
\bar{D}_{0}=\int_{\mathbb{R}^{3}}D_{0}\left( t,\mathbf{x}\right) \psi ^{\ast
}\,\mathrm{d}\mathbf{x}=\,\int_{\mathbb{R}^{3}}-q\varphi ^{\left( 1\right)
}\left( t,\mathbf{x}\right) \psi \psi ^{\ast }\,\mathrm{d}\mathbf{x}
\label{discr1} \\
=\,\int_{\mathbb{R}^{3}}-qa^{-3}\left\vert \mathring{\psi}_{1}\left(
a^{-1}\left\vert \mathbf{x}-\mathbf{r}\right\vert \right) \right\vert
^{2}\varphi _{\mathrm{ex}}^{\left( 1\right) }\left( t,\mathbf{x}\right) \,%
\mathrm{d}\mathbf{x}.  \notag
\end{gather}%
Notice that the relations (\ref{fiex1}) for $\tilde{\varphi}_{\mathrm{ex}%
}\left( t,\mathbf{x}\right) $ and the charge normalization condition (\ref%
{expo8}) imply that a similar integral involving $\tilde{\varphi}_{\mathrm{ex%
}}$ \ equals 
\begin{equation}
\int_{\mathbb{R}^{3}}q\tilde{\varphi}_{\mathrm{ex}}\left( t,\mathbf{x}%
\right) \left\vert \psi \right\vert ^{2}\,\mathrm{d}\mathbf{x}=q\varphi _{%
\mathrm{ex}}\left( t,\mathbf{r}\left( t\right) \right) ,  \label{elenr}
\end{equation}%
which coincides with the potential energy of a point charge $q$ in the
potential $\ \varphi _{\mathrm{ex}}\left( t,\mathbf{x}\right) ,$ therefore
it is natural to compare $\bar{D}_{0}$ with variation of this energy in the
dynamics. To assess typical scales of inhomogeneity of the external field we
introduce the potential variation quantity%
\begin{equation}
\bar{\varphi}_{0,T}=\max_{0\leq t\leq T}\left\vert \varphi _{\mathrm{ex}%
}\left( t,\mathbf{r}\left( t\right) \right) -\varphi _{\mathrm{ex}}\left( 0,%
\mathbf{r}\left( 0\right) \right) \right\vert .  \label{elenr1}
\end{equation}%
Note that $\left\vert q\right\vert \bar{\varphi}_{0,T}$ equals the point
charge potential energy variation on time interval $\left[ 0,T\right] .$ We
also introduce a parameter $\sigma _{\psi }$ which plays a role similar to $%
3\sigma $ for Gaussian probability but with respect to the function $%
\mathring{\psi}_{1}\left( \left\vert z\right\vert \right) ^{2}$\ in (\ref%
{nrac5}), namely%
\begin{equation}
\mathring{\psi}_{1}\left( r\right) \simeq 0\text{ if \ }r\geq \sigma _{\psi
}.  \label{thetpsi}
\end{equation}%
The above approximation means that the discrepancy created by replacing $%
\mathring{\psi}_{1}\left( r\right) $ by zero for large $r$ is smaller than
other discrepancies that appear below.

Let us introduce the following characteristic lengths $R_{\varphi }\left( t,%
\mathbf{r}\right) $ and $R_{\varphi }$ similar to the radius of the
curvature: 
\begin{equation}
\frac{1}{R_{\varphi }^{2}\left( t,\mathbf{r}\right) }=\sup_{0<\left\vert
z\right\vert \leq a\sigma _{\psi }}\frac{\left\vert \varphi _{\mathrm{ex}%
}^{\left( 1\right) }\left( t,\mathbf{r}+\mathbf{z}\right) \right\vert }{%
\mathbf{z}^{2}\left\vert \bar{\varphi}\right\vert },\ \frac{1}{R_{\varphi
}^{2}}=\max_{0\leq t\leq T}\frac{1}{R_{\varphi }^{2}\left( t,\mathbf{r}%
\left( t\right) \right) },  \label{rfi}
\end{equation}%
where%
\begin{equation}
\left\vert \bar{\varphi}\right\vert =\max_{0\leq t\leq T}\max_{0<\left\vert
z\right\vert \leq a\sigma _{\psi }}\left\vert \varphi \left( t,\mathbf{r}%
\left( t\right) +\mathbf{z}\right) -\varphi \left( 0,\mathbf{r}\left(
0\right) \right) \right\vert .  \label{fibar}
\end{equation}%
The quantity $R_{\varphi }$ represents the typical spatial scale at which
the spatially curvilinear component $\varphi _{\mathrm{ex}}^{\left( 1\right)
}\left( t,\mathbf{x}\right) \ $of the external field $\varphi _{\mathrm{ex}%
}\left( t,\mathbf{x}\right) $ changes significantly in a vicinity of $%
\mathbf{r}\left( t\right) $. For small $a$\ the quantity $R_{\varphi }$ is
essentially determined by the maximal eigenvalue $\left\vert \lambda _{\max
}\right\vert $ of the matrix of the second spatial derivatives of $\varphi _{%
\mathrm{ex}}\left( t,\mathbf{x}\right) $ at $\mathbf{x}=\mathbf{r}\left(
t\right) $. In addition to that, $R_{\varphi }\left( t,\mathbf{r}\right)
\rightarrow 1/\left\vert \lambda _{\max }\right\vert $ as $a\rightarrow 0$
where $1/\left\vert \lambda _{\max }\right\vert $ is the minimal curvature
radius of the graph of the normalized potential $\varphi _{\mathrm{ex}%
}^{\left( 1\right) }\left( t,\mathbf{x}\right) /\left\vert \bar{\varphi}%
\right\vert $ at the point $\mathbf{r}\left( t\right) $. It follows from (%
\ref{figr0}) and (\ref{rfi}) that $1/R_{\varphi }^{2}$ is bounded as long as
the curve $\mathbf{r}\left( t\right) $\ is not close to singularities of the
external field $\varphi _{\mathrm{ex}}\left( t,\mathbf{x}\right) $ if any.
Then we estimate the integral discrepancy as follows: 
\begin{align}
\left\vert \bar{D}_{0}\right\vert & =\left\vert q\right\vert \left\vert
\,\int_{\mathbb{R}^{3}}a^{-3}\left\vert \mathring{\psi}_{1}\left(
a^{-1}\left\vert \mathbf{x}-\mathbf{r}\right\vert \right) \right\vert
^{2}\varphi _{\mathrm{ex}}^{\left( 1\right) }\left( t,\mathbf{r}+\mathbf{x}-%
\mathbf{r}\right) \,\mathrm{d}\mathbf{x}\right\vert  \label{d0} \\
& \lesssim \frac{\left\vert q\right\vert \left\vert \bar{\varphi}\right\vert 
}{R_{\varphi }^{2}}\,\int_{\mathbb{R}^{3}}a^{-3}\left\vert \mathring{\psi}%
_{1}\left( a^{-1}\left\vert \mathbf{x}-\mathbf{r}\right\vert \right)
\right\vert ^{2}\left( \mathbf{x}-\mathbf{r}\right) ^{2}\,\mathrm{d}\mathbf{x%
}\leq C_{0}\frac{a^{2}\left\vert q\right\vert \left\vert \bar{\varphi}%
\right\vert }{R_{\varphi }^{2}}.  \notag
\end{align}%
\ Combining inequality (\ref{d0}) with relations (\ref{elenr}), (\ref{elenr1}%
), (\ref{fibar}) we can judge the quality of approximation by requiring the
relative dimensionless discrepancy $\left\vert \bar{D}_{0}\right\vert
/\left( \left\vert q\right\vert \bar{\varphi}_{0,T}\right) $ to be small,
namely 
\begin{equation}
\frac{\left\vert \bar{D}_{0}\right\vert }{\left\vert q\right\vert \bar{%
\varphi}_{0,T}}\lesssim \frac{a^{2}\left\vert \bar{\varphi}\right\vert }{%
R_{\varphi }^{2}\bar{\varphi}_{0,T}}\ll 1\text{ is a requirement for an
accurate approximation.}  \label{d01}
\end{equation}

For further applications we briefly consider an effect on the discrepancy of
a perturbation of the external potential $\varphi _{\mathrm{ex}}\left( t,%
\mathbf{x}\right) $ when it is substituted with a slightly different
potential $\acute{\varphi}_{\mathrm{ex}}\left( t,\mathbf{x},\epsilon \right) 
$ \ with $\epsilon $ being a small perturbation parameter and the
approximate solution is determined based on $\varphi _{\mathrm{ex}}\left( t,%
\mathbf{x}\right) =\acute{\varphi}_{\mathrm{ex}}\left( t,\mathbf{x},0\right) 
$. Supposing the initial data $\mathbf{r}_{0}$, $\mathbf{\dot{r}}_{0}$ and
hence the position function (trajectory) $\mathbf{r}\left( t\right) $
solving the equation of motion (\ref{nlaw}) being fixed we assume that there
exists fixed positive constants $C$, $C_{1}$, $T$ and $\epsilon _{1}$ such
that for any small $\epsilon $ we have 
\begin{equation}
\left\vert \varphi _{\mathrm{ex}}\left( t,\mathbf{x}\right) -\acute{\varphi}%
_{\mathrm{ex}}\left( t,\mathbf{x},\epsilon \right) \right\vert \leq
C_{1}\epsilon ,\mathbf{\ }\left\vert \nabla \varphi _{\mathrm{ex}}\left( t,%
\mathbf{x}\right) -\nabla \acute{\varphi}_{\mathrm{ex}}\left( t,\mathbf{x}%
,\epsilon \right) \right\vert \leq C_{1}\epsilon ,  \label{eeep}
\end{equation}%
for any $t$ and $\mathbf{x}$ such that%
\begin{equation*}
\left\vert \mathbf{x-r}\left( t\right) \right\vert \leq \epsilon _{1},%
\mathbf{\ }0\leq t\leq T.
\end{equation*}%
The above condition simply requires the external perturbed field potential
to be close to the original one in a small vicinity of the trajectory $%
\mathbf{r}\left( t\right) $. Substitution of the original wave-corpuscle
solution $\left\{ \psi ,\varphi \right\} $ defined by (\ref{psil0})
(corresponding to $\epsilon =0$ ) and the original complementary point
charge equations of motion (\ref{nlaw}) for $\mathbf{r}\left( t\right) $
into the equation (\ref{NLSel}) with the external potential $\acute{\varphi}%
_{\mathrm{ex}}\left( t,\mathbf{x},\epsilon \right) $ \ produces the total
discrepancy 
\begin{equation}
\acute{D}_{0}\left( t,\mathbf{x}\right) =D_{0}\left( t,\mathbf{x}\right) +%
\hat{D}_{1}\left( t,\mathbf{x}\right) ,\ \hat{D}_{1}\left( t,\mathbf{x}%
\right) =-\left[ \varphi _{\mathrm{ex}}\left( t,\mathbf{x}\right) -\acute{%
\varphi}_{\mathrm{ex}}\left( t,\mathbf{x},\epsilon \right) \right] \psi .
\label{eeep2}
\end{equation}%
Note that if $a$ is small $a\sigma _{\psi }\leq \epsilon _{1}$ and using (%
\ref{eeep}) we get 
\begin{align}
\left\vert \,\int_{\mathbb{R}^{3}}\hat{D}_{1}\left( t,\mathbf{x}\right) \psi
^{\ast }\,\mathrm{d}\mathbf{x}\right\vert & \lesssim \left\vert
\int_{\left\vert \mathbf{r}-\mathbf{r}\left( t\right) \right\vert \leq
\epsilon _{1}}\left\vert \varphi _{\mathrm{ex}}\left( t,\mathbf{x}\right) -%
\acute{\varphi}_{\mathrm{ex}}\left( t,\mathbf{x},\epsilon \right)
\right\vert \left\vert \psi \right\vert ^{2}\mathrm{d}\mathbf{x}\right\vert
\label{eeep3} \\
& \leq \sup_{\left\vert \mathbf{r}-\mathbf{r}\left( t\right) \right\vert
\leq \epsilon _{1}}\left\vert \varphi _{\mathrm{ex}}\left( t,\mathbf{r}%
\right) -\acute{\varphi}_{\mathrm{ex}}\left( t,\mathbf{r},\epsilon \right)
\right\vert \leq C_{1}\epsilon ,  \notag
\end{align}%
where $C_{1}/\left\vert \bar{\varphi}\right\vert $ is a dimensionless
constant. Combining (\ref{d0}), (\ref{eeep2}), (\ref{eeep3}) we get the
following rough estimate 
\begin{equation}
\frac{\left\vert \hat{D}_{0}\left( t,\mathbf{x}\right) \right\vert
+\left\vert \hat{D}_{1}\left( t,\mathbf{x}\right) \right\vert }{\left\vert 
\bar{\varphi}\right\vert }\lesssim \left( C_{0}\frac{a^{2}}{R_{\varphi }^{2}}%
+\frac{C_{1}\epsilon }{\left\vert \bar{\varphi}\right\vert }\right)
\left\vert q\right\vert .  \label{totd}
\end{equation}

It is instructive to look at the trajectory $\mathbf{r}\left( t\right) $
determined from (\ref{nlaw}) for previously constructed exact and
approximate wave-corpuscle solutions from another point of view. Namely, let
us introduce a moving frame $\mathbf{y}=\mathbf{x}-\mathbf{r}\left( t\right) 
$, where $\mathbf{r}=\mathbf{r}\left( t\right) $, $s_{\mathrm{p}}=s_{\mathrm{%
p}}\left( t\right) $, $\mathbf{v}=\mathbf{v}\left( t\right) $ solve (\ref%
{nlaw}), (\ref{eel}). Notice that the origin of the new frame is at the
center $\mathbf{r}\left( t\right) $ of the wave-corpuscle. \ Let us change
variables in equations (\ref{NLSel}) 
\begin{equation}
\psi \left( t,\mathbf{x}\right) =\exp \left\{ \frac{\mathrm{i}m\mathbf{v}%
\cdot \mathbf{y}}{\chi }+\frac{\mathrm{i}s_{\mathrm{p}}}{\chi }\right\} \hat{%
\psi}\left( t\mathbf{,y}\right) ,\ \varphi \left( \mathbf{x}\right) =\hat{%
\varphi}\left( \mathbf{y}\right) ,
\end{equation}%
where $\hat{\psi}\left( t\mathbf{,y}\right) $ \ is a new unknown function.
Then repeating the above calculations (without using (\ref{nop7})) we obtain
an equivalent equation of the same form as (\ref{NLSel}), namely 
\begin{gather}
\mathrm{i}\chi \partial _{t}\hat{\psi}=-\frac{\chi ^{2}}{2m}\nabla _{y}^{2}%
\hat{\psi}+q\left( \hat{\varphi}+\hat{\varphi}_{\mathrm{ex}}\right) \hat{\psi%
}+\frac{\chi ^{2}}{2m}G^{\prime }\left( \left\vert \hat{\psi}\right\vert
^{2}\right) \acute{\psi},\   \label{NLSel1} \\
\nabla _{y}^{2}\hat{\varphi}=-4\pi q\left\vert \hat{\psi}\right\vert ^{2}, 
\notag
\end{gather}%
with an external potential $\hat{\varphi}_{\mathrm{ex}}\left( t,y\right) $
which satisfies an additional condition $\hat{\varphi}_{\mathrm{ex}}\left(
0\right) =0$,\ $\nabla \hat{\varphi}_{\mathrm{ex}}\left( 0\right) =0$. If
the original potential $\varphi _{\mathrm{ex}}\left( t,\mathbf{x}\right) $
is linear in $\mathbf{x},$ the external potential $\hat{\varphi}_{\mathrm{ex}%
}\left( t,\mathbf{y}\right) $ in the moving frame vanishes, i.e. $\hat{%
\varphi}_{\mathrm{ex}}\left( t,\mathbf{y}\right) =0$ for all $t,\mathbf{y}$.
In this case (\ref{NLSel1}) coincides with the equilibrium condition (\ref%
{nop7}), hence $\hat{\psi}=\mathring{\psi}$ and the wave-corpuscle rests at
the origin of the moving frame.

\subsection{\ Accelerated motion in a general external EM field\label%
{exelmag}}

In this subsection we consider a single charge in a general external EM
field \emph{which can have nonzero magnetic component}. The primary goal of
this section is to show that the wave-corpuscle defined by relations (\ref%
{psil0}) and the complementary point charge equations (\ref{Lor1}), (\ref%
{Lor2}) is an approximate solution to the field equations (\ref{NLS1}), (\ref%
{fis}). We accomplish this by following the method of Section \ref{swcacel})
and showing first that the wave-corpuscle is an exact solution to the
auxiliary field equations (\ref{Schmag1}) and then provide estimations of
the discrepancy between the auxiliary and the original field equations (\ref%
{NLS1}), (\ref{fis}).

\subsubsection{Wave-corpuscle as an exact solution\label{Thexactq}}

In this section we show that the wave-corpuscle defined by relations (\ref%
{psil0}) and the complementary point charge equations (\ref{Lor1}), (\ref%
{Lor2}) is an exact solution to the auxiliary field equations (\ref{Schmag1}%
). One way to do that is to plug in $\psi ,\varphi $ defined by the formulas
(\ref{psil0}) into the auxiliary field equations (\ref{Schmag1}) and, using
the complementary point charge equations (\ref{Lor1}), (\ref{Lor2}), verify
that the equality does hold. An alternative and more instructive, we
believe, way to accomplish the same goal is (i) to assume for the sake of
the argument that the point charge functions $\mathbf{r}\left( t\right) $, $%
\mathbf{v}\left( t\right) $, $s_{\mathrm{p}}\left( t\right) $ are unknown;
(ii) to find out if there is a way to choose those functions so that the
wave-corpuscle fields $\psi ,\varphi $ solves exactly the auxiliary field
equations (\ref{Schmag1}); (iii) verify that the chosen $\mathbf{r}\left(
t\right) $, $\mathbf{v}\left( t\right) $, $s_{\mathrm{p}}\left( t\right) $
satisfy the complementary point charge equations. Following this way we
substitute (\ref{psil0}) into (\ref{Schmag1}) and obtain the following
equation for $\mathbf{r},\mathbf{v},s_{\mathrm{p}}$: 
\begin{gather}
\mathrm{i}\left\{ \mathrm{i}m\left[ \mathbf{\dot{v}}\cdot \left( \mathbf{x}-%
\mathbf{r}\right) -\mathbf{v}\cdot \mathbf{\dot{r}}\right] +\mathrm{i}\dot{s}%
_{\mathrm{p}}\right\} \hat{\psi}-\mathrm{i}\chi \mathbf{\dot{r}}\cdot \hat{%
\psi}^{\prime }\frac{\mathbf{x}-\mathbf{r}}{\left\vert \mathbf{x}-\mathbf{r}%
\right\vert }-  \label{wpex1} \\
-\frac{m\mathbf{v}^{2}\hat{\psi}}{2}+\mathrm{i}\chi \hat{\psi}^{\prime }%
\mathbf{v}\cdot \frac{\mathbf{x}-\mathbf{r}}{\left\vert \mathbf{x}-\mathbf{r}%
\right\vert }+\frac{\chi ^{2}\nabla ^{2}\hat{\psi}}{2m}+  \notag \\
+\left[ \frac{q\mathbf{v}\hat{\psi}}{\mathrm{c}}+\frac{\chi q\hat{\psi}%
^{\prime }}{m\mathrm{ci}}\frac{\mathbf{x}-\mathbf{r}}{\left\vert \mathbf{x}-%
\mathbf{r}\right\vert }\right] \cdot \left[ \mathbf{A}_{\mathrm{ex},0}+\frac{%
\mathbf{B}_{0}\times \left( \mathbf{x}-\mathbf{r}\right) }{2}\right] - 
\notag \\
-q\left( \tilde{\varphi}_{\mathrm{ex}}+\varphi \right) \hat{\psi}-\frac{\chi
^{2}G^{\prime }\hat{\psi}}{2m}=0.  \notag
\end{gather}%
Taking into account the obvious identity $\left( \mathbf{x}-\mathbf{r}%
\right) \cdot \left[ \mathbf{B}_{0}\times \left( \mathbf{x}-\mathbf{r}%
\right) \right] =0$ we recast the\ above equation as follows: 
\begin{gather}
\left[ -m\left( \mathbf{\dot{v}}\cdot \left( \mathbf{x}-\mathbf{r}\right) -%
\mathbf{v}\cdot \mathbf{\dot{r}}\right) -\dot{s}_{\mathrm{p}}\right] \hat{%
\psi}-\mathrm{i}\chi \mathbf{\dot{r}}\cdot \hat{\psi}^{\prime }\frac{\mathbf{%
x}-\mathbf{r}}{\left\vert \mathbf{x}-\mathbf{r}\right\vert }  \notag \\
-\frac{m\mathbf{v}^{2}\hat{\psi}}{2}+\mathrm{i}\chi \hat{\psi}^{\prime }%
\mathbf{v}\cdot \frac{\mathbf{x}-\mathbf{r}}{\left\vert \mathbf{x}-\mathbf{r}%
\right\vert }+\frac{\chi ^{2}\nabla ^{2}\hat{\psi}}{2m}+\frac{\chi q\hat{\psi%
}^{\prime }}{m\mathrm{ci}}\frac{\left( \mathbf{x}-\mathbf{r}\right) }{%
\left\vert \mathbf{x}-\mathbf{r}\right\vert }\cdot \mathbf{A}_{\mathrm{ex}%
,0}+  \label{eq2a} \\
+\frac{q}{\mathrm{c}}\mathbf{v}\cdot \left[ \mathbf{A}_{\mathrm{ex},0}+\frac{%
\mathbf{B}_{0}\times \left( \mathbf{x}-\mathbf{r}\right) }{2}\right] \hat{%
\psi}-q\left( \tilde{\varphi}_{\mathrm{ex}}+\mathring{\varphi}\right) \hat{%
\psi}-\frac{\chi ^{2}}{2m}G^{\prime }\hat{\psi}=0.  \notag
\end{gather}%
Using the charge equilibrium equation (\ref{nop7}) we can eliminate $%
G^{\prime }$ in equation (\ref{eq2a}), obtaining the following equation
equivalent to it:%
\begin{gather}
-\left[ m\left( \mathbf{\dot{v}}\cdot \left( \mathbf{x}-\mathbf{r}\right) +%
\mathbf{v}\cdot \mathbf{\dot{r}}\right) +\dot{s}_{\mathrm{p}}\right] \hat{%
\psi}-  \label{eq2b} \\
-\mathrm{i}\chi \mathbf{\dot{r}}\cdot \frac{\mathbf{x}-\mathbf{r}}{%
\left\vert \mathbf{x}-\mathbf{r}\right\vert }\hat{\psi}^{\prime }-\frac{m%
\mathbf{v}^{2}\hat{\psi}}{2}+\mathrm{i}\chi \hat{\psi}^{\prime }\mathbf{v}%
\cdot \frac{\mathbf{x}-\mathbf{r}}{\left\vert \mathbf{x}-\mathbf{r}%
\right\vert }+  \notag \\
+\frac{q}{\mathrm{c}}\mathbf{v}\cdot \left[ \mathbf{A}_{\mathrm{ex},0}+\frac{%
\mathbf{B}_{0}\times \left( \mathbf{x}-\mathbf{r}\right) }{2}\right] \hat{%
\psi}+  \notag \\
+\frac{\chi q\hat{\psi}^{\prime }}{m\mathrm{ci}}\frac{\mathbf{x}-\mathbf{r}}{%
\left\vert \mathbf{x}-\mathbf{r}\right\vert }\cdot \mathbf{A}_{\mathrm{ex}%
,0}-q\tilde{\varphi}_{\mathrm{ex}}\hat{\psi}=0.  \notag
\end{gather}%
For equation (\ref{eq2b}) to hold we may require the coefficient before $%
\hat{\psi}$ and $\hat{\psi}^{\prime }$ in it to be zero. Executing that by
collecting terms with $\hat{\psi}$ and $\hat{\psi}^{\prime }$ and using $%
\tilde{\varphi}_{\mathrm{ex}}=\varphi _{0,\mathrm{ex}}+\varphi _{0,\mathrm{ex%
}}^{\prime }\cdot \left( \mathbf{x}-\mathbf{r}\right) $ we obtain the
following equations:%
\begin{equation}
\mathbf{\dot{r}}=\mathbf{v}-\frac{q}{m\mathrm{c}}\mathbf{A}_{\mathrm{ex},0},
\label{raa}
\end{equation}%
\begin{gather}
-m\left[ \mathbf{\dot{v}}\cdot \left( \mathbf{x}-\mathbf{r}\right) -\mathbf{v%
}\cdot \mathbf{\dot{r}}\right] -\dot{s}_{\mathrm{p}}-\frac{m\mathbf{v}^{2}}{2%
}+  \label{fira} \\
+\frac{q\mathbf{v}}{\mathrm{c}}\cdot \left[ \mathbf{A}_{\mathrm{ex},0}+\frac{%
1}{2}\mathbf{B}_{0}\times \left( \mathbf{x}-\mathbf{r}\right) \right]  \notag
\\
-q\left( \varphi _{0,\mathrm{ex}}+\varphi _{0,\mathrm{ex}}^{\prime }\cdot
\left( \mathbf{x}-\mathbf{r}\right) \right) =0.  \notag
\end{gather}%
To solve (\ref{fira}) we require the coefficient of $\left( \mathbf{x}-%
\mathbf{r}\right) $ and the remaining one to be zero, and this with the help
of an elementary identity $\mathbf{v}\cdot \left( \mathbf{B}_{0}\times
\left( \mathbf{x}-\mathbf{r}\right) \right) =\left( \mathbf{x}-\mathbf{r}%
\right) \cdot \left( \mathbf{v}\times \mathbf{B}_{0}\right) $ yields the
following pair of equations:%
\begin{gather}
m\mathbf{\dot{v}}=-q\left[ \frac{1}{2\mathrm{c}}\mathbf{B}_{0}\left(
t\right) \times \mathbf{v}+\varphi _{0,\mathrm{ex}}^{\prime }\left( t\right) %
\right] ,  \label{V1aa} \\
m\mathbf{v}\cdot \mathbf{\dot{r}}-\dot{s}_{\mathrm{p}}+\frac{q}{\mathrm{c}}%
\mathbf{v}\cdot \mathbf{A}_{\mathrm{ex},0}-\frac{m\mathbf{v}^{2}}{2}%
-q\varphi _{0,\mathrm{ex}}=0.  \label{v2a}
\end{gather}%
Now being given \textbf{$v$}$\left( 0\right) $ we readily find $\mathbf{v}%
\left( t\right) $ from the linear equation (\ref{V1aa}). Then using $\mathbf{%
v}\left( t\right) $ and being given \textbf{$r$}$\left( 0\right) $ we
immediately find $\mathbf{r}\left( t\right) $ from equation (\ref{raa}).
Combining equations (\ref{raa}) and (\ref{v2a}) we obtain the following
equations for $s_{\mathrm{p}}\left( t\right) $%
\begin{equation}
\dot{s}_{\mathrm{p}}=m\mathbf{v}\cdot \mathbf{\dot{r}}+\frac{q}{\mathrm{c}}%
\mathbf{v}\cdot \mathbf{A}_{\mathrm{ex},0}-\frac{m}{2}\mathbf{v}%
^{2}-q\varphi _{0,\mathrm{ex}}=\frac{m\mathbf{v}^{2}}{2}-q\varphi _{0,%
\mathrm{ex}}.  \label{thetp1a}
\end{equation}%
It remains to verify that the triple $\left\{ \mathbf{r}\left( t\right) ,%
\mathbf{v}\left( t\right) ,s_{\mathrm{p}}\left( t\right) \right\} $
satisfies the complementary point charge equations (\ref{Lor1}), (\ref{Lor2}%
). Indeed, combining the relations (\ref{V1aa}) and (\ref{raa}) we obtain 
\begin{equation}
m\mathbf{\ddot{r}}=-q\left[ \frac{1}{2\mathrm{c}}\mathbf{B}_{0}\times \left(
-\mathbf{\dot{r}}-\frac{q}{m\mathrm{c}}\mathbf{A}_{\mathrm{ex},0}\right)
+\varphi _{0,\mathrm{ex}}^{\prime }+\frac{1}{\mathrm{c}}\partial _{t}\mathbf{%
A}_{\mathrm{ex},0}\right] .  \label{V2aa}
\end{equation}%
A straightforward comparison taking into account (\ref{fiexlin}) shows that
the above equation (\ref{V2aa}) coincides with the point charge equation of
motion (\ref{Lor1}), and the equations (\ref{raa}) and (\ref{thetp1a})
provide the point charge equations (\ref{Lor2}). \emph{With that we
completed the desired verification of the fact the wave-corpuscle does solve
exactly the auxiliary field equations (\ref{Schmag1})}.

As to the \emph{exact solvability} issue let us compare the coefficients of
auxiliary equations (\ref{Schmag1}) and the original field equations (\ref%
{NLS1}), (\ref{fis}). Taking into account the relation (\ref{fiexlin})- (\ref%
{Aexlin}) we find that the EM potentials $\varphi _{\mathrm{ex}}\left( t,%
\mathbf{x}\right) $, $\mathbf{A}_{\mathrm{ex}}\left( t,\mathbf{x}\right) $
in the field equations (\ref{NLS1}), (\ref{fis}) are compatible with the
auxiliary system (\ref{Schmag1}) for the wave-corpuscle defined by (\ref%
{psil0}) and the complementary point charge equations (\ref{Lor1}), (\ref%
{Lor2}) if the following \emph{exact solvability condition} holds 
\begin{gather}
\varphi _{\mathrm{ex}}\left( t,\mathbf{x}\right) =\varphi _{0,\mathrm{ex}%
}+\varphi _{0,\mathrm{ex}}^{\prime }\cdot \left( \mathbf{x}-\mathbf{r}%
\right) -  \label{fifi1} \\
-\frac{q}{2m\mathrm{c}^{2}}\left( \mathbf{A}_{\mathrm{ex},0}+\frac{1}{2}%
\mathbf{B}_{0}\left( t\right) \times \left( \mathbf{x}-\mathbf{r}\left(
t\right) \right) \right) ^{2},  \notag \\
\mathbf{A}_{\mathrm{ex}}\left( t,\mathbf{x}\right) =\mathbf{\tilde{A}}_{%
\mathrm{ex}}\left( t,\mathbf{x}\right) =\mathbf{A}_{\mathrm{ex},0}+\frac{1}{2%
}\mathbf{B}_{0}\left( t\right) \times \left( \mathbf{x}-\mathbf{r}\left(
t\right) \right) .  \notag
\end{gather}%
Note that if $\mathbf{B}_{0}\neq 0$ the electrical field potential $\varphi
_{\mathrm{ex}}$ involves a quadratic term which vanishes at the center of
the wave-corpuscle. One can naturally ask how broad is the class of external
EM fields as in (\ref{fifi1}) for which there are exact solutions as the
wave-corpuscles? The class of such EM field is sufficiently broad in the
sense that for any accelerated motion of a point charge in an arbitrary EM
field there is wave-corpuscle as an exact solution to the field equations
with an external field from the class. To see that let us consider a point
charge in an \emph{arbitrary} external EM field and find its trajectory $%
\mathbf{r}\left( t\right) $. Then we introduce a special EM field defined by
(\ref{fiexlin}), (\ref{Aexlin}) with $\mathbf{B}_{0}=\mathbf{B}\left(
t\right) $ and $\varphi _{0,\mathrm{ex}}^{\prime }\left( t\right) $ defined
by (\ref{ea1}) and for this external field the wave-corpuscle (\ref{psil0})
is an exact solution to the field equations(\ref{NLS1}), (\ref{fis}): its
center moves exactly according to the trajectory $\mathbf{r}\left( t\right) $%
. More than that, an arbitrary vector-function $\mathbf{r}\left( t\right) $
can be obtained as a solution of (\ref{Lor1}) with appropriate choice of $%
\mathbf{\acute{E}}\left( t,\mathbf{r}\right) $, $\mathbf{\acute{B}}\left( t,%
\mathbf{r}\right) $.\ Indeed, let $\mathbf{\acute{E}}\left( t,\mathbf{r}%
\right) =m\mathbf{\ddot{r}}\left( t\right) /q,$\ and such $\mathbf{r}\left(
t\right) $ is a solution of (\ref{Lor1}). Note that for the given $\mathbf{%
\acute{E}}\left( t\right) $ along the trajectory we can take $\mathbf{A}_{%
\mathrm{ex},0}\left( t\right) $ to be arbitrary and determine $\varphi _{0,%
\mathrm{ex}}^{\prime }\left( t\right) $ by the following formula 
\begin{eqnarray}
\varphi _{0,\mathrm{ex}}^{\prime }\left( t\right) &=&-\left\{ \frac{\partial
_{t}\mathbf{A}_{\mathrm{ex},0}\left( t\right) }{\mathrm{c}}-\frac{\mathbf{B}%
_{0}\left( t\right) \times \mathbf{\dot{r}}\left( t\right) }{2\mathrm{c}}+%
\mathbf{\acute{E}}\left( t\right) \right. +  \label{ea1} \\
&&+\left. \frac{q}{m\mathrm{c}^{2}}\left[ \frac{1}{2}\mathbf{B}_{0}\left(
t\right) \times \mathbf{A}_{\mathrm{ex},0}\left( t\right) \right] \right\} .
\notag
\end{eqnarray}%
Thus, we can conclude that the wave-corpuscle (\ref{psil0}) \ as an exact
solution to the field equations(\ref{NLS1}), (\ref{fis}) with an appropriate
choice of the external EM field can model any motion of a point charge.

\subsubsection{de Broglie factor for accelerating charge}

In this subsection we would like to take a look at the de Broglie
exponential factor in the wave-corpuscle defined by (\ref{psil0}) and the
complementary point charge equations (\ref{Lor1}), (\ref{Lor2}). Let $\check{%
\psi}\left( \mathbf{k}\right) =\left[ \mathcal{F}\psi \right] \left( \mathbf{%
k}\right) $ be the Fourier transform of the wave function $\psi \left( 
\mathbf{x}\right) $%
\begin{equation}
\check{\psi}\left( \mathbf{k}\right) =\left[ \mathcal{F}\psi \right] \left( 
\mathbf{k}\right) =\int_{\mathbb{R}^{3}}\mathrm{e}^{-\mathrm{i}\mathbf{%
k\cdot x}}\psi \left( \mathbf{x}\right) \,\mathrm{d}\mathbf{x}.
\label{futran}
\end{equation}%
Then in view of the charge normalization condition (\ref{expo8}) and by the
Parseval theorem $\check{\psi}\left( \mathbf{k}\right) $ satisfies similar
condition, namely 
\begin{equation}
\left( 2\pi \right) ^{-3}\int_{\mathbb{R}^{3}}\left\vert \check{\psi}\left( 
\mathbf{k}\right) \right\vert ^{2}\,\mathrm{d}\mathbf{k}=1,  \label{ndual}
\end{equation}%
\ and we can introduce the center $\mathbf{k}_{\ast }\left( \psi \right) $
for $\check{\psi}\left( \mathbf{k}\right) $ as follows: 
\begin{equation}
\mathbf{k}_{\ast }\left( \psi \right) =\mathbf{k}_{\ast }=\left( 2\pi
\right) ^{-3}\int_{\mathbb{R}^{3}}\mathbf{k}\left\vert \check{\psi}\left( 
\mathbf{k}\right) \right\vert ^{2}\,\mathrm{d}\mathbf{k.}  \label{kstar}
\end{equation}%
Note that the following identity holds%
\begin{equation*}
\mathbf{k}\left\vert \check{\psi}\left( \mathbf{k}\right) \right\vert ^{2}=%
\mathbf{k}\check{\psi}\left( \mathbf{k}\right) \check{\psi}^{\ast }\left( 
\mathbf{k}\right) =\frac{1}{2\mathrm{i}}\left\{ \mathrm{i}\mathbf{k}\check{%
\psi}\left( \mathbf{k}\right) \check{\psi}^{\ast }\left( \mathbf{k}\right) -%
\check{\psi}\left( \mathbf{k}\right) \left[ \mathrm{i}\mathbf{k}\check{\psi}%
\left( \mathbf{k}\right) \right] ^{\ast }\right\} ,
\end{equation*}%
implying together with (\ref{kstar}) the following representation%
\begin{equation}
\mathbf{k}_{\ast }=\int_{\mathbb{R}^{3}}\func{Im}\frac{\nabla \psi \left( 
\mathbf{x}\right) }{\psi \left( \mathbf{x}\right) }\left\vert \psi \left( 
\mathbf{x}\right) \right\vert ^{2}\,\mathrm{d}\mathbf{x}.  \label{kstarp}
\end{equation}%
Observe that the Fourier transform (\ref{futran})\ of the wave-corpuscle
defined by (\ref{psil0}) is 
\begin{equation}
\check{\psi}\left( t,\mathbf{k}\right) =\exp \left\{ \mathrm{i}\mathbf{r}%
\left( t\right) \mathbf{k}-\frac{\mathrm{i}s_{\mathrm{p}}\left( t\right) }{%
\chi }\right\} \left( \mathcal{F}\left[ \mathring{\psi}\right] \right)
\left( \mathbf{k-}\frac{m\mathbf{v}\left( t\right) }{\chi }\right) ,
\label{omk1}
\end{equation}%
and since $\mathring{\psi}=\mathring{\psi}\left( \left\vert \mathbf{x}%
\right\vert \right) $ is a radial function, its Fourier transform $\mathcal{F%
}\left[ \mathring{\psi}\right] \left( \mathbf{k}\right) $ is a radial
function as well. Let us consider $\mathbf{k}_{\ast }$ defined by the
formula (\ref{kstar}) that corresponds to the wave-corpuscle (\ref{omk1}).
Using the fact that $\mathcal{F}\left[ \mathring{\psi}\right] \left( \mathbf{%
k}\right) $ is a radial function and the relations (\ref{Lor2}), (\ref{Lor3}%
) we readily find that%
\begin{equation}
\mathbf{k}_{\ast }=\frac{m\mathbf{v}\left( t\right) }{\chi },\text{ }\mathbf{%
v}\left( t\right) =\mathbf{\dot{r}}\left( t\right) +\frac{q}{m\mathrm{c}}%
\mathbf{A}_{\mathrm{ex},0}=\frac{\mathbf{\mathring{p}}}{m}.  \label{omk1a}
\end{equation}%
The dispersion relation $\omega \left( \mathbf{k}\right) $ for the linear
part of equation (\ref{Schmag1}) and the corresponding group velocity $%
\nabla _{\mathbf{k}}\omega \left( \mathbf{k}\right) $ are, respectively, 
\begin{gather}
\omega \left( \mathbf{k}\right) =\frac{\chi }{2m}\mathbf{k}^{2}-\frac{q}{m%
\mathrm{c}}\mathbf{A}_{\mathrm{ex},0}\cdot \mathbf{k},\text{ }  \label{omk2}
\\
\nabla _{\mathbf{k}}\omega \left( \mathbf{k}\right) =\frac{\chi }{m}\mathbf{k%
}-\frac{q}{m\mathrm{c}}\mathbf{A}_{\mathrm{ex},0}.  \notag
\end{gather}%
Combining relations (\ref{omk2}) and (\ref{omk1a})\ 
\begin{equation}
\nabla _{\mathbf{k}}\omega \left( \mathbf{k}_{\ast }\right) =\frac{\chi }{m}%
\mathbf{k}_{\ast }-\frac{q}{m\mathrm{c}}\mathbf{A}_{\mathrm{ex},0}=\mathbf{v}%
\left( t\right) -\frac{q}{m\mathrm{c}}\mathbf{A}_{\mathrm{ex},0}=\mathbf{%
\dot{r}}\left( t\right) ,  \label{omk3}
\end{equation}%
we find that the charge velocity $\mathbf{v}\left( t\right) $ is identical
with the group velocity $\nabla _{\mathbf{k}}\omega \left( \mathbf{k}_{\ast
}\right) $ indicating the wave origin of the charge motion.

\subsubsection{General external EM field}

The subject of this section and the treatment are similar to the ones in
Section \ref{seinhomo}, but estimates in the presence of external magnetic
field are more involved. Below we provide the most essential estimates
related to this case omitting tedious details. The main result of this
section is that the wave-corpuscle defined by (\ref{psil0}) and the
complementary point charge equations (\ref{Lor1}), (\ref{Lor2}) is an
approximate solution to the field equations (\ref{NLS1}), (\ref{fis}) with a
discrepancy of the order $O\left( \left( a/R_{\mathrm{ex}}\right)
^{2}\right) $ for $a\ll R_{\mathrm{ex}},$ where $a$ is the size parameter
defined by relations (\ref{nrac5}) and $R_{\mathrm{ex}}$ is a typical length
for inhomogeneity of the external field. We assume here that the function $%
\left\vert \mathring{\psi}_{1}\left( s\right) \right\vert ^{2}$ \ decays
exponentially \ as $s\rightarrow \infty $, \ and (\ref{psexp}) holds.

We define the coefficients (\ref{fiexlin})- (\ref{Aexlin}) of the auxiliary
field equations in (\ref{Schmag1}) as follows 
\begin{gather}
\varphi _{0,\mathrm{ex}}\left( t\right) =\varphi _{\mathrm{ex}}\left( 
\mathbf{r}\left( t\right) ,t\right) +\frac{q}{2m\mathrm{c}^{2}}\mathbf{A}_{%
\mathrm{ex}}^{2}\left( \mathbf{r}\left( t\right) ,t\right) ,  \label{fiexa}
\\
\varphi _{0,\mathrm{ex}}^{\prime }\left( t\right) =\nabla _{x}\varphi _{%
\mathrm{ex}}\left( \mathbf{r}\left( t\right) ,t\right) +\frac{q}{2m\mathrm{c}%
^{2}}\left[ \mathbf{A}_{\mathrm{ex},0}\times \mathbf{B}_{0}\left( t\right) %
\right] ,\text{ }  \notag
\end{gather}%
implying%
\begin{gather}
\tilde{\varphi}_{\mathrm{ex}}\left( t,\mathbf{x}\right) =\varphi _{0,\mathrm{%
ex}}\left( t\right) +\varphi _{0,\mathrm{ex}}^{\prime }\left( t\right) \cdot
\left( \mathbf{x}-\mathbf{r}\left( t\right) \right) ,  \label{fico} \\
\mathbf{B}_{0}\left( t\right) =\mathbf{B}\left( \mathbf{r}\left( t\right)
,t\right) ,\mathbf{\ A}_{\mathrm{ex},0}\left( t\right) =\mathbf{A}_{\mathrm{%
ex}}\left( \mathbf{r}\left( t\right) ,t\right) .  \notag
\end{gather}%
Note that the wave-corpuscle defined by (\ref{psil0}) and the complementary
point charge equations (\ref{Lor1}), (\ref{Lor2}) generally speaking is not
an exact solution of (\ref{NLS1}) since the potentials of (\ref{NLS1}) \
satisfy more general relations than (\ref{fifi1}): 
\begin{gather}
\varphi _{\mathrm{ex}}\left( t,\mathbf{x}\right) =\varphi _{0,\mathrm{ex}%
}\left( t\right) +\varphi _{0,\mathrm{ex}}^{\prime }\cdot \left( \mathbf{x}-%
\mathbf{r}\left( t\right) \right) -  \label{fia} \\
-\frac{q}{2m\mathrm{c}^{2}}\left( \mathbf{A}_{\mathrm{ex},0}+\frac{1}{2}%
\mathbf{B}_{0}\left( t\right) \times \left( \mathbf{x}-\mathbf{r}\left(
t\right) \right) \right) ^{2}+\varphi _{\mathrm{ex}}^{\left( 1\right)
}\left( t,\mathbf{x}\right) ,  \notag \\
\mathbf{B}\left( t,\mathbf{x}\right) =\mathbf{B}_{0}\left( t\right) +\mathbf{%
B}_{1}\left( t,\mathbf{x}\right) ,  \notag \\
\mathbf{A}_{\mathrm{ex}}\left( t,\mathbf{x}\right) =\mathbf{A}_{\mathrm{ex}%
,0}\left( t\right) +\frac{1}{2}\mathbf{B}_{0}\left( t\right) \times \left( 
\mathbf{x}-\mathbf{r}\left( t\right) \right) +\mathbf{A}_{\mathrm{ex}%
,1}\left( t,\mathbf{x}\right) ,  \notag
\end{gather}%
with non-zero terms $\varphi _{\mathrm{ex}}^{\left( 1\right) }\left( t,%
\mathbf{x}\right) $, $\mathbf{B}_{1}\left( t,\mathbf{x}\right) $, $\mathbf{A}%
_{\mathrm{ex},1}\left( t,\mathbf{x}\right) $. These extra terms are small in
the vicinity of $\mathbf{r}\left( t\right) $ where $\psi $ is localized
since 
\begin{eqnarray}
\varphi _{\mathrm{ex}}^{\left( 1\right) }\left( x,\mathbf{r}\left( t\right)
\right) &=&0,\ \nabla \varphi _{\mathrm{ex}}^{\left( 1\right) }\left( x,%
\mathbf{r}\left( t\right) \right) =0,  \label{fiaeq0} \\
\mathbf{B}_{1}\left( \mathbf{r}\left( t\right) ,t\right) &=&0,\mathbf{\ A}_{%
\mathrm{ex},1}\left( \mathbf{r}\left( t\right) ,t\right) =0.  \notag
\end{eqnarray}%
To estimate of the discrepancy resulting from magnetic field we introduce
quantities 
\begin{gather}
\left\vert \mathbf{\bar{B}}\right\vert =\max_{0\leq t\leq T}\mathbf{B}\left( 
\mathbf{r}\left( t\right) ,t\right) ,\   \label{rm} \\
\left\vert \mathbf{\bar{A}}\right\vert =\max_{0\leq t\leq
T}\max_{0<\left\vert z\right\vert \leq a\sigma _{\psi }}\left\vert \mathbf{A}%
\left( \mathbf{r}\left( t\right) +\mathbf{z},t\right) -\mathbf{A}\left( 
\mathbf{r}\left( 0\right) ,0\right) \right\vert ,  \notag
\end{gather}%
\emph{\ } 
\begin{gather}
\frac{1}{R_{B}\left( t,\mathbf{r}\right) }=\sup_{0<\left\vert z\right\vert
\leq a\sigma _{\psi }}\frac{\left\vert \mathbf{B}_{1}\left( \mathbf{r}\left(
t\right) +\mathbf{z},t\right) \right\vert }{\left\vert \mathbf{z}\right\vert
\left\vert \mathbf{\bar{B}}\right\vert },  \notag \\
\frac{1}{R_{A}\left( t,\mathbf{r}\right) }=\sup_{0<\left\vert z\right\vert
\leq a\sigma _{\psi }}\frac{\left\vert \mathbf{A}_{\mathrm{ex},1}\left( 
\mathbf{r}\left( t\right) +\mathbf{z},t\right) \right\vert }{\left\vert 
\mathbf{z}\right\vert \left\vert \mathbf{\bar{A}}\right\vert },  \notag \\
\ \frac{1}{R_{M}}=\max_{0\leq t\leq T}\left( \frac{1}{R_{B}\left( t,\mathbf{r%
}\right) },\frac{1}{R_{A}\left( t,\mathbf{r}\right) }\right) .
\end{gather}%
The quantity $R_{A}\left( t,\mathbf{r}\right) $ is a spatial distance at
which the local variation of $\mathbf{A}$ is of the same order as the global
variation $\left\vert \mathbf{\bar{A}}\right\vert $ of $\mathbf{A}$ itself,
and consequently it represents a spatial scale at which the local variation
of $\mathbf{A}$ is not negligible. \ By (\ref{fiaeq0}) the quantity $\
1/R_{M}$\ is bounded. \ We substitute (\ref{psil0}) in (\ref{NLS1})\ and
observe \ that according to Subsection \ref{exelmag}, $\psi ,\varphi $
exactly satisfy (\ref{NLS1}) \ if $\varphi _{\mathrm{ex}}^{\left( 1\right)
}\left( t,\mathbf{x}\right) $, $\mathbf{B}_{1}\left( t,\mathbf{x}\right) $, $%
\mathbf{A}_{\mathrm{ex},1}\left( t,\mathbf{x}\right) $ are identically zero.
If they are not zero, we obtain a discrepancy $D=D_{0}+D_{1}$ with $D_{0}$
as in Section \ref{seinhomo} and 
\begin{equation}
D_{1}=\frac{\chi q}{m\mathrm{ci}}\left[ \mathbf{A}_{\mathrm{ex},1}+\mathbf{B}%
_{1}\left( t,\mathbf{x}\right) \times \left( \mathbf{x}-\mathbf{r}\left(
t\right) \right) \right] \cdot \nabla \psi .  \label{dqb1}
\end{equation}%
\ Similarly to (\ref{discr1}) we introduce the \emph{integral discrepancy }$%
\bar{D}_{1}$\emph{:} 
\begin{align}
\bar{D}_{1}& =\func{Re}\int_{\mathbb{R}^{3}}\frac{\chi q}{m\mathrm{ci}}%
\left( \mathbf{A}_{\mathrm{ex},1}+\mathbf{B}_{1}\left( t,\mathbf{x}\right)
\times \left( \mathbf{x}-\mathbf{r}\left( t\right) \right) \right) \cdot
\nabla \psi \psi ^{\ast }\,\mathrm{d}\mathbf{x}  \label{dqb2} \\
& =\int_{\mathbb{R}^{3}}\frac{\chi q}{m\mathrm{c}}\left[ \mathbf{A}_{\mathrm{%
ex},1}+\mathbf{B}_{1}\left( t,\mathbf{x}\right) \times \left( x-\mathbf{r}%
\left( t\right) \right) \right] \cdot \func{Im}\left( \frac{\nabla \psi }{%
\psi }\right) \left\vert \psi \right\vert ^{2}\,\mathrm{d}\mathbf{x}  \notag
\end{align}%
Note that for solutions of the form (\ref{psil0}) we have 
\begin{equation*}
\func{Im}\left( \frac{\nabla \psi }{\psi }\right) =\func{Im}\nabla \left( 
\frac{\mathrm{i}m}{\chi }\mathbf{v}\cdot \left( \mathbf{x}-\mathbf{r}\right)
+\frac{\mathrm{i}}{\chi }s_{p}\right) =\frac{m}{\chi }\mathbf{v},
\end{equation*}%
implying when combined with (\ref{dqb2})%
\begin{equation*}
\bar{D}_{1}=\frac{q}{\mathrm{c}}\int_{\mathbb{R}^{3}}\left[ \mathbf{A}_{%
\mathrm{ex},1}+\mathbf{B}_{1}\left( t,\mathbf{x}\right) \times \left( 
\mathbf{x}-\mathbf{r}\left( t\right) \right) \right] \cdot \mathbf{v}%
\left\vert \psi \right\vert ^{2}\,\mathrm{d}\mathbf{x},
\end{equation*}%
which, in turn, yields the following estimate%
\begin{gather}
\left\vert \bar{D}_{1}\right\vert \lesssim \frac{\left\vert \mathbf{v}%
\right\vert }{\mathrm{c}}q\int_{\mathbb{R}^{3}}\frac{\left\vert \mathbf{\bar{%
A}}\right\vert \left\vert x-\mathbf{r}\left( t\right) \right\vert \left\vert
\psi _{a}\right\vert ^{2}}{R_{A}\left( t,\mathbf{r}\right) }\,\mathrm{d}%
\mathbf{x}+  \label{dqb3} \\
+\frac{\left\vert \mathbf{v}\right\vert }{\mathrm{c}}q\left\vert \mathbf{%
\bar{B}}\right\vert \int_{\mathbb{R}^{3}}\frac{1}{R_{B}\left( t,\mathbf{r}%
\right) }\left\vert \mathbf{x}-\mathbf{r}\left( t\right) \right\vert
^{2}\left\vert \psi _{a}\right\vert ^{2}\,\mathrm{d}\mathbf{x}\leq  \notag \\
\leq C_{0}\frac{\left\vert \mathbf{v}\right\vert }{\mathrm{c}}\frac{a}{R_{M}}%
\left( \left\vert q\right\vert \left\vert \mathbf{\bar{A}}\right\vert
+qa\left\vert \mathbf{\bar{B}}\right\vert \right) ,\ 0\leq t\leq T.  \notag
\end{gather}%
Combining relation (\ref{d0}) and (\ref{dqb3}) we get 
\begin{equation}
\left\vert \bar{D}_{0}+\bar{D}_{1}\right\vert \lesssim C_{0}\frac{%
a^{2}\left\vert q\right\vert \left\vert \bar{\varphi}\right\vert }{%
R_{\varphi }^{2}}+\frac{\left\vert \mathbf{v}\right\vert }{\mathrm{c}}\frac{a%
}{R_{M}}C_{0}\left( \left\vert q\right\vert \left\vert \mathbf{\bar{A}}%
\right\vert +qa\left\vert \mathbf{\bar{B}}\right\vert \right) .  \label{dqb4}
\end{equation}%
\ yielding the following conditions for the discrepancy to be relatively
small 
\begin{equation}
\frac{\left\vert \mathbf{v}\right\vert }{\mathrm{c}}\frac{a}{R_{M}}\ll 1,\ 
\frac{a^{2}}{R_{\varphi }^{2}}\ll 1.  \label{dqb5}
\end{equation}

\subsection{Stability issues}

A comprehensive analysis of the stability is complex, involved and beyond
the scope of this paper. Nevertheless, we would like to give a concise
consideration to three aspects of stability for well separated charges in
the nonrelativistic regime : (i) no "blow-up" or "collapse"; (ii)\
preservation with high accuracy of the form of a wave-corpuscle solution for
a limited time; (iii) preservation of spatial localization for certain
solutions on long time intervals.

Here is an argument for why there can not be a "blow-up" in finite time. A
"blow-up" is an issue since the nonlinearity $G^{\prime }\left( s\right) $
provides focusing properties with consequent soliton-like solutions $%
\mathring{\psi}^{\ell },$ $\mathring{\varphi}^{\ell }$. In our model the
possibility of "blow up" is excluded when we define $\left[ G_{a}^{\ell }%
\right] ^{\prime }$ to be a constant for large amplitudes of the fields,
namely for $s\geq \left( \mathring{\psi}^{\ell }\right) ^{2}\left( 0\right) $
as in (\ref{intps1}). This factor combined with the charge normalization
condition (\ref{expo8}) implies that the energy is bounded from below.
Indeed, according to (\ref{nama5}),(\ref{nama7}) and (\ref{emfr11a}) the
energy of a free charge can be written in the form%
\begin{equation}
\mathcal{E}\left( \psi ,\varphi \right) =\int_{\mathbb{R}^{3}}\left( w+%
\tilde{u}\right) \,\mathrm{d}\mathbf{x}=\int_{\mathbb{R}^{3}}\frac{%
\left\vert \nabla \varphi \right\vert ^{2}}{8\pi }+\frac{\chi ^{2}}{2m}\left[
\left\vert \nabla \psi \right\vert ^{2}+G\left( \left\vert \psi \right\vert
^{2}\right) \right] \,\mathrm{d}\mathbf{x,}  \label{ene}
\end{equation}%
where $\varphi =\varphi _{\psi }$ \ is determined from (\ref{jco3}). In view
of relations (\ref{intps1}) the nonlinearity derivative $G^{\prime }\left(
s\right) $ is bounded, implying $G\left( \left\vert \psi \right\vert
^{2}\right) \geq -C\left\vert \psi \right\vert ^{2}$ for a constant $C$.
That combined with the charge normalization condition (\ref{expo8}) implies
boundedness of the energy from below, namely 
\begin{equation}
\mathcal{E}\left( \psi ,\varphi \right) \geq -C\text{\ for all }\psi
,\varphi ,\ \left\Vert \psi \right\Vert ^{2}=\int_{\mathbb{R}^{3}}\left\vert
\psi \right\vert ^{2}\,\mathrm{d}\mathbf{x}=1.  \label{boue}
\end{equation}%
A similar argument in the case of many interacting charges also shows that
the energy is bounded from below. Since energy is a conserved quantity,
using the boundedness of the energy from below one can prove along lines of 
\cite{Kato89} the global existence of a unique solution $\psi ^{\ell }\left(
t,\mathbf{x}\right) $, $\varphi ^{\ell }\left( t,\mathbf{x}\right) $ to (\ref%
{NLSj0}), (\ref{delfi})\ for all times $0\leq t<\infty $ for given initial
data $\psi ^{\ell }\left( 0,\mathbf{x}\right) $.

The second aspect of stability is about a preservation of the wave-corpuscle
shape with high accuracy for limited times. A basis for it is provided in
Section \ref{seinhomo}. Since discrepancies in the equations (\ref{auf1})
are of the order $\left\vert q\right\vert \left\vert \bar{\varphi}%
\right\vert a^{2}/R^{2}$ \ for the charge in an external EM field the fields 
$\psi ,\varphi $ have to be close to the wave-corpuscle of the form (\ref%
{psil0}) on time intervals of order $\left\vert q\right\vert \left\vert \bar{%
\varphi}\right\vert a^{2}/\left( \chi R^{2}\right) \ $where $R$ is a spatial
scale of inhomogeneity of the external field, and $\left\vert q\right\vert
\left\vert \bar{\varphi}\right\vert $ is a global variation of the external
field potential energy near the trajectory of the wave-corpuscle.

The third aspect is a stability on very long time intervals which is
understood in a broader sense, namely when a charge maintains its spatial
localization without necessarily preserving the exact form of a
wave-corpuscle It is shown in Section \ref{avq} that such a broad
localization assumption is sufficient to identify the corresponding point
charge trajectory. Now let us consider the following argument for the charge
stability based on properties of the energy. For simplicity let us consider
a single free charge with energy (\ref{ene}). The energy conservation law
implies 
\begin{equation}
\mathcal{E}\left( \psi \left( t\right) ,\varphi \left( t\right) \right) =%
\mathcal{E}\left( \psi \left( 0\right) ,\varphi \left( 0\right) \right) ,%
\text{\ for all }0\leq t<\infty .  \label{encon}
\end{equation}%
Note that the rest solution $\mathring{\psi}$ as in (\ref{nop4}), (\ref{nop5}%
) is a critical point of $\mathcal{E}$ \ defined by (\ref{ene}). \ Let us
assume that the rest solution $\mathring{\psi}$ \ is the global minimum
under the charge normalization constraint, namely 
\begin{equation}
\mathcal{E}\left( \mathring{\psi},\varphi _{\mathring{\psi}}\right)
=\min_{\left\Vert \psi \right\Vert =1}\mathcal{E}\left( \psi ,\varphi _{\psi
}\right) =\mathsf{E}_{0}.  \label{encon1}
\end{equation}%
\ Consider then the initial data $\psi _{0}$ for (\ref{nore3a})-(\ref{nore3b}%
) at $t=0$ that (i) satisfies the charge normalization condition (\ref{expo8}%
); (ii) is close to $\mathring{\psi}$ and has almost the same energy, i.e. $%
\left\vert \mathcal{E}\left( \psi \left( 0\right) ,\varphi \left( 0\right)
\right) -\mathsf{E}_{0}\right\vert \ll 1$.\ Note that since every spatial
shift $\mathring{\psi}\left( \mathbf{x-r}\right) $, $\varphi _{\mathring{\psi%
}}\left( \mathbf{x-r}\right) $ of $\mathring{\psi}\left( \mathbf{x}\right)
,\varphi _{\mathring{\psi}}\left( \mathbf{x}\right) $ produces fields
satisfying the charge normalization condition (\ref{expo8}) and of the the
same energy, the minimum in (\ref{encon1}) has to be degenerate. But if we
assume that all the degeneracy is due to to spatial translations, rotations
and the multiplication by $e^{is},$ then the condition $\left\vert \mathcal{E%
}\left( \psi \left( t\right) ,\varphi \left( t\right) \right) -\mathsf{E}%
_{0}\right\vert \ll 1$ though allowing for spatial translation of $\mathring{%
\psi}\left( \mathbf{x}\right) ,\varphi _{\mathring{\psi}}\left( \mathbf{x}%
\right) $\ to large distances and times, still implies that form of $%
\left\vert \psi \left( t,\mathbf{x-r}\left( t\right) \right) \right\vert $, $%
\varphi \left( t,\mathbf{x-r}\left( t\right) \right) $ has to be almost the
same as the form of $\mathring{\psi}\left( \mathbf{x}\right) ,\mathring{%
\varphi}\left( \mathbf{x}\right) $. The same argument works for a local
minimum that is non-degenerate modulo spatial translations, rotations and
the multiplication by $\mathrm{e}^{\mathrm{i}s}$.

\section{Many interacting charges}

A qualitatively new physical component in the theory of two or more charges
compared to the theory of a single charge is obviously the interaction
between them. In our approach \emph{any individual "bare" charge interacts
directly only with the EM} field, and consequently different charges
interact with each other only indirectly trough the EM field. In this
section we develop the Lagrangian theory for many interacting charges for
the both relativistic and nonrelativistic cases based on Lagrangians for
single charges studies in Sections \ref{freepar}, \ref{snfree}, \ref{ssaccel}%
. The primary focus of our studies on many charges is ways of integration
into our wave theory the point charge mechanics in the regime of remote
interaction when the charges are separated by large distances compare to
their sizes. A system of many charges can have charges of different type,
for instance electrons and protons. In that case we naturally assume that
individual Lagrangians for charges of the same type to have identical
Lagrangians with the same mass $m$, charge $q$, form factor $\mathring{\psi}$
and consequently the same nonlinear self-interaction $G$. We use here
general results of the Lagrangian field theory for many interacting charges
including symmetries, conservation laws and the construction of
gauge-invariant and symmetric energy-momentum tensors described in Section %
\ref{Manych}.

Let us introduce a system of $N$ charges interacting directly only with the
EM field described by its $4$-vector potential $A^{\mu }=\left( \varphi ,%
\mathbf{A}\right) $. The charges are described by their wave functions $\psi
^{\ell }$ with the superscript index $\ell =1,\ldots ,N$ labeling them. In
this section we study the dynamics of the system of charges in the \emph{%
regime of remote interaction}, that is when any two different charges of the
system are well separated so that the distance between them is much larger
compare to their typical sizes. We show here that under the assumption of
remote interaction the charges interact which is other by Lorentz forces and
that in non-relativistic case their dynamics is perfect correspondence with
the dynamics of the corresponding system of point charges. In the
relativistic case the correspondence with the point mechanics is more subtle
because of fundamental limitations. In turns out that in the regime of
remote interaction the nonlinear self-interaction terms associated with
charges do not manifest themselves in any way but making charges to behave
as wave-corpuscles similar to ones studied in Sections \ref{freepar}, \ref%
{snfree}, \ref{ssaccel}.

In non-relativistic case we provide a big picture of interaction and
dynamics via a single charge in an external field and charges interacting
instantaneously via their individual electric fields.

\subsection{Non-relativistic theory of interacting charges\label{nonmany}}

The purpose of this section is to develop a non-relativistic theory of many
interacting charges that would be sufficient for establishing its intimate
relation to the point charges mechanics. Developed here non-relativistic
theory for many interacting charges naturally integrates the theory of
single non-relativistic charge developed in Sections \ref{snfree} and \ref%
{ssaccel}, including the set up for the interaction between the bare charge
and the EM field as described by its electric potential $\varphi $. Our
nonrelativistic Lagrangian $\mathcal{\hat{L}}$ for many charges is
constructed based on (i) individual charges nonrelativistic Lagrangians $%
\hat{L}^{\ell }$ of the form (\ref{nore1b}) and (ii) the assumption that
every charge interacts directly only with the EM field as defined by its
electric potential $\varphi $ , namely%
\begin{equation}
\mathcal{\hat{L}}=\mathcal{\hat{L}}\left( \psi ^{\ell },\psi _{,\mu }^{\ell
},\psi ^{\ell \ast },\psi _{,\mu }^{\ell \ast },\nabla \varphi ,\varphi
,x^{\mu }\right) =\frac{\left\vert \nabla \varphi \right\vert ^{2}}{8\pi }%
+\sum_{\ell }\hat{L}^{\ell }\left( \psi ^{\ell },\psi ^{\ell \ast },\varphi
\right) ,  \label{Lagrqst}
\end{equation}%
where%
\begin{equation*}
\hat{L}^{\ell }=\frac{\chi }{2}\mathrm{i}\left[ \psi ^{\ell \ast }\tilde{%
\partial}_{t}^{\ell }\psi ^{\ell }-\psi ^{\ell }\tilde{\partial}_{t}^{\ell
\ast }\psi ^{\ell \ast }\right] -\frac{\chi ^{2}}{2m^{\ell }}\left\{ \tilde{%
\nabla}\psi ^{\ell }\tilde{\nabla}^{\ast }\psi ^{\ell \ast }+G^{\ell }\left(
\psi ^{\ell \ast }\psi ^{\ell }\right) \right\} ,
\end{equation*}%
\begin{equation*}
\tilde{\partial}_{t}^{\ell }=\partial _{t}+\frac{\mathrm{i}q^{\ell }\left(
\varphi +\varphi _{\mathrm{ex}}\right) }{\chi },\ \tilde{\nabla}^{\ell
}=\nabla -\frac{\mathrm{i}q^{\ell }\mathbf{A}_{\mathrm{ex}}}{\chi c},
\end{equation*}%
where $\left( \varphi _{\mathrm{ex}},\mathbf{A}_{\mathrm{ex}}\right) $ is
the potential of the external EM field. Evidently, according to this
Lagrangian, every charge is coupled to the EM field exactly as if it were a
single charge, but since there is just one EM field all charges are coupled.
The Euler-Lagrange field equations for this Lagrangian are%
\begin{gather}
\chi \mathrm{i}\tilde{\partial}_{t}\psi ^{\ell }=\frac{\chi ^{2}}{2m^{\ell }}%
\left[ -\tilde{\nabla}^{\ell 2}\psi ^{\ell }+\left[ G^{\ell }\right]
^{\prime }\left( \left\vert \psi ^{\ell }\right\vert ^{2}\right) \psi \right]
,  \label{nama1} \\
-\Delta \varphi =4\pi \sum_{\ell }q^{\ell }\left\vert \psi ^{\ell
}\right\vert ^{2},\   \label{nama2}
\end{gather}%
where $\left[ G^{\ell }\right] ^{\prime }\left( s\right) =\partial
_{s}G^{\ell }\left( s\right) ,$ \ and as in the case of a single charge $%
\psi ^{\ell \ast }$ is the complex conjugate to $\psi ^{\ell }$ for all $%
\ell $. The nonlinear self-interaction terms $G^{\ell }$ in (\ref{Lagrqst})
are determined based on the corresponding form factors $\mathring{\psi}%
^{\ell }$ from $\ell $-th charge equilibrium equation (\ref{nop7}).

The Lagrangian $\mathcal{\hat{L}}$ defined by (\ref{Lagrqst}) is gauge
invariant with respect to the first and the second gauge transformations (%
\ref{lagco3}), (\ref{lagco4}) and consequently every $\ell $-th charge has a
conserved current $J^{\ell \mu }=\left( c\rho ^{\ell },\mathbf{J}^{\ell
}\right) $ which can be found from relations (\ref{lagco9}), (\ref{lagco10})
yielding the following formulas similar to (\ref{expo3}) 
\begin{gather}
\ \rho ^{\ell }=q\left\vert \psi ^{\ell }\right\vert ^{2},  \label{nama3a} \\
\ \mathbf{J}^{\ell }=\frac{\mathrm{i}\chi q^{\ell }}{2m^{\ell }}\left[ \psi 
\tilde{\nabla}^{\ell \ast }\psi ^{\ell \ast }-\psi ^{\ell \ast }\tilde{\nabla%
}^{\ell }\psi ^{\ell }\right] =\left( \frac{\chi q^{\ell }}{m^{\ell }}\func{%
Im}\frac{\nabla \psi ^{\ell }}{\psi ^{\ell }}-\frac{q^{\ell 2}\mathbf{A}_{%
\mathrm{ex}}}{m^{\ell }\mathrm{c}}\right) \left\vert \psi ^{\ell
}\right\vert ^{2},  \notag
\end{gather}%
Every current $J^{\ell \mu }$ satisfies the conservation/continuity
equations $\partial _{t}\rho ^{\ell }+\nabla \cdot \mathbf{J}^{\ell }=0$ or
explicitly \ 
\begin{equation}
m^{\ell }\partial _{t}\left\vert \psi ^{\ell }\right\vert ^{2}+\nabla \cdot
\left( \chi \func{Im}\frac{\nabla \psi ^{\ell }}{\psi ^{\ell }}\left\vert
\psi ^{\ell }\right\vert ^{2}-\frac{q^{\ell }}{\mathrm{c}}\mathbf{A}_{%
\mathrm{ex}}\left\vert \psi ^{\ell }\right\vert ^{2}\right) =0.
\label{nama3b}
\end{equation}%
The equations (\ref{nama3b}) imply the conservation of the total $\ell $-th
conserved charge. Similarly to the case of a single charge we require every
total $\ell $-th conserved charge to be exactly $q^{\ell }$ and, hence, to
satisfy the following \emph{charge normalization} condition of the form (\ref%
{norm10})%
\begin{equation}
\dint\nolimits_{\mathbb{R}^{3}}\left\vert \psi ^{\ell }\right\vert ^{2}\,%
\mathrm{d}\mathbf{x}=1\text{ for all }t.  \label{nama3c}
\end{equation}%
Next based on the equation (\ref{nama2}) it is natural to introduce for
every $\ell $-th charge the corresponding potential $\varphi ^{\ell }$ using
the Green function (\ref{gre3}), namely 
\begin{equation}
\varphi ^{\ell }\left( t,\mathbf{x}\right) =q^{\ell }\dint_{\mathbb{R}^{3}}%
\frac{\left\vert \psi ^{\ell }\right\vert ^{2}\left( t,\mathbf{y}\right) }{%
\left\vert \mathbf{y}-\mathbf{x}\right\vert }\,\mathrm{d}\mathbf{y}\
,\varphi =\sum_{\ell }\varphi ^{\ell }.  \label{nama4}
\end{equation}%
Taking into account the expression for the covariant derivatives from (\ref%
{Lagrqst}) we can recast the field equations (\ref{nama1}), (\ref{nama2}) as
(\ref{NLSj0}). The charge conservation equations (\ref{nama3b}) can be
alternatively derived directly from the field equations (\ref{NLSj0}) by
multiplying them and their complex conjugate, respectively, by $\psi ^{\ell
\ast }$, $\psi ^{\ell }$ and subtracting from one another.

\emph{One can see in the integral expression (\ref{nama4}) instantaneous
action-at-a-distance, a feature which occurs in the nonrelativistic point
Lagrangian mechanics}.

Many technical aspects needed for the treatment of many charges in the
regime of remote interaction are already developed in our studies of a
single non-relativistic charge in an external EM field in Section \ref%
{ssaccel}, and to avoid repetition whenever the case we use relevant results
from there. \emph{In fact, an accurate guiding principle for the treatment
of distant interaction of non-relativistic charges is to view every charge
as essentially a single one in the slowly varying in the space and time
external EM field created by other charges}.

To study the motion of the energy and momentum for the involved charges and
the EM field we introduce for every $\ell $-th charge its gauge invariant
energy-momentum tensor $\tilde{T}^{\ell \mu \nu }$ based on the formulas (%
\ref{emna7})-(\ref{emna10}) substituting there $\psi ^{\ell }$, $\psi ^{\ell
\ast }$ in place of $\psi $, $\psi ^{\ast }$ and the covariant derivatives
with the index $\ell $ from (\ref{Lagrqst}) in place of the covariant
derivatives for $\psi $, $\psi ^{\ast }$. That yields the following formulas
for energy and momentum densities for individual charges%
\begin{equation}
\tilde{u}^{\ell }=\frac{\chi ^{2}}{2m}\left[ \tilde{\nabla}^{\ell }\psi
^{\ell }\cdot \tilde{\nabla}^{\ell \ast }\psi ^{\ell \ast }+G^{\ell }\left(
\psi ^{\ell \ast }\psi ^{\ell }\right) \right] ,  \label{nama5}
\end{equation}%
\begin{equation}
\tilde{p}_{j}^{\ell }=\frac{\chi \mathrm{i}}{2}\left( \psi ^{\ell }\tilde{%
\partial}_{j}^{\ell \ast }\psi ^{\ell \ast }-\psi ^{\ell \ast }\tilde{%
\partial}_{j}^{\ell }\psi ^{\ell }\right) ,\ j=1,2,3.  \label{nama6}
\end{equation}%
The gauge invariant energy-momentum tensor $\Theta ^{\mu \nu }$ for the EM
field is defined by (\ref{emna3})-(\ref{emna5}) and, in particular, its
energy, momentum and energy flux densities are%
\begin{equation}
\partial _{0}w=\frac{\mathbf{J}\cdot \nabla \varphi }{\mathrm{c}}=-\frac{%
\mathbf{J}\cdot \mathbf{E}}{\mathrm{c}},\ g_{j}=0,\ s_{j}=0.  \label{nama7}
\end{equation}%
Using the field equations (\ref{nama1}), (\ref{nama2}) and the
representation (\ref{nore1c}) for $F^{\nu \mu }$ we can also verify that
following conservations laws for the individual charges and the EM field: 
\begin{equation}
\partial _{\mu }\tilde{T}^{\ell \mu \nu }=f^{\ell \nu }+f_{\mathrm{ex}%
}^{\ell \nu },\ \partial _{\mu }\Theta ^{\mu \nu }=-\dsum_{\ell }f^{\ell \nu
},\ \text{ }  \label{nama8}
\end{equation}%
where%
\begin{equation*}
f^{\ell \nu }=\frac{1}{\mathrm{c}}J_{\mu }^{\ell }F^{\nu \mu }=\left( \frac{1%
}{\mathrm{c}}\mathbf{J}^{\ell }\cdot \mathbf{E},\rho ^{\ell }\mathbf{E}%
\right) ,
\end{equation*}

\begin{equation*}
f_{\mathrm{ex}}^{\ell \nu }=\frac{1}{\mathrm{c}}J_{\mu }^{\ell }F_{\mathrm{ex%
}}^{\nu \mu }=\left( \frac{1}{\mathrm{c}}\mathbf{J}^{\ell }\cdot \mathbf{E}_{%
\mathrm{ex}},\rho ^{\ell }\mathbf{E}_{\mathrm{ex}}+\frac{1}{\mathrm{c}}%
\mathbf{J}^{\ell }\times \mathbf{B}_{\mathrm{ex}}\right) .
\end{equation*}%
\bigskip

\emph{The energy-momentum conservation equations (\ref{nama8}) can be viewed
as equations of motion in an elastic continuum,} \cite[Section 6.4, (6.56),
(6.57)]{Moller}, similar to the case of kinetic energy-momentum tensor for a
single relativistic particle, \cite[Section 37, (3.24)]{Pauli RT}. \emph{It
is important to remember though that in contrast to the point mechanics the
equations of motion (\ref{nama8}) must always be complemented with the field
equations (\ref{nama1}), (\ref{nama2}) or (\ref{NLSj0}) without which they
do not have to hold and are not alone sufficient to determine the motion}.
We also recognize in $f^{\ell \nu }$ and $f_{\mathrm{ex}}^{\ell \nu }$ in
the equations of motion (\ref{nama8}) respectively the Lorentz force
densities for the charge in the EM field (of charges) and the same for the
external EM field. Observe that equations (\ref{nama8}) satisfy manifestly
the Newton's principle "action equals reaction" for all involved densities
at every point of the space-time.

In the regime of remote interactions it makes sense to introduce dressed
charges and attribute to every charge its EM field via the potential $%
\varphi ^{\ell }$ as defined by relations (\ref{NLSj0}) and (\ref{nama4}).
Based on the potentials $\varphi ^{\ell }$ we define the corresponding
energy-momentum tensor $\Theta ^{\ell \mu \nu }$ by formulas (\ref{emna3})-(%
\ref{emna5}) where we substitute $\varphi ^{\ell }$ and $\mathbf{J}^{\ell }$
defined by equalities (\ref{nama4}), (\ref{nama3a}) in place of $\varphi $
and $\mathbf{J}$. One can verify then that the conservation law (\ref{emna13}%
) for $\Theta ^{\ell \mu \nu }$ takes here the form%
\begin{equation}
\partial _{\mu }\Theta ^{\ell \mu \nu }=-\frac{1}{\mathrm{c}}J_{\mu }^{\ell
}F^{\ell \nu \mu }.  \label{thef1}
\end{equation}%
Now for every $\ell $-th dressed charge we define its energy-momentum tensor 
$\mathrm{T}^{\ell \mu \nu }$ by the formula%
\begin{equation}
\mathrm{T}^{\ell \mu \nu }=\tilde{T}^{\ell \mu \nu }+\Theta ^{\ell \mu \nu }.
\label{thef2}
\end{equation}%
It is also natural and useful to introduce for every $\ell $-th charge the
EM field $\mathbf{E}_{\mathrm{ex}}^{\ell }$ and $F_{\mathrm{ex}}^{\ell \nu
\mu }$ of all other charges $\ell ^{\prime }\neq \ell $ by 
\begin{equation}
\mathbf{E}_{\mathrm{ex}}^{\ell }=\dsum_{\ell ^{\prime }\neq \ell }\mathbf{E}%
^{\ell ^{\prime }},\ F_{\mathrm{ex}}^{\ell \nu \mu }=\dsum_{\ell ^{\prime
}\neq \ell }F^{\ell ^{\prime }\nu \mu },\ \mathbf{E}^{\ell }=-\nabla \varphi
^{\ell }.  \label{thef3}
\end{equation}%
Then combining relations (\ref{nama8}), (\ref{thef1}) with (\ref{nore1c}) we
obtain the following \emph{equations of motion for dressed charges} 
\begin{gather}
\partial _{\mu }\mathrm{T}^{\ell \mu \nu }=\mathrm{f}^{\ell \nu }+f_{\mathrm{%
ex}}^{\ell \nu },  \label{thef4} \\
\mathrm{f}^{\ell \nu }=\frac{1}{\mathrm{c}}J_{\mu }^{\ell }\dsum_{\ell
^{\prime }\neq \ell }F^{\ell ^{\prime }\nu \mu }=\left( \frac{1}{\mathrm{c}}%
\mathbf{J}^{\ell }\cdot \mathbf{E}_{\mathrm{ex}}^{\ell },\rho ^{\ell }%
\mathbf{E}_{\mathrm{ex}}^{\ell }\right) ,  \notag
\end{gather}%
describing the motion of energies and momenta of the dressed charges in the
space-time continuum. Importantly, \emph{the Lorentz force }$\mathrm{f}%
^{\ell \nu }$\emph{\ in the right-hand of (\ref{thef4}) excludes manifestly
the self-interaction} in contrast to the Lorentz force acting upon bare
charge as in (\ref{nama8}) which explicitly includes the self-interaction
term $\frac{1}{\mathrm{c}}J_{\mu }^{\ell }F^{\ell \nu \mu }$. Thus, we can
conclude that when the charge and its EM field are treated as a single
entity, namely dressed charge, there is no self-interaction as signified by
the exact equations (\ref{empa7}).

It follows from (\ref{nama3a}) and (\ref{nama6}) that\emph{\ the charge
gauge invariant momentum density }$\mathbf{P}$ $^{\ell }$\emph{\ equals
exactly the microcurrent density }$\mathbf{J}^{\ell }$\textbf{\ }\emph{%
multiplied by the constant }$m^{\ell }/q^{\ell }$, namely the following
identity holds 
\begin{gather}
\mathbf{P}^{\ell }=\frac{m^{\ell }}{q^{\ell }}\mathbf{J}^{\ell }=\frac{%
\mathrm{i}\chi }{2}\left[ \psi ^{\ell }\tilde{\nabla}^{\ell \ast }\psi
^{\ell \ast }-\psi ^{\ell \ast }\tilde{\nabla}^{\ell }\psi ^{\ell }\right]
\label{thef5} \\
=\left( \chi \func{Im}\frac{\nabla \psi ^{\ell }}{\psi ^{\ell }}-\frac{%
q^{\ell }\mathbf{\bar{A}}}{\mathrm{c}}\right) \left\vert \psi ^{\ell
}\right\vert ^{2},  \notag
\end{gather}%
\emph{that can be viewed as the momentum density kinematic representation}.
We can also recast the above equality as 
\begin{gather}
\mathbf{P}^{\ell }\left( t,\mathbf{x}\right) =m\mathbf{v}^{\ell }\left( t,%
\mathbf{x}\right) ,\ \text{where}  \label{thef6} \\
\mathbf{v}^{\ell }\left( t,\mathbf{x}\right) =\mathbf{J}^{\ell }\left( t,%
\mathbf{x}\right) /q^{\ell }\text{ is the velocity density.}  \notag
\end{gather}

Up to this point we introduced the basic elements of theory of interacting
charges described as fields via the Lagrangian (\ref{Lagrqst}). A natural
question then is in what ways the point charge mechanics is integrated into
this Lagrangian classical field theory? There are two distinct ways to
correspond our field theory to the point charge mechanics: (i) via averaged
quantities in spirit of the well known in quantum mechanics \emph{Ehrenfest\
Theorem, }\cite[Sections 7, 23]{Schiff}; (ii) via a construction of
approximate solutions to the field equations (\ref{nama1}), (\ref{nama2}), (%
\ref{NLSj0}) in terms of radial wave-corpuscles similar to (\ref{psil0}). We
consider these two ways in the next subsections.

\subsubsection{Point mechanics via averaged quantities\label{avq}}

Introducing the total individual momenta $\mathsf{P}^{\ell }$ and energies $%
\mathsf{E}^{\ell }$ for $\ell $-th dressed charge%
\begin{equation}
\mathsf{P}^{\ell }=\int_{\mathbb{R}^{3}}\mathbf{P}^{\ell }\,\mathrm{d}%
\mathbf{x},\ \mathsf{E}^{\ell }=\int_{\mathbb{R}^{3}}\tilde{u}^{\ell }\,%
\mathrm{d}\mathbf{x},  \label{peu1}
\end{equation}%
and using arguments similar to (\ref{crv1})-(\ref{crv3}) combined with
relations (\ref{thef3}), (\ref{thef4}) we obtain the following equations 
\begin{gather}
\frac{\mathrm{d}\mathsf{P}^{\ell }}{\mathrm{d}t}=q^{\ell }\int_{\mathbb{R}%
^{3}}\left[ \left( \dsum\nolimits_{\ell ^{\prime }\neq \ell }\mathbf{E}%
^{\ell ^{\prime }}+\mathbf{E}_{\mathrm{ex}}\right) \left\vert \psi ^{\ell
}\right\vert ^{2}+\frac{1}{\mathrm{c}}\mathbf{v}^{\ell }\times \mathbf{B}_{%
\mathrm{ex}}\right] \,\mathrm{d}\mathbf{x}\text{, where}  \label{peu2} \\
\int_{\mathbb{R}^{3}}\dsum\nolimits_{\ell ^{\prime }\neq \ell }\mathbf{E}%
^{\ell ^{\prime }}\left\vert \psi ^{\ell }\right\vert ^{2}\,\mathrm{d}%
\mathbf{x}=  \label{peu3} \\
=-\dsum\nolimits_{\ell ^{\prime }\neq \ell }q^{\ell ^{\prime }}\dint_{%
\mathbb{R}^{3}\times \mathbb{R}^{3}}\frac{\left( \mathbf{y}-\mathbf{x}%
\right) \left\vert \psi ^{\ell ^{\prime }}\right\vert ^{2}\left( t,\mathbf{y}%
\right) \left\vert \psi ^{\ell }\left( t,\mathbf{x}\right) \right\vert ^{2}}{%
\left\vert \mathbf{y}-\mathbf{x}\right\vert ^{3}}\,\mathrm{d}\mathbf{y}%
\mathrm{d}\mathbf{x},  \notag
\end{gather}%
\begin{equation}
\frac{\mathrm{d}\mathsf{E}^{\ell }}{\mathrm{d}t}=\int_{\mathbb{R}^{3}}%
\mathbf{J}^{\ell }\cdot \mathbf{E}_{\mathrm{ex}}^{\ell }\left( t,\mathbf{x}%
\right) \,\mathrm{d}\mathbf{x}=q^{\ell }\int_{\mathbb{R}^{3}}\mathbf{v}%
^{\ell }\cdot \left( \dsum\nolimits_{\ell ^{\prime }\neq \ell }\mathbf{E}%
^{\ell ^{\prime }}+\mathbf{E}_{\mathrm{ex}}\right) \,\mathrm{d}\mathbf{x}.
\label{peu3a}
\end{equation}%
Let us introduce the $\ell $-th charge position $\mathbf{r}^{\ell }\left(
t\right) $ and velocity $\mathsf{v}^{\ell }\left( t\right) $ as the
following averages%
\begin{equation}
\mathbf{r}^{\ell }\left( t\right) =\int_{\mathbb{R}^{3}}\mathbf{x}\left\vert
\psi ^{\ell }\left( t,\mathbf{x}\right) \right\vert ^{2}\,\mathrm{d}\mathbf{x%
},\ \mathsf{v}^{\ell }\left( t\right) =\frac{1}{q}\int_{\mathbb{R}^{3}}%
\mathbf{J}^{\ell }\left( t,\mathbf{x}\right) \,\mathrm{d}\mathbf{x}.
\label{peu4}
\end{equation}%
Then using the charge conservation law (\ref{nama3b}) we find the following
identities%
\begin{gather}
\frac{\mathrm{d}\mathbf{r}^{\ell }\left( t\right) }{\mathrm{d}t}=\int_{%
\mathbb{R}^{3}}\mathbf{x}\partial _{t}\left\vert \psi ^{\ell }\right\vert
^{2}\,\mathrm{d}\mathbf{x}=  \label{peu5} \\
=\mathbf{-}\frac{1}{q^{\ell }}\int_{\mathbb{R}^{3}}\mathbf{x}\nabla \cdot 
\mathbf{J}^{\ell }\,\mathrm{d}\mathbf{x}=\frac{1}{q^{\ell }}\int_{\mathbb{R}%
^{3}}\mathbf{J}^{\ell }\mathrm{d}\mathbf{x}=\mathsf{v}^{\ell }\left(
t\right) ,  \notag
\end{gather}%
showing the positions and velocities defined by formulas (\ref{peu4}) are
related exactly as in the point charge mechanics. Then utilizing the
momentum density kinematic representation (\ref{thef5})-(\ref{thef6}) and
the fact that the momentum density of the $\ell $-th charge EM field is
identically zero according to (\ref{emna4}) we obtain the following
kinematic representation for charge and, hence, the dressed charge total
momentum%
\begin{equation}
\mathsf{P}^{\ell }\left( t\right) =\frac{m^{\ell }}{q^{\ell }}\int_{\mathbb{R%
}^{3}}\mathbf{J}^{\ell }\left( t,\mathbf{x}\right) \,\mathrm{d}\mathbf{x}%
=m^{\ell }\mathsf{v}^{\ell }\left( t\right) ,  \label{peu6}
\end{equation}%
which also is exactly the same as in point charges mechanics. Combining
relations (\ref{peu2}), (\ref{peu5}) and (\ref{peu6}) we obtain the
following system of equations of motion for $N$ charges:%
\begin{gather}
m^{\ell }\frac{\mathrm{d}^{2}\mathbf{r}^{\ell }\left( t\right) }{\mathrm{d}%
^{2}t}=\frac{\mathrm{d}\mathsf{P}^{\ell }}{\mathrm{d}t}=  \label{peu7} \\
=q^{\ell }\int_{\mathbb{R}^{3}}\left[ \left( \dsum\nolimits_{\ell ^{\prime
}\neq \ell }\mathbf{E}^{\ell ^{\prime }}+\mathbf{E}_{\mathrm{ex}}\right)
\left\vert \psi ^{\ell }\right\vert ^{2}+\frac{1}{\mathrm{c}}\mathbf{v}%
^{\ell }\times \mathbf{B}_{\mathrm{ex}}\right] \,\mathrm{d}\mathbf{x}, 
\notag
\end{gather}%
where%
\begin{equation*}
\text{ }\mathbf{E}^{\ell }\left( t,\mathbf{x}\right) =-\nabla \varphi ^{\ell
}\left( t,\mathbf{x}\right) ,\ \mathbf{E}_{\mathrm{ex}}=-\nabla \varphi _{%
\mathrm{ex}}\left( t,\mathbf{x}\right) .
\end{equation*}%
The above system is analogous to the well known in quantum mechanics \emph{%
Ehrenfest\ Theorem}, \cite[Sections 7, 23]{Schiff}. Notice that the system
of the equations of motion (\ref{peu7}) departs from the corresponding
system for point charges by having the averaged Lorentz density force
instead of the Lorentz force at exact position $\mathbf{r}^{\ell }\left(
t\right) $. Observe, also that the system of equations of motion (\ref{peu7}%
) is consistent with Newton's third law of motion "action equals reaction"
as it follows from the relations (\ref{peu3}).

Let us suppose now that charges and current densities $\left\vert \psi
^{\ell }\right\vert ^{2}$ and $q^{\ell }\mathbf{v}^{\ell }$ are localized
near a point $\mathbf{r}^{\ell }\left( t\right) ,$ and have localization
scales $a^{\ell }$ which are small compared with the typical variation scale 
$R_{\mathrm{EM}}$ of the EM field. Then for a bounded $R_{\mathrm{EM}}$ and $%
a^{\ell }$ converging to $0$ we have \ as $a^{\ell }\rightarrow 0,$ 
\begin{equation}
\left\vert \psi ^{\ell }\right\vert ^{2}\left( t,\mathbf{x}\right)
\rightarrow \delta \left( \mathbf{x}-\mathbf{r}^{\ell }\left( t\right)
\right) ,\ \mathbf{v}^{\ell }\left( t,x\right) \rightarrow \mathsf{v}^{\ell
}\left( t\right) \delta \left( \mathbf{x}-\mathbf{r}^{\ell }\left( t\right)
\right) \text{ }  \label{todel}
\end{equation}%
where the coefficients at the delta-functions are determined from the charge
normalization conditions (\ref{nama3c}) and relations (\ref{peu4}). Using
potential representations (\ref{nama4}) we infer from (\ref{todel}) the
convergence of the potentials $\varphi ^{\ell }$ to the corresponding
Coulomb's potentials, namely%
\begin{equation}
\varphi ^{\ell }\left( t,\mathbf{x}\right) \rightarrow \varphi _{0}^{\ell
}\left( t,\mathbf{x}\right) =\frac{q^{\ell }}{\left\vert \mathbf{x}-\mathbf{r%
}^{\ell }\left( t\right) \right\vert }\text{ as }a^{\ell }\rightarrow 0.
\label{tocou}
\end{equation}%
Hence, we can recast the equations of motion (\ref{peu7}) as the system 
\begin{equation}
m^{\ell }\frac{\mathrm{d}^{2}\mathbf{r}^{\ell }}{\mathrm{d}t^{2}}=\frac{%
\mathrm{d}\mathsf{P}^{\ell }}{\mathrm{d}t}=\mathsf{f}^{\ell }+\mathbf{%
\epsilon }_{\mathsf{P}^{\ell }},\text{ }  \label{peu8}
\end{equation}%
where%
\begin{equation*}
\mathsf{f}^{\ell }=\sum_{\ell ^{\prime }\neq \ell }q^{\ell }\mathbf{E}%
_{0}^{\ell ^{\prime }}+q^{\ell }\mathbf{E}_{\mathrm{ex}}\left( \mathbf{r}%
^{\ell }\right) +\frac{1}{\mathrm{c}}\mathsf{v}^{\ell }\times \mathbf{B}_{%
\mathrm{ex}}\left( \mathbf{r}^{\ell }\right) ,\ \ell =1,...,N,
\end{equation*}%
with small discrepancies $\epsilon _{\mathsf{P}^{\ell }}\rightarrow 0$ as $%
a^{\ell }/R_{\mathrm{EM}}\rightarrow 0$. Notice that terms $q^{\ell }\mathbf{%
E}_{0}^{\ell ^{\prime }}$ in equations (\ref{peu8}) coincide with the
Lorentz forces for the Coulomb's potentials 
\begin{equation}
q^{\ell }\mathbf{E}_{0}^{\ell ^{\prime }}=-q^{\ell }\nabla \varphi
_{0}^{\ell ^{\prime }}\left( t,\mathbf{x}\right) =-\frac{q^{\ell }q^{\ell
^{\prime }}\left( \mathbf{r}^{\ell ^{\prime }}-\mathbf{r}^{\ell }\right) }{%
\left\vert \mathbf{r}^{\ell ^{\prime }}-\mathbf{r}^{\ell }\right\vert ^{3}}.
\label{peu8a}
\end{equation}%
In the case when there is no external EM field the point charges equations
of motion (\ref{peu8}) in the limit $a^{\ell }\rightarrow 0$ are associated
with the following Lagrangian ("static limit, zeroth order in $\left(
v/c\right) $", \cite[Section 12.6]{Jackson}) 
\begin{equation}
\mathcal{L}_{\mathrm{p}}=\sum_{\ell }\frac{m^{\ell }\left( \partial _{t}%
\mathbf{r}^{\ell }\right) ^{2}}{2}-\frac{1}{2}\sum_{\ell ^{\prime }\neq \ell
}\frac{q^{\ell }q^{\ell ^{\prime }}}{\left\vert \mathbf{r}^{\ell ^{\prime }}-%
\mathbf{r}^{\ell }\right\vert },  \label{minor4}
\end{equation}%
and the equations of motion (\ref{peu8}) take the form%
\begin{equation}
m^{\ell }\frac{\mathrm{d}^{2}\mathbf{r}^{\ell }}{\mathrm{d}t^{2}}%
=-\sum_{\ell ^{\prime }\neq \ell }\frac{q^{\ell }q^{\ell ^{\prime }}\left( 
\mathbf{r}^{\ell ^{\prime }}-\mathbf{r}^{\ell }\right) }{\left\vert \mathbf{r%
}^{\ell ^{\prime }}-\mathbf{r}^{\ell }\right\vert ^{3}},\ \ell =1,\ldots ,N.
\label{minor5}
\end{equation}%
Using similar arguments we obtain from (\ref{peu3a})%
\begin{equation}
\frac{\mathrm{d}\mathsf{E}^{\ell }}{\mathrm{d}t}=\mathsf{v}^{\ell }\cdot 
\mathsf{f}^{\ell }+\epsilon _{\mathsf{E}^{\ell }},\text{ }  \label{peu9}
\end{equation}%
with small discrepancies $\epsilon _{\mathsf{E}^{\ell }}\rightarrow 0$ as $%
a^{\ell }/R_{\mathrm{EM}}\rightarrow 0.$ Combining equalities (\ref{peu6}) (%
\ref{peu8}) (\ref{peu9}) we get%
\begin{gather}
\frac{\mathrm{d}m^{\ell }\mathbf{v}^{\ell }\cdot \mathbf{v}^{\ell }}{2%
\mathrm{d}t}=\mathbf{v}^{\ell }\cdot \frac{\mathrm{d}m\mathbf{v}^{\ell }}{%
\mathrm{d}t}=  \label{peu10} \\
=\mathbf{v}^{\ell }\cdot \mathsf{f}^{\ell }+\mathbf{v}^{\ell }\cdot \mathbf{%
\epsilon }_{\mathsf{P}^{\ell }}=\frac{\mathrm{d}\mathsf{E}^{\ell }}{\mathrm{d%
}t}+\mathbf{v}^{\ell }\cdot \mathbf{\epsilon }_{\mathsf{P}^{\ell }}-\epsilon
_{\mathsf{E}^{\ell }},  \notag
\end{gather}%
implying 
\begin{equation*}
\frac{\mathrm{d}}{\mathrm{d}t}\left( \mathsf{E}^{\ell }-\frac{m^{\ell }%
\mathbf{v}^{\ell }\cdot \mathbf{v}^{\ell }}{2}\right) =\epsilon _{\mathsf{E}%
^{\ell }}-\mathbf{v}^{\ell }\cdot \mathbf{\epsilon }_{\mathsf{P}^{\ell }},
\end{equation*}%
which, up to small errors, are kinematic representations for the energies of
individual charges well known from the point charge mechanics. Note that to
obtain point charges equations of motion (\ref{peu8}) it is sufficient to
assume localization only for $\psi ^{\ell }$.

\subsubsection{Point mechanics via wave-corpuscles\label{pmechwc}}

In this section we construct approximate solution of the field equations (%
\ref{nama1}), (\ref{nama2}) (or, equivalently, (\ref{NLSj0})) for $N$
interacting charges in which every charge is a wave-corpuscle defined by (%
\ref{psil0}) with properly chosen complementary point charges equations of
motion . The construction proposed here is valid for any external EM field,
but to avoid involved expressions we consider the case when the external EM
field is purely electric with $\mathbf{A}_{\mathrm{ex}}=0$. The general case
when $\mathbf{A}_{\mathrm{ex}}\neq 0$ \ is treated similarly based on the
results of Section \ref{exelmag}. We assume here that the shape factor $%
\left\vert \mathring{\psi}_{1}^{\ell }\left( \left\vert \mathbf{x}%
\right\vert \right) \right\vert ^{2}$ \emph{decays exponentially} as $%
\left\vert \mathbf{x}\right\vert \rightarrow \infty $ for every $\ell $-th
charge and (\ref{psexp}) holds. The wave-corpuscle approximation (\ref{apsol}%
) is based on trajectories $\mathbf{r}_{0}^{\ell }$ for the wave-corpuscle
centers determined from equations (\ref{Newtel}) which involve the exact
Coulomb's potentials $\mathring{\varphi}_{0}^{\ell }$ corresponding to the
size parameter $a=0$. To show that the approximation is accurate for small $%
a>0$ we use the results obtained for a single charge motion in an external
field. As the first step for an estimate we introduce an auxiliary system of
equations to determine all center trajectories. This system has the
following property. If $\ell $-th charge is singled out and the potentials $%
\mathring{\varphi}_{a}^{\ell ^{\prime }}\left( \mathbf{x}-\mathbf{r}%
_{0}^{\ell ^{\prime }}\left( t\right) \right) $ of remaining charges are
replaced by the linear approximation of $\mathring{\varphi}_{0}^{\ell
^{\prime }}\left( \mathbf{x}-\mathbf{r}_{0}^{\ell ^{\prime }}\left( t\right)
\right) $ based on the position of the $\ell $-th charge, then the exact
wave-corpuscle solution for the $\ell $-th charge is available in so
modified field. In addition to that, the motion of the center $\mathbf{r}%
^{\ell }$ of the exact solution to the auxiliary equation has the same
trajectories $\mathbf{r}_{0}^{\ell }$. Replacing $\mathring{\varphi}%
_{a}^{\ell ^{\prime }}$ by the exact Coulomb's potential $\mathring{\varphi}%
_{0}^{\ell ^{\prime }}$ produces a contribution to the discrepancy, and the
second source of the discrepancy is the field linearization at $\mathbf{r}%
_{0}^{\ell }$.\ To estimate these discrepancies we use the results of
Section \ref{seinhomo}. First, we find trajectories $\mathbf{r}_{0}^{\ell
}\left( t\right) $ from the auxiliary equations 
\begin{equation}
m^{\ell }\frac{\mathrm{d}^{2}\mathbf{r}_{0}^{\ell }}{\mathrm{d}t^{2}}%
=-q^{\ell }\nabla \varphi _{\mathrm{ex},0}^{\ell }\left( \mathbf{r}%
_{0}^{\ell }\right) ,\   \label{pmvc3}
\end{equation}%
with initial data%
\begin{equation*}
\mathbf{r}_{0}^{\ell }\left( 0\right) =\mathbf{\acute{r}}_{0}^{\ell },\ 
\frac{\mathrm{d}\mathbf{r}_{0}^{\ell }}{\mathrm{d}t}\left( 0\right) =\mathbf{%
\acute{v}}_{0}^{\ell },\ \ell =1,...,N.
\end{equation*}%
The electrostatic potential $\varphi _{\mathrm{ex},0}^{\ell }$ in (\ref%
{pmvc3}) is the Coulomb's potential as in (\ref{fiexl0}), and for $a>0$ we
introduce an intermediate external potential for the $\ell $-th charge as
follows: 
\begin{equation}
\mathring{\varphi}_{\mathrm{ex},a}^{\ell }\left( t,\mathbf{x}\right)
=\varphi _{\mathrm{ex}}\left( t,\mathbf{x}\right) +\sum\nolimits_{\ell
^{\prime }\neq \ell }\mathring{\varphi}_{a}^{\ell ^{\prime }}\left( \mathbf{x%
}-\mathbf{r}_{0}^{\ell ^{\prime }}\right) .  \label{pmvc1}
\end{equation}%
We define then an approximate solution $\psi _{\mathrm{ap}}^{\ell }$ to be
of the form of wave-corpuscles (\ref{apsol}), (\ref{plslfi}), namely: 
\begin{equation}
\psi _{\mathrm{ap}}^{\ell }\left( t,\mathbf{x}\right) =\mathrm{e}^{\mathrm{i}%
S/\chi }\mathring{\psi}\left( \left\vert \mathbf{x}-\mathbf{r}_{0}^{\ell
}\right\vert \right) ,\ \varphi _{\mathrm{ap}}^{\ell }\left( t,\mathbf{x}%
\right) =q^{\ell }\mathring{\varphi}_{a}^{\ell }\left( \mathbf{x}-\mathbf{r}%
_{0}^{\ell }\right) ,\text{ }  \label{minor6}
\end{equation}%
where%
\begin{gather*}
\mathbf{p}^{\ell }=m^{\ell }\frac{\mathrm{d}\mathbf{r}_{0}^{\ell }}{\mathrm{d%
}t}, \\
S^{\ell }\left( t,\mathbf{x}\right) =\mathbf{p}^{\ell }\cdot \mathbf{x}%
-\dint_{0}^{t}\frac{\mathbf{p}^{\ell 2}}{2m}\,\mathrm{d}t^{\prime
}-q\dint_{0}^{t}\varphi _{\mathrm{ex},0}\left( t^{\prime },\mathbf{r}%
_{0}^{\ell }\right) \,\mathrm{d}t^{\prime },
\end{gather*}%
and $\mathbf{r}_{0}^{\ell }$\ is solution of (\ref{pmvc3}). Recall that the
dependence on the size parameter $a$ of the form factor $\mathring{\psi}%
^{\ell }=\mathring{\psi}_{a}^{\ell }$ and corresponding potential $\mathring{%
\varphi}^{\ell }=\mathring{\varphi}_{a}^{\ell }$ is given by (\ref{nrac5}).
Notice that according to relations (\ref{ficr}) the interaction force term
in (\ref{pmvc3}) approaches the Lorentz forces based on the Coulomb's
potential, namely%
\begin{equation}
\mathring{\varphi}_{a}^{\ell }\left( \mathbf{x}\right) =q^{\ell }\dint_{%
\mathbb{R}^{3}}\frac{\left\vert \mathring{\psi}_{a}^{\ell }\right\vert
^{2}\left( t,\mathbf{y}\right) }{\left\vert \mathbf{y}-\mathbf{x}\right\vert 
}\,\mathrm{d}\mathbf{y}\rightarrow \mathring{\varphi}_{0}^{\ell }\left( 
\mathbf{x}\right) =\frac{q^{\ell }}{\left\vert \mathbf{x}\right\vert }\text{
as }a\rightarrow 0.  \label{fiaq}
\end{equation}

Let us introduce an auxiliary spatially linear potential $\tilde{\varphi}_{%
\mathrm{ex,0}}^{\ell }\left( t,\mathbf{x}\right) $ for the $\ell $-th charge
by formula (\ref{fiex1}) with$\ \varphi _{\mathrm{ex}}$ replaced by $\varphi
_{\mathrm{ex},0}^{\ell }$. Observe that for every $\ell $ the pair $\left\{
\psi _{\mathrm{ap}}^{\ell },\varphi _{\mathrm{ap}}^{\ell }\right\} $ is an
exact solution to the auxiliary equation (\ref{auf1}) with the external
potential $\tilde{\varphi}_{\mathrm{ex,0}}^{\ell }\left( t,\mathbf{x}\right) 
$ obtained by the linearization of $\varphi _{\mathrm{ex},0}^{\ell }$ at $%
\mathbf{r}_{0}^{\ell }\left( t\right) $ \ according to (\ref{fiex1}). It
remains to examine if the so defined $\left\{ \psi _{\mathrm{ap}}^{\ell
},\varphi _{\mathrm{ap}}^{\ell }\right\} $ yield an approximation to the
field equations (\ref{NLSj0}). Indeed, the $\ell $-th \ wave-corpuscles $%
\left\{ \psi _{\mathrm{ap}}^{\ell },\varphi _{\mathrm{ap}}^{\ell }\right\} $
is an exact solution to equations (\ref{auf1}). \ To obtain from (\ref{auf1}%
) the $\ell $-th equation (\ref{NLSj0}) we have to replace $\tilde{\varphi}_{%
\mathrm{ex,0}}^{\ell }\left( t,\mathbf{x}\right) $\ by $\mathring{\varphi}_{%
\mathrm{ex},a}^{\ell }\left( t,\mathbf{x}\right) $ resulting in a
discrepancy 
\begin{gather*}
\left[ \tilde{\varphi}_{\mathrm{ex,0}}^{\ell }\left( t,\mathbf{x}\right) -%
\mathring{\varphi}_{\mathrm{ex},a}^{\ell }\left( t,\mathbf{x}\right) \right] 
\mathring{\varphi}_{a}^{\ell }\left( \mathbf{x}\right) = \\
=\left[ \tilde{\varphi}_{\mathrm{ex,0}}^{\ell }\left( t,\mathbf{x}\right) -%
\mathring{\varphi}_{\mathrm{ex},0}^{\ell }\left( t,\mathbf{x}\right) \right] 
\mathring{\varphi}_{a}^{\ell }\left( \mathbf{x}\right) +\left[ \mathring{%
\varphi}_{\mathrm{ex},0}^{\ell }\left( t,\mathbf{x}\right) -\mathring{\varphi%
}_{\mathrm{ex},a}^{\ell }\left( t,\mathbf{x}\right) \right] \mathring{\varphi%
}_{a}^{\ell }\left( \mathbf{x}\right) .
\end{gather*}%
The first term of the above discrepancy is the same as in (\ref{auf2}), and
the corresponding integral discrepancy is estimated as in (\ref{d0}). The
second term has the form \ 
\begin{gather}
\left[ \mathring{\varphi}_{\mathrm{ex},0}^{\ell }\left( t,\mathbf{x}\right) -%
\mathring{\varphi}_{\mathrm{ex},a}^{\ell }\left( t,\mathbf{x}\right) \right] 
\mathring{\varphi}_{a}^{\ell }\left( \mathbf{x}\right) =  \label{sect} \\
=\sum\nolimits_{\ell ^{\prime }\neq \ell }q^{\ell }\left[ \mathring{\varphi}%
_{a}^{\ell ^{\prime }}\left( \mathbf{x-r}_{0}^{\ell ^{\prime }}\right) -%
\mathring{\varphi}_{0}^{\ell ^{\prime }}\left( \mathbf{x}-\mathbf{r}%
_{0}^{\ell ^{\prime }}\right) \right] \mathrm{e}^{\mathrm{i}S/\chi }%
\mathring{\psi}_{a}^{\ell }\left( \left\vert \mathbf{x}-\mathbf{r}_{0}^{\ell
}\right\vert \right) .  \notag
\end{gather}%
Taking into account that $\mathring{\psi}_{a}^{\ell }\left( \left\vert 
\mathbf{x}\right\vert \right) $ decays exponentially as $\left\vert \mathbf{x%
}\right\vert \rightarrow \infty $ we find that every term in (\ref{sect}) is
small if $\mathbf{r}_{0}^{\ell }$ is separated from $\mathbf{r}_{0}^{\ell
^{\prime }}$ for $\ell ^{\prime }\neq \ell $. To take into account point
charges separation we introduce a quantity%
\begin{equation}
R_{\mathrm{\min }}=\min_{\ell \neq \ell ^{\prime },\ 0\leq t\leq
T}\left\vert \mathbf{r}_{0}^{\ell }\left( t\right) -\mathbf{r}_{0}^{\ell
^{\prime }}\left( t\right) \right\vert ,  \label{pmvc6}
\end{equation}%
and assume it to be positive, i.e. $R_{\mathrm{\min }}>0$. Under this
condition using (\ref{fiexp}) we obtain 
\begin{gather}
\left\vert \mathring{\varphi}_{a}^{\ell ^{\prime }}\left( \mathbf{x-r}%
_{0}^{\ell ^{\prime }}\right) -\mathring{\varphi}_{0}^{\ell ^{\prime
}}\left( \mathbf{x}-\mathbf{r}_{0}^{\ell ^{\prime }}\right) \right\vert \leq
C\varepsilon \left( a/R_{\min }\right) ,  \label{est2} \\
\text{if }\left\vert \mathbf{x-r}_{0}^{\ell }\right\vert \leq a\sigma _{\psi
}\text{\ }  \notag
\end{gather}%
where 
\begin{equation*}
\text{ }\varepsilon \left( a/R_{\min }\right) \rightarrow 0\text{ as }%
a/R_{\min }\rightarrow 0,
\end{equation*}%
where constant $C$, $\sigma _{\psi }=\sigma _{\psi ^{\ell }}$\ are the same
as in (\ref{thetpsi}) (note that $\varepsilon \left( a/R_{\min }\right)
\rightarrow 0$ exponentially fast if $\mathring{\psi}^{\ell }\left( r\right) 
$ decays exponentially). Hence, we can take $\mathring{\varphi}_{\mathrm{ex}%
,a}^{\ell }\left( t,\mathbf{x}\right) $\ as $\acute{\varphi}_{\mathrm{ex}%
}\left( t,\mathbf{r},\epsilon \right) $ and $\mathring{\varphi}_{\mathrm{ex}%
,0}^{\ell }\left( t,\mathbf{x}\right) $\ as $\varphi _{\mathrm{ex}}\left( t,%
\mathbf{x}\right) $ in (\ref{eeep}) with $\epsilon =\varepsilon \left(
a/R_{\min }\right) $. \ Then we use the integral discrepancy estimate (\ref%
{totd}) which implies 
\begin{equation}
\left\vert \hat{D}_{0}\left( x,t\right) +\hat{D}_{1}\left( x,t\right)
\right\vert \lesssim \left[ C_{0}\frac{a^{2}}{R_{\varphi }^{2}}+\frac{C_{1}}{%
\left\vert \bar{\varphi}\right\vert }\varepsilon \left( \frac{a}{R_{\min }}%
\right) \right] \max_{\ell }\left\vert q^{\ell }\right\vert \left\vert \bar{%
\varphi}^{\ell }\right\vert ,  \label{totd1}
\end{equation}%
where the potential curvature radius $R_{\varphi }^{2}$ is based on $\varphi
_{\mathrm{ex,0}}\left( \mathbf{x},t\right) $ as in (\ref{pmvc1}), $\bar{%
\varphi}^{\ell }$ is based on $\varphi _{\mathrm{ex},0}^{\ell }$ and $%
\mathbf{r}_{0}^{\ell }$. The factors $\left\vert q^{\ell }\right\vert
\left\vert \bar{\varphi}^{\ell }\right\vert $ are bounded uniformly in $a$.
Consequently, we conclude that the integral discrepancy resulting from the
substitution of $\psi _{\mathrm{ap}}^{\ell }\left( t,\mathbf{x}\right) $
into (\ref{NLSj0})\ and given by (\ref{totd1}) tends to zero as $%
a/R\rightarrow 0$ where $R=\min \left( R_{\min },R_{\varphi }\right) $. Note
that for exponentially decaying $\mathring{\psi}^{\ell }$ the function $%
\varepsilon \left( a/R_{\min }\right) $ decays exponentially as $a/R_{\min
}\rightarrow 0$ and hence 
\begin{equation*}
\left\vert \hat{D}_{0}\left( x,t\right) +\hat{D}_{1}\left( x,t\right)
\right\vert \simeq O\left( \frac{a^{2}}{R_{\varphi }^{2}}\right) U,\
U=\max_{\ell }\left\vert q^{\ell }\right\vert \left\vert \bar{\varphi}^{\ell
}\right\vert .
\end{equation*}%
Interestingly, an additional analysis of the exact equations of motion (\ref%
{Couf}) shows that though the integral discrepancy decays as $%
a^{2}/R_{\varphi }^{2},$ the positions $\mathbf{r}^{\ell }\left( t\right) $
given by (\ref{peu4}) are approximated by $\mathbf{r}_{0}^{\ell }\left(
t\right) $ \ with accuracy of the order $a^{3}/R_{\varphi }^{3}$.

\subsection{Relativistic theory of interacting charges}

Relativistic theory of many interacting point particles is known to have
fundamental difficulties. "The invariant formulation of the motion of two or
more interacting particles is complicated by the fact that each particle
will have a different proper time. ... No exact general theory seems to be
available", \cite[Section II.1, System of colliding particles]{Barut}. Some
of these difficulties are analyzed by H. Goldstein in his classical
monograph, \cite[Section 7.10]{Goldstein}: "The great stumbling block
however is the treatment of the type of interaction that is so natural and
common in nonrelativistic mechanics - direct interaction between particles.
... To say that the force on a particle depends upon the positions or
velocities of other particles at the same time implies propagation of
effects with infinite velocity from one particle to another - "action at a
distance." In special relativity, where signals cannot travel faster than
the speed of light, action-at-a-distance seems outlawed. And in a certain
sense this seems to be the correct picture. It has been proven that if we
require certain properties of the system to behave in the normal way (such
as conservation of total linear momentum), then there can be no covariant
direct interaction between particles except through contact forces." Another
argument, due to von Laue, \cite{von Laue}, on the incompatibility of the
relativity with any finite dimensional mechanical system was articulated by
W. Pauli, \cite[Section 45]{Pauli RT}: "...This in itself raised strong
doubts as to the possibility of introducing the concept of a rigid body into
relativistic mechanics$^{\text{247}}$. The final clarification was brought
about in a paper by Laue$^{\text{248}}$, who showed by quite elementary
arguments that the number of kinematic degrees of freedom of a body cannot
be limited, according to the theory of relativity. For, since no action can
be propagated with a velocity greater than that of light, an impulse which
is given to the body simultaneously at $n$ different places, will, \emph{to
start of with}, produce a motion to which at least $n$ degrees of freedom
must be ascribed."

Now we ask ourselves what features of point charges mechanics can be
integrated into a relativistic mechanics of fields? It seems that the above 
\emph{arguments by Goldstein, von Laue and Pauli completely rule out any
Lagrangian mechanics with finitely many degrees of freedom even as an
approximation} because of its incompatibility with a basic relativity
requirement for the signal speed not to exceed the speed of light. On the
constructive side, these arguments suggest that (i) the EM field has to be
an integral part of charges mechanics, (ii) every charge of the system has
to be some kind of elastic continuum coupled to the EM field. We anticipate
though that point mechanics features that can be integrated into a
relativistic field mechanics are limited and have subtler manifestation
compared to the nonrelativistic theory. We expect point mechanics features
to manifest themselves in (i) identification of the energy-momentum tensor
for every individual bare charge; (ii) certain partition of the EM field
into a sum of EM fields attributed to individual charges with consequent
formation of dressed charges, that is bare charges with attached to them EM
fields. \emph{That energy-momentum partition between individual charges must
comply with the Newton's "action equals to reaction" law, the interaction
forces densities have to be of the Lorentzian form and every dressed charge
has not to interact with itself.}

In the theory proposed here we address the above challenges by (i) the
principal departure from the concept of point charge, which is substituted
by a concept of wave-corpuscle described by a complex valued function in the
space-time; (ii) requirement for every charge to interact directly to only
the EM field implying that different charges interact only via the EM field.
With all that in mind we introduce the system Lagrangian $\mathcal{L}$ to be
of the general form as in Section \ref{Manych}%
\begin{gather}
\mathcal{L}\left( \left\{ \psi ^{\ell },\psi _{;\mu }^{\ell }\right\}
,\left\{ \psi ^{\ell \ast },\psi _{;\mu }^{\ell \ast }\right\} ,A^{\mu
}\right) =\dsum_{\ell }L^{\ell }\left( \psi ^{\ell },\psi _{;\mu }^{\ell
},\psi ^{\ell \ast },\psi _{;\mu }^{\ell \ast }\right) -\frac{F^{\mu \nu
}F_{\mu \nu }}{16\pi },  \label{mplag1} \\
F^{\mu \nu }=\partial ^{\mu }A^{\nu }-\partial ^{\nu }A^{\mu },  \notag
\end{gather}%
with every $\ell $-th charge Lagrangian $L^{\ell }$ to be of the form of
single relativistic charge (\ref{fpar1})-(\ref{fpar1a})%
\begin{equation}
L^{\ell }\left( \psi ^{\ell },\psi _{;\mu }^{\ell },\psi ^{\ell \ast },\psi
_{;\mu }^{\ell \ast }\right) =\frac{\chi ^{2}}{2m^{\ell }}\left\{ \psi
_{;\mu }^{\ell \ast }\psi ^{\ell ;\mu }-\kappa ^{\ell 2}\psi ^{\ell \ast
}\psi ^{\ell }-G^{\ell }\left( \psi ^{\ell \ast }\psi ^{\ell }\right)
\right\} ,  \label{mplag2}
\end{equation}%
where%
\begin{equation*}
\kappa ^{\ell }=\frac{\omega ^{\ell }}{\mathrm{c}}=\frac{m^{\ell }\mathrm{c}%
}{\chi },\ \omega ^{\ell }=\frac{m^{\ell }\mathrm{c}^{2}}{\chi },
\end{equation*}%
and $\psi _{;\mu }^{\ell }$ and $\psi _{;\mu }^{\ell \ast }$ are the \emph{%
covariant derivatives} defined by the following formulas 
\begin{equation}
\psi _{;\mu }^{\ell }=\tilde{\partial}^{\ell \mu }\psi ^{\ell },\ \psi
_{;\mu }^{\ell \ast }=\tilde{\partial}^{\ell \mu \ast }\psi ^{\ell },\ 
\tilde{\partial}^{\ell \mu }=\partial ^{\mu }+\frac{\mathrm{i}q^{\ell
}A^{\mu }}{\chi \mathrm{c}},\ \tilde{\partial}^{\ell \mu \ast }=\partial
^{\mu }-\frac{\mathrm{i}q^{\ell }A^{\mu }}{\chi \mathrm{c}},  \label{mplag2a}
\end{equation}%
where $\tilde{\partial}^{\ell \mu }$ and $\tilde{\partial}^{\ell \mu \ast }$
are called the \emph{covariant differentiation operators}. We also assume
that for every $\ell $: (i) $m^{\ell }>0$ is the charge mass; (ii) $q^{\ell
} $ is a real valued (positive or negative) charge; (iii) $\kappa ^{\ell }>0$%
; (iv) $G^{\ell }$ is a nonlinear self-interaction function. Notice that
charges interaction with the EM field enters the Lagrangian $\mathcal{L}$
via the covariant derivatives (\ref{mplag2a}). The Lagrangian $\mathcal{L}$
defined by (\ref{mplag1})-(\ref{mplag3}) is manifestly local, Lorentz
invariant, and gauge invariant with respect to the second-kind (local) gauge
transformation%
\begin{equation}
\psi ^{\ell }\rightarrow \mathrm{e}^{-\frac{\mathrm{i}q^{\ell }\lambda
\left( x\right) }{\chi \mathrm{c}}}\psi ^{\ell },\ A^{\mu }\rightarrow
A^{\mu }+\partial ^{\mu }\lambda \left( x\right) ,  \label{mplag3}
\end{equation}%
as well as with respect to the group of global (the first-kind) gauge
transformations%
\begin{equation}
\psi ^{\ell }\rightarrow \mathrm{e}^{-\mathrm{i}q^{\ell }\lambda ^{\ell
}}\psi ^{\ell },\ A^{\mu }\rightarrow A^{\mu }\text{ }  \label{mplag4}
\end{equation}%
for any real numbers $\lambda ^{\ell }.$ \ The Euler-Lagrange field
equations (\ref{flagr8c})-(\ref{flagr8d}) for the Lagrangian $\mathcal{L}$
defined by (\ref{mplag1})-(\ref{mplag3}) take the form:%
\begin{equation}
\left[ \tilde{\partial}_{\mu }^{\ell }\tilde{\partial}^{\ell \mu }+\kappa
^{\ell 2}+G^{\ell \prime }\left( \left\vert \psi ^{\ell }\right\vert
^{2}\right) \right] \psi ^{\ell }=0,\ \tilde{\partial}^{\ell \mu }=\partial
^{\mu }+\frac{\mathrm{i}q^{\ell }A^{\mu }}{\chi \mathrm{c}},  \label{mplag5}
\end{equation}%
together with the conjugate equation \ for $\psi ^{\ast \ell }$ 
\begin{equation}
\left[ \tilde{\partial}_{\mu }^{\ell \ast }\tilde{\partial}^{\ell \ast \mu
}+\kappa ^{\ell 2}+G^{\ell \prime }\left( \left\vert \psi ^{\ell
}\right\vert ^{2}\right) \right] \psi ^{\ast \ell }=0,\ \tilde{\partial}%
^{\ell \ast \mu }=\partial ^{\mu }-\frac{\mathrm{i}q^{\ell }A^{\mu }}{\chi 
\mathrm{c}},  \label{mplag6}
\end{equation}%
and equations for the EM field 4-potentials 
\begin{equation}
\partial _{\mu }F^{\mu \nu }=\frac{4\pi }{\mathrm{c}}J^{\nu },\text{ }J^{\nu
}=\dsum_{\ell }J^{\ell \nu },\text{ }F^{\mu \nu }=\partial ^{\mu }A^{\nu
}-\partial ^{\nu }A^{\mu },  \label{mplag7}
\end{equation}%
where the $\ell $-th charge \emph{4-vector EM micro-current} $J^{\ell \nu }$
defined by (\ref{flagr9}) takes here the form%
\begin{eqnarray}
J^{\ell \mu \nu } &=&-\mathrm{i}\frac{q^{\ell }\chi \left[ \left( \tilde{%
\partial}^{\ell \nu \ast }\psi ^{\ell \ast }\right) \psi ^{\ell }-\psi
^{\ell \ast }\tilde{\partial}^{\ell \nu }\psi ^{\ell }\right] }{2m^{\ell }}=
\label{mplag8} \\
&=&-\frac{q^{\ell }\chi \left\vert \psi ^{\ell }\right\vert ^{2}}{m^{\ell }}%
\left( \func{Im}\frac{\partial ^{\nu }\psi ^{\ell }}{\psi ^{\ell }}+\frac{%
q^{\ell }A^{\mu }}{\chi \mathrm{c}}\right) .  \notag
\end{eqnarray}%
Observe that the equations (\ref{mplag7}) are the Maxwell equations (\ref%
{maxw4}) with currents $J^{\ell \mu }$. As in the case of the more general
Lagrangian (\ref{flagr6a}) the gauge invariance (\ref{mplag4}) implies that
every individual $\ell $-th charge 4-vector micro-current $J^{\ell \mu }$
satisfies the continuity equation%
\begin{equation}
\partial _{\nu }J^{\ell \nu }=0,\ \partial _{t}\rho ^{\ell }+\nabla \cdot 
\mathbf{J}^{\ell }=0,\ J^{\ell \nu }=\left( \rho ^{\ell }\mathrm{c},\mathbf{J%
}^{\ell }\right) ,  \label{mplag9}
\end{equation}%
under the assumption that the fields $\left\{ \psi ^{\ell },F^{\mu \nu
}\right\} $ satisfy the Euler-Lagrange field equations (\ref{mplag5})-(\ref%
{mplag7}). Notice that in view of (\ref{mplag8})%
\begin{gather}
\rho ^{\ell }=-\frac{q^{\ell }\left\vert \psi ^{\ell }\right\vert ^{2}}{%
m^{\ell }\mathrm{c}^{2}}\left( \chi \func{Im}\frac{\partial _{t}\psi ^{\ell }%
}{\psi ^{\ell }}+q^{\ell }\varphi \right) ,  \label{mplag10} \\
\mathbf{J}^{\ell }=\frac{q^{\ell }\left\vert \psi ^{\ell }\right\vert ^{2}}{%
m^{\ell }}\left( \chi \func{Im}\frac{\nabla \psi ^{\ell }}{\psi ^{\ell }}-%
\frac{q^{\ell }\mathbf{A}}{\mathrm{c}}\right) .  \notag
\end{gather}%
As a consequence of the continuity equations (\ref{mplag9}) \emph{the space
integral of every} $\rho ^{\ell }\left( x\right) $ \emph{is a conserved
quantity which we assign to be exactly} $q^{\ell }$, i.e. we assume the
following \emph{charge normalization} 
\begin{gather}
\dint_{\mathbb{R}^{3}}\frac{\rho ^{\ell }\left( x\right) }{q^{\ell }}\,%
\mathrm{d}\mathbf{x}=  \label{mplag11} \\
=-\frac{1}{m^{\ell }\mathrm{c}^{2}}\dint_{\mathbb{R}^{3}}\left( \chi \func{Im%
}\frac{\partial _{t}\psi ^{\ell }}{\psi ^{\ell }}+q^{\ell }\varphi \right)
\left\vert \psi ^{\ell }\right\vert ^{2}\,\mathrm{d}\mathbf{x}=1,\ \ell
=1,\ldots N.  \notag
\end{gather}

To summarize, the equations (\ref{mplag5})-(\ref{mplag7}) together with the
normalization (\ref{mplag11}) constitute a complete set of equations
describing the state of the all fields $\left\{ \psi ^{\ell },F^{\mu \nu
}\right\} $ in the space-time. Notice that (\ref{mplag5}) and the Maxwell
equations can be recast as%
\begin{equation}
\left[ \mathrm{c}^{-2}\tilde{\partial}_{t}^{\ell 2}-\tilde{\nabla}^{\ell
2}+\kappa ^{\ell 2}+G^{\ell \prime }\left( \left\vert \psi ^{\ell
}\right\vert ^{2}\right) \right] \psi ^{\ell }=0,  \label{mplag13}
\end{equation}%
where%
\begin{equation*}
\tilde{\partial}_{t}^{\ell }=\partial _{t}+\frac{\mathrm{i}q^{\ell }\varphi 
}{\chi },\ \tilde{\nabla}^{\ell }=\nabla -\frac{\mathrm{i}q^{\ell }\mathbf{A}%
}{\chi \mathrm{c}},
\end{equation*}%
\begin{gather}
\nabla \cdot \left( \partial _{t}\mathbf{A}+\nabla \varphi \right) =4\pi
\dsum_{\ell }\frac{q^{\ell }\left\vert \psi ^{\ell }\right\vert ^{2}}{%
m^{\ell }\mathrm{c}^{2}}\left( \chi \func{Im}\frac{\partial _{t}\psi ^{\ell }%
}{\psi ^{\ell }}+q^{\ell }\varphi \right) ,  \label{mplag14} \\
\nabla \times \left( \nabla \times \mathbf{A}\right) +\frac{1}{\mathrm{c}}%
\partial _{t}\left( \partial _{t}\mathbf{A}+\nabla \varphi \right) =
\label{mplag15} \\
=\frac{4\pi }{\mathrm{c}}\dsum_{\ell }\left( \frac{\chi q^{\ell }}{m^{\ell }}%
\func{Im}\frac{\nabla \psi ^{\ell }}{\psi ^{\ell }}-\frac{q^{\ell 2}\mathbf{A%
}}{m^{\ell }\mathrm{c}}\right) \left\vert \psi ^{\ell }\right\vert ^{2}. 
\notag
\end{gather}

In order to see point charge features in the charges described as fields
over the continuum of the space-time we have to identify the energy-momentum
tensor $T^{\ell \mu \nu }$ for every $\ell $-th charge and the EM field
energy-momentum $\Theta ^{\mu \nu }$. Notice that the system Lagrangian $%
\mathcal{L}$ defined by (\ref{mplag1})-(\ref{mplag3}) satisfies the symmetry
condition (\ref{lagsym}) and consequently the general construction of the
symmetric energy-momenta from Section \ref{Manych} applies here. Namely,
using the formulas (\ref{enmom4})-(\ref{enmom5}) we get the following
representation for the energy-momenta%
\begin{gather}
T^{\ell \mu \nu }=\frac{\chi ^{2}}{2m^{\ell }}\left\{ \left( \psi ^{\ell
;\mu \ast }\psi ^{\ell ;\nu }+\psi ^{\ell ;\mu }\psi ^{\ell ;\nu \ast
}\right) -\right.  \label{tmap1} \\
-\left. \left[ \psi _{;\mu }^{\ell \ast }\psi ^{\ell ;\mu }-\kappa ^{\ell
2}\psi ^{\ell \ast }\psi ^{\ell }-G^{\ell }\left( \psi ^{\ell \ast }\psi
^{\ell }\right) \right] \delta ^{\mu \nu }\right\} ,  \notag
\end{gather}%
\begin{equation}
\Theta ^{\mu \nu }=\frac{1}{4\pi }\left( g^{\mu \gamma }F_{\gamma \xi
}F^{\xi \nu }+\frac{1}{4}g^{\mu \nu }F_{\gamma \xi }F^{\gamma \xi }\right) .
\label{tmap2}
\end{equation}%
The above defined energy-momentum tensors satisfy the equations (\ref%
{divten1})-(\ref{divten2}), namely%
\begin{eqnarray}
\partial _{\mu }T^{\ell \mu \nu } &=&f^{\ell \nu },\text{ }  \label{tmap3} \\
\partial _{\mu }\Theta ^{\mu \nu } &=&-f^{\nu },  \label{tmap4}
\end{eqnarray}%
where%
\begin{equation*}
\text{ }f^{\ell \nu }=\frac{1}{\mathrm{c}}J_{\mu }^{\ell }F^{\nu \mu },\
f^{\nu }=\dsum_{\ell }f^{\ell \nu }=\frac{1}{\mathrm{c}}J_{\mu }F^{\nu \mu
},\ J_{\mu }=\dsum_{\ell }J_{\mu }^{\ell }.
\end{equation*}%
\emph{The energy-momentum conservation equations (\ref{tmap3})-(\ref{tmap4})
can be viewed as equations of motion in elastic continuum,} \cite[Section
6.4, (6.56), (6.57)]{Moller}, similar to the case of kinetic energy-momentum
tensor for a single relativistic particle, \cite[Section 37, (3.24)]{Pauli
RT}. We recognize in the 4-vectors $f^{\ell \nu }$ in right-hand side of
conservation equations (\ref{tmap3}) the density of the Lorentz force acting
upon $\ell $-th charge, and we also see the density of the Lorentz force
with the minus sign in the right-hand side of EM energy-momentum
conservation equation (\ref{tmap3}). We remind to the reader that equations (%
\ref{tmap3})-(\ref{tmap4}) hold only under the assumption that the involved
fields satisfy the field equations (\ref{mplag5})-(\ref{mplag8}).
Consequently, in contrast to the case of the point mechanics the
conservation/equations of motion (\ref{tmap3})-(\ref{tmap4}) in the elastic
continuum alone cannot substitute for the field equations and determine the
motion.

Now we would like to identify the EM\ field attributed to every individual
bare charge. That can be accomplished by partitioning the total EM $F^{\mu
\nu }$ defined as a causal solution to the linear Maxwell equation (\ref%
{mplag7}) (see Section \ref{sGreenMax}) with a source $\frac{4\pi }{\mathrm{c%
}}J^{\mu }$ according to the partition of the current $J^{\mu }=\dsum_{\ell
}J^{\ell \mu }$. Namely we introduce the EM potentials $A^{\ell \mu }$ and
the corresponding EM field $F^{\ell \mu \nu }$ for every individual $\ell $%
-th charge as the causal solution of the form (\ref{grmax9}) to the
following Maxwell equation%
\begin{equation}
\partial _{\mu }F^{\ell \mu \nu }=\frac{4\pi }{\mathrm{c}}J^{\ell \nu }.
\label{empa1}
\end{equation}%
In view of the linearity of the Maxwell equation (\ref{mplag7}) we evidently
always have%
\begin{equation}
F^{\mu \nu }=\dsum_{\ell }F^{\ell \mu \nu }.  \label{empa2}
\end{equation}%
Being given individual EM fields $F^{\ell \mu \nu }$ we introduce the
corresponding EM energy-momentum tensor $\Theta ^{\ell \mu \nu }$ via the
general formula (\ref{maxw11}), namely%
\begin{equation}
\Theta ^{\ell \mu \nu }=\frac{1}{4\pi }\left( g^{\mu \gamma }F_{\gamma \xi
}^{\ell }F^{\ell \xi \nu }+\frac{1}{4}g^{\mu \nu }F_{\gamma \xi }^{\ell
}F^{\ell \gamma \xi }\right) .  \label{empa3}
\end{equation}%
Then combining (\ref{empa1}), (\ref{empa3}) with (\ref{maxw17}) we obtain%
\begin{equation}
\partial _{\mu }\Theta ^{\ell \mu \nu }=-\frac{1}{\mathrm{c}}J_{\mu }^{\ell
}F^{\ell \nu \mu }.  \label{empa4}
\end{equation}%
Notice also that from (\ref{empa2}) and (\ref{tmap4}) we have%
\begin{equation}
\partial _{\mu }T^{\ell \mu \nu }=\frac{1}{\mathrm{c}}J_{\mu }^{\ell
}F^{\ell \nu \mu }+\frac{1}{\mathrm{c}}J_{\mu }^{\ell }\dsum_{\ell ^{\prime
}\neq \ell }F^{\ell ^{\prime }\nu \mu }.  \label{empa5}
\end{equation}%
If we introduce now the energy-momentum $\mathrm{T}^{\ell \mu \nu }$ of the
dressed charge, i.e. the charge with its EM field, by the formula 
\begin{equation}
\mathrm{T}^{\ell \mu \nu }=T^{\ell \mu \nu }+\Theta ^{\ell \mu \nu },
\label{empa6}
\end{equation}%
then the sum of two equalities (\ref{empa4})-(\ref{empa5}) readily yields
the following \emph{equations of motion for dressed charges}%
\begin{equation}
\partial _{\mu }\mathrm{T}^{\ell \mu \nu }=\frac{1}{\mathrm{c}}J_{\mu
}^{\ell }\dsum_{\ell ^{\prime }\neq \ell }F^{\ell ^{\prime }\nu \mu },\ \ell
=1,\ldots ,N.  \label{empa7}
\end{equation}%
describing the motion of energies and momenta of the dressed charges in the
space-time continuum. Importantly, \emph{the Lorentz force in the right-hand
of (\ref{empa7}) excludes manifestly the self-interaction} in contrast to
the Lorentz force acting upon a bare charge as in (\ref{empa6}) which
explicitly includes the self-interaction term $\frac{1}{\mathrm{c}}J_{\mu
}^{\ell }F^{\ell \nu \mu }$. Thus, we can conclude that when the charge and
its EM field are treated as a single entity, namely dressed charge, there is
no self-interaction as signified by the exact equations (\ref{empa7}).

We would like to point out certain subtleties related to the individual EM
fields $F^{\ell \nu \mu }$. Namely, the individual currents $J_{\mu }^{\ell
} $ constructed via solutions to the Euler-Lagrange field equations (\ref%
{mplag5})-(\ref{mplag7}) are implicitly related to each other. Those
implicit relations manifest themselves in particular in the fact that the
sum of the energy-momentum tensors of the individual EM fields does not
exactly equal the energy-momentum tensor of the total EM field, i.e.%
\begin{equation}
\Theta ^{\mu \nu }\neq \dsum_{\ell }\Theta ^{\ell \mu \nu },  \label{empa8}
\end{equation}%
since $\Theta ^{\mu \nu }$ defined by (\ref{tmap3}) is a quadratic, and
hence nonlinear, function of the EM $F^{\mu \nu }$. Nevertheless, the
approximate equality holds 
\begin{equation}
\Theta ^{\mu \nu }\approx \dsum_{\ell }\Theta ^{\ell \mu \nu },
\label{empa9}
\end{equation}%
when the supports of charges wave functions $\psi ^{\ell }$ are well
separated in the space for the time period of interest and, hence, 
\begin{equation}
F_{\gamma \xi }^{\ell }F^{\ell ^{\prime }\xi \nu }\approx 0\text{ for }\ell
^{\prime }\neq \ell  \label{empa10}
\end{equation}%
implying (\ref{empa9}). In other words, for well separated charges the above
mentioned subtle correlations between individual currents become
insignificant for their EM energy-momenta.

We can ask now how far one can go in extracting from the equations of motion
(\ref{empa7}) equations of point charges similarly to the non-relativistic
theory in previous subsection. We can argue that in the relativistic theory
in the regime of remote interaction we study, every charge can behave
similarly to a wave-corpuscle but their motion can not be reduced to a
system of differential equations obtained from a conventional
finite-dimensional Lagrangian since it is prohibited by presented above
arguments by Goldstein, von Laue and Pauli. The next option in simplicity
can be the motion governed by a system of integro-differential equations
which can account for retardation effects similar to the Sommerfeld
integro-differential-difference equation of motion for the
nonrelativistically rigid electron and its relativistic generalizations, 
\cite[Sections 8-10]{Pearle1}, but we are not going to pursue this problem
any further in this paper.

\section{Equations in dimensionless form\label{cheqdimf}}

We introduce here changes of variables allowing to recast the original field
equations into a dimensionless form. \ These equations in dimensionless form
allow to clarify three aspects of the theory for a single charge: (i) out of
all the constants involved there is only one parameter of significance
denoted by $\alpha $, and it coincides with the Sommerfeld fine structure
constant $\alpha _{S}\simeq 1/137$ if $\chi =\hbar $ and $q,m$ are the
electron charge and mass respectively; (ii) the non-relativistic Lagrangian (%
\ref{nore1b}) can be obtained from the relativistic one via the
frequency-shifted Lagrangian (\ref{fshif3}) by setting there $\alpha =0$;
(iii) the fulfillment of charge and energy normalization conditions in
relativistic case follows from smallness of $\alpha $.

Recall that the single charge nonrelativistic Lagrangian $\mathring{L}_{0}$
defined by (\ref{nore1b}) is constructed in Section \ref{snfree} based on
the relativistic one via the the frequency\ shifted Lagrangian $L_{\omega
_{0}}$ defined by (\ref{fshif1})-(\ref{fshif3}) (see also Sections \ref%
{sfreq}, \ref{stharm}). The corresponding to $L_{\omega _{0}}$
Euler-Lagrange field equations are 
\begin{gather}
\frac{\tilde{\partial}_{t}^{2}\psi }{\mathrm{c}^{2}}-\frac{\mathrm{i}m}{\chi 
}\left( 2\partial _{t}\psi +2\frac{\mathrm{i}q\bar{\varphi}}{\chi }\hat{\psi}%
\right) -\tilde{\nabla}^{2}\psi +G^{\prime }\left( \left\vert \psi
\right\vert ^{2}\right) \psi =0,  \label{mlagom} \\
\frac{1}{4\pi }\nabla \cdot \left( \frac{\partial _{t}\mathbf{A}}{\mathrm{c}}%
+\nabla \varphi \right) =\left( \frac{\chi q}{m\mathrm{c}^{2}}\func{Im}\frac{%
\partial _{t}\hat{\psi}}{\hat{\psi}}+\frac{q^{2}\bar{\varphi}}{m\mathrm{c}%
^{2}}\right) \left\vert \psi \right\vert ^{2}-q\left\vert \psi \right\vert
^{2},  \notag \\
-\left( \nabla \times \left( \nabla \times \mathbf{A}\right) +\frac{\partial
_{t}}{\mathrm{c}}\left( \frac{\partial _{t}\mathbf{A}}{\mathrm{c}}+\nabla
\varphi \right) \right) =\frac{4\pi }{\mathrm{c}}\left( -\frac{\chi q}{m}%
\func{Im}\frac{\nabla \psi }{\psi }-\frac{q^{2}\mathbf{\bar{A}}}{m\mathrm{c}}%
\right) \left\vert \psi \right\vert ^{2},  \notag
\end{gather}%
where%
\begin{equation}
\tilde{\partial}_{t}=\partial _{t}+\frac{\mathrm{i}q\bar{\varphi}}{\chi },\ 
\bar{\varphi}=\varphi +\varphi _{\mathrm{ex}},\ \mathbf{\bar{A}=A}+\mathbf{A}%
_{\mathrm{ex}}.  \label{mlagom1}
\end{equation}%
Let us introduce the following constants and new variables:%
\begin{equation}
a_{\chi }=\frac{\chi ^{2}}{mq^{2}},\mathbf{\ }\alpha =\frac{q^{2}}{\chi 
\mathrm{c}},\mathbf{\ }\omega _{0}=\frac{m\mathrm{c}^{2}}{\chi }=\frac{%
\mathrm{c}}{\alpha a_{\chi }},  \label{not}
\end{equation}%
\begin{gather}
\alpha ^{2}\omega _{0}t=\tau ,\mathbf{\ x}=a_{\chi }\mathbf{y},
\label{chvar} \\
\psi \left( \mathbf{x}\right) =\frac{1}{a_{\chi }^{3/2}}\Psi \left( \frac{%
\mathbf{x}}{a_{\chi }}\right) ,\mathbf{\ }\varphi \left( \mathbf{x}\right) =%
\frac{q}{a_{\chi }}\Phi \left( \frac{\mathbf{x}}{a_{\chi }}\right) \mathbf{%
,\ A\left( \mathbf{x}\right) =}\frac{q}{a_{\chi }}\mathsf{A}\left( \frac{%
\mathbf{x}}{a_{\chi }}\right) .  \notag
\end{gather}%
In those new variables the field equations (\ref{mlagom}) turn into the
following dimensionless form: 
\begin{gather}
\alpha ^{2}\left( \partial _{\tau }+\mathrm{i}\bar{\Phi}\right) ^{2}\Psi -2%
\mathrm{i}\left( \partial _{\tau }+\mathrm{i}\bar{\Phi}\right) \Psi -\left(
\nabla _{y}-\mathrm{i}\alpha \mathsf{\bar{A}}\right) ^{2}\Psi +\mathsf{G}%
^{\prime }\left( \left\vert \Psi \right\vert ^{2}\right) \Psi =0,
\label{mlagat} \\
\frac{1}{4\pi }\nabla _{y}\cdot \left( \alpha \partial _{\tau }\mathsf{A}%
+\nabla _{y}\Phi \right) =\left( \alpha ^{2}\func{Im}\frac{\partial _{\tau
}\Psi }{\Psi }+\alpha ^{2}\Phi \right) \left\vert \Psi \right\vert
^{2}-\left\vert \Psi \right\vert ^{2},\   \notag \\
-\left( \nabla _{y}\times \left( \nabla _{y}\times \mathsf{A}\right) +\alpha
\partial _{\tau }\left( \alpha \partial _{\tau }\mathsf{A}+\nabla _{y}\Phi
\right) \right) =-4\pi \alpha \left( \func{Im}\frac{\nabla _{y}\Psi }{\Psi }%
+\alpha \mathsf{\bar{A}}\right) \left\vert \Psi \right\vert ^{2},  \notag
\end{gather}

We would like to show that the dimensionless form of the non-relativistic
equations field equations (\ref{NLS1}), (\ref{fis}) can be obtained from the
field equations (\ref{mlagat}) in the limit $\alpha \rightarrow 0$.\ To have
a nonvanishing external magnetic field in the limit $\alpha \rightarrow 0$
we set 
\begin{equation}
\mathsf{A}_{\mathrm{ex}}=\alpha ^{-1}\mathsf{A}_{\mathrm{ex}}^{0}
\label{aaex1}
\end{equation}%
Plugging in the expression (\ref{aaex1}) into the equations (\ref{mlagat})
we obtain in the limit $\alpha \rightarrow 0$\ the following dimensionless
version of the field equations (\ref{NLS1}), (\ref{fis}):\ 
\begin{gather}
\mathrm{i}\partial _{\tau }\Psi =-\frac{1}{2}\left( \nabla _{y}-\mathrm{i}%
\mathsf{A}_{\mathrm{ex}}^{0}\right) ^{2}\Psi +\left( \Phi +\Phi _{\mathrm{ex}%
}\right) \Psi +\frac{1}{2}\mathsf{G}^{\prime }\left( \left\vert \Psi
\right\vert ^{2}\right) \Psi ,  \label{mlagatn} \\
-\nabla _{y}^{2}\cdot \Phi =4\pi \left\vert \Psi \right\vert ^{2},\ \left(
\nabla _{y}\times \left( \nabla _{y}\times \mathsf{A}\right) \right) =0. 
\notag
\end{gather}

To get an insight in the nonrelativistic case as an approximation to the
relativistic one we would like to make a few comments on the relative
magnitude of terms that have to be omitted in equation (\ref{mlagat}) in
order\ to obtain equation (\ref{mlagatn}). The nonrelativistic case is
defined as one when the charge velocity $v$ is much smaller than the speed
of light $\mathrm{c}$, and a careful look at those omitted terms in (\ref%
{mlagat}) that have factors $\alpha $ and $\alpha ^{2}$ shows that they can
be small not only because of $\alpha $, but also because of the smallness of
typical values of velocities compared to the speed of light. Indeed, every
term that has factor $\alpha $ involves time derivatives with only one
exception: the term $\alpha ^{2}\left( \mathrm{i}\bar{\Phi}\right) ^{2}\Psi $%
. An estimation of the magnitude of the omitted terms for solutions of the
form of wave-corpuscles (\ref{psil0}) indicated that every such a term is of
order $\alpha \left\vert \mathbf{v}\right\vert \mathbf{/}\mathrm{c}$ where $%
\mathbf{v}$ is the wave-corpuscle velocity. The only omitted term in (\ref%
{mlagat}) which does not involve time derivatives is $\alpha ^{2}\Phi
^{2}\Psi $ and, in fact, it can be preserved in the nonrelativistic system
which would be similar to (\ref{mlagatn}) with properties analogous to (\ref%
{NLS1}). Analysis of that system is more involved and the treatment is
similar to the one for the rest solution of the relativistic equation
involving that term considered in next Subsection \ref{relone}.

\subsection{Single relativistic charge at rest\label{relone}}

In this section we a consider a single relativistic charge. Using the new
constants and variables (\ref{not}), (\ref{chvar}) we get the following
dimensionless version of the resting charge equations (\ref{psfi1a}), (\ref%
{psfi1b}) and the charge normalization condition (\ref{psfi7})%
\begin{gather}
-\frac{1}{2}\nabla _{y}^{2}\Psi +\left( \Phi -\frac{\alpha ^{2}\Phi ^{2}}{2}%
\right) \Psi +\frac{\hat{G}^{\prime }\left( \Psi ^{2}\right) }{2}\Psi =0,
\label{atomic1} \\
-\nabla ^{2}\Phi =4\pi \left( 1-\alpha ^{2}\Phi \right) \left\vert \Psi
\right\vert ^{2}=0,  \label{atomic}
\end{gather}%
$\ \ $ \ \ 
\begin{equation}
\int_{\mathbb{R}^{3}}\left[ \left( 1-\alpha ^{2}\Phi \right) \left\vert \Psi
\right\vert ^{2}\right] \,\mathrm{d}\mathbf{x}=1.  \label{normat}
\end{equation}%
Setting in the above equations $\alpha =0$ we obtain the dimensionless form
nonrelativistic equilibrium equations (\ref{nop4}), (\ref{nop5}) and the
charge normalization condition (\ref{nop11}), namely 
\begin{gather}
-\frac{1}{2}\nabla _{y}^{2}\Psi +\Phi \Psi +\hat{G}_{0}^{\prime }\left(
\left\vert \Psi \right\vert ^{2}\right) \check{\Psi}=0,\ -\nabla ^{2}\Phi
=4\pi \left\vert \Psi \right\vert ^{2},  \label{atomic0} \\
\int_{\mathbb{R}^{3}}\left\vert \Psi \right\vert ^{2}\,\mathrm{d}\mathbf{x}%
=1.  \label{normc}
\end{gather}

Now using perturbations analysis we argue that for small $\alpha $\ the
solution $\Psi _{\alpha },\Phi _{\alpha }$ to the equations (\ref{atomic})
is close to the solution $\Psi _{0},\Phi _{0}$ of the equations (\ref%
{atomic0}). Indeed, for zero approximation $\Phi _{0}\left( \mathbf{x}%
\right) $ 
\begin{equation*}
\Phi _{0}\left( \mathbf{x}\right) =\int_{\mathbb{R}^{3}}\frac{\left\vert
\Psi _{0}\left( \mathbf{y}\right) \right\vert ^{2}}{\left\vert \mathbf{x}-%
\mathbf{y}\right\vert }\,\mathrm{d}\mathbf{y},
\end{equation*}%
and the first order approximation $\Phi _{1}$ is a solution to%
\begin{equation*}
-\nabla ^{2}\Phi _{1}=4\pi \left( 1-\alpha ^{2}\Phi _{0}\right) \left\vert
\Psi _{0}\right\vert ^{2}.
\end{equation*}%
Using the Maximum principle we get%
\begin{equation}
0<\Phi _{1}\left( \mathbf{x}\right) <\Phi \left( \mathbf{x}\right) <\Phi
_{0}\left( \mathbf{x}\right) =\check{\Phi}\left( \mathbf{x}\right) \text{
for all }\mathbf{x}.  \label{vvest0}
\end{equation}%
Obviously, 
\begin{equation}
\Phi _{1}\left( \mathbf{x}\right) =\Phi _{0}\left( \mathbf{x}\right) +\alpha
^{2}\Phi _{01}\left( \mathbf{x}\right) \text{, where }\nabla ^{2}\Phi
_{01}=4\pi \Phi _{0}\left\vert \Psi _{0}\right\vert ^{2},  \label{fi01}
\end{equation}%
and hence 
\begin{equation*}
\Phi _{01}\left( \mathbf{x}\right) =-\int_{\mathbb{R}^{3}}\frac{\Phi
_{0}\left( \mathbf{y}\right) \left\vert \Psi _{0}\left( \mathbf{y}\right)
\right\vert ^{2}}{\left\vert \mathbf{x}-\mathbf{y}\right\vert }\,\mathrm{d}%
\mathbf{y}.
\end{equation*}%
Consequently, inequalities (\ref{vvest0}) imply explicit estimate%
\begin{equation}
\alpha ^{2}\Phi _{01}\left( \mathbf{x}\right) <\Phi \left( \mathbf{x}\right)
-\Phi _{0}\left( \mathbf{x}\right) <0\text{ for all }\mathbf{x}\in \mathbb{R}%
^{d}.  \label{vvest}
\end{equation}

\subsection{Charge and energy simultaneous normalization\label{simnorm}}

We consider here some technical details related to the size representation (%
\ref{psfi8}), (\ref{psfi9}) and the problem of simultaneous normalization of
the charge and the energy by equations (\ref{psixi2}) and (\ref{psi2d}). Let
function $\Psi _{\theta }\left( \mathbf{y}\right) $ be the dimensionless
version of the function $\psi _{a}$ in (\ref{psfi8}), (\ref{psfi9}), namely%
\begin{gather}
\psi _{a}\left( \mathbf{x}\right) =\frac{\theta ^{3/2}}{a_{\chi }^{3/2}}\psi
_{1}\left( \frac{\theta ^{3/2}\mathbf{x}}{a_{\chi }^{3/2}}\right) =\Psi
_{\theta }\left( \mathbf{y}\right) =C_{\Psi }\theta ^{3/2}\Psi _{1}\left(
\theta \mathbf{y}\right) ,  \label{psikapc} \\
\text{where}\ \theta =\frac{a_{\chi }}{a},\text{ and }\int_{\mathbb{R}%
^{3}}\left\vert \Psi _{1}\right\vert ^{2}\,\mathrm{d}\mathbf{y}=1.  \notag
\end{gather}%
Then the charge and the energy normalization conditions (\ref{psixi2}) and (%
\ref{psi2d}) take the form 
\begin{equation}
\mathcal{N}=\int_{\mathbb{R}^{3}}\left( 1-\alpha ^{2}\Phi _{\theta }\right)
\left\vert \Psi _{\theta }\right\vert ^{2}\,\mathrm{d}\mathbf{y}=1.
\label{catom1}
\end{equation}%
Using the relation (\ref{kab9}) and (\ref{kab13}) we obtain the following
representation for the energy $\mathcal{E}\left( \psi _{a},\varphi
_{a}\right) $, which is a version of the Pohozhaev-Derrik formula (see
Section \ref{stharm}),%
\begin{gather}
\mathcal{E}_{0}\left( \psi _{a},\varphi _{a}\right) =m\mathrm{c}^{2}+%
\mathcal{E}_{0}^{\prime }\left( \psi _{a},\varphi _{a}\right) ,
\label{pohat} \\
\mathcal{E}_{0}^{\prime }\left( \psi _{a},\varphi _{a}\right) =\frac{2}{3}%
\int_{\mathbb{R}^{3}}\left( \frac{\chi ^{2}\left\vert \nabla \psi
_{a}\right\vert ^{2}}{2m}-\frac{\left\vert \nabla \varphi _{a}\right\vert
^{2}}{8\pi }\right) \,\mathrm{d}\mathbf{x}.  \notag
\end{gather}%
Then the energy normalization condition (\ref{psi2d}), namely $\mathcal{E}%
_{0}^{\prime }=0$, turns into the dimensionless variables into the condition%
\begin{equation}
\mathcal{E}_{0}^{\prime }\left( \Psi _{\theta },\Phi _{\theta }\right) =%
\frac{q^{2}}{3a_{\chi }}\int_{\mathbb{R}^{3}}\left( \left\vert \nabla \Psi
_{\theta }\right\vert ^{2}-\frac{\left\vert \nabla \Phi _{\theta
}\right\vert ^{2}}{4\pi }\right) \,\mathrm{d}\mathbf{y}=0.  \label{e20}
\end{equation}

First, let us consider a simple case $\alpha =0$ using for it the notation $%
\check{\Phi}_{\theta }=\left. \Phi _{\theta }\right\vert _{\alpha =0}$, $%
\check{\Psi}_{\theta }=\left. \Psi _{\theta }\right\vert _{\alpha =0}$. In
this case the charge normalization condition (\ref{catom1}), in view of the
normalization condition in (\ref{psikapc}), is satisfied for $C_{\Psi }=1$
and 
\begin{equation}
\check{\Psi}_{\theta }\left( \mathbf{y}\right) =\theta ^{3/2}\check{\Psi}%
_{1}\left( \theta \mathbf{y}\right) ,\ \check{\Phi}_{\theta }\left( \mathbf{y%
}\right) =\theta \check{\Phi}_{1}\left( \theta \mathbf{y}\right) \text{ for }%
\alpha =0.  \label{fit}
\end{equation}%
It follows then from (\ref{e20}) that 
\begin{equation}
\mathcal{E}_{0}^{\prime }\left( \check{\Psi}_{\theta },\check{\Phi}_{\theta
}\right) =\frac{q^{2}}{3a_{\chi }}\int_{\mathbb{R}^{3}}\left( \theta
^{2}\left\vert \nabla \check{\Psi}_{1}\right\vert ^{2}-\frac{\left\vert
\nabla \check{\Phi}_{1}\right\vert ^{2}}{4\pi }\theta \right) \,\mathrm{d}%
\mathbf{y}.  \label{fit1}
\end{equation}%
implying that $\mathcal{E}_{0}^{\prime }\left( \check{\Psi}_{\theta },\check{%
\Phi}_{\theta }\right) $ is a quadratic function of the parameter $\theta $.
Since $\Psi $ is fixed\ we set in (\ref{psikapc}) $C_{\Psi }=1$ \ and $%
\theta =\theta _{0}$, where $\theta _{0}$ is defined by\ 
\begin{equation}
\theta _{0}=\frac{b_{\Phi }}{b_{\Psi }},\ b_{\Phi }=\frac{1}{4\pi }\int_{%
\mathbb{R}^{3}}\left\vert \nabla \check{\Phi}_{1}\right\vert ^{2}\,\mathrm{d}%
\mathbf{y},\ b_{\Psi }=\int_{\mathbb{R}^{3}}\left\vert \nabla \check{\Psi}%
_{1}\right\vert ^{2}\,\mathrm{d}\mathbf{y},  \label{kap0}
\end{equation}%
and obtain the desired energy normalization condition (\ref{e20}), namely 
\begin{equation}
\mathcal{E}_{0}^{\prime }\left( \check{\Psi}_{\theta },\check{\Phi}_{\theta
}\right) =0,\ \int_{\mathbb{R}^{3}}\left\vert \check{\Psi}_{\theta
}\right\vert ^{2}\,\mathrm{d}\mathbf{x}=1.  \label{eqa101}
\end{equation}%
Note that $\theta _{0}$ as in (\ref{kap0}) coincides with $\theta _{\psi }$
as in (\ref{apsi}).

Let us consider now the case $\alpha >0$. We would like to show that for a
given form factor $\Psi $ and small $\alpha >0$\ there exist constants $%
C_{\Psi }$ and $\theta $ such that the following two equation hold 
\begin{equation}
\mathcal{E}_{0}^{\prime }\left( \check{\Psi}_{\theta },\check{\Phi}_{\theta
},\alpha \right) =0,\ \mathcal{N}\left( C_{\Psi },\theta ,\alpha \right)
-1=0,  \label{eqal}
\end{equation}%
\ where $\mathcal{E}_{0}^{\prime }$ and $\mathcal{N}_{1}$ are defined
respectively by relations (\ref{e20}) and (\ref{catom1}). In other words he
need to find two parameters $C_{\Psi }$ \ and $\theta \ $\ from a system of
two nonlinear equations (\ref{eqal}). \ We have already established that for 
$\alpha =0$ :\ the solution $\ $is $C_{\Psi }=1$, $\theta =\theta _{0}$ as
in (\ref{kap0}). A geometrical argument shows\emph{\ }that \emph{for
sufficiently small} $\alpha $ equations (\ref{eqal}) have a solution $%
\left\{ C,\theta \right\} $ \ which is close for small $\alpha $ to the
solution $C_{\Psi }=1$, $\theta =\theta _{0}$. The complete argument is
based on the inequality (\ref{vvest}) but its details are rather technical
and we omit them here.

\section{Hydrogen atom model\label{shydro}}

The purpose of this section is to introduce a hydrogen atom model within the
framework of our theory for two interacting charges: an electron and a
proton. We\ have no intend though to study this model in detail here. Our
modest effort on the subject in this paper is, first, to indicate a
similarity between our and Schr\"{o}dinger's hydrogen atom models and to
contrast it to any kind of Kepler's model. Another point we can make based
on our hydrogen atom model is that our theory does provide a basis for a
regime of close interaction between two charges which differs significantly
from the regime of remote interaction which is the primary focus of this
paper. Clearly the results on interaction of many charges as in Section \ref%
{pmechwc} do not apply to the hydrogen atom model. Indeed in this case the
two charges are in close proximatity and the potentials can vary
significantly over their locations, and hence, other methods have to be
developed for the hydrogen atom model.

To model interaction of two charges \emph{at a short distance} we must
consider the original system (\ref{NLSj0}) for two charges with $%
-q_{1}=q_{2}=q>0$, that is%
\begin{gather}
\mathrm{i}\chi \partial _{t}\psi _{1}=-\frac{\chi ^{2}\nabla ^{2}\psi _{1}}{%
2m_{1}}-q\left( \varphi _{1}+\varphi _{2}\right) \psi _{1}+\frac{\chi
^{2}G_{1}^{\prime }\left( \left\vert \psi _{1}\right\vert ^{2}\right) \psi
_{1}}{2m_{1}},  \label{hyd1} \\
-\nabla ^{2}\varphi _{1}=-4\pi q\left\vert \psi _{1}\right\vert ^{2},  \notag
\\
\mathrm{i}\chi \partial _{t}\psi _{2}=-\frac{\chi ^{2}\nabla ^{2}\psi _{2}}{%
2m_{2}}+q\left( \varphi _{1}+\varphi _{2}\right) \psi _{2}+\frac{\chi
^{2}G_{2}^{\prime }\left( \left\vert \psi _{2}\right\vert ^{2}\right) \psi
_{2}}{2m_{2}},  \notag \\
\nabla ^{2}\varphi _{2}=-4\pi q\left\vert \psi _{2}\right\vert ^{2}.  \notag
\end{gather}%
Note that the model describes proton-electron interaction if $q=e$ is the
electron charge, $\chi =\hbar $ is the Planck constant, $m_{1}$ \ and $m_{2}$
are the electron and the proton masses respectively. Let us look now at
time-harmonic solutions to the system (\ref{hyd1}) in the form 
\begin{equation}
\psi _{1}\left( t,\mathbf{x}\right) =\mathrm{e}^{-\mathrm{i}\omega
_{1}t}u_{1}\left( \mathbf{x}\right) ,\ \psi _{2}=\mathrm{e}^{-\mathrm{i}%
\omega _{2}t}u_{2}\left( \mathbf{x}\right) ,\ \Phi _{1}=-\frac{\varphi _{1}}{%
q},\ \Phi _{2}=\frac{\varphi _{2}}{q}.  \label{hyd4}
\end{equation}%
The system (\ref{hyd1}) for such solutions turns into the following
nonlinear eigenvalue problem: 
\begin{gather}
-\frac{a_{1}}{2}\nabla ^{2}u_{1}+\left( \Phi _{1}-\Phi _{2}\right) u_{1}+%
\frac{a_{1}}{2}G_{1}^{\prime }\left( \left\vert u_{1}\right\vert ^{2}\right)
u_{1}=\frac{\chi }{q^{2}}\omega _{1}u_{1},  \label{H1} \\
\text{where }a_{1}=\frac{\chi ^{2}}{q^{2}m_{1}},  \notag \\
-\frac{a_{2}}{2}\nabla ^{2}u_{2}+\left( \Phi _{2}-\Phi _{1}\right) u_{2}+%
\frac{a_{2}}{2}G_{2}^{\prime }\left( \left\vert u_{2}\right\vert ^{2}\right)
u_{2}=\frac{\chi }{q^{2}}\omega _{2}u_{2},  \notag \\
\text{where }a_{2}=\frac{\chi ^{2}}{q^{2}m_{2}}.  \notag
\end{gather}%
Here $a_{1}$\ coincides with Bohr radius. We seek solutions of (\ref{H1}))
satisfying the charge normalization conditions 
\begin{equation}
\int_{\mathbb{R}^{3}}\left\vert u_{1}\right\vert ^{2}\,\mathrm{d}\mathbf{x}%
=1,\ \int_{\mathbb{R}^{3}}\left\vert u_{2}\right\vert ^{2}\,\mathrm{d}%
\mathbf{x}=1.  \label{C1C2}
\end{equation}%
The potentials $\Phi _{i}$ are presented using (\ref{hyd1}) as follows: 
\begin{equation}
\Phi _{i}=4\pi \left( -\nabla ^{2}\right) ^{-1}\left\vert u_{i}\right\vert
^{2}=4\pi \dint_{\mathbb{R}^{3}}\frac{\left\vert u_{i}\right\vert ^{2}\left( 
\mathbf{y}\right) }{\left\vert \mathbf{y}-\mathbf{x}\right\vert }\,\mathrm{d}%
\mathbf{y},\ i=1,2.  \label{fihyd}
\end{equation}%
Let us introduce now the following energy functional%
\begin{gather}
\mathcal{E}\left( u_{1},u_{2}\right) =  \label{Epe} \\
q^{2}\int_{\mathbb{R}^{3}}\left\{ \frac{a_{1}\left[ \left\vert \nabla
u_{1}\right\vert ^{2}+G_{1}\left( \left\vert u_{1}\right\vert ^{2}\right) %
\right] }{2}+\frac{a_{2}\left[ \left\vert \nabla u_{2}\right\vert
^{2}+a_{2}G_{2}\left( \left\vert u_{2}\right\vert ^{2}\right) \right] }{2}%
\right.  \notag \\
\left. -\left( \Phi _{1}-\Phi _{2}\right) \left( \left\vert u_{2}\right\vert
^{2}-\left\vert u_{1}\right\vert ^{2}\right) -\frac{\left\vert \nabla \left(
\Phi _{1}-\Phi _{2}\right) \right\vert ^{2}}{8\pi }\right\} \,\mathrm{d}%
\mathbf{x},  \notag
\end{gather}%
where $\Phi _{1},\Phi _{2}$ are determined in terms of $u_{1},u_{2}$ by (\ref%
{fihyd}). Notice that the equations (\ref{H1}) describe stationary points of
the functional $\mathcal{E}$\ and can be obtained by its variation under
constraints (\ref{C1C2}) with the frequencies $\omega _{1},\omega _{2}$
being the Lagrange multipliers. Observe that the energy functional and the
constraints are invariant with respect to multiplication by $-1$, and that
allows us to apply the Lusternik-Schnirelman theory which guarantees the
existence of an infinite set of critical points under appropriate
conditions. The critical points are the eigenfunctions of the corresponding
Schr\"{o}dinger operators (see, for example, \cite{Heid}, \cite{Lions}).\
Our preliminary analysis shows that properties of solutions are similar to
those in the spectral theory of the hydrogen atom described by the linear
Schr\"{o}dinger equation. The smallness of the ratio $m_{1}/m_{2}\cong
1/1836 $ of electron to proton masses plays an important role in the
analysis. For the critical points with low energies the potential $\Phi _{2}$
of the proton \ is close to the Coulomb's potential $1/\left\vert \mathbf{x}%
\right\vert $ at spatial scales of the order $a_{1}$. For a properly chosen
nonlinearity the quadratic part of the energy functional and the
corresponding linear Schr\"{o}dinger equation can be used to find discrete
low levels of energy. Rough estimates of the energy levels of approximate
solutions of the nonlinear problem based on the eigenfunctions of the linear
Schr\"{o}dinger operator show good agreement with well-known energy levels
for the hydrogen atom if \ choose \ the size parameter $a_{1}$ 10-20 times
greater then the Bohr radius. It seems reasonable to assume that the wave
function of a free electron contracts when it becomes bound to a positively
charged proton and apperently its size is reduced by an order of magnitude.

\section{Comparison with the Schr\"{o}dinger wave theory\label{scompair}}

We already made some points on common features and differences between our
theory and the Schr\"{o}dinger wave mechanics in Section \ref{scompsum}, and
here we discuss in more detail a few significant differences between the two
theories. Recall that the Schr\"{o}dinger wave mechanics is constructed
based on the classical point particle Hamiltonian%
\begin{equation}
\mathcal{E}=H\left( \mathbf{p},\mathbf{x}\right) =\frac{\mathbf{p}^{2}}{2m}%
+V\left( \mathbf{x}\right)  \label{clin1}
\end{equation}%
by substituting there, \cite[Section 2, 11]{Pauli WM}, \cite[Sections 3, 4]%
{Pauli PWM},%
\begin{equation}
\mathbf{p}\rightarrow -\mathrm{i}\hbar \nabla ,\ \mathcal{E}\rightarrow 
\mathrm{i}\hbar \frac{\partial }{\partial t},  \label{clin2}
\end{equation}%
that yields the celebrated Schr\"{o}dinger equation%
\begin{equation}
\mathrm{i}\hbar \frac{\partial \psi }{\partial t}=-\frac{\hbar ^{2}\nabla
^{2}\psi }{2m}+V\left( \mathbf{x}\right) \psi .  \label{clin3}
\end{equation}%
The substitution (\ref{clin2}) is essentially the quantization procedure
allowing to correspond the classical point Hamiltonian (\ref{clin1}) to the
quantum mechanical wave equation (\ref{clin3}). If we introduce the polar
representation 
\begin{equation}
\psi =\mathrm{e}^{\mathrm{i}\frac{S}{\hbar }}R,\ R\geq 0  \label{clin3a}
\end{equation}%
then the Schr\"{o}dinger equation (\ref{clin3}) can be recast as the
following system of two coupled equations for the real valued phase function 
$S$ and the amplitude $R$, \cite[Section 3.2.1]{Holland},%
\begin{equation}
\frac{\partial S}{\partial t}+\frac{\left( \nabla S\right) ^{2}}{2m}+V-\frac{%
\hbar ^{2}\nabla ^{2}R}{2mR}=0,  \label{clin3b}
\end{equation}%
\begin{equation}
\frac{\partial R^{2}}{\partial t}+\nabla \cdot \frac{R^{2}\nabla S}{m}=0.
\label{clin3c}
\end{equation}%
If we expand the phase $S$ into power series with respect to $\hbar $ i.e. 
\begin{equation}
S=S_{0}+\hbar S_{1}+\hbar ^{2}S_{2}+\ldots .,  \label{clin4}
\end{equation}%
we obtain from the equation (\ref{clin3b}) the so called WKB approximation, 
\cite[Section 12]{Pauli PWM}. In particular, the function $S_{0}$ satisfies
the Hamilton-Jacobi equation 
\begin{equation}
\frac{\partial S_{0}}{\partial t}+\frac{\left( \nabla S_{0}\right) ^{2}}{2m}%
+V=0,  \label{clin5}
\end{equation}%
which shows, in particular, that the Schr\"{o}dinger wave equation (\ref%
{clin3}) does "remember" how it was constructed by "returning back" the
original Hamiltonian $H$ via the Hamilton-Jacobi equation (\ref{clin5}) for $%
S_{0}$.

Our approach works the other way around. We introduce the Lagrangian (\ref%
{fpar1})-(\ref{fpar1b}) and the corresponding field equations as a
fundamental basis and deduce from them the classical Newtonian mechanics as
a certain approximation (see Sections \ref{ssaccel}, \ref{nonmany}). To
appreciate the difference let us look at a system of $N$ charges. The here
introduced wave-corpuscle mechanics would have $N$ wave functions and the EM
fields defined over the same 3 dimensional space, whereas the same system of 
$N$ charges in the Schr\"{o}dinger wave mechanics has a single wave function
defined over a $3N$-dimensional "configuration space".

It is quite instructive to compare the polar representation (\ref{exac1})
for wave-corpuscle $\psi $ with the same for the Schr\"{o}dinger wave
function $\psi $ for the potential $V\left( \mathbf{x}\right) =-q\mathbf{E}_{%
\mathrm{ex}}\cdot \mathbf{x}$ corresponding to a homogeneous external
electric field (the eigenfunctions of the corresponding Schr\"{o}dinger
equations can be expressed in terms of the Airy functions, \cite[Section 26]%
{Pauli WM}). The amplitude $\mathring{\psi}\left( \left\vert \mathbf{x}-%
\mathbf{r}\left( t\right) \right\vert \right) $ of the wave-corpuscle in the
expression (\ref{exac1}) for $\psi $ is a soliton-like field moving exactly
as the point charge described by its position $\mathbf{r}\left( t\right) $
in contrast to the amplitude $R$ of the Schr\"{o}dinger wave function which
describes the location of the charge rather implicitly via the coupled
equations (\ref{clin3c}). The difference between the phases is equally
significant. Indeed, the exponential factor $e^{\mathrm{i}S/\chi }$ for the
accelerating wave-corpuscle is just a plane wave with the phase $S$ that
depends only on the point charge position $r$ and momentum $p$ whereas the
same for the Schr\"{o}dinger equations captures the features of the point
charge only via WKB approximation and the Hamilton-Jacobi equation (\ref%
{clin5}) which holds for the phase $S$ only in the limit $\hbar \rightarrow
0 $.

Let us now take a look at uncertainty relations which constitute a very
important consequence of the Schr\"{o}dinger wave mechanics as a linear wave
theory. Detailed studies of this subject is not in the scope of this paper
but we can already see significant alterations of the uncertainty relations
brought by the nonlinearity. W. Pauli writes in section "The uncertainty
principle", \cite[Section 3]{Pauli WM}): "The kinematics of waves does not
allow the simultaneous specification of the location and the exact
wavelength of a wave. Indeed, one can only speak of the location of a wave
in the case of a spatially localized wave packet. The number of different
wavelengths contained in the Fourier spectrum increases as the wave packet
becomes more localized. A relation of the form $\Delta k_{i}\Delta x_{i}\geq 
\limfunc{constant}$ seems reasonable, and we now want to derive this
relation quantitatively". Then in the same section he derives the well known
Heisenberg uncertainty principle for a wavepacket in the form%
\begin{equation}
\Delta k\Delta x\geq \frac{1}{2},\ \Delta p\Delta x\geq \frac{\hbar }{2},
\label{unce1}
\end{equation}%
where $\Delta x$, $\Delta k$ and $\Delta p=\hbar \Delta k$ are respectively
the spacial range, the wave number range and the momentum range for the
wavepacket $\psi \left( \mathbf{x}\right) $ defined by%
\begin{gather}
\Delta x^{2}=\int_{\mathbb{R}^{3}}\left( \mathbf{x}-\mathbf{\bar{x}}\right)
^{2}\left\vert \psi \left( \mathbf{x}\right) \right\vert ^{2}\,\mathrm{d}%
\mathbf{x},\ \mathbf{\bar{x}}=\int_{\mathbb{R}^{3}}\mathbf{x}\left\vert \psi
\left( \mathbf{x}\right) \right\vert ^{2}\,\mathrm{d}\mathbf{x},
\label{unce1a} \\
\Delta k^{2}=\int_{\mathbb{R}^{3}}\left( \mathbf{k}-\mathbf{\bar{k}}\right)
^{2}\left\vert \hat{\psi}\left( \mathbf{k}\right) \right\vert ^{2}\,\mathrm{d%
}\mathbf{k},\text{ where}  \notag \\
\mathbf{\bar{k}}=\int_{\mathbb{R}^{3}}\psi ^{\ast }\left( \mathbf{x}\right)
\left( -\mathrm{i}\frac{\partial }{\partial x}\right) \psi \left( \mathbf{x}%
\right) \,\mathrm{d}\mathbf{x}=\int_{\mathbb{R}^{3}}\mathbf{k}\left\vert 
\hat{\psi}\left( \mathbf{k}\right) \right\vert ^{2}\,\mathrm{d}\mathbf{k}, 
\notag \\
\hat{\psi}\left( \mathbf{k}\right) =\frac{1}{\left( 2\pi \right) ^{3/2}}%
\int_{\mathbb{R}^{3}}e^{ikx}\psi \left( \mathbf{x}\right) \,\mathrm{d}%
\mathbf{x}.  \notag
\end{gather}%
Importantly, in the orthodox quantum wave mechanics, which is a linear
theory, if $\psi \left( \mathbf{x}\right) $ is the wave function $\left\vert
\psi \left( \mathbf{x}\right) \right\vert ^{2}$ is interpreted as the
probability density for a point particle to be at a location $\mathbf{x}$.
Hence, in this theory the uncertainty is already in the very interpretation
of the wave function, and $\Delta x$ as in (\ref{unce1a}) is an uncertainty
in the location of the point particle with a similar interpretation holding
for $\Delta p$. \emph{Thus the exact form of the Heisenberg uncertainty
relation (\ref{unce1}) is a direct consequence of the fundamental definition
(\ref{clin2}) of the momentum and the definition of the uncertainty as in (%
\ref{unce1a}) based on the probabilistic interpretation of the wave function}%
.

An important feature of the uncertainty relations in the linear theory is
that any freely propagating wavepacket spreads out as a quadratic function
of the time $t$ and such a spread out takes a particularly simple form for a
Gaussian wavepacket, \cite[Section 3]{Pauli WM},%
\begin{equation}
\left( \Delta x\right) ^{2}=\frac{1}{4\left( \Delta k\right) ^{2}}+\frac{%
\hbar ^{2}\left( \Delta k\right) ^{2}}{m^{2}}t^{2}.  \label{unce2}
\end{equation}%
We would like to point out that the very concept of wavepacket is based on
the medium linearity, and the same is true for of the uncertainty relations (%
\ref{unce1}), (\ref{unce2}) as general wave phenomena.

Let us consider now the here proposed wave-corpuscle mechanics from the
uncertainty relations point of view. In wave-corpuscle mechanics we denote
uncertainties in the position $x$ and the momentum $p$ by respectively by $%
\tilde{\Delta}x$ and $\tilde{\Delta}p$ using a different symbol $\tilde{%
\Delta}$ to emphasize the difference in their definition since the
wave-function does not carry a probabilistic interpretation. \emph{We also
limit our considerations of the uncertainties }$\tilde{\Delta}x$\emph{\ and }%
$\tilde{\Delta}p$\emph{\ to special cases when a wave-corpuscle is an exact
solution to either relativistic or nonrelativistic field equations, since in
these cases we can argue more convincingly what constitutes uncertainty
without giving its definition in a general case}.

Notice that a common feature of the dressed charge (wave-corpuscle) and a
wavepacket is the wave origin of their kinematics as manifested by the
equality of the velocity to the group velocity of the underlying linear
medium. But, in contrast to a wavepacket in a linear medium, the free
relativistic dressed charge described by the relations (\ref{mvch1})-(\ref%
{mvch4}) does not disperse as it moves and preserves its shape up the
Lorentz contraction. The de Broglie vector $\mathbf{k}$ and the frequency $%
\omega $ can be determined from the factor $\mathrm{e}^{-\mathrm{i}\left(
\omega t-\mathbf{k}\cdot \mathbf{x}\right) }$ in (\ref{mvch1}), and
consequently the same applies to the total momentum $\mathsf{P}=\hbar 
\mathbf{k}$. In the case of a nonrelativistic wave-corpuscle as defined by
relations (\ref{nexac1}), (\ref{nexac1a}) similarly its total momentum is $%
\mathsf{P}=m\mathbf{v}=\hbar \mathbf{k}$. More than that as we show in
Section \ref{ssaccel} the wave-corpuscle as defined by relations (\ref{psil0}%
) is an exact solution to the field equations and it propagates in space
without dispersion even when accelerates.

To further clarify differences with the uncertainties in Schr\"{o}dinger
wave mechanics and the wave-corpuscle mechanics introduced here notice the
following. The dressed charge described by the relations (\ref{mvch1})-(\ref%
{mvch4}) is a material wave for which we can reasonably assign size $D$ in
relativistic and nonrelativistic cases by formulas%
\begin{equation}
D^{2}\left( t\right) =\int_{\mathbb{R}^{3}}\left( \mathbf{x}-\mathbf{\bar{x}}%
\right) ^{2}\frac{\rho \left( t,\mathbf{x}\right) }{q}\,\mathrm{d}\mathbf{x},
\label{unce2a}
\end{equation}%
where%
\begin{equation*}
\mathbf{\bar{x}}=\int_{\mathbb{R}^{3}}\mathbf{x}\frac{\rho \left( t,\mathbf{x%
}\right) }{q}\,\mathrm{d}\mathbf{x},
\end{equation*}%
and, respectively,%
\begin{equation*}
\frac{\rho \left( t,\mathbf{x}\right) }{q}=q\left( 1-\frac{q\mathring{\varphi%
}\left( \mathbf{x}-\mathbf{v}t\right) }{m\mathrm{c}^{2}}\right) \mathring{%
\psi}^{2}\left( \mathbf{x}-\mathbf{v}t\right) ,
\end{equation*}%
and%
\begin{equation*}
\frac{\rho \left( t,\mathbf{x}\right) }{q}=\mathring{\psi}^{2}\left( \mathbf{%
x}-\mathbf{v}t\right) .
\end{equation*}%
\emph{In the case of a relativistic or nonrelativistic wave-corpuscle we
define "safely" the uncertainty }$\tilde{\Delta}x=D\left( t\right) $\emph{\
and notice that it follows from the definition (\ref{unce1}) and charge
normalization conditions (\ref{fpar11}), (\ref{nrac4}) that in fact }$\tilde{%
\Delta}x=D\left( t\right) =D$\emph{\ does not depend on }$t$. Hence, the
uncertainty $\tilde{\Delta}x=D$ unlike the uncertainty $\Delta x$ from (\ref%
{unce2}) does not grow without bound for large times. As to the charges
momentum $\mathsf{P}=m\mathbf{v}=\hbar \mathbf{k}$ we can argue that $\tilde{%
\Delta}p=0$ for the exact wave-corpuscle solutions. Indeed, for a
wave-corpuscle as defined by relations (\ref{mvch1})-(\ref{mvch4}) or (\ref%
{nexac1}), (\ref{nexac1a}) respectively in the relativistic and
nonrelativistic cases the motion of dressed charge is obtained by
application of \ space translations (or the Lorentz transformations) to a
fixed rest charge, therefore \ "every point" of the dressed charge moves
with exactly the same velocity $\mathbf{v}$ similarly to a rigid body, which
allows naturally define its velocity and momentum without uncertainty.
Summarizing, we can conclude that in the wave-corpuscle mechanics the
Heisenberg uncertainty principle cannot be a universal principle.

\section{Relation to the Quantum Mechanics and hidden variables theories}

Since the wave-corpuscle mechanics (WCM) naturally covers all spatial scales
one can wonder how it relates to the quantum mechanics (QM), including the
probabilistic interpretation of the wave function, the hydrogen atom
frequencies, the double-slit experiment, scattering of a charge by a
potential created by another charge and more. In sections \ref{shydro}, \ref%
{scompair} we already provided comparisons of some features of the WCM and
the QM. But a systematic comparison of all fundamental features of the two
mechanics requires more extensive studies of the WCM, and at this point we
can only formulate hypotheses on some significant elements of the WCM and
their relations to fundamentals of the QM.

A key element of the WCM that has not been studied yet is a regime of time
limited close interaction. More precisely, it is a regime when initially
free moving charge undergoes for a naturally limited time a close
interaction with another charge or an external EM field after which the
charge continues to move freely again but with altered location and
velocity. A typical example of such an interaction is when one moving charge
is scattered by another one, or when a charge passes through a bounded
domain in the space with a strong external EM field. Let us try to imagine
what can happen according to the WCM to a charge during a time limited close
interaction. Recall that in the WCM when charges are far apart every charge
is represented by a particle-like well localized wave function as in (\ref%
{exac1}) as a result of a fine balance of forces including the nonlinear
self-interaction. Importantly, the cohesive action of the nonlinear
self-interaction is very subtle, and by no means it is a brute attractive
force since there is no action at distance in the WCM. Now, when one charge
comes close to another or if it enters a domain with an external EM field
varying at a sufficiently small scale, a fine balance of forces holding the
charge together is disrupted. We can already see consequences of such a
disruption in the WCM hydrogen atom model in section \ref{shydro} where the
electron size reduces by a factor of order 10 compare to that of the free
electron under attractive action of a single proton. A disruption of the
subtle cohesive action of the nonlinear self-interaction can also cause the
charge wave function to spread out substantially and become wave-like. We
can imagine further that during the time of interaction the evolution of the
extended wave function is determined by an interplay of two factors: (i) the
linear Schr\"{o}dinger component of the field equation; (ii) its nonlinear
component due the nonlinear self-interaction. Shortly after the interaction
ends the wave-function of the charge restores the particle-like form but its
position and velocity after the contraction will depend sensitively on
details of the interaction. So, effectively, a limited time interaction
switches one particle-like state of the charge to another.

Based the above hypothetical features of a time limited close interaction
one can explain how the entirely deterministic WCM can conceivably lead to
some of probabilistic aspects of the QM. Suppose for the sake of argument
that the time scale of the interaction process is smaller than an observer
can resolve, and, consequently, he sees the interaction result as a
transition from one particle-like state to another. The interaction process
can alter the total momentum of the charge quite considerably. This momentum
alteration combined with effects of the nonlinear self-interaction can cause
an extreme sensitivity to the initial data and that in turn can make the
transition look like it is random and, hence, a subject to a statistical
theory. An interesting feature of the nonlinear self-interaction in the WCM
that might be relevant to the extreme sensitity is that it is not analytic
and singular for small wave-amplitudes (see examples of the WCM nonlinearity
in \ref{Examnon}). Consequently, small wave amplitudes can play far more
important role in the WCM than in the case of conventional nonlinearities
which are analytic for small amplitudes. Going further we observe that the
WCM field equations are similar to the Schr\"{o}dinger equation. Hence, it
is conceivable that the statistics of the transition will be determined with
certain degree of accuracy by a wave function satisfying an effective linear
Schr\"{o}dinger equation. Some general ideas on the "determinism beneath
quantum mechanics" at the Planck scale were put forward recently by 't Hooft
(see \cite{t Hooft} and references therein).

Let us use the described above hypothetical scenario of interaction to
explain the double-slit experiment. Suppose that a single electron is fired
by a device and moves freely as a particle-like wave toward a double-slit
apparatus. As the electron approaches and interacts with the double-slit
apparatus its wave function spreads out quite substantially, its amplitude
reduces and consequently the electron turns into a "real wave". This
extended wave penetrates through the both slits and over a limited time
leaves the apparatus. Being outside of the apparatus in a free space the
electron wave function contracts back to its particle-like shape and
proceeds toward a sensitive screen until it hits it at a well defined impact
location. Assume for the sake of argument that the initial condition of the
fired electron can not be controlled with a sufficient accuracy (which may
be higher than in the existing experiments). Then, the impact location can
appear to be random with a statistics consistent with well known
interference pattern as described in the modern double-slit experiments, 
\cite[Section 1.1]{Greenstein}.

Another qualitatively important regime is a regime of close interaction for
an extended or even infinite period of time. This regime can occur, in
particular, in complex systems involving many charges such as atoms,
molecules or solids. As we have already indicated in Section \ref{shydro}
the WCM hydrogen atom has features which are very similar the Schr\"{o}%
dinger one. In particular, the primary binding force in that model between
the electron and the nucleus is the EM attraction. As to solids let us
briefly recall basics of their treatment in the QM. As any theory of many
particles, the fundamental QM theory of a solid is of enormous complexity,
but the standard simplified QM treatment of charges in crystalline solids is
based on a free-electron model with the following basics assumptions, \cite[%
Chapter 1]{Ashcroft}: (i) positively charged ionized atoms, consisting of
nuclei and tightly bound to them "core electrons", form an immobile periodic
lattice structure; (ii) "valence electrons" called also conductance
electrons are "allowed to wander far away from their parent atoms"; (iii)
the conductance are non-interacting and independent and the interaction
between a conductance electron and the periodic lattice is modeled via a
periodic potential. Such a simplified QM theory is effectively reduced to
the one-electron theory for the Schr\"{o}dinger operator with the periodic
potential. Consequently, the eigenfunctions of such an operator are of the
Bloch form and are extended over the entire crystal. The fundamental WCM
theory of a solid is of an enormous complexity as well, but similarly to the
QM theory we can introduce a simplified WCM model for a solid based on the
same assumptions as in the QM theory. Hence, as in the QM model there is an
immobile periodic lattice of ionized atoms described by a periodic potential
corresponding to an external electric field. The one-electron WCM model is
similar to the QM one, but it differs from it by the presence of nonlinear
self-interaction. In this non-relativistic WCM\ model a mobile
conductance/valence electron is subjected to (i) attractive forces of the
periodic lattice; (ii) electric force of its own electric field; (iii)
nonlinear-self interaction forces. Since in a solid the distance between
atoms is pretty small, it is of order the atom Bohr radius, the cohesive
action of the nonlinear self-interaction can be disrupted and the wave
function can spread out significantly and even it might resemble a Bloch
eigenmode, in which case the electron would occupy the entire crystal sample.

The above considerations bring us naturally to issues of the charge size and
the WCM theory locality. As the above considerations suggest in the WCM the
electron size can vary significantly depending on whether it is free or if
it is bound in a atom, or if it is a conductance electron in a crystalline
solid. In particular, the size of the electron can increase dramatically
when it undergoes a strong close interaction with an external EM field or a
system of other charges. As to the locality of the WCM it is perfectly local
in one sense but can be non-local in another sense. Namely, the WCM theory
is perfectly local in the sense that there is no action of distance. But the
charge evidently is not perfectly local since in the WCM\ it is not a point
but at the best a localized wave which under indicated above conditions can
spread out significantly. Consequently, it is conceivable within the WCM
that an elementary charge being a spatially distributed quantity can be
simultaneously at two distant spatial locations, and in this sense the WCM
might deviate significantly from being a local theory. If one excepts that,
then the Bell's inequalities \cite[Section 2]{Bell}, \cite[Section 12.3]%
{Gottfried} do not apply to the WCM. A question if a modification of the
Bell's approach is possible for the WCM would require more studies
including, in particular, an introduction of the spin concept in the theory.

One can also wonder what is a relation between the WCM and hidden variables
theories, see \cite[12.2]{Gottfried}, \cite[Section 1.5, 3.7.2]{Holland},
and a review article \cite{Genovese} with references therein. Particularly,
it is interesting to look how does the WCM compare with the Bohmian
Mechanics (BM), \cite{Bohm Hiley}, \cite[Section 3.1, 3.2]{Holland}, a well
known example of hidden variables theories. Even a brief look on the WCM and
the BM shows their significant differences: (i) in the BM an elementary
charge is a point whereas in the WCM it is a distributed quantity, a wave;
(ii) the WCM theory is local in the sense that there is no action at
distance, and it is no so for the BM; (iii) the WCM is a genuine Lagrangian
Mechanics and, consequently, the Third Newton Law is always satisfied, and
it is not so for the BM, \cite[Section 3.3.2]{Holland}. In addition to that,
as we have already indicated above, the WCM might account for the QM
statistics via the dynamic instability approximately, and the verification
of that, including the accuracy of approximation, is a subject of future
studies. But it is absolutely clear already that the statistical predictions
of the WCM can not be precisely the same as those of the QM, since the WCM
field equations might only approximately and under certain conditions
produce the QM evolution equation. The later factor evidently differs the
WCM from the BM in which the Schr\"{o}dinger equation is an exact equation
for the wave function as a part of the BM variables.

\section{Classical field theory\label{sclasfield}}

In this section we discuss important elements of the classical field theory
including the variational principles and the Lagrangian formalism, gauge
invariance and conservation laws. There are many classical references on the
classical field theory: \cite{Barut}, \cite{Goldstein}, \cite{LandauLif F}, 
\cite[Section 3.4]{Morse Feshbach I}, \cite{Pauli RFTh} and more. So we
picked and chose different parts of theory from different sources to
emphasize concepts and constructions important for our own arguments. Often
we gave multiple references to provide different and complementary points of
view on the same subjects. We also extended some aspects of the theory as
needed. In particularly, we did that for an important for us subject of many
charges interacting with the EM.

\subsection{Relativistic Kinematics\label{srelkin}}

Here where provide very basic facts and notations related to the
relativistic kinematics following to \cite[Section 1]{Barut}, \cite[Sections
1.1-1.4, 2]{LandauLif F}, \cite[Section 11.3]{Jackson}, \cite[Section 7]%
{Goldstein}.

The time-space four vector in its contravariant $x^{\mu }$ and covariant $%
x_{\mu }$ forms are represented as follows%
\begin{equation}
x=x^{\mu }=\left( x^{0},x^{1},x^{2},x^{3}\right) =\left( \mathrm{c}t,\mathbf{%
x}\right) ,\ \mu =0,1,2,3;  \label{rkin1}
\end{equation}%
\begin{equation}
x_{\mu }=g_{\mu \nu }x^{\nu }=\left( x^{0},-x^{1},-x^{2},-x^{3}\right) ,
\label{rkin2}
\end{equation}%
with the common convention on the summation of the same indices, and where $%
g_{\mu \nu }$ or $g^{\mu \nu }$, called metric tensor, $4\times 4$ matrix,
are defined by%
\begin{equation}
\left\{ g_{\mu \nu }\right\} =\left\{ g^{\mu \nu }\right\} =\left[ 
\begin{array}{cccc}
1 & 0 & 0 & 0 \\ 
0 & -1 & 0 & 0 \\ 
0 & 0 & -1 & 0 \\ 
0 & 0 & 0 & -1%
\end{array}%
\right] .  \label{rkin2a}
\end{equation}%
Notice also that, \cite[Section 11.6]{Jackson}, \cite[Section 6.1]{Schwabl},%
\begin{equation}
g_{\ \nu }^{\mu }=g^{\mu \sigma }g_{\sigma \nu }=\delta _{\ \nu }^{\mu }%
\text{ where }\delta _{\ \nu }^{\mu }\text{ is the Dirac symbol,}\ \left\{
\delta _{\ \nu }^{\mu }\right\} =\left[ 
\begin{array}{cccc}
1 & 0 & 0 & 0 \\ 
0 & 1 & 0 & 0 \\ 
0 & 0 & 1 & 0 \\ 
0 & 0 & 0 & 1%
\end{array}%
\right] ,  \label{rkin2b}
\end{equation}%
and%
\begin{equation}
\partial _{\mu }=\frac{\partial }{\partial x^{\mu }}=\left( \frac{1}{\mathrm{%
c}}\partial _{t},\nabla \right) ,\ \partial ^{\mu }=\frac{\partial }{%
\partial x_{\mu }}=\left( \frac{1}{\mathrm{c}}\partial _{t},-\nabla \right) .
\label{rkin2c}
\end{equation}

The elementary Lorentz transformation to a moving with a velocity $\mathbf{v}
$ frame is%
\begin{gather}
x^{0\prime }=\gamma \left( x^{0}-\mathbf{\beta }\cdot \mathbf{x}\right) ,\ 
\mathbf{x}^{\prime }=\mathbf{x}+\frac{\gamma -1}{\beta ^{2}}\left( \mathbf{%
\beta }\cdot \mathbf{x}\right) \mathbf{\beta }-\gamma \mathbf{\beta }x^{0},
\label{rkin3} \\
\mathbf{\beta }=\frac{\mathbf{v}}{\mathrm{c}},\ \beta =\left\vert \mathbf{%
\beta }\right\vert ,\ \gamma =\frac{1}{\sqrt{1-\left( \frac{v}{\mathrm{c}}%
\right) ^{2}}}.  \notag
\end{gather}%
If for a space vector $\mathbf{x}$\ we introduce $\mathbf{x}_{\Vert }$ and $%
\mathbf{x}_{\bot }$ so that they are respectively its components parallel
and perpendicular to the velocity $\mathbf{v}$, i.e. $\mathbf{x}=\mathbf{x}%
_{\Vert }+\mathbf{x}_{\bot }$, then (\ref{rkin3}) can be recast as%
\begin{equation}
x^{0\prime }=\gamma \left( x^{0}-\mathbf{\beta }\cdot \mathbf{x}\right) ,\ 
\mathbf{x}_{\Vert }^{\prime }=\gamma \left( \mathbf{x}_{\Vert }-\mathbf{%
\beta }x^{0}\right) ,\ \mathbf{x}_{\bot }^{\prime }=\mathbf{x}_{\bot },
\label{rkin3a}
\end{equation}%
which in the case when $\mathbf{v}$ is parallel to the axis $x^{1}$ turns
into 
\begin{equation}
x^{0\prime }=\gamma \left( x^{0}-\beta x^{1}\right) ,\ x^{1\prime }=\gamma
\left( x^{1}-\beta x^{0}\right) ,\ x^{2\prime }=x^{2},\ x^{3\prime }=x^{3}.
\label{rkin3b}
\end{equation}%
The Lorentz invariance then of a $4$-vector $x$ under the above
transformation reduces to%
\begin{equation}
\left( x^{0\prime }\right) ^{2}-\left\vert \mathbf{x}^{\prime }\right\vert
^{2}=\left( x^{0}\right) ^{2}-\left\vert \mathbf{x}\right\vert ^{2}.
\label{rkin3c}
\end{equation}%
The general infinitesimal form of the inhomogeneous Lorentz transformation
is, \cite[Section 6.1]{Moller},%
\begin{equation}
x^{\prime \mu }=x^{\mu }+\xi ^{\mu \nu }x_{\nu }+a^{\mu },\ \xi ^{\mu \nu
}=-\xi ^{\nu \mu },  \label{ttbe2}
\end{equation}%
where $\xi ^{\mu \nu }$ and $a^{\mu }$ are its ten parameters.

The Lagrangian $L_{\mathrm{p}}$ of the moving point mass is%
\begin{equation}
L_{\mathrm{p}}=-m\mathrm{c}^{2}\sqrt{1-\left( \frac{\mathbf{v}}{\mathrm{c}}%
\right) ^{2}},  \label{rkin4}
\end{equation}%
implying the following nonrelativistic approximation%
\begin{equation}
L_{\mathrm{p}}\cong -m\mathrm{c}^{2}+\frac{m\mathbf{v}^{2}}{2},\text{ for }%
\left\vert \frac{\mathbf{v}}{\mathrm{c}}\right\vert \ll 1.  \label{rkin5}
\end{equation}%
The momentum (the ordinary kinetic momentum) and the energy of the point
mass for the relativistic Lagrangian $L_{\mathrm{p}}$ defined by (\ref{rkin4}%
) are, \cite[Section 2.9]{LandauLif F},%
\begin{equation}
\mathbf{p}=\frac{m\mathbf{v}}{\sqrt{1-\left( \frac{\mathbf{v}}{\mathrm{c}}%
\right) ^{2}}},\ \mathcal{E}=p^{0}\mathrm{c}=\mathbf{p}\cdot \mathbf{v}-L=%
\frac{m\mathrm{c}^{2}}{\sqrt{1-\left( \frac{\mathbf{v}}{\mathrm{c}}\right)
^{2}}}=\mathrm{c}\sqrt{\mathbf{p}^{2}+m^{2}\mathrm{c}^{2}}.  \label{rkin6}
\end{equation}

The relativistic Lagrangian $L_{\mathrm{p}}$ of a point charge $q$ with a
mass $m$ in an external EM field as described by electric potential $\varphi 
$ and vector potential $\mathbf{A}$ is defined by, \cite[Section 12.1]%
{Jackson}%
\begin{equation}
L_{\mathrm{p}}=-m\mathrm{c}^{2}\sqrt{1-\left( \frac{\mathbf{v}}{\mathrm{c}}%
\right) ^{2}}-q\varphi +\frac{q}{\mathrm{c}}\mathbf{v}\cdot \mathbf{A}.
\label{rkin7}
\end{equation}%
For this Lagrangian the ordinary kinetic momentum $\mathbf{p}$, the
canonical (conjugate) momentum $\mathbf{\mathring{p}}$, and the Hamiltonian $%
H_{\mathrm{p}}$ are defined by the following relations%
\begin{equation}
\mathbf{p}=\frac{m\mathbf{v}}{\sqrt{1-\left( \frac{\mathbf{v}}{\mathrm{c}}%
\right) ^{2}}},\ \mathbf{\mathring{p}}=\mathbf{p}+\frac{q}{\mathrm{c}}%
\mathbf{A},  \label{rkin7a}
\end{equation}%
\begin{equation}
H_{\mathrm{p}}=\mathbf{\mathring{p}}\cdot \mathbf{v}-L=\sqrt{\left( \mathrm{c%
}\mathbf{p}-q\mathbf{A}\right) ^{2}+m^{2}\mathrm{c}^{4}}+q\varphi ,
\label{rkin7b}
\end{equation}%
and the Euler-Lagrange equations are%
\begin{equation}
\frac{\mathrm{d}\mathbf{p}}{\mathrm{d}t}=\mathbf{F}=q\mathbf{E}+\frac{q}{%
\mathrm{c}}\mathbf{v}\times \mathbf{B},\ \text{and }\frac{\mathrm{d}\mathcal{%
E}}{\mathrm{d}t}=q\mathbf{v}\cdot \mathbf{E,}\text{ where }\mathbf{E}%
=-\nabla \varphi -\frac{\partial \mathbf{A}}{\partial t},\ \mathbf{B}=\nabla
\times \mathbf{A},  \label{rkin7c}
\end{equation}%
where $\mathbf{F}$ is the \emph{Lorentz force}, and $\mathbf{E}$ and $%
\mathbf{B}$ are respectively the electric field and the magnetic induction.

The nonrelativistic version of the above Lagrangian in view of (\ref{rkin5})
is, \cite[Section 1.5]{Goldstein}%
\begin{equation}
L_{\mathrm{p}}=\frac{m\mathbf{\dot{r}}^{2}}{2}-q\varphi \left( \mathbf{r}%
\right) +\frac{q}{\mathrm{c}}\mathbf{A}\left( \mathbf{r}\right) \cdot 
\mathbf{\dot{r}},\ \mathbf{\dot{r}}=\frac{\mathrm{d}\mathbf{r}}{\mathrm{d}t}.
\label{rkin8}
\end{equation}%
The corresponding ordinary kinetic momentum $\mathbf{p}$, the canonical
(conjugate) momentum $\mathbf{\mathring{p}}$, and the Hamiltonian $H_{%
\mathrm{p}}$ are defined by the following relations%
\begin{equation}
\mathbf{p}=m\mathbf{\dot{r}},\ \mathbf{\mathring{p}}=\mathbf{p}+\frac{q}{%
\mathrm{c}}\mathbf{A},\ H_{\mathrm{p}}=\mathbf{\mathring{p}}\cdot \mathbf{%
\dot{r}}-L=\frac{m\mathbf{\dot{r}}^{2}}{2}+q\varphi .  \label{rkin8a}
\end{equation}%
The canonical Euler-Lagrange equations for the nonrelativistic Lagrangian (%
\ref{rkin8}) take the form%
\begin{equation}
\frac{\mathrm{d}\mathbf{\mathring{p}}}{\mathrm{d}t}=-q\nabla \varphi +\frac{q%
}{\mathrm{c}}\left[ D\mathbf{A}\right] \mathbf{\dot{r}},\ \left[ D\mathbf{A}%
\right] _{j}=\frac{\partial }{\partial x^{j}}\mathbf{A},\ j=1,2,3,
\label{rkinc}
\end{equation}%
or, if use $\mathbf{\mathring{p}}=\mathbf{p}+\frac{q}{\mathrm{c}}\mathbf{A}$
and the identity 
\begin{equation}
\frac{\mathrm{d}\mathbf{A}}{\mathrm{d}t}=\frac{\partial \mathbf{A}}{\partial
t}+\left[ D\mathbf{A}\right] \mathbf{\dot{r}},  \label{rkin8b}
\end{equation}%
we can recast (\ref{rkinc}) as 
\begin{equation}
\frac{\mathrm{d}\mathbf{p}}{\mathrm{d}t}=m\frac{\mathrm{d}^{2}\mathbf{r}}{%
\mathrm{d}^{2}t}=\mathbf{F}=q\mathbf{E}+\frac{q}{\mathrm{c}}\frac{\mathrm{d}%
\mathbf{r}}{\mathrm{d}t}\times \mathbf{B},  \label{rkin9}
\end{equation}%
where the right-hand side of the equation (\ref{rkin9}) is the Lorentz
force. Importantly for what we study in this paper the canonical
Euler-Lagrange equation (\ref{rkinc}) involves the canonical momentum $%
\mathbf{\mathring{p}}$ and the canonical force $-q\nabla \varphi +\frac{q}{%
\mathrm{c}}\left[ D\mathbf{A}\right] \mathbf{\dot{r}}$ which manifestly
depend on the EM potentials $\varphi $ and $\mathbf{A}$ rather than EM
fields $\mathbf{E}$ and $\mathbf{B}$. Consequently, the equation (\ref{rkinc}%
) involves quantities which are not directly measurable in contrast to the
equivalent to it equation (\ref{rkin9}) which is gauge invariant and
involves measurable quantities, namely the kinematic momentum $\mathbf{p}$
and the Lorentz force $q\mathbf{E}+\frac{q}{\mathrm{c}}\mathbf{\dot{r}}%
\times \mathbf{B}$.

\subsection{Lagrangians, field equations and conserved quantities\label%
{senergymom}}

In this section we collect basic well known facts on the Lagrangian
formalism for classical fields following to \cite[Section III.3]{Barut}, 
\cite[Section 3.4]{Morse Feshbach I}, \cite{Pauli RFTh} and other classical
sources. Let us assume that physical systems of interest are described by
fields real-valued $q^{\ell }\left( x\right) $, $\ell =1,\ldots ,N$, with
the Lagrangian density%
\begin{equation}
\mathcal{L}(\left\{ q^{\ell }\left( x\right) \right\} ,\left\{ q_{,\mu
}^{\ell }\left( x\right) \right\} ,x),\ q_{,\mu }^{\ell }\left( x\right)
=\partial _{\mu }q^{\ell }\left( x\right) ,\ \mu =0,1,2,3,  \label{flagr1}
\end{equation}%
According to the Lagrangian formalism the field equation are derived from
the variational principle%
\begin{equation}
\delta \dint_{R}\mathcal{L}\mathrm{\,d}x=0  \label{flagr1a}
\end{equation}%
where $R$ is four-dimensional \emph{space-like region} with a
three-dimensional boundary $\partial R$. Importantly, the variation $\delta $
is such that $\delta q^{\ell }$ vanish on the boundary $\partial R$. Then
the corresponding Euler-Lagrange field equations take the form%
\begin{equation}
\Lambda _{\ell }=\frac{\partial \mathcal{L}}{\partial q^{\ell }}-\partial
_{\mu }\frac{\partial \mathcal{L}}{\partial q_{,\mu }^{\ell }}=0.
\label{flagr2}
\end{equation}%
The equation (\ref{flagr2}) can be recast in a Hamiltonian form as%
\begin{equation}
\frac{\partial \pi _{\ell }}{\partial x_{0}}=F_{\ell },\ \pi _{\ell }=\frac{%
\partial \mathcal{L}}{\partial q_{,0}^{\ell }},\ F_{\ell }=\frac{\partial 
\mathcal{L}}{\partial q^{\ell }}-\dsum\limits_{\mu =1}^{3}\partial _{\mu }%
\frac{\partial \mathcal{L}}{\partial q_{,\mu }^{\ell }},\ x_{0}=\mathrm{c}t,
\label{flagr2a}
\end{equation}%
where $\pi _{\ell }$ is interpreted as the \emph{canonical momentum density}
of the field $q^{\ell }$ and $F_{\ell }$ is the canonical force density
acting upon the field. The term $\frac{\partial L}{\partial q^{\ell }}$ of
the canonical force density $F_{\ell }$ has to do with external forces
acting on the field $q^{\ell }$. \emph{In the view of the last remark, if
the Lagrangian }$L$\emph{\ involves a nonlinear terms }$G_{\ell }\left(
q^{\ell }\right) $\emph{\ its derivative }$\frac{\partial G}{\partial
q^{\ell }}$\emph{\ can be interpreted as a self-force.}

In our considerations the fields $q^{\ell }\left( x\right) $ describe
elementary charges (and in fact they are complex-valued) and the potential
4-vector field $A_{\mu }$ describes the classical EM field. The extension of
the Lagrangian formalism to complex-valued in considered in a following
section.

The \emph{canonical energy-momentum tensor} (also called \emph{stress-energy
tensor} or \emph{stress-tensor}) is defined by the following formula, \cite[%
(3.63)]{Barut}, \cite[Section 32, (32.5)]{LandauLif F}, \cite[Section 13.3,
(13.30)]{Goldstein}, \cite[Section 12.4, (12.4.1)]{Schwabl} 
\begin{equation}
\mathring{T}^{\mu \nu }=\dsum_{\ell }\frac{\partial \mathcal{L}}{\partial
q_{,\mu }^{\ell }}q^{\ell ,\nu }-g^{\mu \nu }\mathcal{L},  \label{flagr5}
\end{equation}%
with the energy conservation laws in the form, \cite[(3.94)]{Barut}, \cite[%
(13.28)]{Goldstein} 
\begin{equation}
\partial _{\mu }\mathring{T}^{\mu \nu }=-\frac{\partial \mathcal{L}}{%
\partial x_{\nu }}.  \label{flagr5e}
\end{equation}%
Notice that in \cite[(3.4.2)]{Morse Feshbach I}, \cite[Section 13.3]%
{Goldstein} the canonical energy-momentum tensor is defined as
matrix-transposed to $\mathring{T}^{\mu \nu }$ defined by (\ref{flagr5}),
namely $\mathring{T}^{\mu \nu }\rightarrow \mathring{T}^{\nu \mu }$. The
conservation laws for the energy-momentum are examples of conservation laws
obtained from the Noether's theorem considered below in Section \ref%
{snoether}.

In the case of the Lagrangian $\mathcal{L}$ does not depend explicitly on $%
x_{\nu }$, in other words invariant under translations $x_{\nu }\rightarrow
x_{\nu }+a_{\nu }$, $\frac{\partial L}{\partial x_{\nu }}=0$ and the
conservation laws (\ref{flagr5e}) turn into the following \emph{continuity
equations}, 
\begin{equation}
\partial _{\mu }\mathring{T}^{\mu \nu }=0.  \label{flagr5a}
\end{equation}%
A typical situation when the general conservation laws (\ref{flagr5e}) apply
rather than (\ref{flagr5a}) is the presence of external forces which can
"drive" our field. For instance, an external EM field driving a charge or an
external "imposed" current which becomes a source for the EM field.

We would like to remind the reader that the canonical tensor of
energy-momentum defined by (\ref{flagr5}) is not the only one that satisfies
the conservation laws (\ref{flagr5e}) or (\ref{flagr5a}). For instance, any
tensor of the form, \cite[(3,73)]{Barut}, \cite[(32.7)]{LandauLif F}, \cite[%
(14)]{Pauli RFTh} 
\begin{equation}
T^{\mu \nu }=\mathring{T}^{\mu \nu }-\partial _{\gamma }f^{\mu \gamma \nu },%
\text{ where }f^{\mu \gamma \nu }=-f^{\gamma \mu \nu },  \label{emten1}
\end{equation}%
would satisfy (\ref{flagr5a}) as long as $\mathring{T}^{\mu \nu }$ does. In
view of the (\ref{emten1}) the energy-momentum tensor $T^{\mu \nu }$ satisfy
the same conservation laws (\ref{flagr5e}) or (\ref{flagr5a}) as $\mathring{T%
}^{\mu \nu }$, namely%
\begin{equation}
\partial _{\mu }T^{\mu \nu }=-\frac{\partial \mathcal{L}}{\partial x_{\nu }},
\label{emten1a}
\end{equation}%
or, if the Lagrangian $L$ does not depend explicitly on $x_{\nu }$ and,
hence, invariant under time and space translations, the above conservation
laws turn into 
\begin{equation}
\partial _{\mu }T^{\mu \nu }=0.  \label{emten1b}
\end{equation}

In fact, this flexibility in choosing the energy-momentum can be used to
define $f^{\gamma \mu \nu }$ and construct a \emph{symmetric energy-momentum
tensor} $T^{\mu \nu }$, i.e. $T^{\mu \nu }=T^{\nu \mu }$, which is a
necessary and sufficient condition for the field angular momentum density to
be represented by the usual formula in terms of the field momentum density, 
\cite[Section 32]{LandauLif F}, \cite[Section III.4]{Barut}. The symmetry of
the energy-momentum tensor for matter fields is also fundamentally important
since it is a source for the gravitational field, \cite[Section 3.8]{Nair}, 
\cite[Section 5.7]{Misner}. As to the uniqueness of the energy-momentum we
can quote \cite[Section 21.3]{Misner}: "... the theory of gravity in the
variational formulation gives a unique prescription for fixing the
stress-energy tensor, a prescription that, besides being symmetric, also
automatically satisfies the laws of conservation of momentum and energy".
This unique form of the symmetric energy-momentum can be derived based on a
variational principle involving charge of boundary, with varied boundary, 
\cite[Section III.3(B)]{Barut}, and under the following assumptions: (i) the
Lagrangian does not depend explicitly on $x$; (ii) the fields $q^{\ell
}\left( x\right) $ satisfy the fields equations (\ref{flagr2}); (iii) the
fields vanish at the spacial infinity sufficiently fast. The result is a
symmetric \emph{Belinfante-Rosenfeld energy-momentum tensor} $T^{\mu \nu }$, 
\cite[(3.73)-(3.75)]{Barut}, \cite[(13a), (13b), (13c), (14)]{Pauli RFTh}, 
\cite{Belinfante1}, \cite{Belinfante2}, \cite{Rosenfeld}, namely%
\begin{gather}
T^{\mu \nu }=\mathring{T}^{\mu \nu }-\partial _{\gamma }f^{\mu \gamma \nu
}=\dsum_{\ell }\frac{\partial \mathcal{L}}{\partial q_{,\mu }^{\ell }}%
q^{\ell ,\nu }-g^{\mu \nu }\mathcal{L}-\partial _{\gamma }f^{\mu \nu \gamma
},\text{ where}  \label{ttbe1} \\
f^{\mu \gamma \nu }=-f^{\gamma \mu \nu }=\frac{1}{2}\left[
\dsum\limits_{\ell \ell ^{\prime }}\left( L_{\ell }^{\mu }S_{\ell ^{\prime
}}^{\ell \gamma \nu }+L_{\ell }^{\gamma }S_{\ell ^{\prime }}^{\ell \nu \mu
}-L_{\ell }^{\nu }S_{\ell ^{\prime }}^{\ell \mu \gamma }\right) q^{\ell
^{\prime }}\right] ,\ L_{\ell }^{\mu }=\frac{\partial \mathcal{L}}{\partial
q_{,\mu }^{\ell }},  \notag
\end{gather}%
where the tensor $S_{\ell ^{\prime }}^{\ell \mu \nu }$ describes the
infinitesimal transformation of the involved fields $q^{\ell }\left(
x\right) \rightarrow q^{\prime \ell }\left( x^{\prime }\right) $ along with
the infinitesimal inhomogeneous Lorentz transformation of the space time
vector $x\rightarrow x^{\prime }$ as described by (\ref{ttbe2}), namely, 
\begin{equation}
x^{\prime \mu }=x^{\mu }+\xi ^{\mu \nu }x_{\nu }+a^{\mu },\ \xi ^{\mu \nu
}=-\xi ^{\nu \mu },  \label{ttbe1a}
\end{equation}%
where $\xi ^{\mu \nu }$ and $a^{\mu }$ are the ten parameters, and%
\begin{equation}
q^{\prime \ell }\left( x^{\prime }\right) =q^{\ell }\left( x\right) +\xi
_{\mu \nu }\dsum\limits_{\ell ^{\prime }}S_{\ell ^{\prime }}^{\ell \mu \nu
}q^{\ell ^{\prime }}\left( x\right) ,\ S_{\ell ^{\prime }}^{\ell \mu \nu
}=-S_{\ell ^{\prime }}^{\ell \nu \mu }.  \label{ttbe3}
\end{equation}%
In particular, \cite[III.4(A)]{Barut},%
\begin{eqnarray}
S_{\ell ^{\prime }}^{\ell \mu \nu } &=&0\text{ if }q\text{ is a scalar field,%
}  \label{ttbe3a} \\
S_{\beta }^{\alpha \mu \nu } &=&g^{\alpha \mu }g_{\beta }^{\nu }-g^{\alpha
\nu }g_{\beta }^{\mu }\text{ if }q\text{ is a vector field.}  \label{ttbe3b}
\end{eqnarray}%
The conserved quantities are, \cite[(3.76)-(3.77)]{Barut}, \cite[(6)]{Pauli
RFTh} 
\begin{equation}
P^{\nu }=\dint_{\sigma }T^{\mu \nu }\,\mathrm{d}\sigma _{\mu },\ J^{\nu
\gamma }=\dint_{\sigma }M^{\mu \nu \gamma }\,\mathrm{d}\sigma _{\mu },\
M^{\mu \nu \gamma }=T^{\mu \nu }x^{\gamma }-T^{\mu \gamma }x^{\nu },
\label{ttbe4}
\end{equation}%
where $\sigma $ is any space-like surface, for instance $x_{0}=\limfunc{const%
}$. $P^{\nu }$ is \emph{four-vector of the total energy-momentum} and $%
J^{\nu \gamma }=-J^{\gamma \nu }$ is \emph{the total angular momentum tensor}%
. The differential form of the conservation laws is%
\begin{equation}
\partial _{\mu }T^{\mu \nu }=0,\ \partial _{\mu }M^{\mu \nu \gamma
}=T^{\gamma \nu }-T^{\nu \gamma }=0.  \label{ttbe5}
\end{equation}%
Observe that the conservation of the angular momentum $M^{\mu \nu \gamma }$
in (\ref{ttbe5}) implies the symmetry of the energy-momentum tensor and, in
view of (\ref{ttbe1}), the following identities, \cite[(3.81$^{\prime }$)]%
{Barut}, 
\begin{equation}
T^{\mu \nu }=T^{\nu \mu },\ \dsum_{\ell }\frac{\partial \mathcal{L}}{%
\partial q_{,\mu }^{\ell }}q^{\ell ,\nu }-\partial _{\gamma }f^{\mu \gamma
\nu }=\dsum_{\ell }\frac{\partial \mathcal{L}}{\partial q_{,\nu }^{\ell }}%
q^{\ell ,\mu }-\partial _{\gamma }f^{\nu \gamma \mu }.  \label{ttbe6}
\end{equation}%
For an alternative insightful derivation of the symmetric energy-momentum
tensor based on kinosthenic (ignorable) variables and the Noether's method
as a way to generate such variables we refer to \cite[Section 3.5, 3.6, 3.10]%
{Lanczos VP}. Interestingly under this approach the conservations laws take
the form of the Euler-Lagrange equations for those kinosthenic variables. 
\emph{We would like to point out that the symmetry of the energy-momentum
tensor and the corresponding identities (\ref{ttbe6}) are nontrivial
relations which hold provided that the involved fields satisfy the field
equations (\ref{flagr2}).}

\emph{Since the symmetric energy-momentum tensor }$T^{\mu \nu }$\emph{\ is
the one used in most of the cases we often refer to it just as the
energy-momentum tensor,} while the tensor $\mathring{T}^{\mu \nu }$ defined
by (\ref{flagr5}) is referred to as the \emph{canonical energy-momentum
tensor}.

The interpretation of the symmetric energy-momentum tensor $T^{\mu \nu }$
entries is as follows, \cite[Section 32]{LandauLif F}, \cite[Sections 13.2,
13.3]{Goldstein}, \cite[Chapter 3.4]{Morse Feshbach I} 
\begin{equation}
T^{\mu \nu }=\left[ 
\begin{array}{cccc}
u & \mathrm{c}p_{1} & \mathrm{c}p_{2} & \mathrm{c}p_{3} \\ 
\mathrm{c}^{-1}s_{1} & -\sigma _{11} & -\sigma _{12} & -\sigma _{13} \\ 
\mathrm{c}^{-1}s_{2} & -\sigma _{21} & -\sigma _{22} & -\sigma _{23} \\ 
\mathrm{c}^{-1}s_{3} & -\sigma _{31} & -\sigma _{32} & -\sigma _{33}%
\end{array}%
\right] ,  \label{emten2}
\end{equation}%
where%
\begin{equation}
\begin{tabular}{|l|l|}
\hline
$u$ & field energy density, \\ \hline
$p_{j},\ j=1,2,3,$ & field momentum density, \\ \hline
$s_{j}=\mathrm{c}^{2}p_{j},\ j=1,2,3,$ & field energy flux density, \\ \hline
$\sigma _{ij}=\sigma _{ji},\ j=1,2,3,$ & field symmetric stress tensor. \\ 
\hline
\end{tabular}
\label{emten3}
\end{equation}%
If we introduce the 4-momentum and 4-flux densities by the formulas 
\begin{equation*}
p^{\nu }=\frac{1}{\mathrm{c}}T^{0\nu }=\left( \frac{1}{\mathrm{c}}u,\mathbf{p%
}\right) ,\ s^{\nu }=\mathrm{c}T^{\nu 0}=\left( \mathrm{c}u,\mathbf{s}%
\right) .
\end{equation*}%
then the energy-momentum conservation law (\ref{emten1a}) can be recast as
follows%
\begin{equation}
\partial _{t}p^{\nu }+\dsum_{j=1,2,3}\partial _{j}T^{j\nu }=-\frac{\partial 
\mathcal{L}}{\partial x_{\nu }},  \label{emten4}
\end{equation}%
or, in other words,%
\begin{equation}
\partial _{t}p_{i}=\dsum_{j=1,2,3}\partial _{j}\sigma _{ji}-\frac{\partial 
\mathcal{L}}{\partial x^{i}},\ \text{where }p_{i}=\frac{1}{\mathrm{c}}%
T^{0i},\ \sigma _{ji}=-T^{ji},\ i,j=1,2,3,  \label{emten5}
\end{equation}%
\begin{equation}
\partial _{t}u+\dsum_{j=1,2,3}\partial _{j}s_{j}=-\frac{\partial \mathcal{L}%
}{\partial t},\ \text{where }u=T^{00},\ s_{i}=\mathrm{c}T^{i0}=\mathrm{c}%
^{2}p_{i},\ i=1,2,3.  \label{emten6}
\end{equation}%
If the Lagrangian $\mathcal{L}$ does not depend explicitly on $x_{\nu }$ the
above energy-momentum conservation laws turn into%
\begin{equation}
\partial _{t}p_{i}=\dsum_{j=1,2,3}\partial _{j}\sigma _{ji},\ \text{where }%
p_{i}=\frac{1}{\mathrm{c}}T^{0i},\ \sigma _{ji}=-T^{ji},\ i,j=1,2,3,
\label{emten7}
\end{equation}%
\begin{equation}
\partial _{t}u+\dsum_{j=1,2,3}\partial _{j}s_{j}=0,\ \text{where }u=T^{00},\
s_{i}=\mathrm{c}T^{i0}=\mathrm{c}^{2}p_{i},\ i=1,2,3.  \label{emten8}
\end{equation}%
Consequently, the \emph{total conserved energy momentum} 4-vector takes the
form%
\begin{equation}
P^{\nu }=\frac{1}{\mathrm{c}}\dint\nolimits_{\mathbb{R}^{3}}T^{0\nu }\left(
x\right) \,\mathrm{d}\mathbf{x},  \label{emten9}
\end{equation}%
and its components, the total energy and momentum are respectively%
\begin{equation}
H=\mathrm{c}P^{0}=\dint\nolimits_{\mathbb{R}^{3}}T^{0\nu }\left( x\right) \,%
\mathrm{d}\mathbf{x},\ P^{j}=\dint\nolimits_{\mathbb{R}^{3}}T^{0j}\left(
x\right) \,\mathrm{d}\mathbf{x},\ j=1,2,3.  \label{emten10}
\end{equation}%
Evidently, the formulas (\ref{emten9}), (\ref{emten10}) are particular cases
of the formulas (\ref{ttbe4}) for the special important choice of $\sigma
=\left\{ x=\left( x_{0},\mathbf{x}\right) :\ \mathbf{x}\in \mathbb{R}%
^{3}\right\} $.

Importantly, \emph{for closed systems the conserved total energy-momentum }$%
P^{\nu }$\emph{\ and }$J^{\nu \gamma }$\emph{\ angular momentum as defined
by formulas (\ref{ttbe4}) and (\ref{emten9}) transform respectively as
4-vector and 4-tensor} under Lorentz transformation and that directly
related to the conservations laws (\ref{ttbe5}), \cite[Section 6.2]{Moller}, 
\cite[Section 12.10 A]{Jackson}. But \emph{for open (not closed) systems
generally the total energy-momentum }$P^{\nu }$\emph{\ and }$J^{\nu \gamma }$%
\emph{\ angular momentum do not transform as respectively 4-vector and
4-tensor}, \cite[Section 7.1, 7.2]{Moller}, \cite[Section 12.10 A, 16.4]%
{Jackson}.

\subsection{Noether's theorem\label{snoether}}

In this section following to \cite[Section 13.7]{Goldstein} we provide basic
information on the Noether's theorem which relates symmetries to
conservation laws based on the Lagrangian formalism. Suppose that the
Lagrangian $\mathcal{L}$ as defined (\ref{flagr1}) does not depend
explicitly on the field variable $q^{\ell }$. Then the Euler-Lagrange
equations (\ref{flagr2}) for that variable turns into%
\begin{equation}
\partial _{\mu }J_{\ell }^{\mu }=0,\ J_{\ell }^{\mu }=\frac{\partial 
\mathcal{L}}{\partial q_{,\mu }^{\ell }},\   \label{noe1}
\end{equation}%
which is a conservation law (continuity equation) for the four-"current" $%
J_{\ell }^{\mu }$. Noether theory interprets the above situation as certain
invariance (symmetry) of the Lagrangian and provides its far reaching
generalization allowing to obtain conservation laws based on the Lagrangian
invariance (symmetry) with respect to a general Lie group of transformation.
Symmetry under coordinate transformation refers to the effects of
infinitesimal transformation of the form%
\begin{equation}
x^{\mu }\rightarrow x^{\prime \mu }=x^{\mu }+\delta x^{\mu },  \label{noe2}
\end{equation}%
where the infinitesimal change $\delta x^{\mu }$ may depend on other $x^{\nu
}$. The field transformations are assumed to be of the form%
\begin{equation}
q^{\ell }\left( x\right) \rightarrow q^{\prime \ell }\left( x^{\prime
}\right) =q^{\ell }\left( x\right) +\delta q^{\ell }\left( x\right) ,
\label{noe3}
\end{equation}%
where $\delta q^{\ell }\left( x\right) $ measures to \emph{total} change of $%
q^{\ell }$ due to both $x$ and $q^{\ell }$ and it can depend on other field
variables $q^{\ell _{1}}$. We also consider a \emph{local} change $\bar{%
\delta}q^{\ell }\left( x\right) $ of $q^{\ell }\left( x\right) $ at a point $%
x$%
\begin{equation}
q^{\prime \ell }\left( x\right) =q^{\ell }\left( x\right) +\bar{\delta}%
q^{\ell }\left( x\right)  \label{noe4}
\end{equation}%
We assume the following three conditions to hold for the transformations (%
\ref{noe2})-(\ref{noe3}): (i) the 4-space (time and the space) is flat; (ii)
the Lagrangian density $\mathcal{L}$ displays the same functional form in
terms of the transformed quantities as it does of the original quantities,
that is,%
\begin{equation}
\mathcal{L}(\left\{ q^{\ell }\left( x\right) \right\} ,\left\{ q_{,\mu
}^{\ell }\left( x\right) \right\} ,x)=\mathcal{L}(\left\{ q^{\prime \ell
}\left( x^{\prime }\right) \right\} ,\left\{ q_{,\mu }^{\prime \ell }\left(
x^{\prime }\right) \right\} ,x^{\prime });  \label{noe5}
\end{equation}%
(iii) The magnitude of the action integral is invariant under the
transformation, that is%
\begin{equation}
\dint\nolimits_{\Omega }\mathcal{L}(\left\{ q^{\ell }\left( x\right)
\right\} ,\left\{ q_{,\mu }^{\ell }\left( x\right) \right\} ,x)\,\mathrm{d}%
x=\dint\nolimits_{\Omega ^{\prime }}\mathcal{L}(\left\{ q^{\prime \ell
}\left( x^{\prime }\right) \right\} ,\left\{ q_{,\mu }^{\prime \ell }\left(
x^{\prime }\right) \right\} ,x^{\prime })\,\mathrm{d}x^{\prime },
\label{noe6}
\end{equation}%
where $\mathbb{\Omega }$ is 4-dimensional region bounded by two space-like
3-dimensional surfaces and $\mathrm{d}x=\sqrt{\left\vert \det g\right\vert }%
\mathrm{d}x_{0}\mathrm{d}x_{1}\mathrm{d}x_{2}\mathrm{d}x_{3}$.

If the three above conditions are satisfied the following conservation law
holds%
\begin{equation}
\partial _{\mu }J^{\mu }=0,\ J^{\mu }=\dsum\nolimits_{\ell }\frac{\partial 
\mathcal{L}}{\partial q_{,\mu }^{\ell }}\bar{\delta}q^{\ell }\left( x\right)
+\mathcal{L}\delta x^{\mu }.  \label{noe7}
\end{equation}%
The above conservation law can be made more specific if we introduce
infinitesimal parameters $\xi _{r}$ related to Lie group of transformations
and represent $\delta x^{\mu }$ and $\delta q^{\ell }$ in their terms,
namely 
\begin{equation}
\delta x^{\mu }=\dsum\nolimits_{r}X_{r}^{\mu }\xi _{r},\ \delta q^{\ell
}=\dsum\nolimits_{r}Q_{r}^{\ell }\xi _{r},  \label{noe8}
\end{equation}%
where the functions $X_{r}^{\mu }$ and $Q_{r}^{\ell }$ may depend upon the
other coordinates and field variables, respectively. Then we for $r$ the
following conservation law holds for the respective Noether's current $%
J_{r}^{\mu }$:%
\begin{equation}
\partial _{\mu }J_{r}^{\mu }=0,\ J_{r}^{\mu }=\left[ \dsum\nolimits_{\ell }%
\frac{\partial \mathcal{L}}{\partial q_{,\mu }^{\ell }}q_{,\nu }^{\ell }-%
\mathcal{L}\delta _{\nu }^{\mu }\right] X_{r}^{\nu }-\dsum\nolimits_{\ell }%
\frac{\partial \mathcal{L}}{\partial q_{,\mu }^{\ell }}Q_{r}^{\ell }.
\label{noe9}
\end{equation}%
For instance, in the case of the group of inhomogeneous Lorentz
transformations defined by (\ref{ttbe2}) there are ten parameters $a^{\mu }$
and $\xi ^{\mu \nu }$ and consequently there are ten corresponding conserved
quantities $P^{\nu }$and $J^{\nu \gamma }=-J^{\gamma \nu }$ defined by (\ref%
{ttbe4}). Another important example is the group of gauge transformation of
the first kind defined by (\ref{flagr8}) below. For this group there is a
conserved current $J^{\ell \nu }$ for every $\ell $ defined by (\ref{flagr9}%
).

\subsection{Electromagnetic fields and the Maxwell equations}

We consider the Maxwell equations%
\begin{equation}
\nabla \cdot \mathbf{E}=4\pi \varrho ,\qquad \nabla \cdot \mathbf{B}=0
\label{maxw1}
\end{equation}%
\begin{equation}
\nabla \times \mathbf{E}+\frac{1}{\mathrm{c}}\partial _{t}\mathbf{B}%
=0,\qquad \nabla \times \mathbf{B}-\frac{1}{\mathrm{c}}\partial _{t}\mathbf{E%
}=\frac{4\pi }{\mathrm{c}}\mathbf{J}.  \label{maxw2}
\end{equation}%
for the EM fields and their covariant form following to \cite[Section 11.9]%
{Jackson}, \cite[Sections 23, 30]{LandauLif EM}, \cite[Sections 7.4, 11.2]%
{Griffiths}, in CGS units. To represent Maxwell equations in a manifestly
Lorentz invariant form it is common to introduce a four-vector potential $%
A^{\mu }$ and a four-vector current density $J^{\nu }$ 
\begin{gather}
A^{\mu }=\left( \varphi ,\mathbf{A}\right) ,\ J^{\mu }=\left( \mathrm{c}%
\varrho ,\mathbf{J}\right) ,  \label{maxw2a} \\
\partial _{\mu }=\frac{\partial }{\partial x^{\mu }}=\left( \frac{1}{\mathrm{%
c}}\partial _{t},\nabla \right) ,\ \partial ^{\mu }=\frac{\partial }{%
\partial x_{\mu }}=\left( \frac{1}{\mathrm{c}}\partial _{t},-\nabla \right) ,
\notag
\end{gather}%
and, then, an antisymmetric second-rank tensor, the "field strength tensor,%
\begin{equation}
F^{\mu \nu }=\partial ^{\mu }A^{\nu }-\partial ^{\nu }A^{\mu },
\label{maxw2b}
\end{equation}%
so that%
\begin{equation}
F^{\mu \nu }=\left[ 
\begin{array}{cccc}
0 & -E_{1} & -E_{2} & -E_{3} \\ 
E_{1} & 0 & -B_{3} & B_{2} \\ 
E_{2} & B_{3} & 0 & -B_{1} \\ 
E_{3} & -B_{2} & B_{1} & 0%
\end{array}%
\right] ,\ F_{\mu \nu }=\left[ 
\begin{array}{cccc}
0 & E_{1} & E_{2} & -E_{3} \\ 
-E_{1} & 0 & -B_{3} & B_{2} \\ 
-E_{2} & B_{3} & 0 & -B_{1} \\ 
-E_{3} & -B_{2} & B_{1} & 0%
\end{array}%
\right] ,  \label{maxw3}
\end{equation}%
and%
\begin{equation}
\mathbf{E}=-\nabla \varphi -\frac{1}{\mathrm{c}}\partial _{t}\mathbf{A},\ 
\mathbf{B}=\nabla \times \mathbf{A}.  \label{maxw3a}
\end{equation}%
Then the two inhomogeneous equations and the two homogeneous equations from
the four Maxwell equations (\ref{maxw1}) take respectively the form%
\begin{equation}
\partial _{\mu }F^{\mu \nu }=\frac{4\pi }{\mathrm{c}}J^{\nu },  \label{maxw4}
\end{equation}%
\begin{equation}
\partial _{\alpha }F_{\beta \gamma }+\partial _{\beta }F_{\gamma \alpha
}+\partial _{\gamma }F_{\alpha \beta }=0,\ \alpha ,\beta ,\gamma =0,1,2,3.
\label{maxw4a}
\end{equation}%
It follows from the asymmetry of $F^{\mu \nu }$, the Maxwell equation (\ref%
{maxw4}) and (\ref{maxw2a})-(\ref{maxw2b}) that the four-vector current $%
J^{\mu }$ must satisfy the \emph{continuity equation} 
\begin{equation}
\partial _{\mu }J^{\mu }=0\text{ or }\partial _{t}\varrho +\nabla \cdot 
\mathbf{A}=0.  \label{maxw5}
\end{equation}%
The Maxwell equations (\ref{maxw4}) turn into the following equations for
the four-vector potential $A^{\mu }$ 
\begin{equation}
\square A^{\nu }-\partial ^{\nu }\partial _{\mu }A^{\mu }=\frac{4\pi }{%
\mathrm{c}}J^{\nu },  \label{maxw6}
\end{equation}%
where 
\begin{equation}
\square =\partial _{\mu }\partial ^{\mu }=\frac{1}{\mathrm{c}^{2}}\partial
_{t}^{2}-\nabla ^{2}\text{ (d'Alembertian operator).}  \label{maxw7}
\end{equation}%
According to \cite[Section 11.10]{Jackson} the electric and magnetic fields
are transformed from one frame to another one moving relatively with the
velocity $\mathbf{v}$ by the following formulas 
\begin{gather}
\mathbf{E}^{\prime }=\gamma \left( \mathbf{E}+\mathbf{\beta }\times \mathbf{B%
}\right) -\frac{\gamma ^{2}}{\gamma +1}\left( \mathbf{\beta }\cdot \mathbf{E}%
\right) \mathbf{\beta ,}  \label{maxw8} \\
\mathbf{B}^{\prime }=\gamma \left( \mathbf{B}-\mathbf{\beta }\times \mathbf{E%
}\right) -\frac{\gamma ^{2}}{\gamma +1}\left( \mathbf{\beta }\cdot \mathbf{B}%
\right) \mathbf{\beta ,}  \notag \\
\mathbf{\beta }=\frac{\mathbf{v}}{\mathrm{c}},\ \beta =\left\vert \mathbf{%
\beta }\right\vert ,\ \gamma =\frac{1}{\sqrt{1-\left( \frac{v}{\mathrm{c}}%
\right) ^{2}}}.  \notag
\end{gather}%
which also can be recast as, \cite[Section 22]{Grainer EM},%
\begin{eqnarray}
\mathbf{E}_{\bot }^{\prime } &=&\gamma \left( \mathbf{E}_{\bot }+\mathbf{%
\beta }\times \mathbf{B}\right) ,\ \mathbf{E}_{\Vert }^{\prime }=\mathbf{E}%
_{\Vert },  \label{maxw9} \\
\mathbf{B}_{\bot }^{\prime } &=&\gamma \left( \mathbf{B}_{\bot }-\mathbf{%
\beta }\times \mathbf{E}\right) ,\ \mathbf{B}_{\Vert }^{\prime }=\mathbf{B}%
_{\Vert }\mathbf{,}  \notag
\end{eqnarray}%
where subindices $\bot $ and $\Vert $ stand for vector components
respectively parallel and perpendicular to $\mathbf{v}$. Observe that for $%
\beta \ll 1$ the formulas (\ref{maxw8}), (\ref{maxw9}) yield the following
approximations with an error proportional to $\beta ^{2}$ where $J_{\mu }$
is an external four-vector current.%
\begin{equation}
\mathbf{E}_{\bot }^{\prime }\cong \mathbf{E}_{\bot }+\mathbf{\beta }\times 
\mathbf{B},\ \mathbf{E}_{\Vert }^{\prime }=\mathbf{E}_{\Vert },\ \mathbf{B}%
_{\bot }^{\prime }\cong \mathbf{B}_{\bot }-\mathbf{\beta }\times \mathbf{E}%
,\ \mathbf{B}_{\Vert }^{\prime }=\mathbf{B}_{\Vert }\mathbf{.}
\label{maxw10}
\end{equation}

The EM field Maxwell Lagrangian is, \cite[Section 12.7]{Jackson}, \cite[%
Section IV.1]{Barut} 
\begin{equation}
L_{\mathrm{em}}\left( A^{\mu }\right) =-\frac{1}{16\pi }F_{\mu \nu }F^{\mu
\nu }-\frac{1}{\mathrm{c}}J_{\mu }A^{\mu },  \label{flagr7}
\end{equation}%
where $J_{\mu }$ is an external (impressed) current. Using (\ref{maxw3}), (%
\ref{maxw3a}) and (\ref{maxw2a}) we can recast (\ref{flagr7}) as%
\begin{gather}
L_{\mathrm{em}}\left( A^{\mu }\right) =\frac{1}{8\pi }\left( \mathbf{E}^{2}-%
\mathbf{B}^{2}\right) -\rho \varphi +\frac{1}{\mathrm{c}}\mathbf{J}\cdot 
\mathbf{A}  \label{flagr7a} \\
=\frac{1}{8\pi }\left[ \left( \nabla \varphi +\frac{1}{\mathrm{c}}\partial
_{t}\mathbf{A}\right) ^{2}-\left( \nabla \times \mathbf{A}\right) ^{2}\right]
-\rho \varphi +\frac{1}{\mathrm{c}}\mathbf{J}\cdot \mathbf{A.}  \notag
\end{gather}%
In particular, if there are no sources the above Lagrangians turn into%
\begin{gather}
L_{\mathrm{em}}\left( A^{\mu }\right) =-\frac{1}{16\pi }F_{\mu \nu }F^{\mu
\nu }=\frac{1}{8\pi }\left( \mathbf{E}^{2}-\mathbf{B}^{2}\right) =
\label{flagr7aa} \\
=\frac{1}{8\pi }\left[ \left( \nabla \varphi +\frac{1}{\mathrm{c}}\partial
_{t}\mathbf{A}\right) ^{2}-\left( \nabla \times \mathbf{A}\right) ^{2}\right]
.  \notag
\end{gather}

The canonical stress (power-momentum) tensor $\mathring{\Theta}^{\mu \nu }$
for the EM field is as follows, \cite[(12.104)]{Jackson}, \cite[Section
III.4.D]{Barut} 
\begin{equation}
\mathring{\Theta}^{\mu \nu }=-\frac{F^{\mu \gamma }\partial ^{\nu }A_{\gamma
}}{4\pi }+g^{\mu \nu }\frac{F^{\xi \gamma }F_{\xi \gamma }}{16\pi },
\label{flagr7b}
\end{equation}%
or, in particular, for $i,j=1,2,3$ 
\begin{gather}
\mathring{\Theta}^{00}=-\frac{\mathbf{E}^{2}-\mathbf{B}^{2}}{8\pi }+\rho
\varphi -\frac{1}{\mathrm{c}}\mathbf{J}\cdot \mathbf{A}-\frac{\partial _{0}%
\mathbf{A}\cdot \mathbf{E}}{4\pi },  \label{flagr7c} \\
\mathring{\Theta}^{0i}=-\frac{\partial _{i}\mathbf{A}\cdot \mathbf{E}}{4\pi }%
,\ \mathring{\Theta}^{i0}=-\frac{E_{i}\partial _{0}\varphi }{4\pi }+\frac{%
\left( \mathbf{B}\times \partial _{0}\mathbf{A}\right) _{i}}{4\pi },  \notag
\\
\mathring{\Theta}^{ij}=-\frac{E_{i}\partial _{j}\varphi }{4\pi }+\frac{%
\left( \mathbf{B}\times \partial _{j}\mathbf{A}\right) _{i}}{4\pi }+\frac{%
\mathbf{E}^{2}-\mathbf{B}^{2}}{8\pi }-\rho \varphi +\frac{1}{\mathrm{c}}%
\mathbf{J}\cdot \mathbf{A},  \notag
\end{gather}%
whereas the symmetric one $\Theta ^{\alpha \beta }$ for the EM field is, 
\cite[Section 12.10, (12.113)]{Jackson}, \cite[Section III.3]{Barut}%
\begin{equation}
\Theta ^{\alpha \beta }=\frac{1}{4\pi }\left( g^{\alpha \mu }F_{\mu \nu
}F^{\nu \beta }+\frac{1}{4}g^{\alpha \beta }F_{\mu \nu }F^{\mu \nu }\right) ,
\label{maxw11}
\end{equation}%
implying the following formulas for the field energy density $w$, the
momentum $\mathrm{c}\mathbf{g}$ and the Maxwell stress tensor $\tau _{ij}$: 
\begin{equation}
w=\Theta ^{00}=\frac{\mathbf{E}^{2}+\mathbf{B}^{2}}{8\pi },\ \mathrm{c}%
g_{i}=\Theta ^{0i}=\Theta ^{i0}=\frac{\mathbf{E}\times \mathbf{B}}{4\pi },
\label{maxw12}
\end{equation}%
\begin{equation}
\Theta ^{ij}=-\frac{1}{4\pi }\left[ E_{i}E_{j}+B_{i}B_{j}-\frac{1}{2}\delta
_{ij}\left( \mathbf{E}^{2}+\mathbf{B}^{2}\right) \right] ,  \label{maxw13}
\end{equation}%
\begin{eqnarray}
\Theta ^{\alpha \beta } &=&%
\begin{bmatrix}
w & \mathrm{c}\mathbf{g} \\ 
\mathrm{c}\mathbf{g} & -\tau _{ij}%
\end{bmatrix}%
,\ \Theta _{\alpha \beta }=%
\begin{bmatrix}
w & -\mathrm{c}\mathbf{g} \\ 
-\mathrm{c}\mathbf{g} & -\tau _{ij}%
\end{bmatrix}%
,  \label{maxw14} \\
\Theta _{\ \beta }^{\alpha } &=&%
\begin{bmatrix}
w & -\mathrm{c}\mathbf{g} \\ 
\mathrm{c}\mathbf{g} & -\tau _{ij}%
\end{bmatrix}%
,\ \Theta _{\alpha }^{\ \beta }=%
\begin{bmatrix}
w & \mathrm{c}\mathbf{g} \\ 
-\mathrm{c}\mathbf{g} & -\tau _{ij}%
\end{bmatrix}%
.  \notag
\end{eqnarray}%
Note that in the special case when the vector potential $\mathbf{A}$
vanishes and the scalar potential $\varphi $ does not depend on time using
the expressions (\ref{maxw3a}) we get the following representation for the
canonical energy density defined by (\ref{flagr7c})%
\begin{gather}
\mathring{\Theta}^{00}=-\frac{\left( \nabla \varphi \right) ^{2}}{8\pi }%
+\rho \varphi \text{ for }\mathbf{A}=\mathbf{0}\text{ and }\partial
_{0}\varphi =0,  \label{maxw14a} \\
\text{whereas }\Theta ^{00}=\frac{\left( \nabla \varphi \right) ^{2}}{8\pi }.
\notag
\end{gather}%
It is instructive to observe a substantial difference between the above
expressions $\mathring{\Theta}^{00}$, which is the Hamiltonian density of
the EM field, and the energy density $\Theta ^{00}$ defined by (\ref{maxw12}%
).

If there no external currents the with differential \emph{conservation law}
takes the form%
\begin{equation}
\partial _{\alpha }\Theta ^{\alpha \beta }=0,  \label{maxw15}
\end{equation}%
and, in particular, the energy conservation law%
\begin{gather}
0=\partial _{\alpha }\Theta ^{\alpha \beta }=\frac{1}{\mathrm{c}}\left( 
\frac{\partial w}{\partial t}+\nabla \cdot \mathbf{S}\right) ,\text{ where }w%
\text{ is the energy density, and}  \label{maxw16} \\
\mathbf{S}=\mathrm{c}^{2}\mathbf{g}=\frac{\mathrm{c}}{4\pi }\mathbf{E}\times 
\mathbf{B}\text{ is the Poynting vector.}  \notag
\end{gather}%
In the presence of external currents the conservation laws take the form, 
\cite[Section 12.10]{Jackson}%
\begin{equation}
\partial _{\alpha }\Theta ^{\alpha \beta }=-f^{\beta },\ f^{\beta }=\frac{1}{%
\mathrm{c}}F^{\beta \nu }J_{\nu },  \label{maxw17}
\end{equation}%
and the time and space components of the equations (\ref{maxw17}) are the
conservation of energy $w$ and momentum $\mathbf{g}$ which can be recast as%
\begin{equation}
\frac{1}{\mathrm{c}}\left( \frac{\partial w}{\partial t}+\nabla \cdot 
\mathbf{S}\right) =-\frac{1}{\mathrm{c}}\mathbf{J}\cdot \mathbf{E},
\label{maxw18}
\end{equation}%
\begin{equation}
\frac{\partial g_{i}}{\partial t}-\sum\limits_{j=1}^{3}\frac{\partial }{%
\partial x^{j}}\tau _{ij}=-\left[ \rho E_{i}+\frac{1}{\mathrm{c}}\left( 
\mathbf{J}\times \mathbf{B}\right) _{i}\right] .  \label{maxw19}
\end{equation}%
The 4-vector $f^{\beta }$ in the conservation law (\ref{maxw17}) is known as
the \emph{Lorentz force density}%
\begin{equation}
f^{\beta }=\frac{1}{\mathrm{c}}F^{\beta \nu }J_{\nu }=\left( \frac{1}{%
\mathrm{c}}\mathbf{J}\cdot \mathbf{E},\rho \mathbf{E}+\frac{1}{\mathrm{c}}%
\mathbf{J}\times \mathbf{B}\right) .  \label{maxw20}
\end{equation}

\subsubsection{Green functions for the Maxwell equations\label{sGreenMax}}

This section in an excerpt from \cite[Section 12.11]{Jackson}. The EM fields 
$F^{\mu \nu }$ arising from an external source $J^{\nu }$ satisfy the
inhomogeneous Maxwell equations 
\begin{equation}
\partial _{\mu }F^{\mu \nu }=\frac{4\pi }{\mathrm{c}}J^{\nu },\ F^{\mu \nu
}=\partial ^{\mu }A^{\nu }-\partial ^{\nu }A^{\mu },  \label{grmax1}
\end{equation}%
which take the following form for the potentials $A^{\nu }$%
\begin{equation}
\square A^{\nu }-\partial ^{\nu }\partial _{\mu }A^{\mu }=\frac{4\pi }{%
\mathrm{c}}J^{\nu }.  \label{grmax2}
\end{equation}%
If the potentials satisfy the Lorentz condition, $\partial _{\mu }A^{\mu }=0$%
, they are then solutions of the four-dimensional wave equation,%
\begin{equation}
\square A^{\nu }=\frac{4\pi }{\mathrm{c}}J^{\nu }  \label{grmax3}
\end{equation}%
The solution of (\ref{grmax3}) is accomplished by finding a Green function $%
D\left( x,x^{\prime }\right) $ for the equation%
\begin{equation}
\square D\left( z\right) =\delta ^{\left( 4\right) }\left( z\right) ,\
D\left( x,x^{\prime }\right) =D\left( x-x^{\prime }\right) ,  \label{grmax4}
\end{equation}%
where $\delta ^{\left( 4\right) }\left( z\right) =\delta \left( z_{0}\right)
\delta \left( \mathbf{z}\right) $ is a four-dimensional delta function. One
can introduce then the so-called \emph{retarded or causal Green function}
solving the above equation (\ref{grmax4}), namely%
\begin{equation}
D_{\mathrm{r}}\left( x-x^{\prime }\right) =\frac{\theta \left(
x_{0}-x_{0}^{\prime }\right) \delta \left( x_{0}-x_{0}^{\prime }-R\right) }{%
4\pi R},\ R=\left\vert \mathbf{x}-\mathbf{x}^{\prime }\right\vert ,
\label{grmax5}
\end{equation}%
where $\theta \left( x_{0}\right) $ is the Heaviside step function. The name
causal or retarded is justified by the fact that the source-point time $%
x_{0}^{\prime }$ is always earlier then the observation-point time $x_{0}$.
Similarly one can introduce the \emph{advanced Green function} 
\begin{equation}
D_{\mathrm{a}}\left( x-x^{\prime }\right) =\frac{\theta \left[ -\left(
x_{0}-x_{0}^{\prime }\right) \right] \delta \left( x_{0}-x_{0}^{\prime
}+R\right) }{4\pi R},\ R=\left\vert \mathbf{x}-\mathbf{x}^{\prime
}\right\vert .  \label{grmax6}
\end{equation}%
These Green functions can be written in the following covariant form%
\begin{eqnarray}
D_{\mathrm{r}}\left( x-x^{\prime }\right) &=&\frac{1}{2\pi }\theta \left(
x_{0}-x_{0}^{\prime }\right) \delta \left[ \left( x-x^{\prime }\right) ^{2}%
\right] ,  \label{grmax7} \\
D_{\mathrm{a}}\left( x-x^{\prime }\right) &=&\frac{1}{2\pi }\theta \left(
x_{0}^{\prime }-x_{0}\right) \delta \left[ \left( x-x^{\prime }\right) ^{2}%
\right] ,\   \notag
\end{eqnarray}%
where $\left( x-x^{\prime }\right) ^{2}=\left( x_{0}-x_{0}^{\prime }\right)
^{2}-\left\vert \mathbf{x}-\mathbf{x}^{\prime }\right\vert ^{2}$ and 
\begin{equation}
\delta \left[ \left( x-x^{\prime }\right) ^{2}\right] =\frac{1}{2R}\left[
\delta \left( x_{0}-x_{0}^{\prime }-R\right) +\delta \left(
x_{0}-x_{0}^{\prime }+R\right) \right] .  \label{grmax8}
\end{equation}%
The solution to the wave equation (\ref{grmax3}) can be written in terms of
the Green functions%
\begin{equation}
A^{\nu }\left( x\right) =A_{\mathrm{in}}^{\nu }\left( x\right) +\frac{4\pi }{%
\mathrm{c}}\dint D_{\mathrm{r}}\left( x-x^{\prime }\right) J^{\nu }\left(
x^{\prime }\right) \,\mathrm{d}x  \label{grmax9}
\end{equation}%
or%
\begin{equation}
A^{\nu }\left( x\right) =A_{\mathrm{out}}^{\nu }\left( x\right) +\frac{4\pi 
}{\mathrm{c}}\dint D_{\mathrm{a}}\left( x-x^{\prime }\right) J^{\nu }\left(
x^{\prime }\right) \,\mathrm{d}x  \label{grmax10}
\end{equation}%
where $A_{\mathrm{in}}^{\nu }\left( x\right) $ and $A_{\mathrm{out}}^{\nu
}\left( x\right) $ are solutions to the homogeneous wave equation. In (\ref%
{grmax9}) the retarded Green function is used. In the limit $%
x_{0}\rightarrow -\infty $, the integral over the sources vanishes, assuming
the sources are localized in space and time, because of the retarded nature
of the Green function, and $A_{\mathrm{in}}^{\nu }\left( x\right) $ can be
interpreted as "\emph{incident}" or "\emph{incoming}" potential, specified
at $x_{0}\rightarrow -\infty $. Similarly, in (\ref{grmax10}) with the
advanced Green function, the homogeneous solution $A_{\mathrm{out}}^{\nu
}\left( x\right) $ is the asymptotic "\emph{outgoing}" potential, specified
at $x_{0}\rightarrow +\infty $. The radiation fields are defined as the
difference between the "outgoing" and "incoming" fields, and their $4$%
-vector potential is%
\begin{gather}
A_{\mathrm{rad}}^{\nu }\left( x\right) =A_{\mathrm{out}}^{\nu }\left(
x\right) -A_{\mathrm{in}}^{\nu }\left( x\right) =\frac{4\pi }{\mathrm{c}}%
\dint D\left( x-x^{\prime }\right) J^{\nu }\left( x^{\prime }\right) \,%
\mathrm{d}x,\text{ where}  \label{grmax11} \\
D\left( x-x^{\prime }\right) =D_{\mathrm{r}}\left( x-x^{\prime }\right) -D_{%
\mathrm{a}}\left( x-x^{\prime }\right) .  \notag
\end{gather}

\subsection{Many charges interacting with the electromagnetic field \label%
{Manych}}

In this section we consider the Lagrange formalism for complex-valued fields 
$\psi ^{\ell }$, $\ell =1,\ldots ,N$ that describe charges following to \cite%
[Part I]{Pauli RFTh} and \cite[Section I.3]{Wentzel}. For every
complex-valued $\psi ^{\ell }$ we always assume the presence of its
conjugates $\psi ^{\ell \ast }$, so the product $\psi ^{\ell }\psi ^{\ell
\ast }$ is real. The Lagrangian is assumed to be real valued and its general
form is 
\begin{equation}
\mathcal{L}=\mathcal{L}\left( \left\{ \psi ^{\ell },\psi _{,\mu }^{\ell
},\psi ^{\ell \ast },\psi _{,\mu }^{\ell \ast },\right\} ,\left\{
V^{g},V_{,\mu }^{g}\right\} ,x^{\mu }\right) ,  \label{flagr3}
\end{equation}%
where $\left\{ V^{g}\right\} $ are real-valued quantities. In the Lagrangian
(\ref{flagr3}) the fields $\psi ^{\ell }$ and their conjugates $\psi ^{\ast
\ell }$ are treated as independent and the corresponding Euler-Lagrange
field equations are, \cite[Section 3.3]{Morse Feshbach I}, \cite[Section
II.3, (3.3)]{Wentzel},%
\begin{gather}
\frac{\partial \mathcal{L}}{\partial \psi ^{\ell }}-\partial _{\mu }\left( 
\frac{\partial \mathcal{L}}{\partial \psi _{,\mu }^{\ell }}\right) =0,
\label{flagr4} \\
\frac{\partial \mathcal{L}}{\partial \psi ^{\ell \ast }}-\partial _{\mu
}\left( \frac{\partial \mathcal{L}}{\partial \psi _{,\mu }^{\ell \ast }}%
\right) =0,\ \frac{\partial \mathcal{L}}{\partial V^{g}}-\partial _{\mu
}\left( \frac{\partial \mathcal{L}}{\partial V_{,\mu }^{g}}\right) =0. 
\notag
\end{gather}%
The canonical energy-momentum tensor for the Lagrangian (\ref{flagr3}) is
similar to the general formula (\ref{flagr5}), namely%
\begin{equation}
\mathcal{\mathring{T}}^{\mu \nu }=\dsum_{\ell }\frac{\partial \mathcal{L}}{%
\partial \psi _{,\mu }^{\ell }}\psi ^{\ell ,\nu }+\frac{\partial \mathcal{L}%
}{\partial \psi _{,\mu }^{\ell \ast }}\psi ^{\ell \ast ,\nu }+\dsum_{g}\frac{%
\partial \mathcal{L}}{\partial V_{,\mu }^{g}}V^{g,\nu }-g^{\mu \nu }\mathcal{%
L}.  \label{flagr4a}
\end{equation}%
In the case when the Lagrangian $\mathcal{L}$ depends on only complex-valued
fields $\psi ^{\ell }$ and $\psi ^{\ell \ast }$ the canonical stress tensor
is symmetric and is of the form, \cite[(3.3.23)]{Morse Feshbach I}, \cite[%
(3.8)]{Wentzel},%
\begin{equation}
\mathcal{\mathring{T}}^{\mu \nu }=\dsum_{\ell }\frac{\partial \mathcal{L}}{%
\partial \psi _{,\mu }^{\ell }}\psi ^{\ell ,\nu }+\frac{\partial \mathcal{L}%
}{\partial \psi _{,\mu }^{\ell \ast }}\psi ^{\ell \ast ,\nu }-g^{\mu \nu }%
\mathcal{L}.  \label{flagr6}
\end{equation}

An important for us special case of the Lagrangian (\ref{flagr3}) is when
there are several charges described by complex valued\ fields $\psi ^{\ell }$
and $\psi ^{\ell \ast }$ interacting with the EM field described by the
real-valued four-potential $A^{\mu }$. For this case we introduce the
Lagrangian of the form%
\begin{gather}
\mathcal{L}\left( \left\{ \psi ^{\ell },\psi _{;\mu }^{\ell },\psi ^{\ell
\ast },\psi _{;\mu }^{\ell \ast },\right\} ,A^{\mu }\right) =
\label{flagr6a} \\
=\dsum_{\ell }L^{\ell }\left( \psi ^{\ell },\psi _{;\mu }^{\ell },\psi
^{\ell \ast },\psi _{;\mu }^{\ell \ast }\right) -\frac{F^{\mu \nu }F_{\mu
\nu }}{16\pi },\ F^{\mu \nu }=\partial ^{\mu }A^{\nu }-\partial ^{\nu
}A^{\mu }.  \notag
\end{gather}%
where we have introduced the so-called \emph{covariant derivatives} $\psi
_{;\mu }^{\ell }$ and $\psi _{;\mu }^{\ell \ast }$ by the following formulas 
\begin{equation}
\psi ^{\ell ;\mu }=\tilde{\partial}^{\ell \mu }\psi ^{\ell },\ \psi ^{\ell
;\mu \ast }=\tilde{\partial}^{\ell \mu \ast }\psi ^{\ell },  \label{flagr6b}
\end{equation}%
where%
\begin{equation*}
\tilde{\partial}^{\ell \mu }=\partial ^{\mu }+\frac{\mathrm{i}q^{\ell }}{%
\chi \mathrm{c}}A^{\mu },\ \tilde{\partial}^{\ell \mu \ast }=\partial ^{\mu
}-\frac{\mathrm{i}q^{\ell }}{\chi \mathrm{c}}A^{\mu }.
\end{equation*}%
In the above formula for every $\ell $ the real number $q^{\ell }$ is the
charge of the $\ell $-th elementary charge and $\tilde{\partial}^{\ell \mu }$
and $\tilde{\partial}^{\ell \mu \ast }$ are called the \emph{covariant
differentiation operators}. The particular forms (\ref{flagr6a})-(\ref%
{flagr6b}) of the multiparticle Lagrangian $\mathcal{L}$ and its $\ell $-the
charge components $L^{\ell }$ originates from the condition of \emph{gauge
invariance}. More precisely, one introduces the \emph{gauge transformation
of the first or the second kind} (known also as respectively \emph{global
and local gauge transformation}) for the fields $\psi ^{\ell }$ and $\psi
^{\ell \ast }$. These transformations are described respectively by the
following formulas namely, \cite[(17), (23a), (23b)]{Pauli RFTh}, \cite[%
Section 11, (11.4)]{Wentzel},%
\begin{equation}
\psi ^{\ell }\rightarrow \mathrm{e}^{\mathrm{i}\gamma ^{\ell }}\psi ^{\ell
},\ \psi ^{\ell \ast }\rightarrow \mathrm{e}^{-\mathrm{i}\gamma ^{\ell
}}\psi ^{\ell \ast },\text{ where }\gamma ^{\ell }\text{ is any real
constant,}  \label{flagr8}
\end{equation}%
\begin{equation}
\psi ^{\ell }\rightarrow \mathrm{e}^{-\frac{\mathrm{i}q^{\ell }\lambda
\left( x\right) }{\chi \mathrm{c}}}\psi ^{\ell },\ \psi ^{\ell \ast
}\rightarrow \mathrm{e}^{\frac{\mathrm{i}q^{\ell }\lambda \left( x\right) }{%
\chi \mathrm{c}}}\psi ^{\ell \ast },\ A^{\mu }\rightarrow A^{\mu }+\partial
^{\mu }\lambda ,  \label{flagr8a}
\end{equation}%
and the Lagrangian is assumed to be invariant with respect to the all gauge
transformations (\ref{flagr8}), (\ref{flagr8a}). Notice that for the
multi-charge Lagrangian $\mathcal{L}$ as defined by (\ref{flagr6a})-(\ref%
{flagr6b}) the following is true: (i) every charge interacts with the EM
field described by the four-potential $A^{\mu }$; (ii) different charges
don't interact directly with each other, but they interact only indirectly
through the EM field.

We also introduce the following symmetry condition on charges Lagrangians $%
L^{\ell }$: 
\begin{equation}
\frac{\partial L^{\ell }}{\partial \psi _{;\mu }^{\ell }}\psi ^{\ell ;\nu }+%
\frac{\partial L^{\ell }}{\partial \psi _{;\mu }^{\ell \ast }}\psi ^{\ell
;\nu \ast }=\frac{\partial L^{\ell }}{\partial \psi _{;\nu }^{\ell }}\psi
^{\ell ;\mu }+\frac{\partial L^{\ell }}{\partial \psi _{;\nu }^{\ell \ast }}%
\psi ^{\ell ;\mu \ast }.  \label{lagsym}
\end{equation}%
\emph{As we show below the symmetry condition (\ref{lagsym}) implies that
energy-momentum assigned to every individual charge is symmetric and gauge
invariant energy-momentum. }A simple sufficient condition for the symmetry
condition (\ref{lagsym}) is a requirement for the Lagrangians $L^{\ell }$ to
depend on the field covariant derivatives only through the combination $\psi
^{\ell ;\mu }\psi _{;\mu }^{\ell \ast }$, in other words if there exist such
functions $K^{\ell }\left( \psi ^{\ell },\psi ^{\ell \ast },b\right) $ that%
\begin{equation}
L^{\ell }\left( \psi ^{\ell },\psi _{;\mu }^{\ell },\psi ^{\ell \ast },\psi
_{;\mu }^{\ell \ast }\right) =K^{\ell }\left( \psi ^{\ell },\psi ^{\ell \ast
},\psi ^{\ell ;\mu }\psi _{;\mu }^{\ell \ast }\right) .  \label{lagsym1}
\end{equation}%
Indeed, in this case%
\begin{equation}
\frac{\partial L^{\ell }}{\partial \psi _{;\mu }^{\ell }}\psi ^{\ell ;\nu }+%
\frac{\partial L^{\ell }}{\partial \psi _{;\mu }^{\ell \ast }}\psi ^{\ell
;\nu \ast }=\frac{\partial K^{\ell }}{\partial b}\left( \psi ^{\ell ;\mu
\ast }\psi ^{\ell ;\nu }+\psi ^{\ell ;\mu }\psi ^{\ell ;\nu \ast }\right) ,
\label{lagsym2}
\end{equation}%
readily implying that the symmetry condition (\ref{lagsym}) does hold.

The field equations for the Lagrangian $\mathcal{L}$ defined by (\ref%
{flagr6a})-(\ref{flagr6b}) are%
\begin{equation}
\frac{\partial L^{\ell }}{\partial \psi ^{\ell }}-\tilde{\partial}_{\mu
}^{\ell \ast }\left[ \frac{\partial L^{\ell }}{\partial \psi _{;\mu }^{\ell }%
}\right] =0,\ \frac{\partial L^{\ell }}{\partial \psi ^{\ell \ast }}-\tilde{%
\partial}_{\mu }^{\ell }\left[ \frac{\partial L^{\ell }}{\partial \psi
_{;\mu }^{\ell \ast }}\right] =0,\   \label{flagr8c}
\end{equation}%
\begin{equation}
\partial _{\mu }F^{\mu \nu }=\frac{4\pi }{\mathrm{c}}J^{\nu },\text{ }F^{\mu
\nu }=\partial ^{\mu }A^{\nu }-\partial ^{\nu }A^{\mu },\ J^{\nu
}=\dsum_{\ell }J^{\ell \nu },  \label{flagr8d}
\end{equation}%
where $J^{\ell \nu }$ is the four-vector current related to the $\ell $-th
charge is defined as follows. Under the gauge invariance conditions (\ref%
{flagr8}), (\ref{flagr8a}) for the Lagrangian $\mathcal{L}$ using the
Noether's theorem and the formula (\ref{noe9}) one can introduce for every
charge $\psi ^{\ell }$ the following \emph{4-vector current}, \cite[(19)]%
{Pauli RFTh}, \cite[(3.11)-(3.13)]{Wentzel}%
\begin{equation}
J^{\ell \nu }=-\mathrm{i}\frac{q^{\ell }}{\chi }\left( \frac{\partial
L^{\ell }}{\partial \psi _{;\nu }^{\ell }}\psi ^{\ell }-\frac{\partial
L^{\ell }}{\partial \psi _{;\nu }^{\ast \ell }}\psi ^{\ast \ell }\right) ,
\label{flagr9}
\end{equation}%
or, since $J^{\nu }=\left( \mathrm{c}\rho ,\mathbf{J}\right) $, 
\begin{gather}
\rho ^{\ell }=-\mathrm{i}\frac{q^{\ell }}{\chi }\left( \frac{\partial
L^{\ell }}{\partial \psi _{;0}^{\ell }}\psi ^{\ell }-\frac{\partial L^{\ell }%
}{\partial \psi _{;0}^{\ast \ell }}\psi ^{\ast \ell }\right) ,
\label{flagr10} \\
\mathbf{J}_{j}^{\ell }=-\mathrm{i}\frac{q^{\ell }}{\chi }\left( \frac{%
\partial L^{\ell }}{\partial \psi _{;j}^{\ell }}\psi ^{\ell }-\frac{\partial
L^{\ell }}{\partial \psi _{;j}^{\ast \ell }}\psi ^{\ast \ell }\right) ,\
j=1,2,3,  \notag
\end{gather}%
which satisfy for every $\ell $ the charge conservation/continuity equations%
\begin{equation}
\partial _{\nu }J^{\ell \nu }=0,\text{ or }\partial _{t}\rho ^{\ell }+\nabla
\cdot \mathbf{J}^{\ell }=0,\ J^{\ell \nu }=\left( \mathrm{c}\rho ^{\ell },%
\mathbf{J}^{\ell }\right) .  \label{flagr11}
\end{equation}%
Notice that in view of the relations (\ref{flagr6a})-(\ref{flagr6b}) the
following alternative representation holds for the four-current $J^{\ell \nu
}$ defined by (\ref{flagr9}) 
\begin{equation}
J^{\ell \nu }=-\mathrm{i}\frac{q^{\ell }}{\chi }\left( \frac{\partial
L^{\ell }}{\partial \psi _{;\nu }^{\ell }}\psi ^{\ell }-\frac{\partial
L^{\ell }}{\partial \psi _{;\nu }^{\ast \ell }}\psi ^{\ast \ell }\right) =-%
\mathrm{c}\frac{\partial L^{\ell }}{\partial A_{\nu }}.  \label{flagr12}
\end{equation}%
\emph{We would like to emphasize here the physical significance of identity (%
\ref{flagr12}) equating two complementary views on the electric current: (i)
as a source current in the Maxwell equations (\ref{flagr8d}); (ii) as the
gauge electric current (\ref{flagr9}) satisfying the continuity equation (%
\ref{flagr11}).} Notice that the equality (\ref{flagr12}) originates from a
particular form of the coupling between the EM field and charges in the
Lagrangian (\ref{flagr6a}), namely the coupling through the covariant
derivatives (\ref{flagr6b}). One may also view the electric currents
identity (\ref{flagr12}) as a physical rational for introducing the coupling
exactly as it is done in the expressions (\ref{flagr6a})-(\ref{flagr6b}).

\subsubsection{Gauge invariant and symmetric energy-momentum tensors\label%
{ginvsym}}

In this subsection we consider a Lagrangian defined by formulas (\ref%
{flagr6a})-(\ref{flagr6b}) and assume it to be gauge invariant with respect
to transformations if the first and the second type. To obtain an expression
for the total symmetric energy-momentum tensor $\mathcal{T}^{\mu \nu }$ for
such a Lagrangian we use the theory described in Section \ref{senergymom},
formulas (\ref{ttbe1}) and (\ref{flagr6}), namely%
\begin{equation}
\mathcal{T}^{\mu \nu }=\mathcal{\mathring{T}}^{\mu \nu }-\partial _{\gamma
}f^{\mu \gamma \nu },\ \mathcal{\mathring{T}}^{\mu \nu }=\mathring{\Theta}%
^{\mu \nu }+\dsum_{\ell }\mathring{T}^{\ell \mu \nu },  \label{enmom1}
\end{equation}%
where the canonical energy-momentum of EM field $\mathring{\Theta}^{\mu \nu
} $ and the energy-momentum tensor $\mathring{T}^{\ell \mu \nu }$ of $\ell $%
-th charge are represented as follows (see \cite[(12.104)]{Jackson}, \cite[%
Section III.4.D]{Barut} for $\mathring{\Theta}^{\mu \nu }$ and (\ref{flagr6}%
) for $\mathring{T}^{\ell \mu \nu }$) 
\begin{eqnarray}
\mathring{T}^{\ell \mu \nu } &=&\frac{\partial L^{\ell }}{\partial \psi
_{;\mu }^{\ell }}\psi ^{\ell ,\nu }+\frac{\partial L^{\ell }}{\partial \psi
_{;\mu }^{\ell \ast }}\psi ^{\ell \ast ,\nu }-g^{\mu \nu }L^{\ell },
\label{enmom2} \\
\mathring{\Theta}^{\mu \nu } &=&-\frac{1}{4\pi }F^{\mu \gamma }\partial
^{\nu }A_{\gamma }+g^{\mu \nu }\frac{F^{\xi \gamma }F_{\xi \gamma }}{16\pi }.
\label{enmom2a}
\end{eqnarray}%
The above canonical energy-momenta tensors are neither gauge invariant nor
symmetric. To find a representation for $f^{\mu \gamma \nu }$ in the formula
(\ref{enmom1}) for $\mathcal{T}^{\mu \nu }$ we use the formulas (\ref{ttbe1}%
) noticing that for the scalar fields $\psi ^{\ell }$ and $\psi ^{\ast \ell
} $ we apply the formula (\ref{ttbe3a}), whereas for the vector field $%
A^{\mu } $ we apply the formula (\ref{ttbe3b}). This yields 
\begin{equation}
f^{\mu \gamma \nu }=-\frac{1}{4\pi }F^{\mu \gamma }A^{\nu },  \label{enmom2b}
\end{equation}%
and, consequently%
\begin{gather}
-\partial _{\gamma }\frac{1}{4\pi }f^{\mu \gamma \nu }=\frac{1}{4\pi }%
\partial _{\gamma }\left( F^{\mu \gamma }\right) A^{\nu }+\frac{1}{4\pi }%
F^{\mu \gamma }\partial _{\gamma }A^{\nu }=  \label{enmom2c} \\
=-\frac{1}{\mathrm{c}}J^{\mu }A^{\nu }+\frac{1}{4\pi }F^{\mu \gamma
}\partial _{\gamma }A^{\nu },  \notag
\end{gather}%
where we used the Maxwell equations (\ref{flagr8d}) producing a term with
the current $J^{\mu }=\dsum_{\ell }J^{\ell \mu }$. We introduce now the
following energy-momentum tensors%
\begin{equation}
T^{\ell \mu \nu }=\mathring{T}^{\ell \mu \nu }-\frac{1}{\mathrm{c}}J^{\ell
\mu }A^{\nu },\ \Theta ^{\mu \nu }=\mathring{\Theta}^{\mu \nu }+\frac{1}{%
4\pi }F^{\mu \gamma }\partial _{\gamma }A^{\nu },  \label{enmom2d}
\end{equation}%
and using relations (\ref{flagr12}) and (\ref{enmom2}) we find that they
have the following representations%
\begin{equation}
\Theta ^{\mu \nu }=\frac{1}{4\pi }\left( g^{\mu \gamma }F_{\gamma \xi
}F^{\xi \nu }+\frac{1}{4}g^{\mu \nu }F_{\gamma \xi }F^{\gamma \xi }\right) ,
\label{enmom4}
\end{equation}%
\begin{equation}
T^{\ell \mu \nu }=\frac{\partial L^{\ell }}{\partial \psi _{;\mu }^{\ell }}%
\psi ^{\ell ;\nu }+\frac{\partial L^{\ell }}{\partial \psi _{;\mu }^{\ell
\ast }}\psi ^{\ell ;\nu \ast }-g^{\mu \nu }L^{\ell }.  \label{enmom5}
\end{equation}%
The formula (\ref{enmom4}) is a well known representation (\ref{maxw11}) for
the symmetric and gauge invariant energy-momentum tensor of the EM field
(see \cite[Section 12.10]{Jackson}, \cite[III.3]{Barut}). Notice also that
each tensors $T^{\ell \mu \nu }$ defined by (\ref{enmom5}) is manifestly
gauge invariant. In the case when symmetry condition (\ref{lagsym}) is
satisfied $T^{\ell \mu \nu }$ is also symmetric.

Using (\ref{enmom4}) and (\ref{enmom5}) we define now the total
energy-momentum tensor by%
\begin{equation}
\mathcal{T}^{\mu \nu }=\Theta ^{\mu \nu }+\dsum_{\ell }T^{\ell \mu \nu },
\label{enmom3}
\end{equation}%
and that it is an admissible since it differs from the canonical one by the
divergence $\partial _{\gamma }f^{\mu \gamma \nu }$.

In the case of the Lagrangian of the form (\ref{lagsym1}) in view of (\ref%
{lagsym2}) the energy-momentum expression (\ref{enmom5}) turns into%
\begin{equation}
T^{\ell \mu \nu }=\frac{\partial K^{\ell }}{\partial b}\left( \psi ^{\ell
;\mu \ast }\psi ^{\ell ;\nu }+\psi ^{\ell ;\mu }\psi ^{\ell ;\nu \ast
}\right) -g^{\mu \nu }K^{\ell }.  \label{enmom5a}
\end{equation}%
Consequently, as expected the energy conservation low for the total system
takes the familiar form 
\begin{equation}
\partial _{\mu }\mathcal{T}^{\mu \nu }=\dsum_{\ell }\partial _{\mu }T^{\ell
\mu \nu }+\partial _{\mu }\Theta ^{\mu \nu }=0.  \label{enmom6}
\end{equation}

\subsubsection{Equations for the energy-momentum tensors\label{eqenmom}}

Notice that using the field equations (\ref{flagr8c})-(\ref{flagr8d}) and
the expression (\ref{enmom4}) for the energy-momentum $\Theta ^{\mu \nu }$
of the EM field we get (see details of the derivation in \cite[Section 12.10C%
]{Jackson}) the following equation 
\begin{equation}
\partial _{\mu }\Theta ^{\mu \nu }=-\frac{1}{\mathrm{c}}J_{\mu }F^{\nu \mu },%
\text{ where }J_{\mu }=\dsum_{\ell }J_{\mu }^{\ell }.  \label{divten1}
\end{equation}%
We show below that the above equation for the energy-momentum $\Theta ^{\mu
\nu }$ is complemented by the following equations for the energy-momenta $%
\mathrm{T}^{\ell \mu \nu }$ defined by (\ref{enmom5}) of individual charges%
\begin{equation}
\partial _{\mu }T^{\ell \mu \nu }=\frac{1}{\mathrm{c}}J_{\mu }^{\ell }F^{\nu
\mu }.  \label{divten2}
\end{equation}%
Observe now that in view of the representation (\ref{maxw20}) for the
Lorentz force the following is true: (i) the right-hand side of the equation
(\ref{divten2}) is the Lorentz force exerted by the EM field on the $\ell $%
-th charge; (ii) the right-hand side of the equation (\ref{divten1}) is the
force exerted by all the charges on the EM field and, as one can expect
based on the Third Newton's Law, "every action has an equal and opposite
reaction", this force it is exactly the negative sum of all the Lorentz
forces for involved charges. In fact, based on general consideration of the
equations for energy-momenta as in relations (\ref{emten2})-(\ref{emten6})
we can view the equations (\ref{divten1})-(\ref{divten2}) with involved
Lorentz forces as a continuum version of classical equations of motion. An
important difference though of the equation (\ref{divten1})-(\ref{divten2})
unlike the equations of motion for point particles do not by themselves
determine the dynamics of all involved fields, and, in fact, they hold only
under an assumption that the field equations (\ref{flagr8c})-(\ref{flagr8d})
are satisfied.

To verify the identities (\ref{divten2}) we, following to \cite[Part I,
Section 2]{Pauli RFTh}, introduce a useful computational tool for dealing
with the covariant differentiation operators $\tilde{\partial}^{\mu }$ and $%
\tilde{\partial}^{\mu \ast }$ as defined in (\ref{flagr6b}). Namely, let us
consider a function $f\left( \psi ,\psi _{;\mu },\psi ^{\ast },\psi _{;\mu
}^{\ast }\right) $, where%
\begin{gather}
\psi _{;\mu }=\tilde{\partial}^{\mu }\psi ,\ \psi _{;\mu }^{\ast }=\tilde{%
\partial}^{\mu \ast }\psi ,  \label{gifu1} \\
\tilde{\partial}^{\mu }=\partial ^{\mu }+\frac{\mathrm{i}q}{\chi \mathrm{c}}%
A^{\mu },\ \tilde{\partial}^{\mu \ast }=\partial ^{\mu }-\frac{\mathrm{i}q}{%
\chi \mathrm{c}}A^{\mu },  \notag
\end{gather}%
which is invariant with respect to the gauge transformations of the first
kind (global) as in (\ref{flagr8}):%
\begin{equation}
\psi \rightarrow \mathrm{e}^{\mathrm{i}\gamma }\psi ,\ \psi ^{\ast
}\rightarrow \mathrm{e}^{-\mathrm{i}\gamma }\psi ^{\ast },\text{ where }%
\gamma \text{ is any real constant.}  \label{gifu2}
\end{equation}%
The invariance of $f$ readily implies the following identity%
\begin{gather}
\left. \frac{d}{d\gamma }\left( f\left( \mathrm{e}^{\mathrm{i}\gamma }\psi ,%
\mathrm{e}^{\mathrm{i}\gamma }\psi _{;\mu },\mathrm{e}^{-\mathrm{i}\gamma
}\psi ^{\ast },\mathrm{e}^{-\mathrm{i}\gamma }\psi _{;\mu }^{\ast }\right)
\right) \right\vert _{\gamma =0}=  \label{gifu3} \\
=\frac{\partial f}{\partial \psi }+\frac{\partial f}{\partial \psi _{;\mu }}%
\tilde{\partial}_{\mu }\psi -\frac{\partial f}{\partial \psi ^{\ast }}-\frac{%
\partial f}{\partial \psi _{;\mu }^{\ast }}\tilde{\partial}_{\mu }^{\ast
}\psi ^{\ast }=0.  \notag
\end{gather}%
Observe also that from the definition (\ref{flagr6b}) of the covariant
differentiation operators $\tilde{\partial}^{\mu }$ and $\tilde{\partial}%
^{\mu \ast }$ we have 
\begin{eqnarray}
\tilde{\partial}^{\mu }\tilde{\partial}^{\nu }-\tilde{\partial}^{\nu }\tilde{%
\partial}^{\mu } &=&\frac{\mathrm{i}q}{\chi \mathrm{c}}\left( \partial ^{\mu
}A^{\nu }-\partial ^{\nu }A^{\mu }\right) =\frac{\mathrm{i}q}{\chi \mathrm{c}%
}F^{\mu \nu }  \label{gifu4} \\
\tilde{\partial}^{\mu \ast }\tilde{\partial}^{\nu \ast }-\tilde{\partial}%
^{\nu \ast }\tilde{\partial}^{\mu \ast } &=&-\frac{\mathrm{i}q}{\chi \mathrm{%
c}}\left( \partial ^{\mu }A^{\nu }-\partial ^{\nu }A^{\mu }\right) =-\frac{%
\mathrm{i}q}{\chi \mathrm{c}}F^{\mu \nu }  \notag
\end{eqnarray}%
Now for a gauge invariant $f$ we have%
\begin{eqnarray}
\partial ^{\nu }f &=&\frac{\partial f}{\partial \psi }\partial ^{\nu }\psi +%
\frac{\partial f}{\partial \psi _{;\mu }}\partial ^{\nu }\tilde{\partial}%
_{\mu }\psi +\frac{\partial f}{\partial \psi ^{\ast }}\partial ^{\nu }\psi
^{\ast }+\frac{\partial f}{\partial \psi _{;\mu }^{\ast }}\partial ^{\nu }%
\tilde{\partial}_{\mu }^{\ast }\psi ^{\ast }=  \label{gifu5} \\
&&\frac{\partial f}{\partial \psi }\tilde{\partial}^{\nu }\psi +\frac{%
\partial f}{\partial \psi _{;\mu }}\tilde{\partial}^{\nu }\tilde{\partial}%
_{\mu }\psi +\frac{\partial f}{\partial \psi ^{\ast }}\tilde{\partial}^{\nu
\ast }\psi ^{\ast }+\frac{\partial f}{\partial \psi _{;\mu }^{\ast }}\tilde{%
\partial}^{\nu \ast }\tilde{\partial}_{\mu }^{\ast }\psi ^{\ast }  \notag \\
&&+\frac{\mathrm{i}q}{\chi \mathrm{c}}A^{\nu }\left( -\frac{\partial f}{%
\partial \psi }\psi -\frac{\partial f}{\partial \psi _{;\mu }}\tilde{\partial%
}_{\mu }\psi +\frac{\partial f}{\partial \psi ^{\ast }}\psi ^{\ast }+\frac{%
\partial f}{\partial \psi _{;\mu }^{\ast }}\tilde{\partial}_{\mu }^{\ast
}\psi ^{\ast }\right)  \notag
\end{eqnarray}%
which together with (\ref{gifu3}) implies the following identity%
\begin{equation}
\partial ^{\nu }f=\frac{\partial f}{\partial \psi }\tilde{\partial}^{\nu
}\psi +\frac{\partial f}{\partial \psi _{;\mu }}\tilde{\partial}^{\nu }%
\tilde{\partial}_{\mu }\psi +\frac{\partial f}{\partial \psi ^{\ast }}\tilde{%
\partial}^{\nu \ast }\psi ^{\ast }+\frac{\partial f}{\partial \psi _{;\mu
}^{\ast }}\tilde{\partial}^{\nu \ast }\tilde{\partial}_{\mu }^{\ast }\psi
^{\ast }.  \label{gifu6}
\end{equation}%
With an argument similar to the above on can verify that if $f$ and $g^{\ast
}$ are functions of $\psi $, $\psi ^{\ast }$, $\tilde{\partial}_{\mu }\psi $%
, $\tilde{\partial}_{\mu }^{\ast }\psi ^{\ast }$ which transform under the
gauge transformations (\ref{gifu2}) as $\mathrm{e}^{\mathrm{i}\gamma }f$ and$%
\ \mathrm{e}^{-\mathrm{i}\gamma }g^{\ast }$ then the following identity
holds 
\begin{equation}
\partial ^{\nu }\left( fg^{\ast }\right) =\left( \tilde{\partial}^{\nu
}f\right) g^{\ast }+f\left( \tilde{\partial}^{\nu \ast }g^{\ast }\right) .
\label{gifu7}
\end{equation}%
One can verify that the function $\mathrm{T}^{\ell \mu \nu }$ defined by (%
\ref{enmom5}) is an expression for which the identities (\ref{gifu6}) and (%
\ref{gifu7}) can be applied. Now applying these identities to $\partial
_{\mu }\mathrm{T}^{\ell \mu \nu }$ and using the field equations (\ref%
{flagr8c}) together with identities (\ref{gifu4}) and the representation (%
\ref{flagr12}) for the current $J_{\nu }^{\ell }$ we obtain%
\begin{gather}
\partial _{\mu }T^{\ell \mu \nu }=\partial _{\mu }\left( \frac{\partial
L^{\ell }}{\partial \psi _{;\mu }^{\ell }}\tilde{\partial}^{\ell \nu }\psi
^{\ell }+\frac{\partial L^{\ell }}{\partial \psi _{;\mu }^{\ell \ast }}%
\tilde{\partial}^{\ast \ell \nu }\psi ^{\ell \ast }\right) -\partial ^{\nu
}L^{\ell }=  \label{gifu7a} \\
=\left[ \tilde{\partial}_{\mu }^{\ell \ast }\left( \frac{\partial L^{\ell }}{%
\partial \psi _{;\mu }^{\ell }}\right) \right] \tilde{\partial}^{\ell \nu
}\psi ^{\ell }+\frac{\partial L^{\ell }}{\partial \psi _{;\mu }^{\ell }}%
\tilde{\partial}_{\mu }^{\ell }\tilde{\partial}^{\ell \nu }\psi ^{\ell }+ 
\notag \\
+\left[ \tilde{\partial}_{\mu }^{\ell }\left( \frac{\partial L^{\ell }}{%
\partial \psi _{;\mu }^{\ell \ast }}\right) \right] \tilde{\partial}^{\ell
\nu \ast }\psi ^{\ell \ast }+\frac{\partial L^{\ell }}{\partial \psi _{;\mu
}^{\ell \ast }}\tilde{\partial}_{\mu }^{\ell \ast }\tilde{\partial}^{\ell
\nu \ast }\psi ^{\ell \ast }-  \notag \\
-\left( \frac{\partial L^{\ell }}{\partial \psi ^{\ell }}\tilde{\partial}%
^{\ell \nu }\psi ^{\ell }+\frac{\partial L^{\ell }}{\partial \psi _{;\mu
}^{\ell }}\tilde{\partial}^{\ell \nu }\tilde{\partial}_{\mu }^{\ell }\psi
^{\ell }+\frac{\partial L^{\ell }}{\partial \psi ^{\ell \ast }}\tilde{%
\partial}^{\ell \nu \ast }\psi ^{\ell \ast }+\frac{\partial L^{\ell }}{%
\partial \psi _{;\mu }^{\ell \ast }}\tilde{\partial}^{\ell \nu \ast }\tilde{%
\partial}_{\mu }^{\ell \ast }\psi ^{\ell \ast }\right) =  \notag \\
=\left( \frac{\partial L^{\ell }}{\partial \psi _{;\mu }^{\ell }}\psi ^{\ell
}-\frac{\partial L^{\ell }}{\partial \psi _{;\mu }^{\ell \ast }}\psi ^{\ell
\ast }\right) \frac{\mathrm{i}q}{\chi \mathrm{c}}F_{\mu }^{\ \nu }=-\frac{1}{%
\mathrm{c}}J_{\mu }^{\ell }F^{\mu \nu }=\frac{1}{\mathrm{c}}J_{\mu }^{\ell
}F^{\nu \mu },  \notag
\end{gather}%
which is the desired equation (\ref{divten2}).

Observe that the equation (\ref{divten1})-(\ref{divten2}) for the
energy-momenta are evidently consistent with the total energy conservation (%
\ref{enmom6}).

\subsubsection{Gauge invariant and partially symmetric energy-momentum
tensors}

In our studies, in particular of non relativistic approximations, we have
Lagrangians which are gauge invariant and invariant with respect space and
time translations but they might not be invariant with respect to the entire
Lorentz group of transformations. This subsection is devoted to this kind of
Lagrangians with the main point that essentially all important results of
the subsections \ref{ginvsym} and \ref{eqenmom} apply to them with the only
difference that the energy-momentum tensor is not fully symmetric but
commonly its space part, the stress tensor, is symmetric.

As in the previous subsection we assume the Lagrangian to be of the form
described by formulas (\ref{flagr6a})-(\ref{flagr6b}) and assume it to be
gauge invariant with respect to transformations if the first and the second
type and invariant with respect space and time translations. A careful
analysis of the arguments in subsections \ref{ginvsym} and \ref{eqenmom}
which produced the expressions (\ref{enmom4}) and (\ref{enmom5}) for
respectively energy-momentum $\Theta ^{\mu \nu }$ of the EM field and
energy-momenta $T^{\ell \mu \nu }$ of charges show the same expressions hold
for gauge and translation invariant Lagrangian even if it is not invariant
with respect to the entire Lorentz group of transformation.

We notice first the field equations and expressions for \emph{conserved
currents} are provided by (\ref{flagr8c}), (\ref{flagr8d}), (\ref{flagr9}), (%
\ref{flagr10}), (\ref{flagr12}), namely%
\begin{equation}
\frac{\partial L^{\ell }}{\partial \psi ^{\ell }}-\tilde{\partial}_{\mu
}^{\ell \ast }\left[ \frac{\partial L^{\ell }}{\partial \psi _{;\mu }^{\ell }%
}\right] =0,\ \frac{\partial L^{\ell }}{\partial \psi ^{\ell \ast }}-\tilde{%
\partial}_{\mu }^{\ell }\left[ \frac{\partial L^{\ell }}{\partial \psi
_{;\mu }^{\ell \ast }}\right] =0,\   \label{ginvp1}
\end{equation}%
\begin{equation}
\partial _{\mu }F^{\mu \nu }=\frac{4\pi }{\mathrm{c}}J^{\nu },\text{ }F^{\mu
\nu }=\partial ^{\mu }A^{\nu }-\partial ^{\nu }A^{\mu },\ J^{\nu
}=\dsum_{\ell }J^{\ell \nu },  \label{ginvp2}
\end{equation}%
where $J^{\ell \nu }$ is the four-vector current related to the $\ell $-th
charge is defined by%
\begin{equation}
J^{\ell \nu }=-\mathrm{i}\frac{q^{\ell }}{\chi }\left( \frac{\partial
L^{\ell }}{\partial \psi _{;\nu }^{\ell }}\psi ^{\ell }-\frac{\partial
L^{\ell }}{\partial \psi _{;\nu }^{\ast \ell }}\psi ^{\ast \ell }\right) =-%
\mathrm{c}\frac{\partial L^{\ell }}{\partial A_{\nu }},  \label{ginvp3}
\end{equation}%
or, since $J^{\nu }=\left( \mathrm{c}\rho ,\mathbf{J}\right) $, 
\begin{gather}
\rho ^{\ell }=-\mathrm{i}\frac{q^{\ell }}{\chi }\left( \frac{\partial
L^{\ell }}{\partial \psi _{;0}^{\ell }}\psi ^{\ell }-\frac{\partial L^{\ell }%
}{\partial \psi _{;0}^{\ast \ell }}\psi ^{\ast \ell }\right) ,
\label{ginvp4} \\
\mathbf{J}_{j}^{\ell }=-\mathrm{i}\frac{q^{\ell }}{\chi }\left( \frac{%
\partial L^{\ell }}{\partial \psi _{;j}^{\ell }}\psi ^{\ell }-\frac{\partial
L^{\ell }}{\partial \psi _{;j}^{\ast \ell }}\psi ^{\ast \ell }\right) ,\
j=1,2,3.  \notag
\end{gather}%
Then \emph{we assign} to the energy-momenta of the EM field $\Theta ^{\mu
\nu }$ and the $\ell $-th charge $T^{\ell \mu \nu }$ respectively
expressions (\ref{enmom4}) and (\ref{enmom5}), namely%
\begin{equation}
\Theta ^{\mu \nu }=\frac{1}{4\pi }\left( g^{\mu \gamma }F_{\gamma \xi
}F^{\xi \nu }+\frac{1}{4}g^{\mu \nu }F_{\gamma \xi }F^{\gamma \xi }\right) ,
\label{ginvp5}
\end{equation}%
\begin{equation}
T^{\ell \mu \nu }=\frac{\partial L^{\ell }}{\partial \psi _{;\mu }^{\ell }}%
\psi ^{\ell ;\nu }+\frac{\partial L^{\ell }}{\partial \psi _{;\mu }^{\ell
\ast }}\psi ^{\ell ;\nu \ast }-g^{\mu \nu }L^{\ell },  \label{ginvp6}
\end{equation}%
The above expressions for the energy-momenta are manifestly gauge invariant.

Looking at the arguments in subsections \ref{ginvsym} and \ref{eqenmom} we
compare the above expressions of the energy-momenta of the EM field $\Theta
^{\mu \nu }$ and the $\ell $-th charge $T^{\ell \mu \nu }$ with their
canonical expression and observe that 
\begin{equation}
T^{\ell \mu \nu }=\mathring{T}^{\ell \mu \nu }-\frac{1}{\mathrm{c}}J^{\ell
\mu }A^{\nu },\ \Theta ^{\mu \nu }=\mathring{\Theta}^{\mu \nu }+\frac{1}{%
4\pi }F^{\mu \gamma }\partial _{\gamma }A^{\nu }.  \label{ginvp7}
\end{equation}%
It remains to verify that the difference between the total energy-momentum
and its canonical value is a 4-divergence. Indeed it follows from (\ref%
{ginvp7}) that%
\begin{gather}
\dsum_{\ell }\left( T^{\ell \mu \nu }-\mathring{T}^{\ell \mu \nu }\right)
+\left( \Theta ^{\mu \nu }-\mathring{\Theta}^{\mu \nu }\right) =
\label{ginvp8} \\
=\frac{1}{4\pi }F^{\mu \gamma }\partial _{\gamma }A^{\nu }-\dsum_{\ell }%
\frac{1}{\mathrm{c}}J^{\ell \mu }A^{\nu }=  \notag \\
=\frac{1}{4\pi }F^{\mu \gamma }\partial _{\gamma }A^{\nu }-\frac{1}{\mathrm{c%
}}J^{\mu }A^{\nu }=-\partial _{\gamma }\frac{1}{4\pi }f^{\mu \gamma \nu }, 
\notag \\
\text{where }f^{\mu \gamma \nu }=-\frac{1}{4\pi }F^{\mu \gamma }A^{\nu }. 
\notag
\end{gather}%
Using the arguments of the subsections \ref{eqenmom} we also find that the
relations (\ref{divten1}) and (\ref{divten2}) hold here, namely%
\begin{equation}
\partial _{\mu }\Theta ^{\mu \nu }=-\frac{1}{\mathrm{c}}J_{\mu }F^{\nu \mu },%
\text{ where }J_{\mu }=\dsum_{\ell }J_{\mu }^{\ell },  \label{ginvp9}
\end{equation}%
\begin{equation}
\partial _{\mu }\mathrm{T}^{\ell \mu \nu }=\frac{1}{\mathrm{c}}J_{\mu
}^{\ell }F^{\nu \mu },  \label{ginvp10}
\end{equation}%
where once again we recognize in the right-hand sides of equalities (\ref%
{ginvp9})-(\ref{ginvp10}) the relevant Lorentz force densities.
Consequently, as expected the energy conservation low for the total system
takes the familiar form 
\begin{equation}
\partial _{\mu }\mathcal{T}^{\mu \nu }=\dsum_{\ell }\partial _{\mu }T^{\ell
\mu \nu }+\partial _{\mu }\Theta ^{\mu \nu }=0.  \label{ginvp11}
\end{equation}%
\emph{The equations (\ref{ginvp9})-(\ref{ginvp11}) reconfirm that our
assignment (\ref{ginvp5})-(\ref{ginvp6}) of energy-momenta to the EM field
and individual charges is physically sound}.

We would like to notice now that even if a Lagrangian of the form (\ref%
{flagr6a})-(\ref{flagr6b}) is not invariant with respect the entire Lorentz
group of transformations it often satisfies a reduced version of the
symmetry condition (\ref{lagsym}) which holds for the space indices only,
namely 
\begin{equation}
\frac{\partial L^{\ell }}{\partial \psi _{;i}^{\ell }}\psi ^{\ell ;j}+\frac{%
\partial L^{\ell }}{\partial \psi _{;i}^{\ell \ast }}\psi ^{\ell ;j\ast }=%
\frac{\partial L^{\ell }}{\partial \psi _{;j}^{\ell }}\psi ^{\ell ;i}+\frac{%
\partial L^{\ell }}{\partial \psi _{;j}^{\ell \ast }}\psi ^{\ell ;i\ast },\
i,j=1,2,3.  \label{ginvp12}
\end{equation}%
Under the reduced symmetry condition (\ref{ginvp12}) the space part of the
energy-momenta $\mathrm{T}^{\ell \mu \nu }$, known as the stress tensor, is
symmetric, namely%
\begin{equation}
\mathrm{T}^{\ell ij}=\mathrm{T}^{\ell ji},\ i,j=1,2,3.  \label{ginvp13}
\end{equation}%
We remind that the symmetry of the stress tensor is a very important
property equivalent to the space angular momentum conservation, see Section %
\ref{senergymom} and, for instance, \cite[Section 6.1, 6.2]{Moller}.

\subsection{A single free charge\label{1part}}

A single free charge interacting with the EM field is evidently a particular
case of considered above system of many charges in Section \ref{Manych} and
the Lagrangian (\ref{flagr6a}) takes the form%
\begin{equation}
\mathcal{L}_{0}=L_{0}\left( \psi ,\psi _{;\mu },\psi ^{\ast },\psi _{;\mu
}^{\ast }\right) -\frac{F^{\mu \nu }F_{\mu \nu }}{16\pi },\ F^{\mu \nu
}=\partial ^{\mu }A^{\nu }-\partial ^{\nu }A^{\mu },  \label{frlag1}
\end{equation}%
where $\psi _{;\mu }$ and $\psi _{;\mu }^{\ast }$ are the covariant
derivatives with respect to the covariant differential operators $\tilde{%
\partial}^{\mu }$ and $\tilde{\partial}^{\mu \ast }$ defined by 
\begin{equation}
\psi _{;\mu }=\tilde{\partial}^{\mu }\psi ,\ \psi _{;\mu }^{\ast }=\tilde{%
\partial}^{\mu \ast }\psi ^{\ast },\ \tilde{\partial}^{\mu }=\partial ^{\mu
}+\frac{\mathrm{i}q}{\chi \mathrm{c}}A^{\mu },\ \tilde{\partial}^{\mu \ast
}=\partial ^{\mu }-\frac{\mathrm{i}q}{\chi \mathrm{c}}A^{\mu }.
\label{frlag2}
\end{equation}%
The Lagrangian is assumed to be Lorentz and gauge invariant with respect to
the gauge transformations of the first and the second type (\ref{flagr8})-(%
\ref{flagr8a}). The field equations are%
\begin{equation}
\frac{\partial L_{0}}{\partial \psi }-\tilde{\partial}_{\mu }^{\ast }\left[ 
\frac{\partial L_{0}}{\partial \psi _{;\mu }}\right] =0,\ \frac{\partial
L_{0}}{\partial \psi ^{\ast }}-\tilde{\partial}_{\mu }\left[ \frac{\partial
L_{0}}{\partial \psi _{;\mu }^{\ast }}\right] =0,\   \label{frlag4}
\end{equation}%
\begin{equation}
\partial _{\mu }F^{\mu \nu }=\frac{4\pi }{\mathrm{c}}J^{\nu },\text{ }F^{\mu
\nu }=\partial ^{\mu }A^{\nu }-\partial ^{\nu }A^{\mu },  \label{frlag5}
\end{equation}%
where $J^{\mu }$ is the four-vector micro-current related to the charge is
defined by%
\begin{equation}
J^{\nu }=-\mathrm{i}\frac{q}{\chi }\left( \frac{\partial L_{0}}{\partial
\psi _{;\nu }}\psi -\frac{\partial L_{0}}{\partial \psi _{;\nu }^{\ast }}%
\psi ^{\ast }\right) =-\mathrm{c}\frac{\partial L_{0}}{\partial A_{\nu }}.
\label{frlag6}
\end{equation}%
or, since $J^{\nu }=\left( \mathrm{c}\rho ,\mathbf{J}\right) $,\emph{\ } 
\begin{gather}
\rho =-\mathrm{i}\frac{q}{\chi }\left( \frac{\partial L_{0}}{\partial \psi
_{;0}}\psi -\frac{\partial L_{0}}{\partial \psi _{;0}^{\ast }}\psi ^{\ast
}\right) ,  \label{frlag7} \\
\mathbf{J}_{j}=-\mathrm{i}\frac{q}{\chi }\left( \frac{\partial L_{0}}{%
\partial \psi _{;j}}\psi -\frac{\partial L_{0}}{\partial \psi _{;j}^{\ast }}%
\psi ^{\ast }\right) ,\ j=1,2,3,  \notag
\end{gather}%
which satisfy the charge conservation/continuity equations%
\begin{equation}
\partial _{\nu }J^{\nu }=0,\text{ or }\partial _{t}\rho +\nabla \cdot 
\mathbf{J}=0,\ J^{\nu }=\left( \mathrm{c}\rho ,\mathbf{J}\right) .
\label{frlag8}
\end{equation}%
The energy-momentum of the charge and the EM field according to the formulas
(\ref{enmom4})-(\ref{enmom5}) are respectively as follows 
\begin{eqnarray}
T^{\mu \nu } &=&\frac{\partial L_{0}}{\partial \psi _{;\mu }}\psi ^{;\nu }+%
\frac{\partial L_{0}}{\partial \psi _{;\mu }^{\ast }}\psi ^{;\nu \ast
}-g^{\mu \nu }L_{0},  \label{frlag9} \\
\Theta ^{\mu \nu } &=&\frac{1}{4\pi }\left( g^{\mu \gamma }F_{\gamma \xi
}F^{\xi \nu }+\frac{1}{4}g^{\mu \nu }F_{\gamma \xi }F^{\gamma \xi }\right) ,
\label{frlag10}
\end{eqnarray}%
and the energy conservation equations (\ref{divten1})-(\ref{divten2}) turn
here into%
\begin{equation}
\partial _{\mu }T^{\mu \nu }=\frac{1}{\mathrm{c}}J_{\mu }F^{\nu \mu },\text{ 
}\partial _{\mu }\Theta ^{\mu \nu }=-\frac{1}{\mathrm{c}}J_{\mu }F^{\nu \mu
}.  \label{frlag11}
\end{equation}

\subsection{A single charge in an external electromagnetic field \label%
{1partext}}

Here we consider a single dressed charge in an external EM field. The very
presence of external forces turns the dressed charge into an open system and
that brings up subtleties in the set up of gauge invariant expressions for
the energy-momentum tensor. One can find signs of those subtleties already
in a simple case of a point charge in an external EM. Indeed for the point
charge model the canonical momentum and force are not gauge invariant as
discussed briefly in Section \ref{srelkin}. The principle source of the
problems lies in the openness of the system with consequent uncertainty of
the energy and the momentum as system changes under action of external
forces. That can be seen, in particular, based on the relativity grounds, 
\cite[Section 7.1, 7.2]{Moller}, when seemingly well defined 4-momenta for a
number of open systems do not transform as 4-vectors, that, in general, can
be taken as a proof of openness of a system.

Coming back to our case we want, first of all, to define a Lagrangian for a
dressed charge in EM field based on (i) our studies in Section \ref{Manych}
of a closed system of many dressed charges and (ii) the Lagrangian of a
singe free charge considered in Section \ref{1part}. We do that by altering
the EM potential $A^{\mu }$ in the expressions (\ref{flagr1})-(\ref{flagr2})
for the Lagrangian of the free single dressed charge with $\bar{A}^{\mu
}=A^{\mu }+A_{\mathrm{ex}}^{\mu }$, where $A_{\mathrm{ex}}^{\mu }$ is the
4-potential of an the external EM field. Namely, we set%
\begin{equation}
L_{0}=L_{0}\left( \psi ,\psi _{;\mu },\psi ^{\ast },\psi _{;\mu }^{\ast
}\right) -\frac{F^{\mu \nu }F_{\mu \nu }}{16\pi },\ F^{\mu \nu }=\partial
^{\mu }A^{\nu }-\partial ^{\nu }A^{\mu },  \label{lagco1}
\end{equation}%
where $\psi _{;\mu }$ and $\psi _{;\mu }^{\ast }$ are the covariant
derivatives with respect to the covariant differential operators $\tilde{%
\partial}^{\mu }$ and $\tilde{\partial}^{\mu \ast }$ defined by 
\begin{gather}
\psi _{;\mu }=\tilde{\partial}^{\mu }\psi ,\ \psi _{;\mu }^{\ast }=\tilde{%
\partial}^{\mu \ast }\psi ^{\ast },  \label{lagco2} \\
\tilde{\partial}^{\mu }=\partial ^{\mu }+\frac{\mathrm{i}q}{\chi \mathrm{c}}%
\bar{A}^{\mu },\ \tilde{\partial}^{\mu \ast }=\partial ^{\mu }-\frac{\mathrm{%
i}q}{\chi \mathrm{c}}\bar{A}^{\mu },\ \bar{A}^{\mu }=A^{\mu }+A_{\mathrm{ex}%
}^{\mu }.  \notag
\end{gather}%
To justify the expressions (\ref{lagco1})-(\ref{lagco2}) for the Lagrangian
let us look at a closed system of many charges studied in Section \ref%
{Manych}. We find there, in particular, that every individual charge with an
index $\ell $ has a conserved current $J^{\ell \nu }$ and the total current
is $J_{\mu }=\dsum_{\ell }J^{\ell \nu }$. Hence, based on the linearity of
Maxwell equation (\ref{flagr8d}) we can introduce individual EM potentials $%
A^{\ell \mu }$ and the corresponding EM fields $F^{\ell \mu \nu }$ as the
causal solutions to the following Maxwell equations 
\begin{equation}
\partial _{\mu }F^{\ell \mu \nu }=\frac{4\pi }{\mathrm{c}}J^{\ell \nu },
\label{scem1}
\end{equation}%
implying%
\begin{equation}
A^{\mu }=\dsum_{\ell }A^{\ell \mu },\ F^{\mu \nu }=\dsum_{\ell }F^{\ell \mu
\nu }.  \label{scem2}
\end{equation}%
Notice that every individual charge satisfies its field equation (\ref%
{flagr8c}) with the EM field entering it via the potential $A^{\mu }$ in the
covariant derivatives (\ref{gifu1}), and we can always represent it as%
\begin{equation}
A^{\mu }=A^{\ell \mu }+A_{\mathrm{ex}}^{\ell \mu },\ A_{\mathrm{ex}}^{\ell
\mu }=\dsum_{\ell ^{\prime }\neq \ell }A^{\ell ^{\prime }\mu }.
\label{scem3}
\end{equation}%
This representation indicates that we can account for the interaction of the 
$\ell $-th charge with remaining charges via an external field as in (\ref%
{scem3}) justifying the expressions (\ref{lagco1})-(\ref{lagco2}) for the
Lagrangian $L_{0}$.

The Lagrangian (\ref{lagco1})-(\ref{lagco2}) is assumed to be invariant with
respect to the first and the second type gauge transformations (\ref{flagr8}%
), (\ref{flagr8a}), which in this case take the form 
\begin{equation}
\psi \rightarrow \mathrm{e}^{\mathrm{i}\gamma }\psi ^{\ell },\ \psi ^{\ast
}\rightarrow \mathrm{e}^{-\mathrm{i}\gamma }\psi ^{\ast },\text{ where }%
\gamma \text{ is any real constant,}  \label{lagco3}
\end{equation}%
\begin{equation}
\psi \rightarrow \mathrm{e}^{-\frac{\mathrm{i}e\lambda \left( x\right) }{%
\chi \mathrm{c}}}\psi ,\ A^{\mu }\rightarrow A^{\mu }+\partial ^{\mu
}\lambda .  \label{lagco4}
\end{equation}%
Similarly to the case of many charges we also assume the charge Lagrangian $%
L_{0}$ to satisfy the following symmetry condition%
\begin{equation}
\frac{\partial L_{0}}{\partial \psi _{;\mu }}\psi ^{;\nu }+\frac{\partial
L_{0}}{\partial \psi _{;\mu }^{\ast }}\psi ^{;\nu \ast }=\frac{\partial L_{0}%
}{\partial \psi _{;\nu }}\psi ^{;\mu }+\frac{\partial L_{0}}{\partial \psi
_{;\nu }^{\ast }}\psi ^{;\mu \ast }.  \label{lagco5}
\end{equation}%
As in already considered case of many charges there is a simple sufficient
condition for the symmetry condition (\ref{lagco5}) to hold. It is when the
Lagrangians $L$ depends on the field covariant derivatives only through the
combination $\psi ^{\ell ;\mu }\psi _{;\mu }^{\ell \ast }$, in other words
if there exist such functions $K\left( \psi ,\psi ^{\ast },b\right) $ that%
\begin{equation}
L_{0}\left( \psi ,\psi _{;\mu }^{\ell },\psi ^{\ast },\psi _{;\mu }^{\ell
\ast }\right) =K\left( \psi ,\psi ^{\ast },\psi ^{;\mu }\psi _{;\mu }^{\ast
}\right) .  \label{lagco6}
\end{equation}

The field equations for the Lagrangian $L_{0}$ defined by (\ref{lagco1})-(%
\ref{lagco2}) are%
\begin{equation}
\frac{\partial L_{0}}{\partial \psi }-\tilde{\partial}_{\mu }^{\ast }\left[ 
\frac{\partial L_{0}}{\partial \psi _{;\mu }}\right] =0,\ \frac{\partial
L_{0}}{\partial \psi ^{\ast }}-\tilde{\partial}_{\mu }\left[ \frac{\partial
L_{0}}{\partial \psi _{;\mu }^{\ast }}\right] =0,  \label{lagco7}
\end{equation}%
\begin{equation}
\partial _{\mu }F^{\mu \nu }=\frac{4\pi }{\mathrm{c}}J^{\nu },\text{ }F^{\mu
\nu }=\partial ^{\mu }A^{\nu }-\partial ^{\nu }A^{\mu },  \label{lagco8}
\end{equation}%
where $J^{\mu }$ is the four-vector current related to the charge is defined
by manifestly gauge invariant expression%
\begin{equation}
J^{\nu }=-\mathrm{i}\frac{q}{\chi }\left( \frac{\partial L_{0}}{\partial
\psi _{;\nu }}\psi -\frac{\partial L_{0}}{\partial \psi _{;\nu }^{\ast }}%
\psi ^{\ast }\right) ,  \label{lagco9}
\end{equation}%
or, since $J^{\nu }=\left( \mathrm{c}\rho ,\mathbf{J}\right) $, the 
\begin{gather}
\rho =-\mathrm{i}\frac{q}{\chi }\left( \frac{\partial L_{0}}{\partial \psi
_{;0}}\psi -\frac{\partial L_{0}}{\partial \psi _{;0}^{\ast }}\psi ^{\ast
}\right) ,  \label{lagco10} \\
\mathbf{J}_{j}=-\mathrm{i}\frac{q}{\chi }\left( \frac{\partial L_{0}}{%
\partial \psi _{;j}}\psi -\frac{\partial L_{0}}{\partial \psi _{;j}^{\ast }}%
\psi ^{\ast }\right) ,\ j=1,2,3,  \notag
\end{gather}%
which satisfy the charge conservation/continuity equations%
\begin{equation}
\partial _{\nu }J^{\nu }=0,\text{ or }\partial _{t}\rho +\nabla \cdot 
\mathbf{J}=0,\ J^{\nu }=\left( \mathrm{c}\rho ,\mathbf{J}\right) .
\label{lagco11}
\end{equation}%
Notice as in the case of many charges the relations (\ref{lagco1})-(\ref%
{lagco2}) imply the following alternative representation for the
four-current $J^{\nu }$ defined by (\ref{lagco9}) 
\begin{gather}
J^{\nu }=-\mathrm{i}\frac{q}{\chi }\left( \frac{\partial L_{0}}{\partial
\psi _{;\nu }}\psi -\frac{\partial L_{0}}{\partial \psi _{;\nu }^{\ast }}%
\psi ^{\ast }\right) =  \label{lagco12} \\
=-\mathrm{c}\frac{\partial L_{0}}{\partial A_{\nu }}=-\mathrm{c}\frac{%
\partial L_{0}}{\partial \bar{A}_{\nu }}=-\mathrm{c}\frac{\partial L_{0}}{%
\partial A_{\mathrm{ex}\nu }}.  \notag
\end{gather}

Now we would like to construct gauge invariant energy-momentum tensors for
the single charge and its EM field. For that we start with their canonical
expressions (\ref{flagr6})-(\ref{flagr7b}) obtained via Noether's theorem%
\begin{gather}
\mathring{T}^{\mu \nu }=\frac{\partial L_{0}}{\partial \psi _{;\mu }}\psi
^{,\nu }+\frac{\partial L_{0}}{\partial \psi _{;\mu }^{\ast }}\psi ^{;\nu
\ast }-g^{\mu \nu }L_{0},  \label{tex1} \\
\mathring{\Theta}^{\mu \nu }=-\frac{F^{\mu \gamma }\partial ^{\nu }A_{\gamma
}}{4\pi }+g^{\mu \nu }\frac{F^{\xi \gamma }F_{\xi \gamma }}{16\pi }.
\label{tex2}
\end{gather}%
The conservation law for the total energy-momentum tensor $\mathring{T}^{\mu
\nu }+\mathring{\Theta}^{\mu \nu }$ in view of the general conservation law (%
\ref{flagr5e}) and the current representation (\ref{lagco12}) take the form 
\begin{equation}
\partial _{\mu }\left( \mathring{T}^{\mu \nu }+\mathring{\Theta}^{\mu \nu
}\right) =-\frac{\partial L_{0}}{\partial x_{\nu }}=-\frac{\partial L_{0}}{%
\partial A_{\mathrm{ex}\mu }}\partial ^{\nu }A_{\mu \mathrm{ex}}=\frac{1}{%
\mathrm{c}}J^{\mu }\partial ^{\nu }A_{\mu \mathrm{ex}}.  \label{tex3}
\end{equation}%
Observe now that the both canonical expressions (\ref{tex1}), (\ref{tex2})
as well as the density of the generalized force $\frac{1}{\mathrm{c}}J^{\mu
}\partial ^{\nu }A_{\mu \mathrm{ex}}$ in (\ref{tex3}) are evidently not
gauge invariant, and this is very similar to what we already observed for
the point charge model in Section \ref{srelkin}. We recall that there is a
general way to alter the canonical energy-momentum tensor as in the relation
(\ref{emten1}), namely%
\begin{equation}
T^{\mu \nu }+\Theta ^{\mu \nu }=\mathring{T}^{\mu \nu }+\mathring{\Theta}%
^{\mu \nu }-\partial _{\gamma }f^{\mu \gamma \nu },\ f^{\mu \gamma \nu
}=-f^{\gamma \mu \nu }.  \label{tex4}
\end{equation}%
But any such alteration alone can not be satisfactory since by its very
construction it would keep unchanged the not gauge invariant density of
generalized force $\frac{1}{\mathrm{c}}J^{\mu }\partial ^{\nu }A_{\mu 
\mathrm{ex}}$ in the right-hand side of (\ref{tex3}). Therefore, a more
profound alteration of the energy-momenta is required that would change the
expression for the force density in the right-hand side of (\ref{tex3}) so
that it becomes gauge invariant. More than that we expect it to produce
exactly the density of the Lorentz force associated with the external EM
potential $A_{\mathrm{ex}}^{\mu }$.

The results of Section \ref{ginvsym} suggest a satisfactory choice for gauge
invariant energy-momentum tensors and it is as in formulas (\ref{enmom2})-(%
\ref{enmom2a}), namely we set%
\begin{gather}
T^{\mu \nu }=\frac{\partial L_{0}}{\partial \psi _{;\mu }}\psi ^{;\nu }+%
\frac{\partial L_{0}}{\partial \psi _{;\mu }^{\ast }}\psi ^{;\nu \ast
}-g^{\mu \nu }L_{0},  \label{tsing1} \\
\Theta ^{\mu \nu }=\frac{1}{4\pi }\left( g^{\mu \gamma }F_{\gamma \xi
}F^{\xi \nu }+\frac{1}{4}g^{\mu \nu }F_{\gamma \xi }F^{\gamma \xi }\right) .
\label{tsing2}
\end{gather}%
Observe that the EM energy-momentum $\Theta ^{\mu \nu }$ is also manifestly
symmetric and the charge energy-momentum $T^{\mu \nu }$ is symmetric if the
the symmetry condition (\ref{lagco5}) is satisfied. The energy-momenta
conservation laws here take the form 
\begin{equation}
\partial _{\mu }\Theta ^{\mu \nu }=\partial _{\mu }\frac{1}{4\pi }\left(
g^{\mu \gamma }F_{\gamma \xi }F^{\xi \nu }+\frac{1}{4}g^{\mu \nu }F_{\gamma
\xi }F^{\gamma \xi }\right) =-\frac{1}{\mathrm{c}}J_{\mu }F^{\nu \mu },
\label{tsing3}
\end{equation}%
\begin{equation}
\partial _{\mu }T^{\mu \nu }=\frac{1}{\mathrm{c}}J_{\mu }\bar{F}^{\nu \mu
},\ \bar{F}^{\nu \mu }=\partial ^{\mu }\bar{A}^{\nu }-\partial ^{\nu }\bar{A}%
^{\mu }.  \label{tsing4}
\end{equation}%
Indeed as in the case of derivation of the similar formula (\ref{divten2})
we observe that the $T^{\mu \nu }$ defined by (\ref{tsing1}) has an
expression for which the identities (\ref{gifu6}) and (\ref{gifu7}) can be
applied. Now we literally repeat the calculation (\ref{gifu7a}). Namely
applying the mentioned identities to $\partial _{\mu }T^{\mu \nu }$ and
using the field equations (\ref{lagco7}) together with identities (\ref%
{gifu4}) and the representation (\ref{flagr12}) for the current $J_{\nu }$
we obtain%
\begin{gather}
\partial _{\mu }T^{\mu \nu }=\partial _{\mu }\left( \frac{\partial L_{0}}{%
\partial \psi _{;\mu }}\tilde{\partial}^{\nu }\psi +\frac{\partial L_{0}}{%
\partial \psi _{;\mu }^{\ast }}\tilde{\partial}^{\ast \nu }\psi ^{\ast
}\right) -\partial ^{\nu }L_{0}=  \label{tsing5} \\
=\left[ \tilde{\partial}_{\mu }^{\ast }\left( \frac{\partial L_{0}}{\partial
\psi _{;\mu }}\right) \right] \tilde{\partial}^{\nu }\psi +\frac{\partial L}{%
\partial \psi _{;\mu }}\tilde{\partial}_{\mu }\tilde{\partial}^{\nu }\psi + 
\notag \\
+\left[ \tilde{\partial}_{\mu }\left( \frac{\partial L_{0}}{\partial \psi
_{;\mu }^{\ast }}\right) \right] \tilde{\partial}^{\nu \ast }\psi ^{\ast }+%
\frac{\partial L_{0}}{\partial \psi _{;\mu }^{\ast }}\tilde{\partial}_{\mu
}^{\ast }\tilde{\partial}^{\nu \ast }\psi ^{\ast }-  \notag \\
-\left( \frac{\partial L_{0}}{\partial \psi }\tilde{\partial}^{\nu }\psi +%
\frac{\partial L_{0}}{\partial \psi _{;\mu }}\tilde{\partial}^{\nu }\tilde{%
\partial}_{\mu }\psi +\frac{\partial L_{0}}{\partial \psi ^{\ast }}\tilde{%
\partial}^{\nu \ast }\psi ^{\ast }+\frac{\partial L_{0}}{\partial \psi
_{;\mu }^{\ast }}\tilde{\partial}^{\nu \ast }\tilde{\partial}_{\mu }^{\ast
}\psi ^{\ast }\right) =  \notag \\
=\left( \frac{\partial L_{0}}{\partial \psi _{;\mu }}\psi -\frac{\partial
L_{0}}{\partial \psi _{;\mu }^{\ell \ast }}\psi ^{\ast }\right) \frac{%
\mathrm{i}q}{\chi \mathrm{c}}\bar{F}_{\mu }^{\ \nu }=-\frac{1}{\mathrm{c}}%
J_{\mu }\bar{F}^{\mu \nu }=\frac{1}{\mathrm{c}}J_{\mu }\bar{F}^{\nu \mu }, 
\notag
\end{gather}%
implying the desired relation (\ref{tsing4}).

Now adding up the equalities (\ref{tsing3}) and (\ref{tsing4}) we get the
conservation law for the total energy-momentum $\mathcal{T}^{\mu \nu
}=T^{\mu \nu }+\Theta ^{\mu \nu }$, i.e. 
\begin{equation}
\partial _{\mu }\mathcal{T}^{\mu \nu }=\partial _{\mu }\left( T_{0}^{\mu \nu
}+\Theta ^{\mu \nu }\right) =\frac{1}{\mathrm{c}}J_{\nu }F_{\mathrm{ex}%
}^{\nu \mu },\ F_{\mathrm{ex}}^{\nu \mu }=\partial ^{\mu }A_{\mathrm{ex}%
}^{\nu }-\partial ^{\nu }A_{\mathrm{ex}}^{\mu }.  \label{tsing6}
\end{equation}%
Notice as we expected we have the density of the Lorentz force $\frac{1}{%
\mathrm{c}}J_{\nu }F_{\mathrm{ex}}^{\nu \mu }$ in the right-hand side of the
conservation laws (\ref{tsing6}) and (\ref{tsing4}).

\subsection{Energy partition for static and time harmonic fields\label%
{senrgypart}}

Let us consider the Lagrangian of the form%
\begin{equation}
\mathcal{L}=\mathcal{L}\left( \left\{ \psi ^{\ell },\psi _{,\mu }^{\ell
},\psi ^{\ell \ast },\psi _{,\mu }^{\ell \ast },V^{g},V_{,\mu }^{g}\right\}
\right) ,  \label{epar1}
\end{equation}%
where $\left\{ V^{g}\right\} $ are real-valued quantities. The corresponding
Euler-Lagrange field equations are%
\begin{gather}
\frac{\partial \mathcal{L}}{\partial \psi ^{\ell }}-\partial _{\mu }\left( 
\frac{\partial \mathcal{L}}{\partial \psi _{,\mu }^{\ell }}\right) =0,
\label{epar2} \\
\frac{\partial \mathcal{L}}{\partial \psi ^{\ell \ast }}-\partial _{\mu
}\left( \frac{\partial \mathcal{L}}{\partial \psi _{,\mu }^{\ell \ast }}%
\right) =0,\ \frac{\partial \mathcal{L}}{\partial V^{g}}-\partial _{\mu
}\left( \frac{\partial \mathcal{L}}{\partial V_{,\mu }^{g}}\right) =0. 
\notag
\end{gather}%
Static regime is characterized as one when the fields $\left\{ \psi ^{\ell
}\right\} $, $\left\{ \psi ^{\ell \ast }\right\} $ and $\left\{
V^{g}\right\} $ are time independent and hence depend only on the space
variable, and we will use the following abbreviated notation for it%
\begin{equation}
\mathrm{stat}\equiv \partial _{t}\psi ^{\ell }=0,\ \partial _{t}\psi ^{\ell
\ast }=0,\ \partial _{t}V^{g}=0.  \label{epar2aa}
\end{equation}%
Then using the canonical energy-momentum $\mathcal{\mathring{T}}^{\mu \nu }$
as defined by (\ref{flagr4a}) we readily obtain the following formulas for
the total energy $\mathcal{E}_{\mathrm{stat}}$ of static field 
\begin{equation}
\mathcal{E}_{\mathrm{stat}}=\int_{\mathbb{R}^{3}}\mathcal{\mathring{U}}\,%
\mathrm{d}x,\text{ where }\mathcal{\mathring{U}}=\left. \mathcal{\mathring{T}%
}^{00}\left( \left\{ \psi ^{\ell },\psi _{,\mu }^{\ell },\psi ^{\ell \ast
},\psi _{,\mu }^{\ell \ast },V^{g},V_{,\mu }^{g}\right\} \right) \right\vert
_{\mathrm{stat}}.  \label{epar2a}
\end{equation}%
The energy $\mathcal{E}_{\mathrm{stat}}$ in a static regime can be
identified with a potential energy. Notice that in view of the formula (\ref%
{flagr4a}) for the canonical energy-momentum $\mathcal{\mathring{T}}^{\mu
\nu }$ the corresponding energy density $\left. \mathcal{\mathring{T}}%
^{00}\right\vert _{\mathrm{stat}}$ is represented by 
\begin{equation}
\mathcal{\mathring{U}}=\left. \mathcal{\mathring{T}}^{00}\right\vert _{%
\mathrm{stat}}=-\left. \mathcal{L}\left( \left\{ \psi ^{\ell },\psi _{,\mu
}^{\ell },\psi ^{\ell \ast },\psi _{,\mu }^{\ell \ast },V^{g},V_{,\mu
}^{g}\right\} \right) \right\vert _{\mathrm{stat}}.  \label{epar3}
\end{equation}%
Consequently, we have the following representation for the potential energy $%
\mathcal{E}_{\mathrm{stat}}$%
\begin{equation}
\mathcal{E}_{\mathrm{stat}}=\int_{\mathbb{R}^{3}}-\left. \mathcal{L}\left(
\left\{ \psi ^{\ell },\psi _{,\mu }^{\ell },\psi ^{\ell \ast },\psi _{,\mu
}^{\ell \ast },V^{g},V_{,\mu }^{g}\right\} \right) \right\vert _{\mathrm{stat%
}}\,\mathrm{d}x.  \label{epar3b}
\end{equation}%
which we use to establish the following variational principle. Based (\ref%
{epar3}) we can conclude that a static solution $\left\{ \psi ^{\ell },\psi
^{\ell \ast },V^{g}\right\} $ to the Euler-Lagrange field equations (\ref%
{epar2}) evidently transforms into a solution to the equation%
\begin{gather}
\frac{\partial \mathcal{\mathring{U}}}{\partial \psi ^{\ell }}%
-\dsum\limits_{j=1,2,3}\partial _{x_{j}}\left( \frac{\partial \mathcal{%
\mathring{U}}}{\partial \psi _{,j}^{\ell }}\right) =0,\ \frac{\partial 
\mathcal{\mathring{U}}}{\partial \psi ^{\ell \ast }}-\dsum\limits_{j=1,2,3}%
\partial _{x_{j}}\left( \frac{\partial \mathcal{\mathring{U}}}{\partial \psi
_{,j}^{\ell \ast }}\right) =0,  \label{epar4} \\
\frac{\partial \mathcal{\mathring{U}}}{\partial V^{g}}-\dsum%
\limits_{j=1,2,3}\partial _{x_{j}}\left( \frac{\partial \mathcal{\mathring{U}%
}}{\partial V_{,j}^{g}}\right) =0,  \notag
\end{gather}%
and, hence, in view of the representation (\ref{epar3b}) it is a stationary
point of the static energy functional $\mathcal{E}_{\mathrm{stat}}$ in the
complete agreement with the \emph{principle of virtual work} for the state
of equilibrium, \cite[Section III.1]{Lanczos VPM}, \cite[Section II.8]%
{Sommerfeld M}.

Let us expand now the potential energy density $\mathcal{\mathring{U}}$
defined by (\ref{epar3}) into the series with respect to the derivatives $%
\nabla \psi ^{\ell }$, $\nabla \psi ^{\ell \ast }$, $\nabla V^{g}$, namely 
\begin{gather}
\mathcal{\mathring{U}}=\sum_{n=0}^{\infty }\mathcal{\mathring{U}}^{\left(
n\right) },\text{ where}  \label{epar6} \\
\mathcal{\mathring{U}}^{\left( n\right) }=\sum_{\sum n_{\ell }+n_{\ell
}^{\ast }+n_{g}=n}\mathcal{\mathring{U}}_{\left\{ n_{\ell },n_{\ell }^{\ast
},n_{g}\right\} }\left( \left\{ \psi ^{\ell },\psi ^{\ell \ast
},V^{g}\right\} \right) \dprod\limits_{\ell ,g}\partial _{j_{\ell
}}^{n_{\ell }}\psi ^{\ell }\partial _{j_{\ell }^{\ast }}^{n_{\ell }^{\ast
}}\psi ^{\ast \ell }\partial _{i_{g}}^{n_{g}}V^{g}.  \notag
\end{gather}%
This expansion via the representation (\ref{epar3b}) for the potential
energy $\mathcal{E}_{\mathrm{stat}}$ readily implies the the corresponding
expansion for $\mathcal{E}_{\mathrm{stat}}$:%
\begin{equation}
\mathcal{E}_{\mathrm{stat}}=\sum_{n=0}^{\infty }\mathcal{E}_{\mathrm{stat}%
}^{\left( n\right) },\text{ where }\mathcal{E}_{\mathrm{stat}}^{\left(
n\right) }=\int_{\mathbb{R}^{3}}\mathcal{\mathring{U}}^{\left( n\right) }\,%
\mathrm{d}x.  \label{epar7}
\end{equation}%
Now being given a static solution $\left\{ \psi ^{\ell },\psi ^{\ell \ast
},V^{g}\right\} $ we use its established above property to be a stationary
point of the functional $\mathcal{E}_{\mathrm{stat}}$ as defined by formula (%
\ref{epar3b}) and (\ref{epar7}). Namely, we introduce the following family
of fields 
\begin{gather}
\psi _{\xi }^{\ell }\left( x\right) =\psi ^{\ell }\left( \xi x\right) ,\
\psi _{\xi }^{\ast \ell }\left( x\right) =\psi ^{\ast \ell }\left( \xi
x\right) ,  \label{epar8} \\
V_{\xi }^{g}\left( x\right) =V^{g}\left( \xi x\right) \text{ where }\xi 
\text{ is real,}  \notag
\end{gather}%
and observe that since $\left\{ \psi ^{\ell },\psi ^{\ell \ast
},V^{g}\right\} $ is a stationary point of the functional $\mathcal{E}_{%
\mathrm{stat}}$ we have 
\begin{gather}
\left. \frac{d}{d\xi }\mathcal{E}_{\mathrm{stat}}\left( \left\{ \psi _{\xi
}^{\ell },\psi _{\xi }^{\ell \ast },V_{\xi }^{g},\nabla \psi _{\xi }^{\ell
},\nabla \psi _{\xi }^{\ell \ast },\nabla V_{\xi }^{g}\right\} \right)
\right\vert _{\xi =1}=\left. \sum_{n=0}^{\infty }\xi ^{n-3}\mathcal{E}_{%
\mathrm{stat}}^{\left( n\right) }\right\vert _{\xi =1}  \label{epar9} \\
=\sum_{n=0}^{\infty }\left( n-3\right) \mathcal{E}_{\mathrm{stat}}^{\left(
n\right) }\left( \left\{ \psi ^{\ell },\psi ^{\ell \ast },V^{g},\nabla \psi
^{\ell },\nabla \psi ^{\ell \ast },\nabla V^{g}\right\} \right) =0  \notag
\end{gather}%
In other words, for a static solution its energy components $\mathcal{E}_{%
\mathrm{stat}}^{\left( n\right) }$ always satisfy the identity 
\begin{equation}
\sum_{n=0}^{\infty }\left( n-3\right) \mathcal{E}_{\mathrm{stat}}^{\left(
n\right) }=0.  \label{epar10}
\end{equation}%
Very often the density $\mathcal{\mathring{U}}$ of the potential energy
depends on the field derivatives so that 
\begin{gather}
\mathcal{\mathring{U}}\left( \left\{ \psi ^{\ell },\psi ^{\ell \ast
},V^{g},\nabla \psi ^{\ell },\nabla \psi ^{\ell \ast },\nabla V^{g}\right\}
\right) =  \label{epar11} \\
=\mathcal{\mathring{U}}^{\left( 2\right) }\left( \left\{ \psi ^{\ell },\psi
^{\ell \ast },V^{g},\nabla \psi ^{\ell },\nabla \psi ^{\ell \ast },\nabla
V^{g}\right\} \right) +\mathcal{\mathring{U}}^{\left( 0\right) }\left(
\left\{ \psi ^{\ell },\psi ^{\ell \ast },V^{g}\right\} \right) .  \notag
\end{gather}%
where $U^{\left( 2\right) }$ satisfies the following identity for any real $%
\theta $%
\begin{gather}
\mathcal{\mathring{U}}^{\left( 2\right) }\left( \left\{ \psi ^{\ell },\psi
^{\ell \ast },V^{g},\theta \nabla \psi ^{\ell },\theta \nabla \psi ^{\ell
\ast },\theta \nabla V^{g}\right\} \right) =  \label{epar11a} \\
=\theta ^{2}\mathcal{\mathring{U}}^{\left( 2\right) }\left( \left\{ \psi
^{\ell },\psi ^{\ell \ast },V^{g},\nabla \psi ^{\ell },\nabla \psi ^{\ell
\ast },\nabla V^{g}\right\} \right) .  \notag
\end{gather}%
In this case the identity (\ref{epar10}) turns into the following important
identity for the two constituting components $\mathcal{E}_{\mathrm{stat}%
}^{\left( 2\right) }$ and $\mathcal{E}_{\mathrm{stat}}^{\left( 0\right) }$
of the total potential energy $\mathcal{E}_{\mathrm{stat}}$: 
\begin{equation}
\mathcal{E}_{\mathrm{stat}}=\mathcal{E}_{\mathrm{stat}}^{\left( 2\right) }+%
\mathcal{E}_{\mathrm{stat}}^{\left( 0\right) },\ \mathcal{E}_{\mathrm{stat}%
}^{\left( 0\right) }=-\frac{1}{3}\mathcal{E}_{\mathrm{stat}}^{\left(
2\right) }\text{ implying }\mathcal{E}_{\mathrm{stat}}=\frac{2}{3}\mathcal{E}%
_{\mathrm{stat}}^{\left( 2\right) }.  \label{epar12}
\end{equation}%
The significance of the above identity for our goals is that in the cases of
interest the energy component $\mathcal{E}_{\mathrm{stat}}^{\left( 0\right)
} $ accounts for the energy of nonlinear self-interactions and the formula $%
\mathcal{E}_{\mathrm{stat}}=\frac{2}{3}\mathcal{E}_{\mathrm{stat}}^{\left(
2\right) }$ shows the \emph{total energy has a representation that does not
depend explicitly on the nonlinear self-interactions}. \emph{This is one
among other properties allowing us to characterize the introduced nonlinear
self-interactions as stealthy}.

The identity (\ref{epar12}) for a single field is known as the \emph{%
Pokhozhaev-Derrick identity}, \cite{Pokhozhaev}, \cite{Derrick} (see also 
\cite{Kapitanskii} and \cite[Section 2.4]{Coleman 1}). It was often used to
prove the nonexistence of nonzero solutions to the corresponding field
equations in situations when a priory the both energies $\mathcal{E}_{%
\mathrm{stat}}^{\left( 2\right) }$ and $\mathcal{E}_{\mathrm{stat}}^{\left(
0\right) }$ are nonnegative and vanish for the zero field. Indeed if the
nonnegativity of the energy components is combined with the identity (\ref%
{epar12}) the both energies $\mathcal{E}_{\mathrm{stat}}^{\left( 2\right) }$
and $\mathcal{E}_{\mathrm{stat}}^{\left( 0\right) }$ must vanish implying
that the field must vanish as well.

\subsubsection{Time-harmonic fields\label{stharm}}

The above statements for static fields can be generalized for the case when
complex valued fields $\psi ^{\ell }$ are time harmonic, namely when%
\begin{equation}
\psi ^{\ell }=\mathrm{e}^{-\mathrm{i}\omega ^{\ell }t}\tilde{\psi}^{\ell },\
\psi ^{\ell \ast }=\mathrm{e}^{\mathrm{i}\omega ^{\ell }t}\tilde{\psi}^{\ell
\ast },  \label{thar1}
\end{equation}%
where $\tilde{\psi}^{\ell }$ and $\tilde{\psi}^{\ell \ast }$ are static,
i.e. time independent. We provide such a generalization for the Lagrangian
of the form (\ref{flagr6a})-(\ref{flagr6b}) describing many charges coupled
with the EM field $F^{\mu \nu }$, i.e.%
\begin{equation}
\mathcal{L}\left( \left\{ \psi ^{\ell },\psi _{;\mu }^{\ell },\psi ^{\ell
\ast },\psi _{;\mu }^{\ell \ast }\right\} ,A^{\mu }\right) =\dsum_{\ell
}L^{\ell }\left( \psi ^{\ell },\psi _{;\mu }^{\ell },\psi ^{\ell \ast },\psi
_{;\mu }^{\ell \ast }\right) -\frac{F^{\mu \nu }F_{\mu \nu }}{16\pi },
\label{thar1a}
\end{equation}%
with an additional assumption on the charge Lagrangians $L^{\ell }$ to be of
the form%
\begin{equation}
L^{\ell }\left( \psi ^{\ell },\psi _{;\mu }^{\ell },\psi ^{\ell \ast },\psi
_{;\mu }^{\ell \ast }\right) =K^{\ell }\left( \psi ^{\ell \ast }\psi ^{\ell
},\psi _{;\mu }^{\ell \ast }\psi ^{\ell ;\mu }\right) ,  \label{thar2}
\end{equation}%
where $K^{\ell }\left( a,b\right) $ is a function of real variables $a$ and $%
b$. In the case of interest represented by the Lagrangian (\ref{thar1a})-(%
\ref{thar2}) we add to the assumption (\ref{thar1}) an assumption that the
EM field is static namely%
\begin{equation}
\partial _{t}\varphi =0,\ \mathbf{A}=\mathbf{0}.  \label{thar2aa}
\end{equation}

Treating the equalities (\ref{thar1}) as variables change let us recast the
charges Lagrangians in the new variable $\tilde{\psi}^{\ell }$ and $\tilde{%
\psi}^{\ell \ast }$. Notice first that%
\begin{equation}
\psi ^{\ell \ast }\psi ^{\ell }=\tilde{\psi}^{\ell \ast }\tilde{\psi}^{\ell
}.  \label{thar2a}
\end{equation}%
Then using (\ref{rkin1}), (\ref{rkin2}) and (\ref{rkin2c}) and (\ref{maxw2a}%
) we obtain%
\begin{gather}
\psi _{;0}^{\ell \ast }=\tilde{\partial}_{\mu }^{\ell \ast }\psi ^{\ell \ast
}=\mathrm{e}^{\mathrm{i}\omega ^{\ell }t}\left( \frac{\partial _{t}}{\mathrm{%
c}}+\mathrm{i}\frac{\omega ^{\ell }}{\mathrm{c}}-\frac{\mathrm{i}q^{\ell
}\varphi }{\chi \mathrm{c}}\right) \tilde{\psi}^{\ell \ast }=\mathrm{e}^{%
\mathrm{i}\omega ^{\ell }t}\left( \partial _{0}+\mathrm{i}\frac{\omega
^{\ell }}{\mathrm{c}}-\frac{\mathrm{i}q^{\ell }\varphi }{\chi \mathrm{c}}%
\right) \tilde{\psi}^{\ell \ast },  \label{thar2b} \\
\psi ^{\ell ;0}=\tilde{\partial}^{\ell \mu }\psi ^{\ell }=\mathrm{e}^{-%
\mathrm{i}\omega ^{\ell }t}\left( \frac{\partial _{t}}{\mathrm{c}}-\mathrm{i}%
\frac{\omega ^{\ell }}{\mathrm{c}}+\frac{\mathrm{i}q^{\ell }\varphi }{\chi 
\mathrm{c}}\right) \tilde{\psi}^{\ell }=\mathrm{e}^{-\mathrm{i}\omega ^{\ell
}t}\left( \partial _{0}-\mathrm{i}\frac{\omega ^{\ell }}{\mathrm{c}}+\frac{%
\mathrm{i}q^{\ell }\varphi }{\chi \mathrm{c}}\right) \tilde{\psi}^{\ell }, 
\notag \\
\psi _{;j}^{\ell \ast }=\tilde{\partial}_{j}^{\ell \ast }\psi ^{\ell \ast }=%
\mathrm{e}^{\mathrm{i}\omega ^{\ell }t}\left( \partial _{j}-\frac{\mathrm{i}%
q^{\ell }A_{j}}{\chi \mathrm{c}}\right) \tilde{\psi}^{\ell \ast },\ \psi
^{\ell ;j}=\tilde{\partial}^{\ell \mu }\psi ^{\ell }=\mathrm{e}^{-\mathrm{i}%
\omega ^{\ell }t}\left( \partial ^{j}+\frac{\mathrm{i}q^{\ell }A_{j}}{\chi 
\mathrm{c}}\right) \tilde{\psi}^{\ell }.  \notag
\end{gather}%
Observe that in the case when there is just a single charge the expressions (%
\ref{thar2b}) show that the time derivatives $\psi _{;0}^{\ell \ast }$ and $%
\psi ^{\ell ;0}$ are modified so as the potential $\varphi $ is added a
constant, namely 
\begin{equation}
\varphi \rightarrow \varphi -\frac{\chi \omega }{q}.  \label{thar2bb}
\end{equation}%
Substituting (\ref{thar2a}) and (\ref{thar2b}) into the Lagrangian $L^{\ell
} $ we get the Lagrangian which denote $L^{\omega \ell }$ as a function of
the variables $\tilde{\psi}^{\ell }$ and $\tilde{\psi}^{\ell \ast }$ and we
obtain%
\begin{gather}
L_{\omega }^{\ell }=K^{\ell }\left( \tilde{\psi}^{\ell \ast }\tilde{\psi}%
^{\ell },\psi _{;\mu }^{\ell \ast }\psi ^{\ell ;\mu }\right) ,\text{ where }%
\psi _{;\mu }^{\ell \ast }\psi ^{\ell ;\mu }=  \label{thar2c} \\
=\left( \frac{\partial _{t}}{\mathrm{c}}+\frac{\mathrm{i}\omega ^{\ell }}{%
\mathrm{c}}-\frac{\mathrm{i}q^{\ell }\varphi }{\chi \mathrm{c}}\right) 
\tilde{\psi}^{\ell \ast }\left( \frac{\partial _{t}}{\mathrm{c}}-\frac{%
\mathrm{i}\omega ^{\ell }}{\mathrm{c}}+\frac{\mathrm{i}q^{\ell }\varphi }{%
\chi \mathrm{c}}\right) \tilde{\psi}^{\ell }-  \label{thar2d} \\
-\left( \nabla +\frac{\mathrm{i}q^{\ell }\mathbf{A}}{\chi \mathrm{c}}\right) 
\tilde{\psi}^{\ell \ast }\cdot \left( \nabla -\frac{\mathrm{i}q^{\ell }%
\mathbf{A}}{\chi \mathrm{c}}\right) \tilde{\psi}^{\ell }.  \notag
\end{gather}%
We can apply now to the Lagrangian (\ref{thar1a}), (\ref{thar2d}) as a
function of the fields $\tilde{\psi}^{\ell }$, $\tilde{\psi}^{\ell \ast }$
and $A^{\mu }$ the obtained above results for the static fields taking into
account also the assumption (\ref{thar2aa}) for the EM field to static. In
this case the static regime is characterized by the time independence of the
charge fields $\tilde{\psi}^{\ell }$, $\tilde{\psi}^{\ell \ast }$ and the
assumption (\ref{thar2aa}) on the potential $A^{\mu }$ and these conditions
which are abbreviated by the symbol $\mathrm{stat}$: 
\begin{equation}
\mathrm{stat}\equiv \partial _{t}\tilde{\psi}^{\ell }=0,\ \partial _{t}%
\tilde{\psi}^{\ell \ast }=0,\ \partial _{t}\varphi =0,\ \mathbf{A}=\mathbf{0}%
,\ A^{\mu }=\left( \varphi ,\mathbf{A}\right) .  \label{thar2da}
\end{equation}%
Hence the Lagrangian of interest now is%
\begin{equation}
\mathcal{L}_{\omega }\left( \left\{ \psi ^{\ell },\psi _{;\mu }^{\ell },\psi
^{\ell \ast },\psi _{;\mu }^{\ell \ast }\right\} ,A^{\mu }\right)
=\dsum_{\ell }L_{\omega }^{\ell }\left( \psi ^{\ell },\psi _{;\mu }^{\ell
},\psi ^{\ell \ast },\psi _{;\mu }^{\ell \ast }\right) -\frac{F^{\mu \nu
}F_{\mu \nu }}{16\pi }.  \label{thar2db}
\end{equation}

Now applying the formula (\ref{epar3}) to the Lagrangian $\mathcal{L}%
_{\omega }$, as defined by (\ref{thar2db}), (\ref{thar2}), (\ref{thar2c})-(%
\ref{thar2d}), and the formula (\ref{flagr7aa}) for the Lagrangian of EM
field we obtain the following expression for the energy density $\mathcal{%
\mathring{U}}_{\omega \,\mathrm{stat}}$ of the system of charges and EM
field:%
\begin{gather}
\mathcal{\mathring{U}}_{\omega \,\mathrm{stat}}=\left. \mathcal{\mathring{T}}%
_{\omega }^{00}\right\vert _{\mathrm{stat}}=-\frac{\left( \nabla \varphi
\right) ^{2}}{8\pi }-  \label{thar2e} \\
-\sum_{\ell }K^{\ell }\left( \tilde{\psi}^{\ell \ast }\tilde{\psi}^{\ell
},\left( \frac{\omega ^{\ell }}{\mathrm{c}}-\frac{q^{\ell }\varphi }{\chi 
\mathrm{c}}\right) ^{2}\tilde{\psi}^{\ell \ast }\tilde{\psi}^{\ell }-\nabla 
\tilde{\psi}^{\ell \ast }\cdot \nabla \tilde{\psi}^{\ell }\right) ,  \notag
\end{gather}%
Let us take now the function $K^{\ell }\left( a,b\right) $ to be of a more
special form%
\begin{gather}
K^{\ell }\left( a,b\right) =k_{2}^{\ell }\left( a\right) b+k_{0}^{\ell
}\left( a\right) ,\text{ implying}  \label{kab1} \\
L_{\omega }^{\ell }=K^{\ell }\left( \psi ^{\ell \ast }\psi ^{\ell },\psi
_{;\mu }^{\ell \ast }\psi ^{\ell ;\mu }\right) =k_{2}^{\ell }\left( \psi
^{\ell \ast }\psi ^{\ell }\right) \psi _{;\mu }^{\ell \ast }\psi ^{\ell ;\mu
}+k_{0}^{\ell }\left( \psi ^{\ell \ast }\psi ^{\ell }\right) .  \label{kab2}
\end{gather}%
Notice that the term $k_{0}^{\ell }\left( a\right) $ in the cases of
interest contains the nonlinear self-interaction. In the special case (\ref%
{kab1}) the expression (\ref{thar2e})\ for the total energy density of the
charges and the EM field takes the form%
\begin{gather}
\mathcal{\mathring{U}}_{\omega \,\mathrm{stat}}=\mathcal{\mathring{U}}%
_{\omega \,\mathrm{stat}}^{\left( 2\right) }+\mathcal{\mathring{U}}_{\omega
\,\mathrm{stat}}^{\left( 0\right) }\text{, where}  \label{kab3} \\
\mathcal{\mathring{U}}_{\omega \,\mathrm{stat}}^{\left( 2\right) }=-\frac{%
\left( \nabla \varphi \right) ^{2}}{8\pi }+\sum_{\ell }k_{2}^{\ell }\left( 
\tilde{\psi}^{\ell \ast }\tilde{\psi}^{\ell }\right) \nabla \tilde{\psi}%
^{\ell \ast }\cdot \nabla \tilde{\psi}^{\ell },  \notag \\
\mathcal{\mathring{U}}_{\omega \,\mathrm{stat}}^{\left( 0\right)
}=-\sum_{\ell }\left[ k_{2}^{\ell }\left( \tilde{\psi}^{\ell \ast }\tilde{%
\psi}^{\ell }\right) \left( \frac{\omega ^{\ell }}{\mathrm{c}}-\frac{q^{\ell
}\varphi }{\chi \mathrm{c}}\right) ^{2}\tilde{\psi}^{\ell \ast }\tilde{\psi}%
^{\ell }+k_{0}^{\ell }\left( \tilde{\psi}^{\ell \ast }\tilde{\psi}^{\ell
}\right) \right] .  \notag
\end{gather}%
The corresponding expression for the total energy is%
\begin{gather}
\mathcal{E}_{\omega \,\mathrm{stat}}=\mathcal{E}_{\omega \,\mathrm{stat}%
}^{\left( 2\right) }+\mathcal{E}_{\omega \,\mathrm{stat}}^{\left( 0\right) },%
\text{ where}  \label{kab4} \\
\mathcal{E}_{\omega \,\mathrm{stat}}^{\left( 2\right) }=\int_{\mathbb{R}^{3}}%
\left[ -\frac{\left( \nabla \varphi \right) ^{2}}{8\pi }+\sum_{\ell
}k_{2}^{\ell }\left( \tilde{\psi}^{\ell \ast }\tilde{\psi}^{\ell }\right)
\nabla \tilde{\psi}^{\ell \ast }\cdot \nabla \tilde{\psi}^{\ell }\right] \,%
\mathrm{d}x,  \label{kab5} \\
\mathcal{E}_{\omega \,\mathrm{stat}}^{\left( 0\right) }=-\sum_{\ell }\int_{%
\mathbb{R}^{3}}\left[ k_{2}^{\ell }\left( \tilde{\psi}^{\ell \ast }\tilde{%
\psi}^{\ell }\right) \left( \frac{\omega ^{\ell }}{\mathrm{c}}-\frac{q^{\ell
}}{\chi \mathrm{c}}\varphi \right) ^{2}\tilde{\psi}^{\ell \ast }\tilde{\psi}%
^{\ell }+k_{0}^{\ell }\left( \tilde{\psi}^{\ell \ast }\tilde{\psi}^{\ell
}\right) \right] \,\mathrm{d}x.  \label{kab6}
\end{gather}%
Applying now the formula (\ref{epar12}) we get%
\begin{equation}
\mathcal{E}_{\omega \,\mathrm{stat}}^{\left( 0\right) }=-\frac{1}{3}\mathcal{%
E}_{\omega \,\mathrm{stat}}^{\left( 2\right) }.  \label{kab7}
\end{equation}%
implying the following representation for the total system energy%
\begin{equation}
\mathcal{E}_{\omega \,\mathrm{stat}}=\frac{2}{3}\mathcal{E}_{\omega \,%
\mathrm{stat}}^{\left( 2\right) }=\frac{2}{3}\int_{\mathbb{R}^{3}}\left[ -%
\frac{\left( \nabla \varphi \right) ^{2}}{8\pi }+\sum_{\ell }k_{2}^{\ell
}\left( \psi ^{\ell \ast }\psi ^{\ell }\right) \nabla \tilde{\psi}^{\ell
\ast }\cdot \nabla \tilde{\psi}^{\ell }\right] \,\mathrm{d}x.  \label{kab8}
\end{equation}%
In the case of a single charge the above formula turns into%
\begin{equation}
\mathcal{E}_{\omega \,\mathrm{stat}}=\frac{2}{3}\mathcal{E}_{\omega \,%
\mathrm{stat}}^{\left( 2\right) }=\frac{2}{3}\int_{\mathbb{R}^{3}}\left[ -%
\frac{\left( \nabla \varphi \right) ^{2}}{8\pi }+k_{2}\left( \psi ^{\ast
}\psi \right) \nabla \tilde{\psi}^{\ast }\cdot \nabla \tilde{\psi}\right] \,%
\mathrm{d}x.  \label{kab9}
\end{equation}%
We want to emphasize once more the importance of the representation (\ref%
{kab8}) in comparison with the original formula (\ref{kab4})-(\ref{kab6}),
which shows that the \emph{total energy of the system of charges interacting
with EM field does not explicitly depend on the terms} $k_{0}^{\ell }\left(
\psi ^{\ell \ast }\psi ^{\ell }\right) $\emph{\ which include the nonlinear
self-interactions}.

We would like to point out now that so far we carried computation for the
energy computation for the Lagrangian $\mathcal{L}_{\omega }$, as defined by
(\ref{thar2db}), (\ref{thar2}), (\ref{thar2c})-(\ref{thar2d}). But, in fact,
what we really need is the energy for fields of the form (\ref{thar1}) under
static conditions (\ref{thar2da}) for the initial Lagrangian $\mathcal{L}$
as defined (\ref{thar1a})-(\ref{thar2}). In turns out, as one may expect,
the difference between the two is just the sum of the rest energies. Indeed,
using once more the formula (\ref{flagr4a}) for the canonical energy $%
\mathcal{\mathring{T}}^{00}$ under static conditions (\ref{thar2da}), the
formulas (\ref{epar3}), (\ref{thar2e}) together with the formulas (\ref%
{flagr9}), (\ref{fpar10}) for microcharge density $\rho ^{\ell }$ we obtain%
\begin{gather}
\mathcal{\mathring{U}}\left( \mathrm{e}^{-\mathrm{i}\omega ^{\ell }t}\tilde{%
\psi}^{\ell },\ \mathrm{e}^{\mathrm{i}\omega ^{\ell }t}\tilde{\psi}^{\ell
\ast },\varphi \right) =  \label{kab10} \\
=\dsum_{\ell }\frac{\partial \mathcal{L}}{\partial \psi _{,0}^{\ell }}\left(
-\mathrm{i}\frac{\omega ^{\ell }}{\mathrm{c}}\tilde{\psi}^{\ell }\right) +%
\frac{\partial \mathcal{L}}{\partial \psi _{,\mu }^{\ell \ast }}\left( 
\mathrm{i}\frac{\omega ^{\ell }}{\mathrm{c}}\tilde{\psi}^{\ell \ast }\right)
-\mathcal{L}\left( \mathrm{e}^{-\mathrm{i}\omega ^{\ell }t}\tilde{\psi}%
^{\ell },\ \mathrm{e}^{\mathrm{i}\omega ^{\ell }t}\tilde{\psi}^{\ell \ast
},\varphi \right)  \notag \\
=\dsum_{\ell }-\mathrm{i}\frac{\omega ^{\ell }}{\mathrm{c}}\left( \frac{%
\partial \mathcal{L}}{\partial \psi _{,0}^{\ell }}\tilde{\psi}^{\ell }-\frac{%
\partial \mathcal{L}}{\partial \psi _{,\mu }^{\ell \ast }}\tilde{\psi}^{\ell
\ast }\right) +\left. \mathcal{\mathring{T}}_{\omega }^{00}\right\vert _{%
\mathrm{stat}}=  \notag \\
=\dsum_{\ell }\frac{\omega ^{\ell }\chi }{q}\rho ^{\ell }+\mathcal{\mathring{%
U}}_{\omega \,\mathrm{stat}}=\dsum_{\ell }\frac{m^{\ell }\mathrm{c}^{2}}{q}%
\rho ^{\ell }+\mathcal{\mathring{U}}_{\omega \,\mathrm{stat}}  \notag
\end{gather}%
Now integrating the above density over the entire space and using the
micro-charge normalization condition (\ref{psfi6}) obtain 
\begin{equation}
\mathcal{E}\left( \mathrm{e}^{-\mathrm{i}\omega ^{\ell }t}\tilde{\psi}^{\ell
},\ \mathrm{e}^{\mathrm{i}\omega ^{\ell }t}\tilde{\psi}^{\ell \ast },\varphi
\right) =\dsum_{\ell }m^{\ell }\mathrm{c}^{2}+\mathcal{E}_{\omega \,\mathrm{%
stat}}.  \label{kab11}
\end{equation}%
For the special case (\ref{kab1}) combining the last formula with formulas (%
\ref{kab8}), (\ref{kab9}) we obtain the following important formulas for
respectively many charges and a single charge 
\begin{gather}
\mathcal{E}\left( \mathrm{e}^{-\mathrm{i}\omega ^{\ell }t}\tilde{\psi}^{\ell
},\ \mathrm{e}^{\mathrm{i}\omega ^{\ell }t}\tilde{\psi}^{\ell \ast },\varphi
\right)  \label{kab12} \\
=\dsum_{\ell }m^{\ell }\mathrm{c}^{2}+\frac{2}{3}\int_{\mathbb{R}^{3}}\left[
-\frac{\left( \nabla \varphi \right) ^{2}}{8\pi }+\sum_{\ell }k_{2}^{\ell
}\left( \psi ^{\ell \ast }\psi ^{\ell }\right) \nabla \tilde{\psi}^{\ell
\ast }\cdot \nabla \tilde{\psi}^{\ell }\right] \,\mathrm{d}x  \notag
\end{gather}%
\begin{gather}
\mathcal{E}\left( \mathrm{e}^{-\mathrm{i}\omega t}\tilde{\psi},\ \mathrm{e}^{%
\mathrm{i}\omega t}\tilde{\psi}^{\ast },\varphi \right) =  \label{kab13} \\
=m\mathrm{c}^{2}+\frac{2}{3}\int_{\mathbb{R}^{3}}\left[ -\frac{\left( \nabla
\varphi \right) ^{2}}{8\pi }+k_{2}\left( \psi ^{\ast }\psi \right) \nabla 
\tilde{\psi}^{\ast }\cdot \nabla \tilde{\psi}\right] \,\mathrm{d}x.  \notag
\end{gather}%
The formulas (\ref{kab12}) and (\ref{kab13}) give important representation
for the energy of time harmonic fields which does not explicitly involve the
nonlinear self-interactions.

\subsection{Compressional waves in nonviscous compressible fluid\label%
{scompress}}

In this section following to \cite[Section 3.3]{Morse Feshbach I} and \cite[%
Section 6.2]{Morse Ingard} we consider here compressional waves in
nonviscous and compressible fluid which are described by the pressure field $%
p$ and velocity field $\mathbf{v}$ and governed by the following system of
equations 
\begin{equation}
\rho \partial _{t}\mathbf{v}=-\nabla p,\ \kappa \partial _{t}p=-\nabla \cdot 
\mathbf{v},\ \mathrm{c}^{2}=\frac{1}{\rho \kappa }  \label{compr1}
\end{equation}%
where $\rho $ and $\kappa $ are respectively uniform constant mass density
and compressibility (adiabatic) of the fluid at equilibrium and $\mathrm{c}$
is the velocity of wave propagation. We also have 
\begin{equation}
\frac{1}{2}\rho \mathbf{v}\cdot \mathbf{v}\text{ is the kinetic energy and }%
\frac{1}{2}\kappa p^{2}\text{ is the potential energy}  \label{compr2}
\end{equation}%
Then we if introduce the velocity potential $\psi $ so that 
\begin{equation}
p=\rho \partial _{t}\psi ,\ \mathbf{v}=-\nabla \psi ,  \label{compr3}
\end{equation}%
it immediately follows from (\ref{compr3}) that $\psi $ satisfies the
classical wave equation%
\begin{equation}
\frac{1}{\mathrm{c}^{2}}\partial _{t}^{2}\psi -\nabla ^{2}\psi =0.
\label{compr4}
\end{equation}%
The compressional waves have the following Lagrangian density%
\begin{equation}
L=\frac{1}{2}\rho \mathbf{v}\cdot \mathbf{v}-\frac{1}{2}\kappa p^{2}=\frac{1%
}{2}\rho \left[ \frac{1}{\mathrm{c}^{2}}\left( \partial _{t}\psi \right)
^{2}-\left( \nabla \psi \right) ^{2}\right]  \label{compr5}
\end{equation}%
and the following canonical energy-momentum tensor%
\begin{gather}
T^{\mu \nu }=\left[ 
\begin{array}{cccc}
T^{00} & \rho \partial _{0}\psi \partial _{1}\psi & \rho \partial _{0}\psi
\partial _{2}\psi & \rho \partial _{0}\psi \partial _{3}\psi \\ 
\rho \partial _{1}\psi \partial _{0}\psi & T^{11} & -\rho \partial _{1}\psi
\partial _{2}\psi & -\rho \partial _{1}\psi \partial _{3}\psi \\ 
\rho \partial _{2}\psi \partial _{0}\psi & -\rho \partial _{2}\psi \partial
_{1}\psi & T^{22} & -\rho \partial _{2}\psi \partial _{3}\psi \\ 
\rho \partial _{3}\psi \partial _{0}\psi & -\rho \partial _{3}\psi \partial
_{1}\psi & -\rho \partial _{3}\psi \partial _{2}\psi & T^{33}%
\end{array}%
\right] ,\ \partial _{0}=\frac{1}{\mathrm{c}}\partial _{t},  \label{compr6}
\\
T^{00}=\frac{\rho }{2}\left[ \left( \partial _{0}\psi \right) ^{2}+\left(
\nabla \psi \right) ^{2}\right] ,\ T^{jj}=\frac{\rho }{2}\left[ \left(
\nabla \psi \right) ^{2}-2\left( \partial _{j}\psi \right) ^{2}-\left(
\partial _{0}\psi \right) ^{2}\right] .  \notag
\end{gather}

\subsection{Klein-Gordon equation and Yukawa potential\label{skleingordon}}

Klein-Gordon equation is well known model for a free charge, \cite[Section 18%
]{Pauli PWM}. In particular, its certain modification describes a charge
interacting with an external EM field, \cite[Section 8.1]{Schwabl}. Here we
follow to \cite[Section 1.5.2]{Martin}. If the spin is neglected a freely
propagating particle $X$ of the rest mass $m_{X}$ is described by a
complex-valued wave function $\varphi \left( \mathbf{r}\right) $ satisfying
the \emph{Klein-Gordon equation}%
\begin{equation}
-\frac{1}{\mathrm{c}^{2}}\partial _{t}^{2}\varphi =\left\{ -\Delta +\left( 
\frac{m_{X}\mathrm{c}}{\hbar }\right) ^{2}\right\} \varphi .  \label{kgor1}
\end{equation}%
This equation is obtained from the fundamental relativistic mass-energy
relation%
\begin{equation}
\frac{E^{2}}{\mathrm{c}^{2}}=\mathbf{p}^{2}+m_{X}\mathrm{c}^{2},
\label{kgor2}
\end{equation}%
where $E$ is the particle energy, $\mathbf{p}$ is the three-dimensional
space momentum, by the substitution $E=\hbar \partial _{t}$ and $\mathbf{p}%
=-\hbar \nabla _{\mathbf{r}}$. A static solution $V$ to the Klein-Gordon
equation (\ref{kgor1}) with a $\delta $-function source, i.e.%
\begin{equation}
\left\{ -\Delta +\mu ^{2}\right\} V=-g^{2}\delta \left( \mathbf{x}\right) ,
\label{kgor4}
\end{equation}%
is called the \emph{Yukawa potential}%
\begin{equation}
V\left( \left\vert \mathbf{x}\right\vert \right) =-\frac{g^{2}}{4\pi }\frac{%
\mathrm{e}^{-\mu \left\vert \mathbf{x}\right\vert }}{\left\vert \mathbf{x}%
\right\vert }=-\left( \mu ^{2}-\Delta \right) ^{-1}g^{2}\delta \left( 
\mathbf{x}\right) ,\ \mu =\frac{m_{X}\mathrm{c}}{\hbar },\ .  \label{kgor3}
\end{equation}%
The quantity $\mu ^{-1}=\frac{\hbar }{m\mathrm{c}}$ is called the \emph{%
range of the potential} $V$, is and it is also known as the \emph{Compton
wavelength} of the relativistic particle of the mass $m_{X}$. The constant $%
g $ is a so-called coupling constant representing the basic strength of the
interaction.

There is an interpretation of Klein-Gordon equation as a flexible string
with additional stiffness forces provided by the medium surrounding it.
Namely, if the string is embedded in a thin sheet of rubber or if it is
along the axis of a cylinder of rubber whose outside surface kept fixed, 
\cite[Section 2.1]{Morse Feshbach I}.

\subsection{Schr\"{o}dinger Equation\label{sschrodinger}}

The Schr\"{o}dinger equation with the potential $V$ is 
\begin{equation}
\hbar \mathrm{i}\partial _{t}\psi =-\frac{\hbar ^{2}}{2m}\nabla ^{2}\psi
+V\psi .  \label{schr1}
\end{equation}%
It is the Euler-Lagrange field equation (together with its conjugate) for
the following Lagrangian, \cite[(3.3.20)]{Morse Feshbach I} 
\begin{equation}
L=\mathrm{i}\frac{\hbar }{2}\left( \psi ^{\ast }\partial _{t}\psi -\partial
_{t}\psi ^{\ast }\psi \right) -\frac{\hbar ^{2}}{2m}\nabla \psi ^{\ast
}\cdot \nabla \psi -\psi ^{\ast }V\psi .  \label{schr2}
\end{equation}%
The stress-tensor here%
\begin{equation}
\mathring{T}^{\mu \nu }=\frac{\partial L}{\partial \psi _{,\mu }}\psi ^{,\nu
}+\frac{\partial L}{\partial \psi _{,\mu }^{\ast }}\psi ^{\ast ,\nu }-\delta
^{\mu \nu }L,  \label{schr3}
\end{equation}%
implying the following formula for the energy density%
\begin{equation}
H=\mathring{T}^{00}=-\frac{\hbar ^{2}}{2m}\nabla \psi ^{\ast }\cdot \nabla
\psi +\psi ^{\ast }V\psi .  \label{schr4}
\end{equation}%
The energy flow vector $\mathbf{S}$, the momentum density vector $\mathbf{P}$
and the current density vector $\mathbf{J}$ for the Schr\"{o}dinger equation
(\ref{schr1}) are respectively, \cite[(3.3.25), (3.3.26)]{Morse Feshbach I},%
\begin{gather}
\mathbf{S}=-\frac{\hbar ^{2}}{2m}\left[ \partial _{t}\psi ^{\ast }\cdot
\nabla \psi +\partial _{t}\psi \cdot \nabla \psi ^{\ast }\right] ,
\label{schr5} \\
\mathbf{P}=\mathrm{i}\frac{\hbar }{2}\left[ \psi ^{\ast }\cdot \nabla \psi
-\psi \cdot \nabla \psi ^{\ast }\right] ,\ \mathbf{J}=-\frac{q}{m}\mathbf{P},
\notag
\end{gather}%
with the equation of continuity $\partial _{t}H+\nabla \cdot \mathbf{S}=0$.

Quantum mechanical charged particle in an external EM field with the
4-potential $A^{\mu }=\left( \varphi ,\mathbf{A}\right) $ is described by
the following Schr\"{o}dinger equation, \cite[(2.6.47)]{Morse Feshbach I} 
\begin{equation}
\hbar \mathrm{i}\partial _{t}\psi =\frac{1}{2m}\left( \frac{\hbar }{\mathrm{i%
}}\nabla -\frac{q}{\mathrm{c}}\mathbf{A}\right) \cdot \left( \frac{\hbar }{%
\mathrm{i}}\nabla -\frac{q}{\mathrm{c}}\mathbf{A}\right) \psi +q\varphi \psi
\label{schr6}
\end{equation}%
or%
\begin{equation}
\hbar \mathrm{i}\partial _{t}\psi =-\frac{\hbar ^{2}}{2m}\nabla ^{2}\psi +%
\mathrm{i}\frac{q\hbar }{2m\mathrm{c}}\mathbf{A}\cdot \nabla \psi +\left[ 
\frac{q^{2}\left\vert \mathbf{A}\right\vert ^{2}}{2m\mathrm{c}^{2}}+q\varphi %
\right] \psi ,  \label{schr7}
\end{equation}%
with the charge density $\rho =q\psi \psi ^{\ast }$ and the current $\mathbf{%
J}$ as follows, \cite[(2.6.46)]{Morse Feshbach I}%
\begin{equation}
\rho =q\psi \psi ^{\ast },\ \mathbf{J}=\mathrm{i}\frac{q\hbar }{2m}\left(
\psi ^{\ast }\nabla \psi -\psi \nabla \psi ^{\ast }\right) .  \label{schr8}
\end{equation}%
The quantities $\rho $ and $\mathbf{J}$ satisfy the continuity equation%
\begin{equation}
\partial _{t}\rho +\nabla \cdot \mathbf{J}=0.  \label{schr9}
\end{equation}

\section{Appendix: Fourier transforms and Green functions}

The polar coordinates representation of the Laplace operator in $\mathbb{R}%
^{3}$, \cite[(4.56)]{Taylor1}, is%
\begin{equation}
\Delta =\Delta _{r}+\frac{1}{r^{2}}\Delta _{s}=\frac{d^{2}}{dr^{2}}+\frac{2}{%
r}\frac{d}{dr}+\frac{1}{r^{2}}\Delta _{s},\ \mathbf{x}\in \mathbb{R}^{3},\
r=\left\vert \mathbf{x}\right\vert ,  \label{gre2}
\end{equation}%
where $\Delta _{s}$ is the Laplace operator on the unit sphere $\mathbb{S}%
^{2}$. We also have, \cite[(5,53)-(5,59)]{Taylor1},%
\begin{equation}
\left( \kappa ^{2}-\Delta \right) ^{-1}\delta \left( \mathbf{x}\right) =%
\frac{\mathrm{e}^{-\kappa \left\vert \mathbf{x}\right\vert }}{4\pi
\left\vert \mathbf{x}\right\vert },\ \mathbf{x}\in \mathbb{R}^{3},\ \kappa
\geq 0.  \label{gre3}
\end{equation}%
Notice that the action of the operator $\Delta $ on radial functions $%
g\left( r\right) $, i.e. functions depending on $r=\left\vert x\right\vert $%
, is reduced to the action of $\Delta _{r}$ only for smooth functions, i.e.%
\begin{equation}
\Delta g\left( r\right) =\Delta _{r}g\left( r\right) \text{ if }g\left(
r\right) \text{ is continuos and smooth for }r\geq 0.  \label{gre3a}
\end{equation}%
Indeed, in view of (\ref{gre3})%
\begin{equation}
\Delta _{r}\frac{1}{r}=0,\text{ whereas }\Delta \frac{1}{r}=-4\pi \delta
\left( \mathbf{x}\right) .  \label{gre3b}
\end{equation}%
Let us consider the Fourier transform of radial functions following to \cite[%
Section 3.6]{Taylor1}: 
\begin{gather}
\hat{f}\left( \mathbf{k}\right) =\hat{f}\left( \left\vert \mathbf{k}%
\right\vert \right) =  \label{gre4} \\
=\frac{1}{\left( 2\pi \right) ^{\frac{3}{2}}}\dint\nolimits_{0}^{\infty
}f\left( r\right) \psi _{d}\left( r\left\vert \mathbf{k}\right\vert \right)
r^{n-1}\,\mathrm{d}r,\ \psi _{d}\left( \left\vert \mathbf{k}\right\vert
\right) =\dint\nolimits_{\left\vert \mathbf{x}\right\vert =1}\mathrm{e}^{-%
\mathrm{i}\mathbf{k}\cdot \mathbf{x}}\,\mathrm{d}s.  \notag
\end{gather}%
Then the following identity holds%
\begin{equation}
\hat{f}\left( \mathbf{k}\right) =\left( \frac{2}{\pi }\right) ^{\frac{1}{2}}%
\frac{1}{\left\vert \mathbf{k}\right\vert }\dint\nolimits_{0}^{\infty
}f\left( r\right) \sin \left( r\left\vert \mathbf{k}\right\vert \right) \,%
\mathrm{d}r.  \label{gre6}
\end{equation}%
Let $w\left( \mathbf{x}\right) $, $\mathbf{x}\in \mathbb{R}^{3}$ be a real
function satisfying%
\begin{equation}
0\leq w\left( \mathbf{x}\right) \leq w_{\infty }<\infty .  \label{gre7}
\end{equation}%
Then the Green function $G\left( \mathbf{x},\mathbf{y}\right) =\left(
-\Delta +w\right) ^{-1}\left( \mathbf{x},\mathbf{y}\right) $ defined as a
fundamental solution to the equation%
\begin{equation}
\left( -\Delta +w\right) G\left( \mathbf{x},\mathbf{y}\right) =\delta \left( 
\mathbf{x}-\mathbf{y}\right) ,  \label{gre8}
\end{equation}%
satisfies the following inequalities%
\begin{gather}
\frac{\mathrm{e}^{-\sqrt{w_{\infty }}\left\vert \mathbf{x}-\mathbf{y}%
\right\vert }}{4\pi \left\vert \mathbf{x}-\mathbf{y}\right\vert }\leq \left(
-\Delta +w_{\infty }\right) ^{-1}\left( \mathbf{x},\mathbf{y}\right) \leq
\left( -\Delta +w\right) ^{-1}\left( \mathbf{x},\mathbf{y}\right) \leq
\label{gre9} \\
\leq \left( -\Delta \right) ^{-1}\left( \mathbf{x},\mathbf{y}\right) =\frac{1%
}{4\pi \left\vert \mathbf{x}-\mathbf{y}\right\vert },  \notag
\end{gather}%
which follow from the Feynman-Kac formula for the heat kernel, \cite[Section
8.2]{Oksendal} , applied to the operator $-\Delta +w$.

\textbf{Acknowledgment.} The research was supported through Dr. A. Nachman
of the U.S. Air Force Office of Scientific Research (AFOSR), under grant
number FA9550-04-1-0359. The authors are very grateful to Michael Kiessling
for reading the manuscript and suggestions which helped to improve the
exposition.

\end{document}